\documentclass[12pt]{article}
\usepackage{slashed}
\usepackage{mathrsfs}
\makeatletter \@addtoreset{equation}{section} \makeatother
\renewcommand{\theequation}{\thesection.\arabic{equation}}
\newcommand{\ba}{\begin{array}} \newcommand{\ea}{\end{array}}
\newcommand{\beq}{\begin{equation}} \newcommand{\eeq}{\end{equation}}
\newcommand{\bea}{\begin{eqnarray}} \newcommand{\eea}{\end{eqnarray}}
 
 \def\bce{\begin{center}} \def\ece{\end{center}}
 \def\nonu{\nonumber} 
\def\pa{\partial}  \def\be{\beta} 
 \def\de{\delta} \def\De{\Delta} \def\ep{\epsilon}

   \def\si{\sigma}

     \def\eps6{{\displaystyle
    \mathop{\epsilon}^{6}}{}} \def\g6{{\displaystyle
    \mathop{g}^{6}}{}}  \def\nab6{{\displaystyle
    \mathop{\nabla}^{6}}{}}

 \def\0{{\sst{(0)}}} \def\1{{\sst{(1)}}}
\def\2{{\sst{(2)}}} \def\3{{\sst{(3)}}} \def\4{{\sst{(4)}}}
\def\5{{\sst{(5)}}} \def\6{{\sst{(6)}}} \def\7{{\sst{(7)}}}
\def\8{{\sst{(8)}}}

\def\ba{\begin{array}} \def\ea{\end{array}} \def\beq{\begin{equation}}
\def\eeq{\end{equation}} \def\be{\begin{equation}}
\def\ee{\end{equation}}

  \def\eps{\epsilon}

\def\ba{\begin{array}} \def\ea{\end{array}} \def\beq{\begin{equation}}
\def\eeq{\end{equation}} \def\be{\begin{equation}}
\def\ee{\end{equation}}

  \def\eps{\epsilon}

      \def\eps6{{\displaystyle
    \mathop{\epsilon}^{6}}{}}  \def\nab6{{\displaystyle
    \mathop{\nabla}^{6}}{}}

\newcommand{\bean}{\begin{eqnarray*}}
\newcommand{\eean}{\end{eqnarray*}}

\usepackage{amsmath}

\usepackage{kotex}

\usepackage{amsmath} \usepackage{graphicx} \usepackage{amssymb}
\usepackage{amsthm} \usepackage{grffile} \usepackage{simplewick}
\usepackage{tikz}

\usepackage{nccmath} \usepackage{physics} \usepackage{wrapfig}
\usepackage{mathtools} \usepackage{stackengine}

\usetikzlibrary{intersections,decorations.markings}

\allowdisplaybreaks

\begin{document}
\thispagestyle{empty} \addtocounter{page}{-1}
   \begin{flushright}
\end{flushright}

  
\centerline{ \Large \bf
Multi-Particle Contributions to the Celestial
Algebra}
\vspace*{0.3cm}
\centerline{ \Large \bf  
in the  ${\cal N}=8$ Supergravity} 
\vspace*{1.5cm}
\centerline {\bf
Changhyun Ahn$^{\dagger,\star,}$\footnote{CA is a visitor to
Seoul
National University of Science and Technology
and a professor emeritus at
Kyungpook National University.  }
}
\vspace*{1.0cm} 
\centerline{\it
$\dagger$
Institute for Convergent Fundamental Studies,
}
\centerline{\it
Seoul
National University of Science and Technology, Seoul
01811, Korea} 
\centerline{\it
$\star$
Department of Physics, Kyungpook National University, Daegu
41566, Korea} 
\vspace*{0.5cm}
\vspace*{0.5cm}
\centerline{\tt ahn@knu.ac.kr
} 
\vskip2cm

\centerline{\bf Abstract}
\vspace*{0.5cm}

Recently, Calkins and Pate
calculated the multi-particle operator product expansions (OPEs) of
single-particle celestial operators with two-particle
celestial operators in pure
Einstein gravity. We apply their construction to
the graviton, gravitinos, graviphotons, graviphotinos
and scalars in ${\cal N}=8$ supergravity.
By performing various contour integrals on these multi-particle OPEs,
we obtain ninety-five (anti)commutators for the single-particle
contributions from the celestial soft current
algebra (derived from the single-particle OPEs)
in a two-dimensional boundary.
Moreover, these  ninety-five multi-particle OPEs
provide nontrivial relations between the corresponding celestial amplitudes.
The triple-collinear limits between these bosonic and fermionic particles
from the four-dimensional bulk are also described as a double check.
Finally, we implicitly propose  both the multi-particle OPEs of
single-particle celestial operators with $(N-1)$-particle
celestial operators  and the corresponding (anti)commutators
for the modes of celestial operators that have linear terms
(or single-particle exchange terms due to the sum of $N$-particle factorization
channels from amplitudes)
on the right-hand sides.


\baselineskip=18pt
\newpage
\renewcommand{\theequation}
{\arabic{section}\mbox{.}\arabic{equation}}

\tableofcontents

\section{
Introduction}

In celestial holography \cite{Strominger1703,Raclariu,Pasterski,PPR},
the four-dimensional scattering amplitudes
(or $S$-matrix) are related to the two-dimensional
conformal correlators (or celestial amplitudes) on the
celestial sphere at null infinity via the Mellin transform
of the external particle energies.
This transform converts the momentum-space scattering
into a basis of Lorentz eigenstates,
making scattering amplitudes transform as two-dimensional conformal
field theory operators.
The collinear limit of bulk scattering amplitudes
corresponds to the operator product expansion (OPE)
of the boundary two-dimensional (single-particle) operators,
where the splitting functions are identified with the OPE
coefficients (i.e., the Euler beta functions).

In \cite{ESW},
by recognizing that the product of Wilson coefficients in two dimensions
(after substituting the operator product expansions (OPEs)
for the $N$-point correlator)
plays the role of the multi-collinear splitting functions
in four dimensions, they studied-for example-the multi-graviton OPE for the
$n$ positive-helicity gravitons.
This was achieved recursively from the two-point defining OPE of
the positive-helicity graviton with itself.
They used
the Mellin transform on
the multi-collinear splitting functions
(which depend on the $n$ different energies and $2n$ complex coordinates)
in momentum space
and identified those
splitting functions with the above OPE coefficient result
from the celestial conformal
field theory.

In \cite{BHP}, the $n$-point amplitude in momentum space
for the specific (holomorphic) triple-collinear limit
\footnote{
\label{multi}
The multi-collinear limits involve more than
two asymptotic particles.} is factorized
as two three-point amplitudes, two propagators and an $(n-2)$-point
amplitude. The overall factor in front of the $(n-2)$-point amplitude
depends on the three energies and six complex coordinates 
of three (collinear) particles.
Then the splitting function appearing in the coefficient of
the $(n-2)$-point amplitude on the right-hand side
is obtained by summing over
three possible consecutive collinear limits, contrary to the approach of
\cite{ESW}.
By applying this formula for the splitting function to
three positive-helicity gravitons and using the Mellin transform
to these three terms of the splitting function,
they have constructed the celestial OPE of three
operators. This depends on i) the generalized Euler beta functions
\footnote{
\label{generalizedbeta}
This can be written as
$B(x,y,z)\equiv \frac{\Gamma(x) \Gamma(y) \Gamma(z)}{\Gamma(x+y+z)}$
in terms of Gamma functions.
We have the relation $B(x,y,z)=B(x,y)\, B(x+y,z)$.
In other words, once one of the arguments in the
second Euler beta function is the same as the sum of the two
arguments of the first Euler beta function, then the product of
two Euler beta functions can be written as the generalized Euler
beta function.},
ii) the three conformal dimensions, iii) the above six complex coordinates,
iv) other factors from the Taylor expansion over two new variables
in the Mellin transform
and v) the infinite sum over the antiholomorphic derivatives acting on
the positive graviton. 

In \cite{GHP},
by defining two-particle celestial operators
\footnote{They are given by the normal-ordered product
of two celestial operators by considering
the additional antiholomorphic
contour integral (for given
holomorphic contour integral in the standard
conformal field theory).

The normal-ordered product is a regularization method
for composite operators where singular short-distances
in the OPE are removed. In the normal-ordered products
of three operators or more, one should specify how
the normal ordering is carried out between the successive pairs
of operators. The precise convention we are using is given by
\cite{BBSS}. This is so-called fully normal-ordered and denoted as
round brackets. In the context of celestial holography,
the celestial operators depend on the antiholomorphic
coordinates in addition to the holomorphic coordinates.
By considering the contour integral for the antiholomorphic
coordinates further,
we define the normal-ordered product of
two operators \cite{GHP,CP}
\bea
({\cal O}_1 \, {\cal O}_2)(z_2, \bar{z}_2) \equiv
\oint_{z_2} \, \frac{d z_1}{2\pi i}\, \frac{1}{z_{12}}
\oint_{\bar{z}_2} \, \frac{d \bar{z}_1}{2\pi i}\, \frac{1}{\bar{z}_{12}}\,
{\cal O}_1(z_1,\bar{z}_1) \, {\cal O}_2(z_2, \bar{z}_2).
\label{NORMAL}
\eea
The OPE $
{\cal O}_1(z_1,\bar{z}_1) \, {\cal O}_2(z_2, \bar{z}_2)$
can be expanded in terms of $z_{12}^n \, \bar{z}_{12}^{\bar{n}}$
together with some operator which depends on $(z_2, \bar{z}_2)$.
Here $n$ and $\bar{n}$ are any integers.
The normal-ordered product does not depend on both $z_{12}$
and $\bar{z}_{12}$. In other words, the operator having
$z_{12}^0 \, \bar{z}_{12}^{0}$ on the right-hand side of the OPE
is the normal-ordered product
$({\cal O}_1 \, {\cal O}_2)(z_2, \bar{z}_2)$.
},
they have described the OPEs of the single-particle operators
with those two-particle celestial operators in the gluon amplitudes.
The second-order poles in the antiholomorphic
coordinates-case one, where
the three particles have the negative helicities and
case two, where particles $1$ and $2$ have the negative helicities
and particle $3$ has positive-helicity,
provide the single-particle exchanges
(with the rest of the celestial correlator function) and the
first-order pole
of this case-one gives two-particle exchanges.
Similarly, the second-order pole in the holomorphic coordinates
(the case-three where particle $1$ has the negative helicity
and particles $2$ and $3$ have the positive helicities)
provides the single-particle exchanges \footnote{
They expect that the higher-particle celestial operators
(by generalizing the above two-particle celestial operators)
can give the higher-order poles in the OPE
of the single-particle celestial operators with
the  higher-particle celestial operators
with the (anti)holomorphic collinear
limits.}.

In \cite{KP},
the massless eigenstate can be described as the inverse
Mellin transform (the integral
over the dummy variable, the conformal dimension)
of the celestial operator acting on the vacuum for one-particle
states.
Similarly, the two-particle states (i.e.,
the simplest multi-particle states along the same lines
as the footnote \ref{multi}) can be written in terms of
two creation operators acting on the vacuum,
which can be described as the inverse Mellin transform
(the integrals over the two conformal dimensions)
of the product of celestial operators acting on the vacuum.
They show that there should be multi-particle operators
appearing in the celestial OPE.

In \cite{CP},
they have found the OPE of the single-particle celestial operators
with the two-particle celestial operators explicitly in pure Einstein
gravity \footnote{Equivalently, we denote this by the multi-particle
OPE in this paper.}.
The second-order pole in the holomorphic coordinates
provides the (infinite number of)
antiholomorphic derivatives acting on the
(linear) gravitons, corresponding to single-particle
exchanges above, while
the first-order pole leads to i) the normal-ordered product
between the right-hand side of the OPE of the single-particle
operators and themselves (particles $1$ and $2$) and particle $3$
and ii)
the normal-ordered product
between particle $2$ and
the right-hand side of the OPE of the single-particle
operators  and themselves (particles $1$ and $3$) \footnote{
Recall that the right-hand side of the OPE
of single-particle operators has the (infinite number of)
antiholomorphic derivatives
on the celestial operators.}.
The first-order pole is related to the two-particle exchanges
(with  the remainder of the celestial correlator function) above.
The main observation of \cite{CP} is that
the above second-order pole appears from the second contraction
between particle $I$ (in the process $12 \rightarrow I$)
which is {\it the leading (or first) antiholomorphic descendant } 
and particle $3$
\footnote{
\label{geometric}
The singular term at the first contraction behaves as
$\frac{\bar{z}_{12}^2}{z_{12}}$ where $z_{12} \equiv z_1-z_2$
and $\bar{z}_{12} \equiv \bar{z}_1-\bar{z}_2$.  
Then the fractional-holomorphic
factor $\frac{1}{z_{12}}$ can be written as
$\frac{1}{(z_{13}-z_{23})}$ where $z_{13} \equiv z_1-z_3$ and
$z_{23} \equiv z_2-z_3$ and
the first-order pole in the holomorphic
coordinates is given by $\sum_{n=0}^{\infty} \,
\frac{z_{23}^n}{z_{13}^{n+1}}$ using a geometric series
($|z_{23}| < |z_{13}|$).
Their observation is that the single-particle exchanges
appear at the $n=1$ term in order to have
the nonzero value in the $z_2$ contour integration
after the second contraction. The antiholomorphic
factor $\bar{z}_{12}^2$
is equal to $(\bar{z}_{13}-\bar{z}_{23})^2$.
Now we can apply the Cauchy integral formula to the $\bar{z}_2$
variable and we are left with $\bar{z}_{13}^2$ after
$\bar{z}_2$ integration. By combining the previous factor
$\frac{1}{z_{13}^2}$ with $\bar{z}_{13}^2$, then
we obtain the second-order pole $\frac{\bar{z}_{13}^2}{
z_{13}^2}$ in the holomorphic coordinates.}.
Furthermore, the generalized Euler beta function
in the footnote \ref{generalizedbeta} appear as the structure
constant.

\begin{itemize}
\item[]  
Our goal in this paper is to apply
the construction of \cite{CP} to the ${\cal N}=8$ supergravity \cite{dF}.
We should obtain the multi-particle
OPEs of the single-particle celestial operators
with two-particle celestial operators. 
As in the footnote \ref{geometric}, in general, there exist
the second and first-order poles in the holomorphic complex coordinates.
However, some OPEs do not contain the second-order poles,
compared to the results for pure Einstein gravity in
\cite{CP}, because
the second contractions described above do not have any singular
terms. This is  because the condition on the sum of the helicities
of the three particles 
$s_1+s_2+s_3 \geq 2$ (which is more restrictive than
$s_1+s_2+s_3 \geq 0$ for pure Einstein gravity
case \cite{CP}) is not satisfied.
For the nonzero
first-order pole classified as i) in the previous paragraph,
the condition $s_1+s_2 \geq 0$ should be satisfied and similarly,
for the 
nonzero
first-order pole classified as ii) in the previous paragraph,
the condition $s_1+s_3 \geq 0$ must also be
satisfied \footnote{
\label{25s2s3}
For the pure
Einstein gravity studied in \cite{CP}, these three conditions
are satisfied for the three possibilities for the helicities
of the particles, $(s_1,s_2,s_3)=
(+2,+2,+2)$, $(+2,+2,-2)$ and $(-2,+2,+2)$.
For the ${\cal N}=8$ supergravity,
the possible helicity combinations of $s_2$ and $s_3$ are as
follows \cite{AK2509}:
$(s_2,s_3)=(+2,+2)$, $(+2,+\frac{3}{2})$, $(+2,+1)$,
$(+2,+\frac{1}{2})$, $(+2,0)$, $(+2,-\frac{1}{2})$, $(+2,-1)$,
$(+2,-\frac{3}{2})$, $(+2,-2)$, $(+\frac{3}{2},+\frac{3}{2})$,
$(+\frac{3}{2},+1)$,
$(+\frac{3}{2},+\frac{1}{2})$, $(+\frac{3}{2},0)$,
$(+\frac{3}{2},-\frac{1}{2})$, $(+\frac{3}{2},-1)$,
$(+\frac{3}{2},-\frac{3}{2})$,  $(+1,+1)$,
$(+1,+\frac{1}{2})$, $(+1,0)$, $(+1,-\frac{1}{2})$, $(+1,-1)$,
$(+\frac{1}{2},+\frac{1}{2})$, $(+\frac{1}{2},0)$,
$(+\frac{1}{2},-\frac{1}{2})$ and $(0,0)$.
Moreover, the helicity $s_1$ of particle $1$ can be either $\pm 2$,
$\pm \frac{3}{2}$, $\pm 1$, $\pm \frac{1}{2}$ or $0$.
For the first contraction between particle $1$ and particle $2$,
the  helicity combination
$s_I=s_1+s_2-2$ appears for $p_{12I}=1$. After that,
for
the second contraction  between particle $I$ and particle $3$,
the  helicity combination
$s_J=s_I+s_3-2=s_1+s_2+s_3-4$ appears for $p_{I3J}=1$.
This $s_J$ is greater than or equal to $-2$, and we arrive at
the above condition $s_1+s_2+s_3 \geq 2$.
For the single contractions, we have $s_1+s_2-2 \geq -2$
 leading to $s_1+s_2 \geq 0$
or  $s_1+s_3-2 \geq -2$ which is equivalent to $s_1+s_3 \geq 0$.}.

The main observations in \cite{GHPS,Strominger} are
that the infinite poles of
Euler beta functions appearing in the structure constants
of the OPEs of the single-particle operators with themselves
can be absorbed, via the residue of Gamma function,
by introducing the conformally soft currents
obtained by multiplying both sides of the OPEs by the same factor.
For the first-order poles of the above multi-particle OPEs, 
the analysis done in \cite{GHPS,Strominger} can be applied
and for the second-order poles, due to the property appearing in 
footnote \ref{generalizedbeta}, we expect that
the residues of Gamma functions can be used at four places.
When we compute the corresponding 
(anti)commutators for the multi-particle OPEs,
the four contour integrals where the dummy variables
are given by the two (i.e., holomorphic and antiholomorphic)
complex coordinates and other two
complex coordinates appearing on the left-hand sides of these OPEs,
should be multiplied on both sides.
Then the above second and first-order poles
in the holomorphic coordinates of the multi-particle OPEs
will act on these contour integrals
nontrivially.
As in the single-particle OPEs \footnote{
These stand for the OPEs of the single-particle celestial operators
with themselves.},
the dummy variable related to the number of
antiholomorphic derivatives acting on the descendants is summed over
$0$ to $\infty$ and this will be highly constrained and the infinite summation
will reduce to a finite sum. 

The splitting functions in the triple-collinear limit
for the pure Einstein gravity
have been found in \cite{BHP,CP}. They depend on the three energies,
three helicities (which are integers)
and six complex coordinates
of three particles in the momentum space.
It is natural to {\it interpolate} 
these splitting functions for the half-integers.
In other words, we assume that their explicit formula for the splitting
functions \cite{BHP} holds for the half-integer helicities.
From the relation between the $n$-point celestial amplitude
and the $(n-2)$-point celestial amplitude, by following the work of
\cite{CP},
we obtain the OPE for the three celestial operators and
the multi-particle OPEs for {\it generic}
helicities $s_1,s_2,s_3 = \pm 2$, $\pm \frac{3}{2}$, $\pm 1$,
$\pm \frac{1}{2}$ and $0$ in ${\cal N}=8$ supergravity.

At least, we can consider the higher-particle celestial operators
\footnote{That is, there are three-particle celestial operators,
four-particle celestial operators and so on.}
and the corresponding multi-particle OPEs (i.e., the OPEs
of the single-particle celestial operators with
the higher-particle celestial operators) in two-dimensional spacetimes.
We can perform the analysis given in the footnote \ref{geometric}
starting from the OPEs of the single-particle celestial operators with
three-particle celestial operators. Then we observe that there exists
a third-order pole $\frac{\bar{z}_{14}^3}{z_{14}^3}$
in the holomorphic coordinates of the multi-particle OPEs.
We can apply the descriptions in the
paragraph preceding the previous one to these OPEs.
We continue to calculate the multi-particle
OPEs of the single-particle celestial
operators with the four-particle celestial operators,
$\cdots$, $(N-1)$-particle celestial
operators.

\end{itemize}

In section 2, we calculate all the (anti)commutators
for the particles in the ${\cal N}=8$ supergravity 
corresponding to the second-order poles of the
above multi-particle OPEs mainly. We also briefly describe
the structures of the (anti)commutators corresponding to
the first-order poles. 

In section 3, by following the works of \cite{BHP,CP},
we obtain the various splitting functions for the possible
helicities described in the footnote  \ref{25s2s3} in the
four-dimensional spacetimes.
We also determine the OPE for the three celestial operators
via Mellin transform and the second-order poles
of the multi-particle OPEs described in previous section
are identified explicitly.

In section 4,
we summarize the main results of this paper and
comment on the future directions related to this paper.
Moreover, we briefly sketch  the multi-particle OPEs (those
of the single-particle celestial operators
with $(N-1)$-particle celestial operators) and their
(anti)commutators.

In Appendices $A, B, C, D,$ and $E$, some of the details of sections $2,3$
and $4$ are given.

\section{
The (anti)commutators corresponding to the multi-particle
OPEs
}

\subsection{The single-particle contributions to the
 multi-particle OPEs
\label{2point1}}

In \cite{CP},
the single-particle contributions to the
multi-particle OPEs
\footnote{The simplest multi-particle OPEs are
the OPEs of the single-particle celestial operators
with the two-particle celestial operators. In general,
the multi-particle OPEs are
the OPEs of the multi-particle celestial operators
with the other multi-particle celestial operators.}
in Einstein gravity are given by
\bea
&&
G^{s_1}_{\Delta_1}(z_1, \bar{z}_1) \, (G^{s_2}_{\Delta_2} \,
G^{s_3}_{\Delta_3})(z_3, \bar{z}_3)
\nonu \\
&& = \frac{\bar{z}_{13}^2}{z_{13}^2}
\sum_{m=0}^{\infty}\,
\frac{\bar{z}_{13}^m}{m!}\, B(\Delta_1-s_1+2+m,\Delta_2-s_2+1,
\Delta_3-s_3+1)\, 
\bar{\pa}^m \, G^{\rm{min}(s_1,s_2,s_3)}_{\Delta_1+\Delta_2+\Delta_3}(z_3, \bar{z}_3).
\label{ggope}
\eea
Here the helicities $s_1,s_2$ and $s_3$ are $\pm 2$ but
we allow them to be $\pm \frac{3}{2}, \pm 1, \pm \frac{1}{2}$ or
$0$. Later we will  put the $SU(8)$ indices on the various operators
appearing in the OPEs.
Let us multiply
\bea \lim_{\Delta_1 \rightarrow j, \Delta_2 \rightarrow k,
\Delta_3\rightarrow l} (\Delta_1 -j)(\Delta_2 -k) (\Delta_3-l) \,
\label{threeDelta}
\eea
of
both sides of (\ref{ggope}). Then the contribution from
the right-hand side of
(\ref{ggope}) can be written as
\bea
&& \lim_{\Delta_1 \rightarrow j,
\Delta_2 \rightarrow k, \Delta_3\rightarrow l} (\Delta_1
-j)(\Delta_2 -k) (\Delta_3-l)  \,
\frac{\bar{z}_{13}^2}{z_{13}^2} \nonu \\
&& \times \, \sum_{m=0}^{\infty}\,
\frac{\bar{z}_{13}^m}{m!}\,
B(\Delta_1-s_1+2+m,\Delta_2-s_2+1, \Delta_3-s_3+1)\,
\bar{\pa}^m \, G^{\rm{min}(s_1,s_2,s_3)}_{\Delta_1+\Delta_2+\Delta_3}(z_3, \bar{z}_3)
\nonu \\
&& = \lim_{\Delta_1 \rightarrow j, \Delta_2 \rightarrow k,
\Delta_3\rightarrow l} (\Delta_1 -j)(\Delta_2 -k) (\Delta_3-l)
\, \frac{\bar{z}_{13}^2}{z_{13}^2} \nonu \\
&& \times  \sum_{m=0}^{\infty}\,\frac{\bar{z}_{13}^m}{m!}
\frac{\Gamma(\Delta_1-s_1+2+m) \,
\Gamma(\Delta_2-s_2+1)\, \Gamma(\Delta_3-s_3+1)
}{\Gamma(\Delta_1+\Delta_2+\Delta_3-s_1-s_2-s_3+4+m)}\, \bar{\pa}^m 
G^{\rm{min}(s_1,s_2,s_3)}_{\Delta_1+\Delta_2+\Delta_3}(z_3, \bar{z}_3),
\label{inter2}
\eea
where the expression of $(2.25)$ of \cite{CP} is used
\footnote{The combinations of spins
$p_{12I}\equiv s_1+s_2-s_I-1$ in the process
$12 \rightarrow I$  and
$p_{I3J} \equiv s_I+s_3-s_J-1$ in the process
$I3 \rightarrow J$  
are given by $1$ and $1$ respectively.  
Moreover, the values of
the sum of helicities 
$s_1+s_2+s_3$ on the left-hand sides of the OPEs
in the ${\cal N}=8$ supergravity we are
considering is greater than $+2$.}.  Then we
can use the residues of the Gamma function in (\ref{inter2})
as follows:
\bea
&& \lim_{\Delta_1
\rightarrow j} (\Delta_1 -j) \, \Gamma(\Delta_1-s_1+2+m) =
\frac{(-1)^{s_1-2-m-j}}{(s_1-2-m-j)!}, \nonu \\ &&\lim_{\Delta_2 \rightarrow k}
(\Delta_2 -k) \, \Gamma(\Delta_2-s_2+1) =
\frac{(-1)^{s_2-1-k}}{(s_2-1-k)!},
\nonu \\
&& \lim_{ \Delta_3\rightarrow l} (\Delta_3-l)\,
\Gamma(\Delta_3-s_3+1)
= \frac{(-1)^{s_3-1-l}}{(s_3-1-l)!}, \nonu \\ && \lim_{\Delta_1 \rightarrow j,
\Delta_2 \rightarrow k, \Delta_3 \rightarrow l} \,
(\Delta_1+\Delta_2+\Delta_3 -j-k-l) \,
\Gamma(\Delta_1+\Delta_2+\Delta_3-s_1-s_2-s_3+4+m) \nonu \\ && =
\frac{(-1)^{s_1+s_2+s_3-4-j-k-l-m}}{(s_1+s_2+s_3-4-j-k-l-m)!}.
\label{relation}
\eea

Then the OPE
of a conformally soft particle with
soft multi-particle, after using (\ref{relation}),
can be written as
\bea &&
H^{j}(z_1, \bar{z}_1) \, H^{k,l}(z_3, \bar{z}_3)\nonu \\
&& =
\frac{\bar{z}_{13}^2}{z_{13}^2} \, \frac{1}{(s_2-1-k)!(s_3-1-l)!} \,
\sum_{m=0}^{\infty}\,\frac{\bar{z}_{13}^m}{m!}\, \frac{(
s_1+s_2+s_3-4-j-k-l-m)!
}{(-m-j+s_1-2)!} \,  \nonu \\
&& \times \bar{\pa}^m \, H^{j+k+l}(z_3, \bar{z}_3)
+ \cdots,
\label{HH}
\eea
where
the abbreviated part denoted by
$+ \cdots$ is the regular terms in $z_{13}$ and $\bar{z}_{13}$
and contains the normal-ordered product
$(H^{j} \, H^{k,l})(z_3, \bar{z}_3)$.
The soft operators in the presence of (\ref{threeDelta})
are defined as
\bea H^{j}(z_1,\bar{z}_1)
&\equiv& \lim_{\Delta_1\rightarrow j} (\Delta_1-j)
(G_{\Delta_1}^{s_1})(z_1,\bar{z}_1), \nonu \\
H^{k,l}(z_3,\bar{z}_3)
&\equiv & \lim_{\Delta_2 \rightarrow k, \Delta_3 \rightarrow l} (\Delta_2-k) (\De_3-l)
(G_{\Delta_2+\De+3}^{s_2+s_3})(z_3,\bar{z}_3),
\nonu \\
H^{j+k+l}(z_3,\bar{z}_3)
&\equiv &
\lim_{\De_1 \rightarrow j,
\Delta_2 \rightarrow k, \Delta_3 \rightarrow l} (\De_1-j)
(\Delta_2-k) (\De_3-l)
(G_{\De_1+\Delta_2+\De+3}^{s_1+s_2+s_3})(z_3,\bar{z}_3).
\label{Hdef}
\eea
Of course, the particles from the
${\cal N}=8$ supergravity with $SU(8)$ indices
can be defined similarly to (\ref{Hdef}).
The dimension of right-weight on the LHS of
(\ref{HH}) is given by
\bea
\frac{1}{2}(j-s_1) + \frac{1}{2}(k-s_2+l-s_3)
\label{LHS1}
\eea
while the one on the RHS
is
\bea
\frac{1}{2}(j+k+l-s_4) -2.
\label{RHS1}
\eea
The $-2$ comes from the factor $\bar{z}_{13}^2$.
This implies, by equating (\ref{LHS1}) and (\ref{RHS1}), that
the helicity for the $s_J=s_4$
is constrained and satisfies
$s_1+s_2+s_3-s_4=4$ \footnote{
\label{leftweight}
The dimension of left-weight on the LHS of
(\ref{HH}) is given by
$
\frac{1}{2}(j+s_1) + \frac{1}{2}(k+s_2+l+s_3)
$
and the one on the RHS
is
$
\frac{1}{2}(j+k+l+s_4) +2$.
The $+2$ comes from the factor $\frac{1}{{z}_{13}^2}$.}.
Furthermore,
let us consider the dimension of $(1-\bar{h})$
where $\bar{h}$ is the right-weight.
The LHS has
\bea
1-\frac{1}{2}(j-s_1) + 1-\frac{1}{2}(k-s_2+l-s_3)
\label{oneminushbar}
\eea
while the RHS
has
\bea
1-\frac{1}{2}(j+k+l-s_4)=1-\frac{1}{2}(j+k+l-s_1-s_2-s_3+4),
\label{oneminushbar1}
\eea
where we used the constraint for the helicity $s_4$
in terms of other three helicities.
The numerical value on the LHS is $+2$
while the numerical value on the RHS is
$-1$.
This indicates, by comparing (\ref{oneminushbar})
with (\ref{oneminushbar1}), there should appear the factor
$\frac{1}{\bar{z}_{13}^3}$ which plays the role of
$(1-\bar{h}) =+3$.

Recall that the modes of $H^{j}(z,\bar{z})$ are given by
\bea
(H^{j})_{n,\bar{n}}=\oint \frac{d z}{2 \pi i}\, \oint \frac{d
\bar{z}}{2 \pi i} \, z^{n+h-1}\, \bar{z}^{\bar{n}+\bar{h}-1} H^j(z,
\bar{z}), \qquad h =\frac{1}{2} (j+s_1),
\bar{h} =\frac{1}{2} (j-s_1).
\label{Hmodes}
\eea
In (\ref{Hmodes}),
the left and right-weights $(h, \bar{h})$ are given by
the conformal dimension $j$ and helicity $s_1$.
Moreover, we have
\footnote{
The modes for  the normal-ordered product
can be obtained from the usual CFT technique (for example,
the page $39$
of \cite{CFT}) by considering the holomorphic and antiholomorphic
contour integration
\bea
({\cal O}_1 \, {\cal O}_2)_{n,\bar{n}} &=&
\oint \, \frac{d z}{2\pi i}\, 
\oint \, \frac{d \bar{z}}{2\pi i}\,
z^{n+h_1+h_2-1} \, \bar{z}^{\bar{n}+\bar{h}_1+\bar{h}_2-1}\,
({\cal O}_1 \, {\cal O}_2)(z, \bar{z}) \nonu \\
& = &
\sum_{k \leq -h_1, \, \bar{l} \leq -\bar{h}_1} \,
({\cal O}_1)_{k,\bar{l}} \, ({\cal O}_2)_{n-k,\bar{n}-\bar{l}}
+ \sum_{k > -h_1, \, \bar{l} > -\bar{h}_1} \,
({\cal O}_2)_{n-k,\bar{n}-\bar{l}} \, ({\cal O}_1)_{k,\bar{l}},
\nonu
\eea
where we use the definition of (\ref{NORMAL}).
The $h_1$ and $h_2$ are the left-weights for the
celestial operator one and operator two respectively.
The $\bar{h}_1$ and $\bar{h}_2$ are the corresponding
right-weights for them  respectively.
Note that our notation for the normal ordering
$({\cal O}_1 \, {\cal O}_2)(z,\bar{z})$ is equal to
$N({\cal O}_2 \, {\cal O}_1)(z,\bar{z})$ of \cite{CFT}.
When one of the operators
${\cal O}_1$, or ${\cal O}_2$ contains the (anti)holomorphic
derivatives, further description will appear later.}
\bea (H^{k,l})_{n',\bar{n}'} & =& \oint
\frac{d z}{2 \pi i}\, \oint \frac{d \bar{z}}{2 \pi i} \, z^{n'+h-1}\,
\bar{z}^{\bar{n}'+\bar{h}-1} H^{k,l}(z, \bar{z}), \nonu \\
h
& = &
\frac{1}{2} (k+l+s_2+s_3), \qquad
\bar{h} =\frac{1}{2} (k+l-s_2-s_3).
\label{HHmodes}
\eea
The conformal dimension and helicity
for the quadratic operator in (\ref{HHmodes})
are used in the left and right-weights.
Let us calculate the commutator corresponding to
the OPE (\ref{HH}), by following
the work of \cite{GHPS} and performing the four contour integrals
explicitly,
\bea
&&
\bigg[ (H^{j})_{2-h,\bar{n}},
(H^{k,l})_{n',\bar{n}'} \bigg] = \oint \frac{d z_1}{2 \pi i}\,
\oint \frac{d \bar{z}_1}{2 \pi i} \, z_1^{(2-h)+h-1}\,
\bar{z}_1^{\bar{n}+\bar{h}-1} \, \oint \frac{d z_3}{2 \pi i}\, \oint
\frac{d \bar{z}_3}{2 \pi i} \, z_3^{n'+h-1}\,
\bar{z}_3^{\bar{n}'+\bar{h}-1} \nonu \\
&& \times  H^j(z_1,
\bar{z}_1)\,  H^{k,l}(z_3, \bar{z}_3) \nonu \\
&& = \oint \frac{d
z_1}{2 \pi i}\, \oint \frac{d \bar{z}_1}{2 \pi i} \,
z_1^{(2-h)+h-1}\, \bar{z}_1^{\bar{n}+\bar{h}-1} \, \oint \frac{d
z_3}{2 \pi i}\, \oint \frac{d \bar{z}_3}{2 \pi i} \, z_3^{n'+h-1}\,
\bar{z}_3^{\bar{n}'+\bar{h}-1} \nonu \\
&& \times 
\frac{\bar{z}_{13}^2}{z_{13}^2} \,
\frac{1}{(s_2-1-k)!(s_3-1-l)!} \,
\sum_{m=0}^{\infty}\,\frac{\bar{z}_{13}^m}{m!}\,
\frac{(s_1+s_2+s_3-4-j-k-l-m)!
}{(-m-j+s_1-2)!} \,
\nonu \\
&& \times \bar{\pa}^m \, H^{j+k+l}(z_3, \bar{z}_3).
\label{HHcomm}
\eea
The right-hand side of (\ref{HHcomm}), after $z_1$ integration,
becomes
\footnote{
\label{ncondition}
We are using the Cauchy integral formula
\bea \oint_{|z_{13}| < \ep} \frac{d z_1}{2 \pi i}\,
\frac{z_1}{z_{13}^2}=1.
\nonu
\eea
Note that the holomorphic mode in (\ref{HHcomm})
is fixed by $n=(2-h)$ where $h$ is given by (\ref{Hmodes}). }
\bea && \oint \frac{d \bar{z}_1}{2 \pi i} \,
\bar{z}_1^{\bar{n}+\bar{h}-1} \, \oint \frac{d z_3}{2 \pi i}\, \oint
\frac{d \bar{z}_3}{2 \pi i} \, z_3^{n'+h-1}\,
\bar{z}_3^{\bar{n}'+\bar{h}-1} \,  \bar{z}_{13}^2 \,
\frac{1}{(s_2-1-k)!(s_3-1-l)!} \nonu \\
&& \times
\sum_{m=0}^{\infty}\,\frac{\bar{z}_{13}^m}{m!}\, \frac{(s_1+s_2
+s_3-4-j-k-l-m)!
}{(-m-j+s_1-2)!} \,  \bar{\pa}^m \, H^{j+k+l}(z_3, \bar{z}_3).
\label{HHcomm1}
\eea
Again from the identity \footnote{
We have
\bea \oint_{|\bar{z}_1| < \ep} \frac{d
\bar{z}_1}{2 \pi i} \, \bar{z}_1^{\bar{n}+\frac{1}{2}(j-s_1)-1} \,
\bar{z}_{13}^{m+2} = \frac{(m+2)! \, (-\bar{z}_3)^{m+2+\bar{n}+
\frac{1}{2}(j-s_1)}}{(-\bar{n}- \frac{1}{2}(j-s_1))!(m+2+\bar{n}+
\frac{1}{2}(j-s_1))!}
\label{iden}
\eea
after using the binomial theorem for the $\bar{z}_{13}^{m+2}=
(\bar{z}_1-\bar{z}_3)^{m+2}$
in terms of finite sum and applying the
Cauchy integral formula over the variable $\bar{z}_1$ again.
Note that the $\bar{h}$ in (\ref{Hmodes})
is substituted explicitly and appears on the right-hand side of
(\ref{iden}).}
in \cite{GHPS},
the above right-hand side of (\ref{HHcomm}) or
the equation (\ref{HHcomm1}),
where the left and right weights
in (\ref{HHmodes}) are used, is, after $\bar{z}_1$
integration, given by
\bea && \oint \frac{d z_3}{2 \pi i}\, \oint \frac{d \bar{z}_3}{2 \pi
i} \, z_3^{n'+\frac{1}{2}(k+l+s_2+s_3)-1}\,
\bar{z}_3^{\bar{n}'+\frac{1}{2}(k+l-s_2-s_3)-1}\, \nonu \\
&& \times
\sum_{m=-\bar{n}- \frac{1}{2}(j-s_1)-2}^{-2-(j-s_1)}\,\frac{1}{m!}\, \,
\frac{(s_1+s_2+s_3-4-j-k-l-m)! }{(-m-j+s_1-2)!}  \,
\frac{1}{(s_2-1-k)!(s_3-1-l)!} \, \nonu
\\
&& \times \frac{(m+2)! \, (-\bar{z}_3)^{m+2+\bar{n}+
\frac{1}{2}(j-s_1)}}{(-\bar{n}- \frac{1}{2}(j-s_1))!(m+2+\bar{n}+
\frac{1}{2}(j-s_1))!}  \bar{\pa}^m \, H^{j+k+l}(z_3, \bar{z}_3).
\label{doubleint}
\eea
The upper bound for the summation over $m$
in (\ref{doubleint})
can be fixed by $\bar{\pa}^{-2-(j-s_1)} \, H^{j}(z_1,\bar{z}_1)=0$.
This is consistent with the fact that
the argument in the denominator of
$\frac{1}{(-m-j+s_1-2)!}$ above
should be greater than or equal to zero.
The maximum value of $m$ is given by $m=(-j+s_1-2)$
\footnote{The summation over $m$ from $0$ to
$-\bar{n}- \frac{1}{2}(j-s_1)-3$ vanishes from the
result of (\ref{iden}). This is the reason why the lowest value
for the $m$ is given by $-\bar{n}- \frac{1}{2}(j-s_1)-2$. }.

It is known that
\bea
H^{j+k+l}(z_3, \bar{z}_3) &=&
\sum_{n'',\bar{n}''}\,
\frac{(H^{j+k+l})_{n'',\bar{n}''}}{z_3^{n''+h}\,
\bar{z}_3^{\bar{n}''+\bar{h}}},
\nonu \\
h &= & \frac{1}{2}(j+k+l+s_4), \bar{h}
=\frac{1}{2}( j+k+l-s_4).
\label{Gexp2}
\eea
The nonzero contribution from the ${z}_3$ integration arises at
the condition
\bea
n'+\frac{1}{2}(k+l+s_2+s_3)-1-n''-\frac{1}{2}(j+k+l+s_4)=-1.
\label{n'}
\eea
This
implies, by moving $n''$ in (\ref{n'}) to the right, 
\bea n''=n'-\frac{1}{2}(j-s_2-s_3+s_4)=(\frac{1}{2}(s_1+s_2+s_3-s_4)-h)+n', \qquad
h=\frac{1}{2}(j+s_1).
\label{n''}
\eea
In other words, the holomorphic mode of the right-hand
side $n''$ in (\ref{n''}) is, after using the constraint
for the helicity $s_4$ associated with 
footnote \ref{leftweight},
the sum of those on the left-hand side, $(2-h)+n'$:
the conservation of scaling dimensions for the holomorphic modes.
See also the footnote \ref{ncondition}.

Acting $\bar{\pa}^m$ on the equation (\ref{Gexp2}), the nonzero
contribution from the $\bar{z}_3$ integration arises at the condition
\bea
\bar{n}'+\frac{1}{2}(k+l-s_2-s_3)-1 +
m+2+\bar{n}+ \frac{1}{2}(j-s_1)
-\bar{n}''-\bar{h}-m =-1.
\label{n'bar}
\eea
By putting $\bar{n}''$ in (\ref{n'bar}) to the right,
this implies
\bea
\bar{n}''=\bar{n}+\bar{n}'+\frac{1}{2}(-s_1-s_2-s_3+s_4)+2.
\label{conditionbarn''}
\eea
This, after using
the conservation of scaling dimensions for the antiholomorphic modes,
$\bar{n}''=\bar{n}+\bar{n}'$ (\ref{conditionbarn''}),
leads to the constraint for the helicity $s_4$ described before
\bea
s_1+s_2+s_3-s_4 =4.
\label{s1234}
\eea

Finally, we obtain the
right-hand side
\footnote{
We use
\bea && \oint \frac{d \bar{z}_3}{2 \pi i} \,
\bar{z}_3^{\bar{n}'+\bar{h}-1+m+2+\bar{n}+ \frac{1}{2}(j-s_1)}\,
\bar{\pa}^m \, H^{j+k+l}(z_3, \bar{z}_3) =
\nonu
\\ &&
\frac{(-\bar{n}-\bar{n}'-\frac{1}{2}(
j+k+l-s_4))!}{(-\bar{n}-\bar{n}'-\frac{1}{2}( j+k+l-s_4)-m)!} \,
(H^{j+k+l})_{(2-h)+n',\bar{n}+\bar{n}'}
\nonu
\eea
by performing the antiholomorphic differential operator
on (\ref{Gexp2}) with respect to $\bar{z}_3$ coordinate. }
\bea
&& \bigg[ (H^{j})_{2-h,\bar{n}},
(H^{k,l})_{n',\bar{n}'} \bigg] = \sum_{m=-\bar{n}-
\frac{1}{2}(j-s_1)-2}^{-2-(j-s_1)}\,\frac{1}{m!}\, \,  \frac{
(s_1+s_2+s_3-4-j-k-l-m)!}{(-m-j+s_1-2)!} \nonu \\
&& \times \frac{1}{(s_2-1-k)!(s_3-1-l)!} \,
\frac{(m+2)! \, (-1)^{m+2+\bar{n}+
\frac{1}{2}(j-s_1)}}{
(-\bar{n}- \frac{1}{2}(j-s_1))!(m+2+\bar{n}+
\frac{1}{2}(j-s_1))!}  \nonu \\
&& \times
\frac{(-\bar{n}-\bar{n}'-\frac{1}{2}(
j+k+l-s_4))!}{(-\bar{n}-\bar{n}'-\frac{1}{2}( j+k+l-s_4)-m)!} \,
(H^{j+k+l})_{(2-h)+n',\bar{n}+\bar{n}'} \nonu \\ &&=
\frac{(-1)^{\bar{n}+ \frac{1}{2}(j-s_1)}(-\bar{n}-\bar{n}'-\frac{1}{2}(
j+k+l-s_4))!}{(s_2-1-k)!(s_3-1-l)!
(-\bar{n}- \frac{1}{2}(j-s_1))!} \,
(H^{j+k+l})_{(2-h)+n',\bar{n}+\bar{n}'} \nonu \\ && \times
\sum_{m=-\bar{n}- \frac{1}{2}(j-s_1)-2}^{-2-(j-s_1)}\, \Bigg[
\frac{1}{m!}\, \,  \frac{ (s_1+s_2+s_3-4-j-k-l-m)!}{(-m-j+s_1-2)!}  \, \frac{(m+2)! \,
(-1)^{m}}{(m+2+\bar{n}+ \frac{1}{2}(j-s_1))!}  \nonu \\ && \times
\frac{1}{(-\bar{n}-\bar{n}'-\frac{1}{2}( j+k+l-s_4)-m)!}  \Bigg].
\label{interHH}
\eea
We collect $m$ dependence at the final step of (\ref{interHH}).
The summation over $m$ in (\ref{interHH}) from the
Mathematica \cite{mathematica} can be written as
\bea 
&& \frac{i^{-(j-s_1)} (-1)^{-\bar{n}} (j-s_1+2 \bar{n})
(2+j-s_1+2 \bar{n})
\left(-\frac{j-s_1}{2}-k+s_2-l+s_3+\bar{n}-2\right)!}{
4(-\frac{j-s_1}{2}+\bar{n})!
(\frac{1}{2} (-k+s_2-l+s_3-2 \bar{n}'+4-(s_1+s_2+s_3-s_4)))!}
\label{msummation}
\\ && \times \, {}_3 F_2
\Bigg[
\begin{array}{c}
\frac{k-s_2}{2}+\frac{l-s_3}{2}+\bar{n}'-2+\frac{1}{2}(s_1+s_2+s_3-s_4),\qquad 1-\frac{j-s_1}{2}
-\bar{n},\qquad \frac{j-s_1}{2}-\bar{n}
\\-1 -\frac{j-s_1}{2}-\bar{n},
\qquad
2+\frac{j-s_1}{2}+k-s_2+l-s_3-\bar{n}
\end{array} ; 1
\Bigg].
\nonu
\eea
Note that the whole expression (\ref{msummation})
depends on the right-weights, $(j-s_1)$, $(k-s_2)$,
$(l-s_3)$ and the modes $\bar{n}$ and $\bar{n}'$
after using the relation (\ref{s1234}).

Then the final commutator, after using (\ref{msummation}),
can be written as
\bea && \bigg[
(H^{j})_{2-h,\bar{n}}, (H^{k,l})_{n',\bar{n}'} \bigg]=
\label{COMM1} \\ &&
\Bigg( \frac{(-\frac{j-s_1}{2}-k+s_2-l+s_3+\bar{n}-2)!
(-\frac{j}{2}-\frac{k}{2}-\frac{l}{2}-\bar{n}'-\bar{n}+
\frac{s_4}{2})!}  {4(s_2-1-k)!
(s_3-1-l)! (-\frac{j-s_1}{2}-\bar{n}-2)!  (-\frac{j-s_1}{2}
+\bar{n})!
(-\frac{k-s_2}{2}-\frac{l-s_3}{2}-\bar{n}')!} \Bigg) \nonu \\ && \times \,
 {}_3 F_2
\Bigg[
\begin{array}{c}
\frac{k-s_2}{2}+\frac{l-s_3}{2}+\bar{n}'-2+\frac{1}{2}(s_1+s_2+s_3-s_4),\qquad 1-\frac{j-s_1}{2}
-\bar{n},\qquad \frac{j-s_1}{2}-\bar{n}
\\-1 -\frac{j-s_1}{2}-\bar{n},
\qquad
2+\frac{j-s_1}{2}+k-s_2+l-s_3-\bar{n}
\end{array} ; 1
\Bigg]
\nonu \\
&& \times \, (H^{j+k+l})_{(2-h)+n',\bar{n}+\bar{n}'}.
\nonu \eea
Again the right-hand side of (\ref{COMM1})
depends on the three right-weights and two modes
with (\ref{s1234})
as before.
Here
the generalized hypergeometric function
\footnote{
\label{property3F2}
When at least one of its numerator parameters
is non-positive ($-n$ where $n=0,1,2, \cdots$.), then
the standard infinite series reduces to a finite
polynomial sum of $(n+1)$ terms
\bea
{}_3 F_2
\Bigg[
\begin{array}{c}
-n, a_2, a_3
\\ b_1, b_2
\end{array} ; 1
\Bigg]
&=& \sum_{k=0}^{n}\,
\frac{(-n)_k \, (a_2)_k \, (a_3)_k}{(b_1)_k \, (b_2)_k}
\, \frac{1}{k!} \nonu \\
&=&
\sum_{k=0}^{n}\, \frac{\Gamma(-n+k)}{\Gamma(-n)}\,
\frac{\Gamma(a_2+k)}{\Gamma(a_2)}\,
\frac{\Gamma(a_3+k)}{\Gamma(a_3)}\,
\frac{\Gamma(b_1)}{\Gamma(b_1+k)}\,
\frac{\Gamma(b_2)}{\Gamma(b_2+k)}\,
 \frac{1}{k!}.
\nonu
\eea
Note that $(-n)_k=(-1)^k\, (n-k+1)_k$. The notation of $(a)_n$ is for the
Pochhammer symbol and is given by
$(a)_n =\frac{\Gamma(a+n)}{\Gamma(a)}$.}
is
\bea
&&
{}_3 F_2
\Bigg[
\begin{array}{c}
\frac{k-s_2}{2}+\frac{l-s_3}{2}+\bar{n}'-2+\frac{1}{2}(s_1+s_2+s_3-s_4),\qquad 1-\frac{j-s_1}{2}
-\bar{n},\qquad \frac{j-s_1}{2}-\bar{n}
\\-1 -\frac{j-s_1}{2}-\bar{n},
\qquad
2+\frac{j-s_1}{2}+k-s_2+l-s_3-\bar{n}
\end{array} ; 1
\Bigg]
\nonu \\
&& =  \Bigg( \frac{(\frac{1}{2} (k-s_2+l-s_3)-
\bar{n}'-1)!  (\frac{j-s_1}{2}+k-s_2+l-s_3-\bar{n}+1)! }
{(j-s_1+2
\bar{n}+2) (j-s_1+2 \bar{n}) (k-s_2+l-s_3+1)!
} \Bigg) \nonu \\
&& \times \frac{1}{(\frac{1}{2} (j-s_1+k-s_2+l-s_3-2
(\bar{n}'+\bar{n}))+1)!} \nonu \\
&& \times \,
\bigg(
-8 (j-s_1+1)(k-s_2+l-s_3+1) \bar{n}' \bar{n}
 +4 (j-s_1)
(j-s_1+1) \bar{n}'^2 \nonu \\
&& +(j-s_1)
(k-s_2+l-s_3) (j-s_1+k-s_2+l-s_3+2)\nonu \\
&& +
4(k-s_2+l-s_3) (k-s_2+l-s_3+1) \bar{n}^2 
\bigg).
\label{modequadratic}
\eea
The last three lines of (\ref{modequadratic})
contain the mode dependence $\bar{n}$ and $\bar{n}'$
in addition to the mode independent terms.
The coefficients in front of
mode dependent terms and the
mode independent terms itself depend on
the two $\bar{h}$'s in (\ref{Hmodes}) and (\ref{HHmodes}).
Compared to \cite{Strominger} which contains
the linear dependence on the modes, the degree of the polynomial
is increased by $1$. 

The commutator, by using (\ref{modequadratic}), leads to
\bea && \bigg[ (\hat{H}^{j})_{2-h,\bar{n}},
(\hat{H}^{k,l})_{n',\bar{n}'} \bigg]  =
\frac{1}{(s_2-1-k)! (s_3-1-l)! (1+k-s_2+l-s_3)!}
\nonu \\
&& \times \Bigg(
\frac{
(-\frac{j-s_1}{2}-\bar{n})!
(\frac{j-s_1}{2}+k-s_2+l-s_3-\bar{n}+1)!
(-\frac{j-s_1}{2}-k+s_2-l+s_3+\bar{n}-2)!}{
(-\frac{j-s_1}{2}-\bar{n})!(\frac{j-s_1}{2}-\bar{n}-1)!
(-\frac{j-s_1}{2}+\bar{n})! } \Bigg)
\nonu \\
&& \times \,
\bigg(
-2 (j-s_1+1)(k-s_2+l-s_3+1) \bar{n}' \bar{n}
 + (j-s_1)
(j-s_1+1) \bar{n}'^2 \nonu \\
&& +\frac{1}{4}\, (j-s_1)
(k-s_2+l-s_3) (j-s_1+k-s_2+l-s_3+2)\nonu \\
&& +
(k-s_2+l-s_3) (k-s_2+l-s_3+1) \bar{n}^2 
\bigg)\,
(\hat{H}^{j+k+l})_{(2-h)+n',\bar{n}+\bar{n}'},
\label{COMM2}
\eea
where the
redefined modes are given by
\bea (\hat{H}^{j})_{2-h,\bar{n}}  &
\equiv &  \frac{  (-\frac{j-s_1}{2}-\bar{n})!}{
(\frac{j-s_1}{2}-\bar{n}-1)!
} (H^{j})_{2-h,\bar{n}}, \nonu \\
(\hat{H}^{k,l})_{n',\bar{n}'}
& \equiv & \frac{ (-\frac{k-s_2}{2}-\frac{l-s_3}{2}-\bar{n}')!}
{(\frac{1}{2} (k-s_2+l-s_3-2\bar{n}'-2)!  }
(H^{k,l})_{n',\bar{n}'} 
\label{redefined}
\\ (\hat{H}^{j+k+l})_{(2-h)+n',\bar{n}+\bar{n}'} & \equiv &
\frac{(\frac{1}{2} (-j+s_1-k+s_2-l+s_3-2 \bar{n}'-2 \bar{n}-4))!}{
(\frac{1}{2} (j-s_1+k-s_2+l-s_3-2 (\bar{n}'+\bar{n})+2))!}
(H^{j+k+l})_{(2-h)+n',\bar{n}+\bar{n}'}.
\nonu
\eea
We obtain the coefficient of the third relation
in (\ref{redefined}) from the the coefficient of the first
relation by taking $j \rightarrow (j+k+l)$, $s_1 \rightarrow s_4$
with the constraint (\ref{s1234})
and $\bar{n} \rightarrow \bar{n}+\bar{n}'$.
In this sense, the two redefined relations, the first and the third,
in (\ref{redefined}) are consistent with each other.
They depend on their own $\bar{h}$'s in (\ref{Hmodes}) and
(\ref{Gexp2}).
The coefficient in the second relation of (\ref{redefined})
depends on the $\bar{h}$ in (\ref{HHmodes}).
The factor in the second line of (\ref{COMM2}),
which still depends on the modes, becomes $\pm 1$
depending on the even $k l$ or odd $k l$ because
the following relation holds
\bea && \frac{
(-\frac{j-s_1}{2}-\bar{n})!
(\frac{j-s_1}{2}+k-s_2+l-s_3-\bar{n}+1)!
(-\frac{j-s_1}{2}-k+s_2-l+s_3+\bar{n}-2)!}{
(-\frac{j-s_1}{2}-\bar{n})!(\frac{j-s_1}{2}-\bar{n}-1)!
(-\frac{j-s_1}{2}+\bar{n})! }=
\nonu \\
&& \sin (\pi  (j-s_1+N)) \csc (\pi
(j-s_1+k-s_2+l-s_3+N)),
\label{sincsc}
\eea
after substituting $\bar{n} =
-\frac{(j-s_1)}{2}-N$ for positive integer or half integer $N$
\footnote{The modes for the soft particle are constrained and
satisfy $ \bar{h} \leq \bar{n} \leq -\bar{h}$ \cite{Strominger,Tropper}.}.

Moreover, it is useful to relate
the mode dependent terms in (\ref{COMM2}) to the one in \cite{PRS}
\bea
&&
\bigg(
-2 (j-s_1+1)(k-s_2+l-s_3+1) \bar{n}' \bar{n}
+ (j-s_1)
(j-s_1+1) \bar{n}'^2 \nonu \\
&& +\frac{1}{4}\, (j-s_1)
(k-s_2+l-s_3) (j-s_1+k-s_2+l-s_3+2)\nonu \\
&& +
(k-s_2+l-s_3) (k-s_2+l-s_3+1) \bar{n}^2 
\bigg)= N_1^{1-\frac{1}{2}(j-s_1),1-\frac{1}{2}(k-s_2+l-s_3)}(\bar{n},
\bar{n}').
\label{Nquadratic}
\eea
The mode dependent function is given by
\bea
N^{h_1,h_2}_{h}(m,n)
\!&
\equiv \!&
\sum_{l=0 }^{h+1}(-1)^l
\left(\begin{array}{c}
h+1 \\  l \\
\end{array}\right)
[h_1-1+m]_{h+1-l}[h_1-1-m]_l
\nonu \\
\!& \times \!& [h_2-1+n]_l [h_2-1-n]_{h+1-l}.
\label{Ndef}
\eea
The Pochhammer symbol $[a]_n \equiv \frac{\Gamma(a+1)}{\Gamma(a-n+1)}$.
Note that the two upper indices
on the right-hand side of (\ref{Nquadratic})
indicate the two dimensions of $(1-\bar{h})$
of the left-hand side respectively described in
(\ref{oneminushbar}).
We don't have any information on the
dimension of $(1-\bar{h})$ for the right-hand side of the commutator
from the lower index appearing on the right-hand side of
(\ref{Nquadratic}).

Then we have the simple commutator as
\bea && \bigg[
(\hat{H}^{j})_{2-h,\bar{n}}, \frac{1}{(s_2-1-k)! (s_3-1-l)! (1+k-s_2+l-s_3)!}
\, (\hat{H}^{k,l})_{n',\bar{n}'} \bigg]  =
\nonu \\ && \pm
N_1^{1-\frac{1}{2}(j-s_1),1-\frac{1}{2}(k-s_2+l-s_3)}(\bar{n},
\bar{n}')
\,
(\hat{H}^{j+k+l})_{(2-h)+n',\bar{n}+\bar{n}'}.
\label{comm1}
\eea
It is obvious to introduce the normalization factor for the quadratic
operator on the left-hand side
\footnote{
\label{pandother}
The mode dependent function in \cite{CFT} (See also
\cite{Blumenhagenetal}) is related to the one in (\ref{Ndef})
as follows:
\bea
&& p_{1-\frac{1}{2}(j-s_1),1-\frac{1}{2}(k-s_2+l-s_3),
1-\frac{1}{2}(j+k+l-s_4)}(\bar{n},\bar{n}') \nonu \\
&& =
\frac{1}{2(1+j-s_1+k-s_2+l-s_3)(2+j-s_1+k-s_2+l-s_3)}\,
N_1^{1-\frac{1}{2}(j-s_1),1-\frac{1}{2}(k-s_2+l-s_3)}(\bar{n},
\bar{n}').
\label{pNrelation}
\eea
Compared to the equation (\ref{Ndef}),
the mode dependent function in \cite{CFT} has the information on
the dimension of $(1-\bar{h})$ for the
right-hand side in the commutator.
Then we can express the commutator in terms of the corresponding
OPE
\bea
&& \tilde{H}^{j}(\bar{z})
\,
\tilde{H}^{k,l}(\bar{w})   = 
\pm
p_{1-\frac{1}{2}(j-s_1),1-\frac{1}{2}(k-s_2+l-s_3),
1-\frac{1}{2}(j+k+l-s_4)}(\bar{\pa}_{\bar{z}},\bar{\pa}_{\bar{w}})
\,
\Bigg[ \frac{\tilde{H}^{j+k+l}(\bar{w})}{(\bar{z}-\bar{w})} \Bigg]
\nonu \\
&&= \frac{1}{(\bar{z}-\bar{w})^3}\, \tilde{H}^{j+k+l}(\bar{w})
\nonu \\
&&+  \frac{1}{(\bar{z}-\bar{w})^2}\,
\Bigg[ \frac{(1+j-s_1)}{(2+j-s_1+k-s_2+l-s_3)} \Bigg]
\bar{\pa}_{\bar{w}}\, \tilde{H}^{j+k+l}(\bar{w})
\nonu\\
&&+
\frac{1}{(\bar{z}-\bar{w})}\,
\Bigg[
\frac{(j-s_1)(1+j-s_1)}{(1+j-s_1+k-s_2+l-s_3)(2+j-s_1+k-s_2+l-s_3)}
\Bigg] 
\bar{\pa}^2_{\bar{w}}\,
\tilde{H}^{j+k+l}(\bar{w}) + \cdots.
\label{HtHt}
\eea
The $\tilde{H}^{j}(\bar{z})$ is the $\bar{z}$ dependent factor
in the holomorphic and antiholomorphic modes expansion
of $\hat{H}^{j}(z,\bar{z})$ where there exists
the $\frac{1}{z^2}$ factor from the holomorphic mode
$2-h=2-\frac{1}{2}(j+s_1)$.
Similarly, the $\tilde{H}^{k,l}(\bar{w})$ is the
$\bar{w}$ dependent factor (together with the overall factor
appearing in (\ref{comm1}) and (\ref{pNrelation}))
in the expansion
of $\hat{H}^{k,l}(w,\bar{w})$ for $n'=-h=-
\frac{1}{2}(k+l+s_2+s_3)$.
The $\tilde{H}^{j+k+l}(\bar{w})$ is the
$\bar{w}$ dependent factor in the expansion of 
$\tilde{H}^{j+k+l}(w,\bar{w})$ where there is no $w$ dependent factor.
Note that the difference of $(1-\bar{h})$
on the RHS and the LHS is
\bea
1-\frac{1}{2}(j+k+l-s_4)
-\Bigg(1-\frac{1}{2}(j-s_1)+1-\frac{1}{2}(k-s_2+l-s_3)\Bigg)=-3,
\nonu
\eea
which implies that the highest singular term is given by the
third-order pole.
The differential operator
for the polynomial $p$ in (\ref{HtHt}) can be obtained
from the replacements $\bar{n} \rightarrow \bar{\pa}_{\bar{z}}$
and  $\bar{n}' \rightarrow \bar{\pa}_{\bar{w}}$ on each term
having a degree of polynomial.
In (\ref{Nquadratic}),
the mode independent terms do not contribute because they are
zeroth order.
The structure constants appearing in the second and the
first-order poles of (\ref{HtHt}) can be obtained from the formula
(for example, the page $34$ 
in \cite{CFT}) or we can check them from the first line of
(\ref{HtHt}).}.

\subsection{The total ninety-five (anti)commutators
corresponding to the multi-particle OPEs}

Let us recall that the right-weights of the following particles
are given by Table \ref{hbar1}.
\begin{table}[tbp]
\centering
\renewcommand{\arraystretch}{1.7}
\begin{tabular}{|c|c|c|c|c| }
\hline
The particles & $\bar{h}$ &
$(1-\bar{h})$
\\
\hline
\hline
$\Phi^{(h_1)}_{+2}$ &   $(-1-h_1)=\frac{1}{2}(j-2)$
&
$(2+h_1)=\frac{1}{2}(-j+4)$
\\
\hline
$\Phi^{(h_1),A}_{+\frac{3}{2}}$ & $(-\frac{1}{2}-h_1)=
\frac{1}{2}(j-\frac{3}{2})$
& $(\frac{3}{2}+h_1)=
\frac{1}{2}(-j+\frac{7}{2})$
\\
\hline
$ \Phi^{(h_1),AB}_{+1} $ & $-h_1 =\frac{1}{2}(j-1)$
& $(1+h_1) =\frac{1}{2}(-j+3)$
\\
\hline
$\Phi^{(h_1),ABC}_{+\frac{1}{2}}$ & $(\frac{1}{2}-h_1)=\frac{1}{2}(j-
\frac{1}{2})$
& $(\frac{1}{2}+h_1)=\frac{1}{2}(-j+
\frac{5}{2})$
\\
\hline
$
\Phi^{(h_1),ABCD}_{0} $ & $ (1-h_1) =\frac{1}{2}j$
& $h_1 =\frac{1}{2}(-j+2)$
\\
\hline
$
\Phi^{(h_1)}_{ABC,-\frac{1}{2}}$ & $(\frac{3}{2}-h_1)=\frac{1}{2}(j+
\frac{1}{2})$
& $(-\frac{1}{2}+h_1)=\frac{1}{2}(-j+
\frac{3}{2})$
\\
\hline
$
\Phi^{(h_1)}_{AB,-1} $ & $ (2-h_1)=\frac{1}{2}(j+1)$
& $(-1+h_1)=\frac{1}{2}(-j+1)$
\\
\hline
$
\Phi^{(h_1)}_{A,-\frac{3}{2}} $ & $ (\frac{5}{2}-h_1)=
\frac{1}{2}(j+\frac{3}{2}) $
& $(-\frac{3}{2}+h_1)=
\frac{1}{2}(-j+\frac{1}{2})$
\\
\hline
$ 
\Phi^{(h_1)}_{-2}$ &  $ (3-h_1)=\frac{1}{2}(j+2)$
& $(-2+h_1)=-\frac{1}{2}j$
\\
\hline
\end{tabular}
\caption{
The right-weights $\bar{h}$ and
one minus right-weights $(1-\bar{h})$
of the single-particle operators. }
\label{hbar1}
\end{table}
Then the upper indices $(h_1)$ in the ${\cal N}=8$ multiplet
are written in terms of the conformal dimension $j$ and
the numerical values.
That is, the numerical constant values except $-\frac{j}{2}$
in the last column in Table \ref{hbar1}
are given by $(1+\frac{s_1}{2})$. Note that
the helicity $s_1$ in the ${\cal N}=8$ multiplet
is given by the  numerical values in the lower indices.

Similarly,
the right-weights and one minus right-weights
of the following composite operators, for convenience, 
are given by Tables \ref{hbar2}, and \ref{hbar2-1}.
\begin{table}[tbp]
\centering
\renewcommand{\arraystretch}{1.7}
\begin{tabular}{|c|c|c|c|c| }
\hline
The particles & $\bar{h}$ &
$(1-\bar{h})$
\\
\hline
\hline
$(\Phi_{+2}^{(h_2)}\, \Phi_{+2}^{(h_3)})$ & $(-2-h_2-h_3)=
\frac{1}{2}(k+l-4)$ & $ (3+h_2+h_3)=\frac{1}{2}(-k-l+6)$
\\
\hline
$(\Phi_{+2}^{(h_2)}\,
\Phi_{+\frac{3}{2}}^{(h_3),A}) $ & $ (-\frac{3}{2}-h_2-h_3)=
\frac{1}{2}(k+l-\frac{7}{2})$ & $(\frac{7}{2}+h_2+h_3)=\frac{1}{2}(-k-l+\frac{11}{2})$
\\
\hline
$
(\Phi_{+2}^{(h_2)}\, \Phi_{+1}^{(h_3),AB})$ & $(-1-h_2-h_3)=
\frac{1}{2}(k+l-3)$ & $(2+h_2+h_3)=\frac{1}{2}(-k-l+5)$
\\
\hline
$
(\Phi_{+2}^{(h_2)}\,
\Phi_{+\frac{1}{2}}^{(h_3),ABC}) $ & $(-\frac{1}{2}-h_2-h_3)=
\frac{1}{2}(k+l-
\frac{5}{2})$ & $(\frac{3}{2}+h_2+h_3)=\frac{1}{2}(-k-l+\frac{9}{2})$
\\
\hline
$
(\Phi_{+2}^{(h_2)}\, \Phi_{0}^{(h_3),ABCD})$ & $ (-h_2-h_3)=
\frac{1}{2}(k+l-
2)$ & $ (1+h_2+h_3)=\frac{1}{2}(-k-l+4)$
\\
\hline
$(\Phi_{+2}^{(h_2)}\,
\Phi_{ABC,-\frac{1}{2}}^{(h_3)})$ & $
(\frac{1}{2}-h_2-h_3)=\frac{1}{2}(k+l-
\frac{3}{2})$ & $(\frac{1}{2}+h_2+h_3)=\frac{1}{2}(-k-l+\frac{7}{2})$
\\
\hline
$ (\Phi_{+2}^{(h_2)}\, \Phi_{AB,-1}^{(h_3)})$ & $  (1-h_2-h_3)=
\frac{1}{2}(k+l-
1)$ & $ (h_2+h_3)=\frac{1}{2}(-k-l+3)$
\\
\hline
$(\Phi_{+2}^{(h_2)}\, \Phi_{A,-\frac{3}{2}}^{(h_3)})$ & $
(\frac{3}{2}-h_2-h_3)=\frac{1}{2}(k+l-
\frac{1}{2})$ &  $(-\frac{1}{2}+h_2+h_3)=\frac{1}{2}(-k-l+\frac{5}{2})$
\\
\hline
$ (\Phi_{+2}^{(h_2)}\, \Phi_{-2}^{(h_3)})$ & $
(2-h_2-h_3)=\frac{1}{2}(k+l) $ & $(-1+h_2+h_3)=\frac{1}{2}(-k-l+2)$
\\
\hline
$ (\Phi_{+\frac{3}{2}}^{(h_2),A}\, \Phi_{+\frac{3}{2}}^{(h_3),B})
$ & $ 
(-1-h_2-h_3)=\frac{1}{2}(k+l-
3)$ & $ (2+h_2+h_3)=\frac{1}{2}(-k-l+5)$
\\
\hline
$ (\Phi_{+\frac{3}{2}}^{(h_2),A}\, \Phi_{+1}^{(h_3),BC})$ & $
(-\frac{1}{2}-h_2-h_3)=\frac{1}{2}(k+l-
\frac{5}{2}) $ & $(\frac{3}{2}+h_2+h_3)=\frac{1}{2}(-k-l+\frac{9}{2})$
\\
\hline
$ (\Phi_{+\frac{3}{2}}^{(h_2),A}\, \Phi_{+\frac{1}{2}}^{(h_3),BCD}) $ &
$(-h_2-h_3)=\frac{1}{2}(k+l-
2)$ & $ (1+h_2+h_3)=\frac{1}{2}(-k-l+4)$
\\
\hline
$(\Phi_{+\frac{3}{2}}^{(h_2),A}\, \Phi_{0}^{(h_3),BCDE})$ &
$(\frac{1}{2}-h_2-h_3)=\frac{1}{2}(k+l-
\frac{3}{2})$ & $(\frac{1}{2}+h_2+h_3)=\frac{1}{2}(-k-l+\frac{7}{2})$
\\
\hline
$(\Phi_{+\frac{3}{2}}^{(h_2),A}\, \Phi_{BCD,-\frac{1}{2}}^{(h_3)})$ & $
(1-h_2-h_3)=\frac{1}{2}(k+l-
1)$ & $ (h_2+h_3)=\frac{1}{2}(-k-l+3)$
\\
\hline
$(\Phi_{+\frac{3}{2}}^{(h_2),A}\, \Phi_{BC,-1}^{(h_3)})$ &
$ (\frac{3}{2}-h_2-h_3)=\frac{1}{2}(k+l-
\frac{1}{2})$ & $ (-\frac{1}{2}+h_2+h_3)=\frac{1}{2}(-k-l+\frac{5}{2})$
\\
\hline
\end{tabular}
\caption{
The right-weights $\bar{h}$ and
one minus right-weights $(1-\bar{h})$
of the first fifteen two-particle operators. }
\label{hbar2}
\end{table}

\begin{table}[tbp]
\centering
\renewcommand{\arraystretch}{1.7}
\begin{tabular}{|c|c|c|c|c| }
\hline
The particles & $\bar{h}$ &
$(1-\bar{h})$
\\
\hline
\hline
$
(\Phi_{+\frac{3}{2}}^{(h_2),A}\, \Phi_{B,-\frac{3}{2}}^{(h_3)})$ &
$(2-h_2-h_3)=\frac{1}{2}(k+l)$ & $(-1+h_2+h_3)=\frac{1}{2}(-k-l+2)$
\\
\hline
$(\Phi_{+1}^{(h_2),AB}\, \Phi_{+1}^{(h_3),CD})$ & $
(-h_2-h_3)=\frac{1}{2}(k+l-
2)$ & $ (1+h_2+h_3)= \frac{1}{2}(-k-l+4)$
\\
\hline
$ (\Phi_{+1}^{(h_2),AB}\, \Phi_{+\frac{1}{2}}^{(h_3),CDE})$ & $
(\frac{1}{2}-h_2-h_3)=\frac{1}{2}(k+l-
\frac{3}{2})$ & $ (\frac{1}{2}+h_2+h_3)=\frac{1}{2}(-k-l+\frac{7}{2})$
\\
\hline
$ (\Phi_{+1}^{(h_2),AB}\, \Phi_{0}^{(h_3),CDEF})$ & $
(1-h_2-h_3)=\frac{1}{2}(k+l-
1)$ & $(h_2+h_3)=  \frac{1}{2}(-k-l+3)$
\\
\hline
$(\Phi_{+1}^{(h_2),AB}\, \Phi_{CDE,-\frac{1}{2}}^{(h_3)})$ & $
(\frac{3}{2}-h_2-h_3)=\frac{1}{2}(k+l-
\frac{1}{2})$  & $(-\frac{1}{2}+h_2+h_3)=\frac{1}{2}(-k-l+\frac{5}{2})$
\\
\hline
$
(\Phi_{+1}^{(h_2),AB}\, \Phi_{CD,-1}^{(h_3)})$ &
$(2-h_2-h_3)=\frac{1}{2}(k+l)$  & $ (-1+h_2+h_3)=\frac{1}{2}(-k-l+2)$
\\
\hline
$
(\Phi_{+\frac{1}{2}}^{(h_2),ABC}\, \Phi_{+\frac{1}{2}}^{(h_3),DEF})$ & $
(1-h_2-h_3)=\frac{1}{2}(k+l-
1)$ &  $ (h_2+h_3)=\frac{1}{2}(-k-l+3)$
\\
\hline
$(\Phi_{+\frac{1}{2}}^{(h_2),ABC}\, \Phi_{0}^{(h_3),DEFG})$ & $
(\frac{3}{2}-h_2-h_3)=\frac{1}{2}(k+l-
\frac{1}{2})$ & $ (-\frac{1}{2}+h_2+h_3)= \frac{1}{2}(-k-l+\frac{5}{2})$
\\
\hline
$(\Phi_{+\frac{1}{2}}^{(h_2),ABC}\, \Phi_{DEF,-\frac{1}{2}}^{(h_3)})$ & $
(2-h_2-h_3)=\frac{1}{2}(k+l)$ & $ (-1+h_2+h_3)= \frac{1}{2}(-k-l+2)$
\\
\hline
$(\Phi_{0}^{(h_2),ABCD}\, \Phi_{0}^{(h_3),EFGH})$ & $
(2-h_2-h_3)=\frac{1}{2}(k+l)$ & $ (-1+h_2+h_3)= \frac{1}{2}(-k-l+2)$
\\
\hline
\end{tabular}
\caption{
The right-weights $\bar{h}$ and
one minus right-weights $(1-\bar{h})$
of the remaining ten two-particle operators. }
\label{hbar2-1}
\end{table}

As described before, the
dimension $(1-\bar{h})$ on the LHS is given by
(\ref{oneminushbar}) together with Tables \ref{hbar1},
\ref{hbar2}, and \ref{hbar2-1}.
Each dimension appears in the two superscripts in the
mode-dependent function (\ref{comm1}).
Due to the fact that
\bea
1-\frac{1}{2}(k-s_2+l-s_3) \neq
1-\frac{1}{2}(k-s_2) +1 -\frac{1}{2}(l-s_3),
\label{inequality}
\eea
the sum of $3$ (not $4$) and the
dimension $(1-\bar{h})$ on the RHS
is equal to
the dimension $(1-\bar{h})$ on the LHS.
Note that the left-hand side of
(\ref{inequality}) appears in (\ref{oneminushbar}). 
Originally, the OPE (\ref{HH}) contains
the dimension $\bar{h}$ of $-2$ due to the factor
$\bar{z}_{13}^2$. By taking the negative dimension for this factor
which leads to $\frac{1}{\bar{z}_{13}^2}$ and adding
dimension $1$, this factor becomes $\frac{1}{\bar{z}_{13}^3}$
\footnote{For the linear operators appearing on the LHS
with second argument, the sum of $2$ and the
dimension $(1-\bar{h})$ on the RHS is equal to
the dimension $(1-\bar{h})$ on the LHS.
The corresponding OPE has the dimension $\bar{h}$ of $-1$
from the factor $\bar{z}_{12}$. By following the above procedure
in this case, this factor leads to
the highest singular term $\frac{1}{\bar{z}_{12}^2}$.}. 

We calculate the OPEs of
the nine single operators in Table \ref{hbar1}
with twenty-five composite operators in Table \ref{hbar2} and Table
\ref{hbar2-1}
by taking the linear terms on the right-hand sides.
At the first step,
we can compute the OPEs of
the single operators with the
first operators in the composite operators.
The right-hand sides contain the infinite number of descendants
with respect to the antiholomorphic derivative.
At the next step, we can compute the
OPEs of these descendants with the second operators
in the composite operators by calculating the differentiation
of the OPE of the single operator with other single operator
with respect to the antiholomorphic coordinate
associated with the above descendant operators.
In this procedure, the quantity $\frac{1}{z_{12}}$
can be expressed in terms of $z_{13}$ and $z_{23}$
where $z_{12} \equiv z_{13}-z_{23}$ (as in the footnote \ref{geometric})
by using the geometric series.
Among the infinite terms, the single-particle contribution to the
multi-particle OPE appears at a particular term.
The singular term behaves as $\frac{1}{z_{13}^2}$
with the second to the infinite powers of $\bar{z}_{13}$
in the numerator of the right-hand sides.
The final ninety-five OPEs are summarized in Appendix A
\footnote{Among $165$ outputs
satisfying $s_1+s_2+s_3-s_4=4$ where $s_1,s_2,s_3,s_4=
\pm 2, \pm \frac{3}{2}, \pm 1, \pm \frac{1}{2}$ and $0$
in Mathematica,
the independent $95$ outputs appear after removing
the $70$ redundancies in the different helicities $s_2$ and $s_3$.
In other words, we don't differentiate the composite
operator with helicities $s_2$ and $s_3$ from
the composite operator with helicities $s_3$ and $s_2$.
The latter can be obtained from the former by adding
holomorphic or antiholomorphic derivative terms.}.

The $SU(8)$ indices in Appendix A appear in the operators
and the coefficients on the right-hand sides.
The dependence on the
holomorphic and antiholomorphic
complex coordinates, generalized Euler beta functions,
the summation over the dummy variable $m$ and
antiholomorphic derivatives is common.
We follow the procedure done in the subsection \ref{2point1}
for each case by putting the $SU(8)$ indices properly.

\begin{itemize}
\item[]
i) First of all,
we need to express  all upper indices $h_1, h_2$,
and $h_3$ in the
${\cal N}=8$ supermultiplet
in terms of the conformal dimensions, $j,k$, and $l$
by using Table \ref{hbar1}.\\
ii)  In the mode-dependent function (\ref{Ndef})
which depends on $(\bar{n}, \bar{n}')$,
we should specify
$1-\bar{h}=1-\frac{1}{2}(j-s_1)$
and $1-\bar{h}=1-\frac{1}{2}(k-s_2+l-s_3)$
by using Tables \ref{hbar1},
\ref{hbar2}, and \ref{hbar2-1}. \\
iii) The $(1-\bar{h})$ on the right-hand sides is implicit.
They can be read off from Table
\ref{hbar1} on the nine kinds of particles.
Or according to (\ref{oneminushbar1}),
the ($j,k$, and $l$ independent) numerical value
is given by $1+\frac{s_4}{2}$.
The helicity $s_4$ is fixed by the constraint (\ref{s1234}).\\
iv) From ii),
the ($j,k$, and $l$ independent) sum of numerical values
is
$2+\frac{1}{2} (s_1+s_2+s_3)$
which becomes $4+\frac{s_4}{2}$ from (\ref{s1234}).
Compared to iii),
the $(1-\bar{h})$ of both sides is conserved
after adding $3$ to the right-hand sides.
This conservation should hold for the whole ninety-five (anti)commutators. \\
v) The modes on the left-hand sides
appear as $(2-h,\bar{n})$ and $(n',\bar{n}')$
while they on the right-hand sides appear as
$((2-h)+n',\bar{n}+\bar{n}')$. \\
vi) The product of couplings $\kappa_{s_1,s_2,s_I}\, \kappa_{s_I,s_3,s_J}$
appears on the right-hand sides.
\end{itemize}

The twenty-five (anti)commutators, by applying
(\ref{comm1}) and (\ref{redefined}) to the case of
${\cal N}=8$ supergravity, are given by
\bea
&& \bigg[
(\hat{\Phi}_{+2}^{(-\frac{j}{2})})_{2-h,\bar{n}},
(\hat{\Phi}_{+2}^{(-\frac{k}{2})}\,
\hat{\Phi}_{+2}^{(-\frac{l}{2})}
)_{n',\bar{n}'} \bigg]  =
\pm  \frac{\kappa_{+2,+2,-2}^2}
{ (1-k)! (1-l)! (-3+k+l)!}  \nonu \\ && \times
N^{2-\frac{j}{2},\frac{1}{2} (-k-l+6)}_{1} (\bar{n},\bar{n}') \,
(\hat{\Phi}_{+2}^{(-\frac{j}{2}-\frac{k}{2}
-\frac{l}{2})})_{(2-h)+n',\bar{n}+\bar{n}'},
\nonu \\
&& \bigg[
(\hat{\Phi}_{+2}^{(-\frac{j}{2})})_{2-h,\bar{n}},
(\hat{\Phi}_{+2}^{(-\frac{k}{2})}\,
\hat{\Phi}_{+\frac{3}{2}}^{(-\frac{l}{2}+\frac{1}{4}),A}
)_{n',\bar{n}'} \bigg]  =
\pm  \frac{\kappa_{+2,+2,-2}\,
\kappa_{+2,+\frac{3}{2},-\frac{3}{2}}}{ (1-k)! (\frac{1}{2}-l)! (k+l-\frac{5}{2})!}  \nonu \\ && \times
N^{2-\frac{j}{2},\frac{1}{2} (-k-l+\frac{11}{2})}_{1} (\bar{n},\bar{n}') \,
(\hat{\Phi}_{+\frac{3}{2}}^{(-\frac{j}{2}-\frac{k}{2}
-\frac{l}{2}+\frac{1}{4}),A})_{(2-h)+n',\bar{n}+\bar{n}'},
\nonu \\
&& \bigg[
(\hat{\Phi}_{+2}^{(-\frac{j}{2})})_{2-h,\bar{n}},
(\hat{\Phi}_{+2}^{(-\frac{k}{2})}\,
\hat{\Phi}_{+1}^{(-\frac{l}{2}+\frac{1}{2}),AB}
)_{n',\bar{n}'} \bigg]  =
\pm  \frac{
\kappa_{+2,+2,-2}\,
\kappa_{+2,+1,-1}}{ (1-k)! (-l)! (k+l-2)!}  \nonu \\&& \times
N^{2-\frac{j}{2},\frac{1}{2} (-k-l+5)}_{1} (\bar{n},\bar{n}') \,
(\hat{\Phi}_{+1}^{(-\frac{j}{2}-\frac{k}{2}
-\frac{l}{2}+\frac{1}{2}),AB})_{(2-h)+n',\bar{n}+\bar{n}'},
\nonu \\
&& \bigg[
(\hat{\Phi}_{+2}^{(-\frac{j}{2})})_{2-h,\bar{n}},
(\hat{\Phi}_{+2}^{(-\frac{k}{2})}\,
\hat{\Phi}_{+\frac{1}{2}}^{(-\frac{l}{2}+\frac{3}{4}),ABC}
)_{n',\bar{n}'} \bigg]  =
\pm  \frac{
\kappa_{+2,+2,-2}\,
\kappa_{+2,+\frac{1}{2},-\frac{1}{2}}}{ (1-k)! (-\frac{1}{2}-l)! (k+l-\frac{3}{2})!}  \nonu \\ && \times
N^{2-\frac{j}{2},\frac{1}{2} (-k-l+\frac{9}{2})}_{1} (\bar{n},\bar{n}') \,
(\hat{\Phi}_{+\frac{1}{2}}^{(-\frac{j}{2}-\frac{k}{2}
-\frac{l}{2}+\frac{3}{4}),ABC})_{(2-h)+n',\bar{n}+\bar{n}'},
\nonu \\
&& \bigg[
(\hat{\Phi}_{+2}^{(-\frac{j}{2})})_{2-h,\bar{n}},
(\hat{\Phi}_{+2}^{(-\frac{k}{2})}\,
\hat{\Phi}_{0}^{(-\frac{l}{2}+1),ABCD}
)_{n',\bar{n}'} \bigg]  =
\pm  \frac{
\kappa_{+2,+2,-2}\,
\kappa_{+2,0,0}}{ (1-k)! (-1-l)! (k+l-1)!}  \nonu \\ && \times
N^{2-\frac{j}{2},\frac{1}{2} (-k-l+4)}_{1} (\bar{n},\bar{n}') \,
(\hat{\Phi}_{0}^{(-\frac{j}{2}-\frac{k}{2}
-\frac{l}{2}+1),ABCD})_{(2-h)+n',\bar{n}+\bar{n}'},
\nonu \\
&& \bigg[
(\hat{\Phi}_{+2}^{(-\frac{j}{2})})_{2-h,\bar{n}},
(\hat{\Phi}_{+2}^{(-\frac{k}{2})}\,
\hat{\Phi}_{ABC,-\frac{1}{2}}^{(-\frac{l}{2}+\frac{5}{4})}
)_{n',\bar{n}'} \bigg]  =
\pm  \frac{
\kappa_{+2,+2,-2}\,
\kappa_{+2,-\frac{1}{2},+\frac{1}{2}}}{ (1-k)! (-\frac{3}{2}-l)! (k+l-\frac{1}{2})!}  \nonu \\ && \times
N^{2-\frac{j}{2},\frac{1}{2} (-k-l+\frac{7}{2})}_{1} (\bar{n},\bar{n}') \,
(\hat{\Phi}_{ABC,-\frac{1}{2}}^{(-\frac{j}{2}-\frac{k}{2}
-\frac{l}{2}+\frac{5}{4})})_{(2-h)+n',\bar{n}+\bar{n}'},
\nonu \\
&& \bigg[
(\hat{\Phi}_{+2}^{(-\frac{j}{2})})_{2-h,\bar{n}},
(\hat{\Phi}_{+2}^{(-\frac{k}{2})}\,
\hat{\Phi}_{AB,-1}^{(-\frac{l}{2}+\frac{3}{2})}
)_{n',\bar{n}'} \bigg]  =
\pm  \frac{
\kappa_{+2,+2,-2}\,
\kappa_{+2,-1,+1}}{ (1-k)! (-2-l)! (k+l)!}  \nonu \\ && \times
N^{2-\frac{j}{2},\frac{1}{2} (-k-l+3)}_{1} (\bar{n},\bar{n}') \,
(\hat{\Phi}_{AB,-1}^{(-\frac{j}{2}-\frac{k}{2}
-\frac{l}{2}+\frac{3}{2})})_{(2-h)+n',\bar{n}+\bar{n}'},
\nonu \\
&& \bigg[
(\hat{\Phi}_{+2}^{(-\frac{j}{2})})_{2-h,\bar{n}},
(\hat{\Phi}_{+2}^{(-\frac{k}{2})}\,
\hat{\Phi}_{A,-\frac{3}{2}}^{(-\frac{l}{2}+\frac{7}{4})}
)_{n',\bar{n}'} \bigg]  =
\pm  \frac{\kappa_{+2,+2,-2}\,
\kappa_{+2,-\frac{3}{2},+\frac{3}{2}}
}{ (1-k)! (-\frac{5}{2}-l)! (\frac{1}{2}+k+l)!}  \nonu \\ && \times
N^{2-\frac{j}{2},\frac{1}{2} (-k-l+\frac{5}{2})}_{1} (\bar{n},\bar{n}') \,
(\hat{\Phi}_{A,-\frac{3}{2}}^{(-\frac{j}{2}-\frac{k}{2}
-\frac{l}{2}+\frac{7}{4})})_{(2-h)+n',\bar{n}+\bar{n}'},
\nonu \\
&& \bigg[
(\hat{\Phi}_{+2}^{(-\frac{j}{2})})_{2-h,\bar{n}},
(\hat{\Phi}_{+2}^{(-\frac{k}{2})}\,
\hat{\Phi}_{-2}^{(-\frac{l}{2}+2)}
)_{n',\bar{n}'} \bigg]  =
\pm  \frac{
\kappa_{+2,+2,-2}\,
\kappa_{+2,-2,+2}}{ (1-k)! (-3-l)! (1+k+l)!}  \nonu \\ && \times
N^{2-\frac{j}{2},\frac{1}{2} (-k-l+2)}_{1} (\bar{n},\bar{n}') \,
(\hat{\Phi}_{-2}^{(-\frac{j}{2}-\frac{k}{2}
-\frac{l}{2}+2)})_{(2-h)+n',\bar{n}+\bar{n}'},
\nonu \\
&& \bigg[
(\hat{\Phi}_{+2}^{(-\frac{j}{2})})_{2-h,\bar{n}},
(\hat{\Phi}_{+\frac{3}{2}}^{(-\frac{k}{2}+\frac{1}{4}),A}\,
\hat{\Phi}_{+\frac{3}{2}}^{(-\frac{l}{2}+\frac{1}{4}),B}
)_{n',\bar{n}'} \bigg]  =
\pm  \frac{
\kappa_{+2,+\frac{3}{2},-\frac{3}{2}}\,
\kappa_{+\frac{3}{2},+\frac{3}{2},-1}}{ (\frac{1}{2}-k)! (\frac{1}{2}-l)! (-2+k+l)!}  \nonu \\ && \times
N^{2-\frac{j}{2},\frac{1}{2} (-k-l+5)}_{1} (\bar{n},\bar{n}') \,
(\hat{\Phi}_{+1}^{(-\frac{j}{2}-\frac{k}{2}
-\frac{l}{2}+\frac{1}{2}),AB})_{(2-h)+n',\bar{n}+\bar{n}'},
\nonu \\
&& \bigg[
(\hat{\Phi}_{+2}^{(-\frac{j}{2})})_{2-h,\bar{n}},
(\hat{\Phi}_{+\frac{3}{2}}^{(-\frac{k}{2}+\frac{1}{4}),A}\,
\hat{\Phi}_{+1}^{(-\frac{l}{2}+\frac{1}{2}),BC}
)_{n',\bar{n}'} \bigg]  =
\pm  \frac{
\kappa_{+2,+\frac{3}{2},-\frac{3}{2}}\,
\kappa_{+\frac{3}{2},+1,-\frac{1}{2}}}{ (\frac{1}{2}-k)! (-l)! (-\frac{3}{2}+k+l)!}
\nonu \\ && \times
N^{2-\frac{j}{2},\frac{1}{2} (-k-l+\frac{9}{2})}_{1} (\bar{n},\bar{n}') \,
(\hat{\Phi}_{+\frac{1}{2}}^{(-\frac{j}{2}-\frac{k}{2}
-\frac{l}{2}+\frac{3}{4}),ABC})_{(2-h)+n',\bar{n}+\bar{n}'},
\nonu \\
&& \bigg[
(\hat{\Phi}_{+2}^{(-\frac{j}{2})})_{2-h,\bar{n}},
(\hat{\Phi}_{+\frac{3}{2}}^{(-\frac{k}{2}+\frac{1}{4}),A}\,
\hat{\Phi}_{+\frac{1}{2}}^{(-\frac{l}{2}+\frac{3}{4}),BCD}
)_{n',\bar{n}'} \bigg]  =
\pm  \frac{
\kappa_{+2,+\frac{3}{2},-\frac{3}{2}}\,
\kappa_{+\frac{3}{2},+\frac{1}{2},0}}{ (\frac{
1}{2}-k)! (-\frac{1}{2}-l)! (-1+k+l)!}
\nonu \\ && \times
N^{2-\frac{j}{2},\frac{1}{2} (-k-l+4)}_{1} (\bar{n},\bar{n}') \,
(\hat{\Phi}_{0}^{(-\frac{j}{2}-\frac{k}{2}
-\frac{l}{2}+1),ABCD})_{(2-h)+n',\bar{n}+\bar{n}'},
\nonu \\
&& \bigg[
(\hat{\Phi}_{+2}^{(-\frac{j}{2})})_{2-h,\bar{n}},
(\hat{\Phi}_{+\frac{3}{2}}^{(-\frac{k}{2}+\frac{1}{4}),A}\,
\hat{\Phi}_{0}^{(-\frac{l}{2}+1),BCDE}
)_{n',\bar{n}'} \bigg]  =
\pm  \frac{
\kappa_{+2,+\frac{3}{2},-\frac{3}{2}}\,
\kappa_{+\frac{3}{2},0,\frac{1}{2}}}{ (\frac{1}{2}-k)! (-1-l)! (-\frac{1}{2}+k+l)!}
\nonu \\ && \times
N^{2-\frac{j}{2},\frac{1}{2} (-k-l+\frac{7}{2})}_{1} (\bar{n},\bar{n}') \,
\frac{1}{3!} \, \epsilon^{ABCDEFGH}\, 
(\hat{\Phi}_{FGH,-\frac{1}{2}}^{(-\frac{j}{2}-\frac{k}{2}
-\frac{l}{2}+\frac{5}{4})})_{(2-h)+n',\bar{n}+\bar{n}'},
\nonu \\
&& \bigg[
(\hat{\Phi}_{+2}^{(-\frac{j}{2})})_{2-h,\bar{n}},
(\hat{\Phi}_{+\frac{3}{2}}^{(-\frac{k}{2}+\frac{1}{4}),A}\,
\hat{\Phi}_{BCD,-\frac{1}{2}}^{(-\frac{l}{2}+\frac{5}{4})}
)_{n',\bar{n}'} \bigg]  =
\pm  \frac{
\kappa_{+2,+\frac{3}{2},-\frac{3}{2}}\,
\kappa_{+\frac{3}{2},-\frac{1}{2},+1}}{ (\frac{1}{2}-k)! (-\frac{3}{2}-l)! (k+l)!}
\nonu \\ && \times
N^{2-\frac{j}{2},\frac{1}{2} (-k-l+3)}_{1} (\bar{n},\bar{n}') \,
3\, \delta^{A}_{[D}
(\hat{\Phi}_{BC],-1}^{(-\frac{j}{2}-\frac{k}{2}
-\frac{l}{2}+\frac{3}{2})})_{(2-h)+n',\bar{n}+\bar{n}'},
\nonu \\
&& \bigg[
(\hat{\Phi}_{+2}^{(-\frac{j}{2})})_{2-h,\bar{n}},
(\hat{\Phi}_{+\frac{3}{2}}^{(-\frac{k}{2}+\frac{1}{4}),A}\,
\hat{\Phi}_{BC,-1}^{(-\frac{l}{2}+\frac{3}{2})}
)_{n',\bar{n}'} \bigg]  =
\pm  \frac{
\kappa_{+2,+\frac{3}{2},-\frac{3}{2}}\,
\kappa_{+\frac{3}{2},-1,+\frac{3}{2}}}{ (\frac{1}{2}-k)! (-2-l)! (\frac{1}{2}+k+l)!}
\nonu \\ && \times
N^{2-\frac{j}{2},\frac{1}{2} (-k-l+\frac{5}{2})}_{1} (\bar{n},\bar{n}') \,
2\, \delta^{A}_{[B}
(\hat{\Phi}_{C],-\frac{3}{2}}^{(-\frac{j}{2}-\frac{k}{2}
-\frac{l}{2}+\frac{7}{4})})_{(2-h)+n',\bar{n}+\bar{n}'},
\nonu \\
&& \bigg[
(\hat{\Phi}_{+2}^{(-\frac{j}{2})})_{2-h,\bar{n}},
(\hat{\Phi}_{+\frac{3}{2}}^{(-\frac{k}{2}+\frac{1}{4}),A}\,
\hat{\Phi}_{B,-\frac{3}{2}}^{(-\frac{l}{2}+\frac{7}{4})}
)_{n',\bar{n}'} \bigg]  =
\pm  \frac{
\kappa_{+2,+\frac{3}{2},-\frac{3}{2}}\,
\kappa_{+\frac{3}{2},-\frac{3}{2},+2}}{ (\frac{1}{2}-k)! (-\frac{5}{2}-l)! (1+k+l)!}
\nonu \\ && \times
N^{2-\frac{j}{2},\frac{1}{2} (-k-l+2)}_{1} (\bar{n},\bar{n}') \,
\delta^{A}_{B}
(\hat{\Phi}_{-2}^{(-\frac{j}{2}-\frac{k}{2}
-\frac{l}{2}+2)})_{(2-h)+n',\bar{n}+\bar{n}'},
\nonu \\
&& \bigg[
(\hat{\Phi}_{+2}^{(-\frac{j}{2})})_{2-h,\bar{n}},
(\hat{\Phi}_{+1}^{(-\frac{k}{2}+\frac{1}{2}),AB}\,
\hat{\Phi}_{+1}^{(-\frac{l}{2}+\frac{1}{2}),CD}
)_{n',\bar{n}'} \bigg]  =
\pm  \frac{
\kappa_{+2,+1,-1}\,
\kappa_{+1,+1,0}}{(-k)! (-l)! (-1+k+l)!}
\nonu \\ && \times
N^{2-\frac{j}{2},\frac{1}{2} (-k-l+4)}_{1} (\bar{n},\bar{n}') \,
(\hat{\Phi}_{0}^{(-\frac{j}{2}-\frac{k}{2}
-\frac{l}{2}+1),ABCD})_{(2-h)+n',\bar{n}+\bar{n}'},
\nonu \\
&& \bigg[
(\hat{\Phi}_{+2}^{(-\frac{j}{2})})_{2-h,\bar{n}},
(\hat{\Phi}_{+1}^{(-\frac{k}{2}+\frac{1}{2}),AB}\,
\hat{\Phi}_{+\frac{1}{2}}^{(-\frac{l}{2}+\frac{3}{4}),CDE}
)_{n',\bar{n}'} \bigg]  =
\pm  \frac{
\kappa_{+2,+1,-1}\,
\kappa_{+1,+\frac{1}{2},+\frac{1}{2}}}{(-k)! (-\frac{1}{2}-l)! (-\frac{1}{2}+k+l)!}
\nonu \\ && \times
N^{2-\frac{j}{2},\frac{1}{2} (-k-l+\frac{7}{2})}_{1} (\bar{n},\bar{n}') \,
\frac{1}{3!}\,
\epsilon^{ABCDEFGH} \,
(\hat{\Phi}_{FGH,-\frac{1}{2}}^{(-\frac{j}{2}-\frac{k}{2}
-\frac{l}{2}+\frac{5}{4})})_{(2-h)+n',\bar{n}+\bar{n}'},
\nonu \\
&& \bigg[
(\hat{\Phi}_{+2}^{(-\frac{j}{2})})_{2-h,\bar{n}},
(\hat{\Phi}_{+1}^{(-\frac{k}{2}+\frac{1}{2}),AB}\,
\hat{\Phi}_{0}^{(-\frac{l}{2}+1),CDEF}
)_{n',\bar{n}'} \bigg]  =
\pm  \frac{
\kappa_{+2,+1,-1}\,
\kappa_{+1,0,+1}}{(-k)! (-1-l)! (k+l)!}
\nonu \\ && \times
N^{2-\frac{j}{2},\frac{1}{2} (-k-l+3)}_{1} (\bar{n},\bar{n}') \,
\frac{1}{2!}\,
\epsilon^{ABCDEFGH} \,
(\hat{\Phi}_{GH,-1}^{(-\frac{j}{2}-\frac{k}{2}
-\frac{l}{2}+\frac{3}{2})})_{(2-h)+n',\bar{n}+\bar{n}'},
\nonu \\
&& \bigg[
(\hat{\Phi}_{+2}^{(-\frac{j}{2})})_{2-h,\bar{n}},
(\hat{\Phi}_{+1}^{(-\frac{k}{2}+\frac{1}{2}),AB}\,
\hat{\Phi}_{CDE,-\frac{1}{2}}^{(-\frac{l}{2}+\frac{5}{4})}
)_{n',\bar{n}'} \bigg]  =
\pm  \frac{
\kappa_{+2,+1,-1}\,
\kappa_{+1,-\frac{1}{2},+\frac{3}{2}}}{(-k)! (-\frac{3}{2}-l)! (\frac{1}{2}+k+l)!}
\nonu \\ && \times
N^{2-\frac{j}{2},\frac{1}{2} (-k-l+\frac{5}{2})}_{1} (\bar{n},\bar{n}') \,
3!\, \delta^{A}_{[E} \,
(\hat{\Phi}_{C,-\frac{3}{2}}^{(-\frac{j}{2}-\frac{k}{2}
-\frac{l}{2}+\frac{7}{4})})_{(2-h)+n',\bar{n}+\bar{n}'} \, \delta^{B}_{D]},
\nonu \\
&& \bigg[
(\hat{\Phi}_{+2}^{(-\frac{j}{2})})_{2-h,\bar{n}},
(\hat{\Phi}_{+1}^{(-\frac{k}{2}+\frac{1}{2}),AB}\,
\hat{\Phi}_{CD,-1}^{(-\frac{l}{2}+\frac{3}{2})}
)_{n',\bar{n}'} \bigg]  =
\pm  \frac{
\kappa_{+2,+1,-1}\,
\kappa_{+1,-1,+2}}{(-k)! (-2-l)! (1+k+l)!}
\nonu \\ && \times
N^{2-\frac{j}{2},\frac{1}{2} (-k-l+2)}_{1} (\bar{n},\bar{n}') \,
\delta^{AB}_{CD} \,
(\hat{\Phi}_{-2}^{(-\frac{j}{2}-\frac{k}{2}
-\frac{l}{2}+2)})_{(2-h)+n',\bar{n}+\bar{n}'},
\nonu \\
&& \bigg[
(\hat{\Phi}_{+2}^{(-\frac{j}{2})})_{2-h,\bar{n}},
(\hat{\Phi}_{+\frac{1}{2}}^{(-\frac{k}{2}+\frac{3}{4}),ABC}\,
\hat{\Phi}_{+\frac{1}{2}}^{(-\frac{l}{2}+\frac{3}{4}),DEF}
)_{n',\bar{n}'} \bigg]  =
\pm  \frac{
\kappa_{+2,+\frac{1}{2},-\frac{1}{2}}\,
\kappa_{+\frac{1}{2},+\frac{1}{2},+1}}{(-\frac{1}{2}-k)! (-\frac{1}{2}-l)! (k+l)!}
\nonu \\ && \times
N^{2-\frac{j}{2},\frac{1}{2} (-k-l+3)}_{1} (\bar{n},\bar{n}') \,
\frac{1}{2!}\,
\epsilon^{ABCDEFGH} \,
(\hat{\Phi}_{GH,-1}^{(-\frac{j}{2}-\frac{k}{2}
-\frac{l}{2}+\frac{3}{2})})_{(2-h)+n',\bar{n}+\bar{n}'},
\nonu  \\
&& \bigg[
(\hat{\Phi}_{+2}^{(-\frac{j}{2})})_{2-h,\bar{n}},
(\hat{\Phi}_{+\frac{1}{2}}^{(-\frac{k}{2}+\frac{3}{4}),ABC}\,
\hat{\Phi}_{0}^{(-\frac{l}{2}+1),DEFG}
)_{n',\bar{n}'} \bigg]  =
\pm  \frac{
\kappa_{+2,+\frac{1}{2},-\frac{1}{2}}\,
\kappa_{+\frac{1}{2},0,+\frac{3}{2}}}{(-\frac{1}{2}-k)! (-1-l)! (\frac{1}{2}+k+l)!}
\nonu \\ && \times
N^{2-\frac{j}{2},\frac{1}{2} (-k-l+\frac{5}{2})}_{1} (\bar{n},\bar{n}') \,
\epsilon^{ABCDEFGH} \,
(\hat{\Phi}_{H,-\frac{3}{2}}^{(-\frac{j}{2}-\frac{k}{2}
-\frac{l}{2}+\frac{7}{4})})_{(2-h)+n',\bar{n}+\bar{n}'},
\nonu    \\
&& \bigg[
(\hat{\Phi}_{+2}^{(-\frac{j}{2})})_{2-h,\bar{n}},
(\hat{\Phi}_{+\frac{1}{2}}^{(-\frac{k}{2}+\frac{3}{4}),ABC}\,
\hat{\Phi}_{DEF,-\frac{1}{2}}^{(-\frac{l}{2}+\frac{5}{4})}
)_{n',\bar{n}'} \bigg]  =
\pm  \frac{
\kappa_{+2,+\frac{1}{2},-\frac{1}{2}}\,
\kappa_{+\frac{1}{2},-\frac{1}{2},+2}}{(-\frac{1}{2}-k)! (-\frac{3}{2}-l)! (1+k+l)!}
\nonu \\ && \times
N^{2-\frac{j}{2},\frac{1}{2} (-k-l+2)}_{1} (\bar{n},\bar{n}') \,
\frac{1}{5!}\,
\epsilon^{ABCGHIJK} \,
\epsilon_{DEFGHIJK}\,
(\hat{\Phi}_{-2}^{(-\frac{j}{2}-\frac{k}{2}
-\frac{l}{2}+2)})_{(2-h)+n',\bar{n}+\bar{n}'},
\nonu    \\
&& \bigg[
(\hat{\Phi}_{+2}^{(-\frac{j}{2})})_{2-h,\bar{n}},
(\hat{\Phi}_{0}^{(-\frac{k}{2}+1),ABCD}\,
\hat{\Phi}_{0}^{(-\frac{l}{2}+1),EFGH}
)_{n',\bar{n}'} \bigg]  =
\pm  \frac{
\kappa_{+2,0,0}\,
\kappa_{0,0,+2}}{(-1-k)! (-1-l)! (1+k+l)!}
\nonu \\ && \times
N^{2-\frac{j}{2},\frac{1}{2} (-k-l+2)}_{1} (\bar{n},\bar{n}') \,
\epsilon^{ABCDEFGH} \,
(\hat{\Phi}_{-2}^{(-\frac{j}{2}-\frac{k}{2}
-\frac{l}{2}+2)})_{(2-h)+n',\bar{n}+\bar{n}'}.
\label{25comm}
\eea
Note that the $(k,l)$ dependent factors
appearing on the right-hand sides of (\ref{25comm})
originate from the first line of (\ref{COMM2})
for general helicities $s_2$ and $s_3$.
These overall factors can be absorbed in the quadratic operators
of the left-hand sides, similar to (\ref{comm1}).
The overall $\pm$ signs come from the quantity appearing in
(\ref{sincsc}) \footnote{The $SU(8)$ tensorial structure
of the right-hand sides for the first nine commutators follow
from the ones of the third particles on the left-hand sides.
The contractions of the particle $1$ and particle $2$
provide the singlets ${\bf 1}$ and further contractions of
these and the particle $3$ lead to the $SU(8)$ representations
of the particle $3$. For the remaining
$16$ commutators, because the contractions between
the particle $1$ and the particle $2$ allow us to give
the $SU(8)$ representations of the particle $2$,
now further contractions between these and the particle $3$ provide
the tensor products between the
representations of the particle $2$ and
the ones of the particle $3$. Then the $SU(8)$ structure
of the right-hand sides of (\ref{25comm})
follow from the ones of last sixteen (anti)commutators of $(4.1)$ of
\cite{AK2509}.

For the single contractions for the OPEs of the single-particle
operators with the two-particle operators,
where the two-particle exchanges appear, the $SU(8)$ tensorial
structure of the right-hand sides is given by the tensor product
between the $SU(8)$ representations of the particle $2$
and the ones of the particle $3$.
For example, for $s_1=+2$, $s_2= +\frac{3}{2}$ and $s_3=+1$
(the eleventh relation of (\ref{25comm})), the quadratic terms
on the right-hand side have $SU(8)$ structure,
${\bf 8}\otimes {\bf 28}$.
In general,
the $SU(8)$ tensorial
structure of the right-hand sides is given by the sum of
i) the tensor product
between the $SU(8)$ representations coming from
the tensor product between the particle $1$ and the particle $2$
and the ones of the particle $3$ and ii)
the tensor product
between the $SU(8)$ representations
of the particle $2$ and the ones
coming from
the tensor product between the particle $1$ and the particle $3$.
For (\ref{25comm}), the particle $1$
is the singlets ${\bf 1}$ of $SU(8)$
and this general feature reduces to the previous simple ones above.
See Appendix A.}.

The modes of redefined composite operators
appearing in (\ref{25comm})
from (\ref{redefined}) are
\bea
(\hat{\Phi}_{+2}^{(-\frac{k}{2})}\,
\hat{\Phi}_{+2}^{(-\frac{l}{2})}
)_{n',\bar{n}'} &\equiv&
\frac{(\frac{1}{2} (- k- l-2 \bar{n}'+4))!
}{(\frac{1}{2} ( k+ l-2 \bar{n}'-6))!} \,
(\Phi_{+2}^{(-\frac{k}{2})}\,
\Phi_{+2}^{(-\frac{l}{2})}
)_{n',\bar{n}'},
\nonu \\
(\hat{\Phi}_{+2}^{(-\frac{k}{2})}\,
\hat{\Phi}_{+\frac{3}{2}}^{(-\frac{l}{2}+\frac{1}{4}),A}
)_{n',\bar{n}'} &\equiv&
\frac{(\frac{1}{4} (-2 k-2 l-4 \bar{n}'+7))!
}{(\frac{1}{4} (2 k+2 l-4 \bar{n}'-11))!} \,
(\Phi_{+2}^{(-\frac{k}{2})}\,
\Phi_{+\frac{3}{2}}^{(-\frac{l}{2}+\frac{1}{4}),A}
)_{n',\bar{n}'},
\nonu \\
(\hat{\Phi}_{+2}^{(-\frac{k}{2})}\,
\hat{\Phi}_{+1}^{(-\frac{l}{2}+\frac{1}{2}),AB}
)_{n',\bar{n}'} &\equiv& \frac{(\frac{1}{2} (-k-l-2 \bar{n}'+3))!}
{(\frac{1}{2} (k+l-2 \bar{n}'-5))!} \,
(\Phi_{+2}^{(-\frac{k}{2})}\,
\Phi_{+1}^{(-\frac{l}{2}+\frac{1}{2}),AB}
)_{n',\bar{n}'},
\nonu \\
(\hat{\Phi}_{+2}^{(-\frac{k}{2})}\,
\hat{\Phi}_{+\frac{1}{2}}^{(-\frac{l}{2}+\frac{3}{4}),ABC}
)_{n',\bar{n}'} & \equiv & \frac{(\frac{1}{4} (-2 k-2 l-4 \bar{n}'+5))!}
{(\frac{1}{4} (2 k+2 l-4 \bar{n}'-9))!}
\,(\Phi_{+2}^{(-\frac{k}{2})}\,
\Phi_{+\frac{1}{2}}^{(-\frac{l}{2}+\frac{3}{4}),ABC}
)_{n',\bar{n}'},
\nonu \\
(\hat{\Phi}_{+2}^{(-\frac{k}{2})}\,
\hat{\Phi}_{0}^{(-\frac{l}{2}+1),ABCD}
)_{n',\bar{n}'}  & \equiv &  \frac{(\frac{1}{2} (-k-l-2 \bar{n}'+2))!}{(\frac{1}{2} (k+l-2 (\bar{n}'+2)))!} \,
(\Phi_{+2}^{(-\frac{k}{2})}\,
\Phi_{0}^{(-\frac{l}{2}+1),ABCD}
)_{n',\bar{n}'}, \nonu \\
(\hat{\Phi}_{+2}^{(-\frac{k}{2})}\,
\hat{\Phi}_{ABC,-\frac{1}{2}}^{(-\frac{l}{2}+\frac{5}{4})}
)_{n',\bar{n}'}  & \equiv & \frac{(\frac{1}{4} (-2 k-2 l-4 \bar{n}'+3))!}
{(\frac{1}{4} (2 k+2 l-4 \bar{n}'-7))!} \, (\Phi_{+2}^{(-\frac{k}{2})}\,
\Phi_{ABC,-\frac{1}{2}}^{(-\frac{l}{2}+\frac{5}{4})}
)_{n',\bar{n}'},
\nonu \\
(\hat{\Phi}_{+2}^{(-\frac{k}{2})}\,
\hat{\Phi}_{AB,-1}^{(-\frac{l}{2}+\frac{3}{2})}
)_{n',\bar{n}'}  & \equiv & \frac{(\frac{1}{2} (-k-l-2 \bar{n}'+1))!}
{(\frac{1}{2} (k+l-2 \bar{n}'-3))!} \, (\Phi_{+2}^{(-\frac{k}{2})}\,
\Phi_{AB,-1}^{(-\frac{l}{2}+\frac{3}{2})}
)_{n',\bar{n}'},
\nonu \\
(\hat{\Phi}_{+2}^{(-\frac{k}{2})}\,
\hat{\Phi}_{A,-\frac{3}{2}}^{(-\frac{l}{2}+\frac{7}{4})}
)_{n',\bar{n}'}  & \equiv & \frac{(\frac{1}{4} (-2 k-2 l-4 \bar{n}'+1))!}
{(\frac{1}{4} (2 k+2 l-4 \bar{n}'-5))!} \, (\Phi_{+2}^{(-\frac{k}{2})}\,
\Phi_{A,-\frac{3}{2}}^{(-\frac{l}{2}+\frac{7}{4})}
)_{n',\bar{n}'},
\nonu \\
(\hat{\Phi}_{+2}^{(-\frac{k}{2})}\,
\hat{\Phi}_{-2}^{(-\frac{l}{2}+2)}
)_{n',\bar{n}'}  & \equiv & \frac{(\frac{1}{2} (-k-l-2 \bar{n}'))!}{(\frac{1}{2} (k+l-2 (\bar{n}'+1)))!} \, (\Phi_{+2}^{(-\frac{k}{2})}\,
\Phi_{-2}^{(-\frac{l}{2}+2)}
)_{n',\bar{n}'},
\nonu \\
(\hat{\Phi}_{+\frac{3}{2}}^{(-\frac{k}{2}+\frac{1}{4}),A}\,
\hat{\Phi}_{+\frac{3}{2}}^{(-\frac{l}{2}+\frac{1}{4}),B}
)_{n',\bar{n}'}  & \equiv & \frac{(\frac{1}{2} (-k-l-2 \bar{n}'+3))!}{(\frac{1}{2} (k+l-2 \bar{n}'-5))!}
\, (\Phi_{+\frac{3}{2}}^{(-\frac{k}{2}+\frac{1}{4}),A}\,
\Phi_{+\frac{3}{2}}^{(-\frac{l}{2}+\frac{1}{4}),B}
)_{n',\bar{n}'},
\nonu \\
(\hat{\Phi}_{+\frac{3}{2}}^{(-\frac{k}{2}+\frac{1}{4}),A}\,
\hat{\Phi}_{+1}^{(-\frac{l}{2}+\frac{1}{2}),BC}
)_{n',\bar{n}'}  & \equiv & \frac{(\frac{1}{4} (-2 k-2 l-4 \bar{n}'+5))!}{(\frac{1}{4} (2 k+2 l-4 \bar{n}'-9))!} \, (\Phi_{+\frac{3}{2}}^{(-\frac{k}{2}+\frac{1}{4}),A}\,
\Phi_{+1}^{(-\frac{l}{2}+\frac{1}{2}),BC}
)_{n',\bar{n}'},
\nonu \\
(\hat{\Phi}_{+\frac{3}{2}}^{(-\frac{k}{2}+\frac{1}{4}),A}\,
\hat{\Phi}_{+\frac{1}{2}}^{(-\frac{l}{2}+\frac{3}{4}),BCD}
)_{n',\bar{n}'}  & \equiv & \frac{(\frac{1}{2} (-k-l-2 \bar{n}'+2))!}{(\frac{1}{2} (k+l-2 (\bar{n}'+2)))!}
\, (\Phi_{+\frac{3}{2}}^{(-\frac{k}{2}+\frac{1}{4}),A}\,
\Phi_{+\frac{1}{2}}^{(-\frac{l}{2}+\frac{3}{4}),BCD}
)_{n',\bar{n}'},
\nonu \\
(\hat{\Phi}_{+\frac{3}{2}}^{(-\frac{k}{2}+\frac{1}{4}),A}\,
\hat{\Phi}_{0}^{(-\frac{l}{2}+1),BCDE}
)_{n',\bar{n}'}  & \equiv & \frac{(\frac{1}{4} (-2 k-2 l-4 \bar{n}'+3))!}{(\frac{1}{4} (2 k+2 l-4 \bar{n}'-7))!}
\, (\Phi_{+\frac{3}{2}}^{(-\frac{k}{2}+\frac{1}{4}),A}\,
\Phi_{0}^{(-\frac{l}{2}+1),BCDE}
)_{n',\bar{n}'},
\nonu \\
(\hat{\Phi}_{+\frac{3}{2}}^{(-\frac{k}{2}+\frac{1}{4}),A}\,
\hat{\Phi}_{BCD,-\frac{1}{2}}^{(-\frac{l}{2}+\frac{5}{4})}
)_{n',\bar{n}'}  & \equiv & \frac{(\frac{1}{2} (-k-l-2 \bar{n}'+1))!}{(\frac{1}{2} (k+l-2 \bar{n}'-3))!}
\, (\Phi_{+\frac{3}{2}}^{(-\frac{k}{2}+\frac{1}{4}),A}\,
\Phi_{BCD,-\frac{1}{2}}^{(-\frac{l}{2}+\frac{5}{4})}
)_{n',\bar{n}'},
\nonu \\
(\hat{\Phi}_{+\frac{3}{2}}^{(-\frac{k}{2}+\frac{1}{4}),A}\,
\hat{\Phi}_{BC,-1}^{(-\frac{l}{2}+\frac{3}{2})}
)_{n',\bar{n}'}  & \equiv & \frac{(\frac{1}{4} (-2 k-2 l-4 \bar{n}'+1))!}{(\frac{1}{4} (2 k+2 l-4 \bar{n}'-5))!}
\, (\Phi_{+\frac{3}{2}}^{(-\frac{k}{2}+\frac{1}{4}),A}\,
\Phi_{BC,-1}^{(-\frac{l}{2}+\frac{3}{2})}
)_{n',\bar{n}'},
\nonu \\
(\hat{\Phi}_{+\frac{3}{2}}^{(-\frac{k}{2}+\frac{1}{4}),A}\,
\hat{\Phi}_{B,-\frac{3}{2}}^{(-\frac{l}{2}+\frac{7}{4})}
)_{n',\bar{n}'}  & \equiv & \frac{(\frac{1}{2} (-k-l-2 \bar{n}'))!}{(\frac{1}{2} (k+l-2 (\bar{n}'+1)))!}
\, (\Phi_{+\frac{3}{2}}^{(-\frac{k}{2}+\frac{1}{4}),A}\,
\Phi_{B,-\frac{3}{2}}^{(-\frac{l}{2}+\frac{7}{4})}
)_{n',\bar{n}'},
\nonu \\
(\hat{\Phi}_{+1}^{(-\frac{k}{2}+\frac{1}{2}),AB}\,
\hat{\Phi}_{+1}^{(-\frac{l}{2}+\frac{1}{2}),CD}
)_{n',\bar{n}'}  & \equiv & \frac{(\frac{1}{2} (-k-l-2 \bar{n}'+2))!}{(\frac{1}{2} (k+l-2 (\bar{n}'+2)))!}
\, (\Phi_{+1}^{(-\frac{k}{2}+\frac{1}{2}),AB}\,
\Phi_{+1}^{(-\frac{l}{2}+\frac{1}{2}),CD}
)_{n',\bar{n}'},
\nonu \\
(\hat{\Phi}_{+1}^{(-\frac{k}{2}+\frac{1}{2}),AB}\,
\hat{\Phi}_{+\frac{1}{2}}^{(-\frac{l}{2}+\frac{3}{4}),CDE}
)_{n',\bar{n}'}  & \equiv & \frac{(\frac{1}{4} (-2 k-2 l-4 \bar{n}'+3))!}{(\frac{1}{4} (2 k+2 l-4 \bar{n}'-7))!}
\, (\Phi_{+1}^{(-\frac{k}{2}+\frac{1}{2}),AB}\,
\Phi_{+\frac{1}{2}}^{(-\frac{l}{2}+\frac{3}{4}),CDE}
)_{n',\bar{n}'},
\nonu \\
(\hat{\Phi}_{+1}^{(-\frac{k}{2}+\frac{1}{2}),AB}\,
\hat{\Phi}_{0}^{(-\frac{l}{2}+1),CDEF}
)_{n',\bar{n}'}  & \equiv & \frac{(\frac{1}{2} (-k-l-2 \bar{n}'+1))!}{(\frac{1}{2} (k+l-2 \bar{n}'-3))!}
\, (\Phi_{+1}^{(-\frac{k}{2}+\frac{1}{2}),AB}\,
\Phi_{0}^{(-\frac{l}{2}+1),CDEF}
)_{n',\bar{n}'},
\nonu \\
(\hat{\Phi}_{+1}^{(-\frac{k}{2}+\frac{1}{2}),AB}\,
\hat{\Phi}_{CDE,-\frac{1}{2}}^{(-\frac{l}{2}+\frac{5}{4})}
)_{n',\bar{n}'}  & \equiv & \frac{(\frac{1}{4} (-2 k-2 l-4 \bar{n}'+1))!}{(\frac{1}{4} (2 k+2 l-4 \bar{n}'-5))!}
\, (\Phi_{+1}^{(-\frac{k}{2}+\frac{1}{2}),AB}\,
\Phi_{CDE,-\frac{1}{2}}^{(-\frac{l}{2}+\frac{5}{4})}
)_{n',\bar{n}'},
\nonu \\
(\hat{\Phi}_{+1}^{(-\frac{k}{2}+\frac{1}{2}),AB}\,
\hat{\Phi}_{CD,-1}^{(-\frac{l}{2}+\frac{3}{2})}
)_{n',\bar{n}'}  & \equiv & \frac{(\frac{1}{2} (-k-l-2 \bar{n}'))!}{(\frac{1}{2} (k+l-2 (\bar{n}'+1)))!}
\, (\Phi_{+1}^{(-\frac{k}{2}+\frac{1}{2}),AB}\,
\Phi_{CD,-1}^{(-\frac{l}{2}+\frac{3}{2})}
)_{n',\bar{n}'},
\nonu \\
(\hat{\Phi}_{+\frac{1}{2}}^{(-\frac{k}{2}+\frac{3}{4}),ABC}\,
\hat{\Phi}_{+\frac{1}{2}}^{(-\frac{l}{2}+\frac{3}{4}),DEF}
)_{n',\bar{n}'}  & \equiv & \frac{(\frac{1}{2} (-k-l-2 \bar{n}'+1))!}
{(\frac{1}{2} (k+l-2 \bar{n}'-3))!}
\, (\Phi_{+\frac{1}{2}}^{(-\frac{k}{2}+\frac{3}{4}),ABC}\,
\Phi_{+\frac{1}{2}}^{(-\frac{l}{2}+\frac{3}{4}),DEF}
)_{n',\bar{n}'},
\nonu \\
(\hat{\Phi}_{+\frac{1}{2}}^{(-\frac{k}{2}+\frac{3}{4}),ABC}\,
\hat{\Phi}_{0}^{(-\frac{l}{2}+1),DEFG}
)_{n',\bar{n}'}  & \equiv & \frac{(\frac{1}{4} (-2 k-2 l-4 \bar{n}'+1))!}{(\frac{1}{4} (2 k+2 l-4 \bar{n}'-5))!}
\nonu \\
&\times & (\Phi_{+\frac{1}{2}}^{(-\frac{k}{2}+\frac{3}{4}),ABC}\,
\Phi_{0}^{(-\frac{l}{2}+1),DEFG}
)_{n',\bar{n}'},
\label{REDEFINITION}
\\
(\hat{\Phi}_{+\frac{1}{2}}^{(-\frac{k}{2}+\frac{3}{4}),ABC}\,
\hat{\Phi}_{DEF,-\frac{1}{2}}^{(-\frac{l}{2}+\frac{5}{4})}
)_{n',\bar{n}'}  & \equiv & \frac{(\frac{1}{2} (-k-l-2 \bar{n}'))!}
{(\frac{1}{2} (k+l-2 (\bar{n}'+1)))!}
\, (\Phi_{+\frac{1}{2}}^{(-\frac{k}{2}+\frac{3}{4}),ABC}\,
\Phi_{DEF,-\frac{1}{2}}^{(-\frac{l}{2}+\frac{5}{4})}
)_{n',\bar{n}'},
\nonu \\
(\hat{\Phi}_{0}^{(-\frac{k}{2}+1),ABCD}\,
\hat{\Phi}_{0}^{(-\frac{l}{2}+1),EFGH}
)_{n',\bar{n}'}  & \equiv & \frac{(\frac{1}{2} (-k-l-2 \bar{n}'))!}{(\frac{1}{2} (k+l-2 (\bar{n}'+1)))!}
\, (\Phi_{0}^{(-\frac{k}{2}+1),ABCD}\,
\Phi_{0}^{(-\frac{l}{2}+1),EFGH}
)_{n',\bar{n}'}.
\nonu
\eea
Note that the $(k,l,\bar{n}')$ independent numerical values
in the numerator and the denominator of the coefficients
on the right-hand sides of (\ref{REDEFINITION}) are
given by $\frac{1}{2}(s_2+s_3)$ and $-1-\frac{1}{2}(s_2+s_3)$, respectively.
We present the remaining $(95-25)=70$
(anti)commutators in Appendix B \footnote{The modes
of redefined single operators are written as
\bea
(\hat{\Phi}_{+2}^{(-\frac{j}{2})})_{2-h,\bar{n}} &\equiv&
\frac{(\frac{1}{2} (-j-2 \bar{n}+2))!
}{(\frac{1}{2} (j-2 \bar{n}-4))!} \,
(\Phi_{+2}^{(-\frac{j}{2})}
)_{2-h,\bar{n}}, \nonu \\
(\hat{\Phi}_{+\frac{3}{2}}^{(\frac{1}{4}-\frac{j}{2}),P})_{2-h,\bar{n}} &\equiv&
\frac{(\frac{1}{2} (-j-2 \bar{n}+\frac{3}{2}))!
}{(\frac{1}{2} (j-2 \bar{n}-\frac{7}{2}))!} \,
(\Phi_{+\frac{3}{2}}^{(\frac{1}{4}-\frac{j}{2}),P}
)_{2-h,\bar{n}},
\nonu \\
(\hat{\Phi}_{+1}^{(\frac{1}{2}-\frac{j}{2}),PQ})_{2-h,\bar{n}} &\equiv&
\frac{(\frac{1}{2} (-j-2 \bar{n}+1))!
}{(\frac{1}{2} (j-2 \bar{n}-3))!} \,
(\Phi_{+1}^{(\frac{1}{2}-\frac{j}{2}),PQ}
)_{2-h,\bar{n}},\nonu \\
(\hat{\Phi}_{+\frac{1}{2}}^{(\frac{3}{4}-\frac{j}{2}),PQR})_{2-h,\bar{n}} &\equiv&
\frac{(\frac{1}{2} (-j-2 \bar{n}+\frac{1}{2}))!
}{(\frac{1}{2} (j-2 \bar{n}-\frac{5}{2}))!} \,
(\Phi_{+\frac{1}{2}}^{(\frac{3}{4}-\frac{j}{2}),PQR}
)_{2-h,\bar{n}},
\nonu \\
(\hat{\Phi}_{0}^{(1-\frac{j}{2}),PQRS})_{2-h,\bar{n}} &\equiv&
\frac{(\frac{1}{2} (-j-2 \bar{n}))!
}{(\frac{1}{2} (j-2 \bar{n}-2))!} \,
(\Phi_{0}^{(1-\frac{j}{2}),PQRS}
)_{2-h,\bar{n}},
\nonu \\
(\hat{\Phi}_{PQR,-\frac{1}{2}}^{(\frac{5}{4}-\frac{j}{2})})_{2-h,\bar{n}} &\equiv&
\frac{(\frac{1}{2} (-j-2 \bar{n}-\frac{1}{2}))!
}{(\frac{1}{2} (j-2 \bar{n}-\frac{3}{2}))!} \,
(\Phi_{PQR,-\frac{1}{2}}^{(\frac{5}{4}-\frac{j}{2})}
)_{2-h,\bar{n}},
\nonu \\
(\hat{\Phi}_{PQ,-1}^{(\frac{3}{2}-\frac{j}{2})})_{2-h,\bar{n}} &\equiv&
\frac{(\frac{1}{2} (-j-2 \bar{n}-1))!
}{(\frac{1}{2} (j-2 \bar{n}-1))!} \,
(\Phi_{PQ,-1}^{(\frac{3}{2}-\frac{j}{2})}
)_{2-h,\bar{n}},
\nonu \\
(\hat{\Phi}_{P,-\frac{3}{2}}^{(\frac{7}{4}-\frac{j}{2})})_{2-h,\bar{n}} &\equiv&
\frac{(\frac{1}{2} (-j-2 \bar{n}-\frac{3}{2}))!
}{(\frac{1}{2} (j-2 \bar{n}-\frac{1}{2}))!} \,
(\Phi_{P,-\frac{3}{2}}^{(\frac{7}{2}-\frac{j}{2})}
)_{2-h,\bar{n}},
\nonu \\
(\hat{\Phi}_{-2}^{(2-\frac{j}{2})})_{2-h,\bar{n}} &\equiv&
\frac{(\frac{1}{2} (-j-2 \bar{n}-2))!
}{(\frac{1}{2} (j-2 \bar{n}))!} \,
(\Phi_{-2}^{(2-\frac{j}{2})}
)_{2-h,\bar{n}}.
\label{single}
\eea
For the single-particle operators appearing on the right-hand sides
of the (anti)commutators, the coefficients
can be obtained from those in (\ref{single})
by taking $j \rightarrow j+k+l$ and $s_1 \rightarrow s_4$
as explained in (\ref{redefined}) before.
Also note that the two-particle operators in (\ref{REDEFINITION})
come from the normal-ordered products appearing in the
OPEs corresponding to equation $(4.1)$ of
\cite{AK2509}.}.
Furthermore, we can write down the corresponding
twenty-five OPEs in the antiholomorphic coordinates,
similar to (\ref{HtHt}).
In a different basis of the OPEs, we also present
the fourth-order poles of the ninety-five OPEs
in the (anti)holomorphic coordinates 
in the basis of
\cite{AK2509} in Appendix C by using the Thielemans package
\cite{Thielemans}.

\subsection{The two-particle contributions to the multi-particle OPEs}

So far, we have considered the single-particle contributions.
There are also  the two-particle contributions
to the multi-particle OPEs.
We multiply the factor (\ref{threeDelta}) into both sides of
the OPEs in Appendix A by focusing on the quadratic terms
on the right-hand
sides. Then we have the similar expression, compared to
(\ref{inter2}). The overall factor $\frac{\bar{z}_{13}}{z_{13}}$
plays an important role. Inside of the summation over the dummy variable
$m$, the corresponding ratio of Gamma functions appears.
We perform the residues of the Gamma functions described in
(\ref{relation}). Moreover, we can write down
the OPE similar to (\ref{HH}) with a little modification
by introducing the similar soft operators as in (\ref{Hdef}).
After repeating the procedure from (\ref{LHS1}) to (\ref{HHmodes})
where $s_4$ can be written as $s_4 = s_2 +s_4'$ or $s_4 = s_3 + s_4'$,
we can compute the (anti)commutators.
The main difference is that
the holomorphic mode for particle one on the left-hand side 
should be equal to $(1-h)$, related to
the similar analysis in the footnote \ref{ncondition},
where the left-weight is given by
(\ref{Hmodes}). We have the explicit expression for the
integral over $\bar{z}_1$ similar to (\ref{iden}).
Moreover, we can apply the standard mode expansions
done in (\ref{Gexp2}) to the quadratic terms
$(\bar{\pa}^m \, H^{j+k} \, H^{l})(z_3, \bar{z}_3)$
or $(H^k \, \bar{\pa}^m \, H^{j+l})(z_3,\bar{z}_3)$.

How do we extract the $m$ dependence from these
operators when we compute the integral over $\bar{z}_3$?
From the analysis of page $40$ in \cite{CFT}, the normal-ordered
product of two operators where one of the operators
has the (anti)holomorphic derivatives can be expressed explicitly
\footnote{
By performing the antiholomorphic derivatives $m$ times,  
the following relations are satisfied:
\bea
(\bar{\pa}^m \, {\cal O}_1 \, {\cal O}_2)_{n,\bar{n}} &=&
\sum_{k \leq -h_1, \, \bar{l} \leq -\bar{h}_1-m} \,
\frac{(-\bar{l}-\bar{h}_1)!}{(-\bar{l}-\bar{h}_1-m)!}\,
({\cal O}_1)_{k,\bar{l}} \, ({\cal O}_2)_{n-k,\bar{n}-\bar{l}}
\nonu \\
& + & \sum_{k > -h_1, \, \bar{l} > -\bar{h}_1-m} \,
\frac{(-\bar{l}-\bar{h}_1)!}{(-\bar{l}-\bar{h}_1-m)!}\,
({\cal O}_2)_{n-k,\bar{n}-\bar{l}} \, ({\cal O}_1)_{k,\bar{l}}
\nonu \\
&
\equiv &
\sum_{k,\bar{l}} \,
\frac{(-\bar{l}-\bar{h}_1)!}{(-\bar{l}-\bar{h}_1-m)!}\,
(({\cal O}_1)_{k,\bar{l}} \, ({\cal O}_2)_{n-k,\bar{n}-\bar{l}}), 
\nonu \\
( {\cal O}_1 \, \bar{\pa}^m \,{\cal O}_2)_{n,\bar{n}} &=&
\sum_{k \leq -h_1, \, \bar{l} \leq -\bar{h}_1} \,
\frac{(-\bar{n}+\bar{l}-\bar{h}_2)!}{(-\bar{n}+\bar{l}-\bar{h}_2-m)!}\,
({\cal O}_1)_{k,\bar{l}} \, ({\cal O}_2)_{n-k,\bar{n}-\bar{l}}
\nonu \\
& + & \sum_{k > -h_1, \, \bar{l} > -\bar{h}_1} \,
\frac{(-\bar{n} +\bar{l}-\bar{h}_2)!}{(-\bar{n}+\bar{l}-\bar{h}_2-m)!}\,
({\cal O}_2)_{n-k,\bar{n}-\bar{l}} \, ({\cal O}_1)_{k,\bar{l}}
\nonu \\
& \equiv &
\sum_{k , \bar{l}} \,
\frac{(-\bar{n}+\bar{l}-\bar{h}_2)!}{(-\bar{n}+\bar{l}-\bar{h}_2-m)!}\,
(({\cal O}_1)_{k,\bar{l}} \, ({\cal O}_2)_{n-k,\bar{n}-\bar{l}}).
\label{quadnandnbar}
\eea
The $m$ dependence of the boundary on the dummy variable
$\bar{l}$ comes from the right-weight of $\bar{\pa}^m$.
Now we see that the $m$ dependence for the second case of
(\ref{quadnandnbar}) appears at the overall factor inside the
summation.
For the first case, the $m$ dependence
arises also in the boundary of the antiholomorphic
right-weight and therefore the $m$ dependence appears
at the antiholomorphic modes of the operators.
We also use the simplified notation in (\ref{quadnandnbar})
for the equation (\ref{NONLINEAR})}.

Then the intermediate result for the (anti)commutator is 
\bea
&& \bigg[ (H^{j})_{1-h,\bar{n}},
(H^{k,l})_{n',\bar{n}'} \bigg] =
\frac{(-1)^{1+\bar{n}+ \frac{1}{2}(j-s_1)}}{(s_2-1-k)!
(-\bar{n}- \frac{1}{2}(j-s_1))!} \,
\nonu \\ && \times
\sum_{m=-\bar{n}- \frac{1}{2}(j-s_1)-1}^{-1-(j-s_1)}\, \Bigg[
\frac{1}{m!}\, \,  \frac{ (s_1+s_2-2-j-k-m)!}{(-m-j+s_1-1)!}  \,
\frac{(m+1)! \,
(-1)^{m}}{(m+1+\bar{n}+ \frac{1}{2}(j-s_1))!}  \nonu \\
&& \times
\sum_{k_1, \bar{l}_1}\, \frac{(-\bar{l}_1-\frac{1}{2}(
j-s_1+k-s_2+2))!}{(-\bar{l}_1-\frac{1}{2}( j-s_1+k-s_2+2)-m)!} \,
\big((H^{j+k})_{k_1,\bar{l}_1} \, (H^l)_{(1-h)+n'-k_1,\bar{n}+\bar{n}'-\bar{l}_1}\big)
\Bigg]
\nonu \\
&&+ \frac{(-1)^{1+\bar{n}+ \frac{1}{2}(j-s_1)}}{(s_3-1-l)!
(-\bar{n}- \frac{1}{2}(j-s_1))!} \,
\nonu \\ && \times
\sum_{m=-\bar{n}- \frac{1}{2}(j-s_1)-1}^{-1-(j-s_1)}\, \Bigg[
\frac{1}{m!}\, \,  \frac{ (s_1+s_3-2-j-l-m)!}{(-m-j+s_1-1)!}  \,
\frac{(m+1)! \,
(-1)^{m}}{(m+1+\bar{n}+ \frac{1}{2}(j-s_1))!}
\label{NONLINEAR}
\\ && \times
\sum_{k_2, \bar{l}_2}\, \frac{(-\bar{n}-\bar{n}'+\bar{l}_2-\frac{1}{2}(
j-s_1+l-s_3+2))!}{(-\bar{n}-\bar{n}'+\bar{l}_2-\frac{1}{2}(
j-s_1+l-s_3+2)-m)!} \,
\big((H^{k})_{k_2,\bar{l}_2} \,
(H^{j+l})_{(1-h)+n'-k_2,\bar{n}+\bar{n}'-\bar{l}_2} \big) \Bigg].
\nonu
\eea
When we perform the summation over the dummy variable
$m$ on the second case,
the corresponding  expression in (\ref{NONLINEAR})
becomes the linear combination of two hypergeometric functions.
By using the similar property for the
hypergeometric function described in the footnote \ref{property3F2},
we can write down them in terms of $j$, $l$, $s_1$, $s_3$,
$\bar{n}$, $\bar{n}'$, $k_2$ and $\bar{l}_2$.
By considering the overall factors,
we are left with
\bea
&& \bigg[ (H^{j})_{1-h,\bar{n}},
(H^{k,l})_{n',\bar{n}'} \bigg] =
\frac{(-1)^{1+\bar{n}+ \frac{1}{2}(j-s_1)}}{(s_2-1-k)!
(-\bar{n}- \frac{1}{2}(j-s_1))!} \,
\nonu \\ && \times
\sum_{m=-\bar{n}- \frac{1}{2}(j-s_1)-1}^{-1-(j-s_1)}\, \Bigg[
\frac{1}{m!}\, \,  \frac{ (s_1+s_2-2-j-k-m)!}{(-m-j+s_1-1)!}  \,
\frac{(m+1)! \,
(-1)^{m}}{(m+1+\bar{n}+ \frac{1}{2}(j-s_1))!}  \nonu \\
&& \times
\sum_{k_1, \bar{l}_1}\, \frac{(-\bar{l}_1-\frac{1}{2}(
j-s_1+k-s_2+2))!}{(-\bar{l}_1-\frac{1}{2}( j-s_1+k-s_2+2)-m)!} \,
\big((H^{j+k})_{k_1,\bar{l}_1} \, (H^l)_{(1-h)+n'-k_1,\bar{n}+\bar{n}'-\bar{l}_1}\big)
\Bigg]
\nonu \\
&&+
\frac{\sin (\pi  (l-s_3)) \csc \left(\frac{1}{2}
\pi  (j+2 l-2 \bar{n}-s_1-2 s_3)\right)}{
\left(-\frac{j}{2}+\bar{n}+\frac{s_1}{2}\right)!
\left(\frac{1}{2} (-j-2 \bar{n}+s_1+2)-1\right)!}
\nonu \\
&&
\times \Bigg[
\sum_{k_2,\bar{l}_2}\,
\frac{\left(\frac{l}{2}+\bar{l}_2-\bar{n}'-
\frac{s_3}{2}-1\right)!
\left(\frac{1}{2} (-j-l+2 \bar{l}_2-2 \bar{n}'-
2 \bar{n}+s_1+s_3)-1\right)!}{
\left(-\frac{l}{2}+\bar{l}_2-\bar{n}'+
\frac{s_3}{2}\right)!
\left(\frac{1}{2} (j+l+2 \bar{l}_2-2 \bar{n}'-2 \bar{n}-
s_1-s_3+2)-1\right)!
}
\nonu \\
&& \times 
N^{1-\frac{(j-s_1)}{2},1-\frac{1}{2} (l-s_3)}_{0} (\bar{n},\bar{n}'-\bar{l}_2) \,
\big((H^{k})_{k_2,\bar{l}_2} \, (H^{j+l})_{(1-h)+n'-k_2,
\bar{n}+\bar{n}'-\bar{l}_2}\big)
\Bigg].
\label{nonlinearcase}
\eea
The mode-dependent function in (\ref{nonlinearcase})
given by (\ref{Ndef}) has the linear dependence on the modes.
Due to the modes for the quadratic operators on the right-hand side
of (\ref{nonlinearcase}), there are summations over
holomorphic and antiholomorphic modes.
We expect that the dependence of trigonometric functions
appearing in the fourth line of (\ref{nonlinearcase})
can have numerical values as described before when we fix
the conformal dimensions, the helicities or mode.
The denominators appearing in the fourth line
can be absorbed into particle one on the left-hand side.
The fractional factor
in the fifth line inside the summations over two dummy variables
can be absorbed into the modes on the right-hand side
\footnote{The dummy variables
$k_1$, $k_2$, $\bar{l}_1$ and $\bar{l}_2$ have nothing to do with
the left- and right-weights of particle one and particle two.}.

Based on the result of (\ref{nonlinearcase}),
we obtain the corresponding (anti)commutators from the
expressions of Appendix A by introducing the $SU(8)$ indices appropriately.
Compared to the single-particle contributions
which always accompany the two-particle contributions, from Appendix A,
the two-particle contributions alone can appear also because
the double contractions between particle one and particle two
vanish. 

\section{The splitting functions, the celestial amplitudes
and the celestial OPEs for three operators 
}

\subsection{The splitting functions}

By substituting the $s_1, s_2, s_3, s_I$ and $s_J$ into
$D_{1,2}+D_{2,3}+D_{1,3}$ \cite{BHP}
which is the sum over factorized channels, one can check that
the ninety-five triple collinear splitting functions
can be summarized by
a single one \footnote{The splitting function of $s_1=s_2=s_3=+2$
in Einstein gravity was found in \cite{BHP,CP} and that of
$s_1=s_2=-s_3=+2$ was checked in \cite{CP}.
The remaining nontrivial splitting function
of $-s_1=s_2=s_3=+2$ can be obtained by using  
(\ref{SPLIT}). \label{Twocases}}
\bea
&& \mbox{Split}[1^{s_1}\, 2^{s_2}\, 3^{s_3} \rightarrow J^{s_J}]=
\nonu \\
&& \frac{1}{\epsilon^2} \, \frac{ (\omega_1+\omega_2+\omega_3)^{s_1+s_2+s_3-4}}
{\omega_1^{s_1-1} \, \omega_2^{s_2-1} \,
\omega_3^{s_3-1}} \, \Bigg( \omega_1 \,
\frac{\bar{z}_{12} \, \bar{z}_{13}}{(1-\eta)} +
\omega_2 \,
\frac{\bar{z}_{12} \, \bar{z}_{23}}{\eta \, (1-\eta)}
+\omega_3 \,
\frac{\bar{z}_{13} \, \bar{z}_{23}}{\eta}\Bigg).
\label{SPLIT}
\eea
See also Appendix D.
These splitting functions appear
in the coefficient of $(n-2)$-particle amplitude
for the $n$-particle amplitude.
The energies $\omega_1, \omega_2$ and $\omega_3$
of each particle
appear at the null momenta in four-dimensional Minkowski space
as the overall factors respectively.
Of course, one can reexpress  (\ref{SPLIT})
in terms of $s_I$ and $s_J$.
In the process $12 \rightarrow I$,
the helicity of $I$-th particle $s_I= s_1+s_2-1-p_{12I}$
with $p_{12I}=d_V-4$.
Here the $d_V$ is the scaling dimension of
three-point vertex in this process.
Moreover, the helicity of $J$-th particle
$s_J= s_I+s_3-1-p_{I3J}=s_1+s_2+s_3-2-p_{12I}-p_{I3J}$ \footnote{In this paper,
we are focusing on the simplest case with $p_{12I}=p_{I3J}=1$.
Then the helicity $s_J=s_4=s_1+s_2+s_3-4$.}.
The holomorphic coordinates in the holomorphic triple collinear limits
between  $z_1, z_2$ and $z_3$ are introduced \cite{BHP,CP}
\bea
z_1 = z_3 -\ep, \qquad
z_2 = z_3 -\eta \, \ep.
\label{holomorphic}
\eea
The triple collinear limit corresponds to taking $\ep \rightarrow 0$.
The $\eta$ variable denotes the relative collinearity
\footnote{
Let us introduce the three coefficients
appearing in front of the $(n-2)$-point amplitude
on the right-hand sides of the $n$-point amplitude
\bea
d_{1,2} &\equiv& \frac{\omega_1^{s_2-s_I-1}
\omega_2^{s_1-s_I-1}
\omega_3^{s_I-s_J}
\bar{z}_{12}^{s_1+s_2-s_I-1}
(\omega_1 \bar{z}_{13}+\omega_2
\bar{z}_{23})^{s_3+s_I-s_J}}
{\ep^2 (1-\eta)
(\omega_1+\omega_2+\omega_3)^{-s_J}
\big((1-\eta) \omega_1 \omega_2
\bar{z}_{12}+\eta \omega_2
\omega_3 \bar{z}_{23}+\omega_1 \omega_3 \bar{z}_{13}\big)},
\nonu \\
d_{2,3} &\equiv&
\frac{
\omega_1^{s_I-s_J}
\omega_2^{s_3-s_I-1}
\omega_3^{s_2-s_I-1}
\bar{z}_{23}^{s_2+s_3-s_I-1}
(-\omega_2 \bar{z}_{12}-\omega_3
\bar{z}_{13})^{s_1+s_I-s_J}}{\ep^2 \eta
(\omega_1+\omega_2+\omega_3)^{-s_J}
\big((1-\eta) \omega_1 \omega_2
\bar{z}_{12}+\eta \omega_2
\omega_3 \bar{z}_{23}+\omega_1 \omega_3 \bar{z}_{13}\big)},
\nonu \\
d_{1,3} &\equiv&
\frac{
\omega_1^{s_3-s_I-1}
\omega_2^{s_I-s_J}
\omega_3^{s_1-s_I-1}
\bar{z}_{13}^{s_1+s_3-s_I-1}
(\omega_1 \bar{z}_{12}-\omega_3
\bar{z}_{23})^{s_2+s_I-s_J}}
{\ep^2 (\omega_1+\omega_2+
\omega_3)^{-s_J} \big((1-\eta)
\omega_1 \omega_2 \bar{z}_{12}+
\eta \omega_2 \omega_3 \bar{z}_{23}+\omega_1
\omega_3 \bar{z}_{13}\big)}.
\label{threed}
\eea
Then by computing
$
\Big(d_{1,2}\Big|_{s_I=s_1+s_2-2}+
d_{2,3}\Big|_{s_I=s_2+s_3-2} + d_{1,3}\Big|_{s_I=s_1+s_3-2}\Big)
\Bigg|_{s_J=s_1+s_2+s_3-4}$,
we obtain (\ref{SPLIT}). Note that each $D_{i,j}$ contains
$d_{i,j}$.}.

\subsection{The celestial amplitudes}

The $n$ particle amplitude after Mellin transform is
given by
\bea
&& \widetilde{\cal A}_n(\Delta_1, s_1, z_1, \bar{z}_1;\Delta_2, s_2, z_2,
\bar{z}_2;\Delta_3, s_3, z_3, \bar{z}_3; \cdots)
\nonu \\
&& =\Bigg( \prod_{i=1}^n \int_0^{\infty} \, d \omega_i \, \omega_i^{\Delta_i-1}
\Bigg) {\cal A}_n(1^{s_1}\, 2^{s_2} \, 3^{s_3} \, \cdots \, n^{s_n}).
\label{Mellin}
\eea
After integrating over the $\omega_1, \omega_2, \cdots, \omega_n$,
the Mellin transformed amplitude depends on $\De_i$ with
$i=1,2, \cdots, n$ as well as the $2n$ complex coordinates.
The tree level $n$ particle amplitude in momentum space can factorize
in the triple collinear limit
and the splitting function is the $\frac{1}{\epsilon^2}$ coefficient\
of the $(n-2)$-particle amplitude
\bea
{\cal A}_n(1^{s_1}\, 2^{s_2} \, 3^{s_3} \, \cdots \, n^{s_n})=
\sum_J \, \mbox{Split}[1^{s_1}\, 2^{s_2} \, 3^{s_3} \rightarrow
J^{s_J}] \, {\cal A}_{n-2}(J^{s_J} \, \cdots \, n^{s_n})+
{\cal O}\Big(\frac{1}{\epsilon}\Big).
\label{Amp}
\eea
The previous sum over factorized channels
$D_{1,2}+D_{2,3}+D_{1,3}$ is the nontrivial term (the summation
over $J$ in (\ref{Amp})) on the right-hand side.

Let us introduce other variables
from the original variables $(\omega_1, \omega_2, \omega_3)$
\cite{BHP,CP}
\bea
\omega \equiv \omega_1 +\omega_2+\omega_3,
\qquad
\si_1 \equiv \frac{\omega_1}{( \omega_1 +\omega_2+\omega_3)},
\qquad
\si_2 \equiv \frac{\omega_2}{( \omega_1 +\omega_2+\omega_3)}.
\label{NEW}
\eea
Then there appears the additional $\omega^2$ factor
in the volume element $d \si_1 \, d \si_2 \, d \omega $.
We also introduce a new splitting function
\cite{BHP,CP}
\bea
\widehat{
\mbox{Split}}[1^{s_1}\, 2^{s_2} \, 3^{s_3} \rightarrow
J^{s_J}] \equiv \omega^{-(s_1+s_2+s_3-s_J-4)}\,
\mbox{Split}[1^{s_1}\, 2^{s_2} \, 3^{s_3} \rightarrow
J^{s_J}].
\label{newsplit}
\eea
Now we perform the Mellin transform on the equation (\ref{Amp})
with (\ref{Mellin})
\bea
&& \widetilde{\cal A}_n(\Delta_1, s_1, z_1, \bar{z}_1;\Delta_2, s_2, z_2,
\bar{z}_2;\Delta_3, s_3, z_3, \bar{z}_3; \cdots)
\nonu \\
&& =\int_{0}^{1}\, d \si_1 \,
\int_{0}^{1-\si_1}\, d \si_2 \,
\si_1^{\Delta_1-1}\, \si_2^{\Delta_2-1} \, (1-\si_1-\si_2)^{\Delta_3-1}
\, \sum_J \, \widehat{
\mbox{Split}}[1^{s_1}\, 2^{s_2} \, 3^{s_3} \rightarrow
J^{s_J}]
\nonu \\
&& \times
\widetilde{\cal A}_{n-2}(\Delta_1+\Delta_2+\Delta_3+s_1+s_2+s_3-s_J-4,
s_J, z_3, \bar{z}_3+
\si_1 \, \bar{z}_{13}+\si_2 \, \bar{z}_{23}
; \cdots) \nonu \\
&& +
{\cal O}(\frac{1}{\epsilon}).
\label{AMP}
\eea
In this calculation,
the power of $\omega$ inside the $d \omega$
is given by
the power of $\omega_i$, $(\De_i-1)$,
the $\omega$ factor in (\ref{newsplit}) and
$+2$ in the Jacobian between the variables (\ref{NEW})
as follows:
\bea
&& \De_1-1+\De_2-1+\De_3-1 +(s_1+s_2+s_3-s_J-4) +2=
\nonu \\
&& \Delta_1+\Delta_2+\Delta_3+s_1+s_2+s_3-s_J-5.
\label{power}
\eea
Then the $\omega$ integral with the integrand
$\omega$ to the power of (\ref{power}) factor
is replaced by the Mellin transformed $(n-2)$-particle amplitude
in (\ref{AMP}).
Moreover, the $\si_1$ and $\si_2$ dependence
comes from 
\bea
\bar{z}_J=
\si_1 \, \bar{z}_1 + \si_2 \, \bar{z}_2
+(1-\si_1-\si_2)\, \bar{z}_3
=\bar{z}_3 + \si_1 \, \bar{z}_{13}+\si_2 \, \bar{z}_{23}
\label{deformation}
\eea
according to the constraint from the
relation $n$-point momentum conservation
and $(n-2)$-point momentum conservation.
In order to obtain the complete factorized form,
the Taylor expansion around $\si_1 \, \bar{z}_{13}+\si_2\,
\bar{z}_{23}$ (\ref{deformation})
can be used and the $n$-point celestial amplitude
in terms of $(n-2)$-point celestial amplitude is summarized by
\bea
&& \widetilde{\cal A}_n(\Delta_1, s_1, z_1, \bar{z}_1;\Delta_2, s_2, z_2,
\bar{z}_2;\Delta_3, s_3, z_3, \bar{z}_3; \cdots)
\nonu \\
&& =\int_{0}^{1}\, d \si_1 \,
\int_{0}^{1-\si_1}\, d \si_2 \,
\si_1^{\Delta_1-1}\, \si_2^{\Delta_2-1} \, (1-\si_1-\si_2)^{\Delta_3-1}
\, \sum_J \, \widehat{
\mbox{Split}}[1^{s_1}\, 2^{s_2} \, 3^{s_3} \rightarrow
J^{s_J}]
\nonu \\
&& \times
\sum_{m=0}^{\infty}\,
\frac{(\si_1 \, \bar{z}_{13}+\si_2 \, \bar{z}_{23})^m}{m!}\,
\pa_{\bar{z}_3}^m\,
\widetilde{\cal A}_{n-2}(\Delta_1+\Delta_2+\Delta_3+s_1+s_2+s_3-s_J-4,
s_J, z_3, \bar{z}_3; \cdots)
\nonu \\
&& +
{\cal O}(\frac{1}{\epsilon}).
\label{FACT}
\eea
Note that the conformal dimension for the $J$-th particle
in the $(n-2)$-point celestial amplitude consists of
the particular combination of $\De_i$, $s_i$ ($i=1,2,3$) and $s_J$. 

\subsection{The celestial OPEs for three operators}

From the above relation (\ref{FACT}),
the operator product expansion for the three operators
is given by \footnote{For the two cases described in the footnote
\ref{Twocases}, the corresponding OPEs are found in \cite{BHP}
and \cite{CP} and are contained in (\ref{generalGGG}).
See also the earlier version in \cite{ESW}.}
\bea
&&
G^{s_1}_{\Delta_1}(z_1, \bar{z}_1) \, G^{s_2}_{\Delta_2}(z_2, \bar{z}_2) \,
G^{s_3}_{\Delta_3}(z_3, \bar{z}_3)=
\nonu \\
&& \sum_{m=0}^{\infty}\, \sum_{l=0}^{m} \,
\frac{\bar{z}_{13}^l \, \bar{z}_{23}^{m-l}}{l! \, (m-l)!}\,
\bar{\pa}^m \, G^{\rm{min}(s_1,s_2,s_3)}_{\Delta_1+\Delta_2+\Delta_3}
(z_3,\bar{z}_3) \nonu \\
&& \times \, \Bigg[ B(\Delta_1+l+2-s_1,\Delta_2+m-l+1-s_2,
\Delta_3+1-s_3)\, \frac{\bar{z}_{12}\, \bar{z}_{13}}{z_{12}\, z_{13}}
\nonu \\
&& +B(\Delta_1+l+1-s_1,\Delta_2+m-l+2-s_2,
\Delta_3+1-s_3)\, \frac{\bar{z}_{12}\, \bar{z}_{23}}{z_{12}\, z_{23}}
\nonu \\
&& +B(\Delta_1+l+1-s_1,\Delta_2+m-l+1-s_2,
\Delta_3+2-s_3)\, \frac{\bar{z}_{13}\, \bar{z}_{23}}{z_{13}\, z_{23}} \Bigg].
\label{generalGGG}
\eea
Note that
the general term inside the $l$ summation
in the $(\si_1 \, \bar{z}_{13}+\si_2 \, \bar{z}_{23})^m$
(\ref{FACT})
contains $\si_1^l \, \si_2^{m-l}$ as well as the binomial coefficient
and $\bar{z}_{13}^l \, \bar{z}_{23}^{m-l}$.
Let us describe how we obtain this result.

\begin{itemize}
\item[]
Note that the contribution from the first term in (\ref{SPLIT})
leads to the fact that \\
i)
the power of $\si_1$
is given by three contributions,
the power of $\si_1$, $(\De_1-1)$,
the power of $\si_1$ in the binomial expansion,
$l$, and the power $\omega_1$ (or $\si_1$) of the splitting function,
$(2-s_1)$, leading to
$(\Delta_1+l-1+2-s_1)$, \\
ii) the power of $\si_2$
is given by
 three contributions,
the power of $\si_2$, $(\De_2-1)$,
the power of $\si_2$ in the binomial expansion,
$(m-l)$, and
the power of $\omega_2$ (or $\si_2$) of the splitting function,
$(1-s_2)$,
leading to $(\Delta_2+m-l-1+1-s_2)$, and
\\
iii) the power of
$(1-\si_1-\si_2)$ is given by
two contributions,
the power of $(1-\si_1-\si_2)$, $(\De_3-1)$,
and the power of $\omega_3$ (or $(1-\si_1-\si_2))$
of the splitting function, $(1-s_3)$,
leading to $(\Delta_3-1+1-s_3)$.\\
They appear in the arguments of the first
generalized Euler beta function
of (\ref{generalGGG}) with some shifts.
\end{itemize}

A similar analysis
for the second and the third terms of (\ref{SPLIT})
can be done and they appear in the arguments of the second
and the third generalized Euler beta functions respectively
\footnote{For the second term, from the expression of
splitting functions, the final result can be obtained
from the result of the first term by replacing
$s_1 \rightarrow s_1+1$ and $s_2 \rightarrow s_2-1$.
For the third term,  the final result can be obtained
from the result of the first term by replacing
$s_1 \rightarrow s_1+1$ and $s_3 \rightarrow s_3-1$.}.
We are using
the relations
\bea
\epsilon^2 \, \eta  =  z_{13} \, z_{23},
\qquad
\epsilon^2 \, (1-\eta)  =  z_{12} \, z_{13},
\qquad
\epsilon^2 \, \eta \, (1-\eta)  =  z_{12} \, z_{23},
\label{somerel}
\eea
which can be obtained from (\ref{holomorphic}).
Then the $\ep$ and $\eta$ dependence in (\ref{SPLIT})
can be written in terms of the holomorphic coordinates.

It is also useful to rewrite
(\ref{generalGGG}) in terms of
the right-weight because the arguments in the
generalized Euler beta function depends on
the combination $(\De_i-s_i)$
which is nothing but $2 \bar{h}_i$
and we arrive at, after using (\ref{somerel}),
\bea
&& G^{s_1}_{\Delta_1}(z_1, \bar{z}_1) \, G^{s_2}_{\Delta_2}(z_2, \bar{z}_2) \,
G^{s_3}_{\Delta_3}(z_3, \bar{z}_3)=
\nonu \\
&&  \sum_{m=0}^{\infty}\, \sum_{l=0}^{m} \,
\frac{\bar{z}_{13}^l \, \bar{z}_{23}^{m-l}}{l! \, (m-l)!}\,
\bar{\pa}^m \, G^{\rm{min}(s_1,s_2,s_3)}_{\Delta_1+\Delta_2+\Delta_3}
(z_3,\bar{z}_3) \nonu \\
&& \times \, \Bigg[ B(2\bar{h}_1+l+2,2\bar{h}_2+m-l+1,
2\bar{h}_3+1)\, \frac{\bar{z}_{12}\, \bar{z}_{13}}{z_{12}\, z_{13}}
\nonu \\
&& +B(2\bar{h}_1+l+1,2\bar{h}_2+m-l+2,
2\bar{h}_3+1)\, \frac{\bar{z}_{12}\, \bar{z}_{23}}{z_{12}\, z_{23}}
\nonu \\
&& +B(2 \bar{h}_1+l+1,2 \bar{h}_2+m-l+1,
2\bar{h}_3+2)\, \frac{\bar{z}_{13}\, \bar{z}_{23}}{z_{13}\, z_{23}} \Bigg],
\label{threeOPE}
\eea
which is equivalent to the equation $(3.17)$ of \cite{CP}
\footnote{The left-hand side of (\ref{threeOPE}) is equivalent to
$G^{s_2}_{\Delta_2}(z_2, \bar{z}_2) \, G^{s_1}_{\Delta_1}(z_1, \bar{z}_1) \,
G^{s_3}_{\Delta_3}(z_3, \bar{z}_3)$. For the time being,
we do not assume they are both fermionic operators.
Then we don't have to worry about the extra minus sign
between the interchange of these operators. Then
the right-hand side of (\ref{threeOPE}) should satisfy
the following property. That is,
by taking $z_1 \leftrightarrow z_2$,
$\bar{z}_1 \leftrightarrow \bar{z}_2$,
$\De_1 \leftrightarrow \De_2$,
and $s_1 \leftrightarrow s_2$, the right-hand side is invariant
by using the proper change of dummy variables, the property of
the generalized Euler beta function and the simple relations
in the complex coordinates.
We can also apply the similar descriptions on the
$G^{s_1}_{\Delta_1}(z_1, \bar{z}_1) \,
G^{s_3}_{\Delta_3}(z_3, \bar{z}_3)\,
G^{s_2}_{\Delta_2}(z_2, \bar{z}_2)$,
$
G^{s_3}_{\Delta_3}(z_3, \bar{z}_3)\,
G^{s_2}_{\Delta_2}(z_2, \bar{z}_2) \,  G^{s_1}_{\Delta_1}(z_1, \bar{z}_1)$,
$G^{s_3}_{\Delta_3}(z_3, \bar{z}_3) \,
G^{s_1}_{\Delta_1}(z_1, \bar{z}_1)\,
G^{s_2}_{\Delta_2}(z_2, \bar{z}_2)$, or
$G^{s_2}_{\Delta_2}(z_2, \bar{z}_2) \,
G^{s_3}_{\Delta_3}(z_3, \bar{z}_3)\, 
G^{s_1}_{\Delta_1}(z_1, \bar{z}_1)$.
}.

Then how do we extract the single-particle
contributions to the multi-particle OPE?
As described in \cite{CP},
we can decompose $z_{12}$ into $z_{12}=(z_{13}-z_{23})$
and  $\bar{z}_{12}$ into $\bar{z}_{12}=(\bar{z}_{13}-\bar{z}_{23})$
and expand the whole expression in (\ref{threeOPE}) in terms of
$z_{23}$ and $\bar{z}_{23}$.
We collect the term of $z_{23}^0 \, \bar{z}_{23}^0$ because
we take the limit of $z_{23} \rightarrow 0$ and $\bar{z}_{23} \rightarrow
0$ at the final stage.
It is obvious to see that the contributions from the second and the
third terms in (\ref{threeOPE}) vanish as we take $\bar{z}_{23}
\rightarrow 0$.
From the first term of (\ref{threeOPE}), the $\frac{1}{z_{12}}$
contains the power of $z_{23}$ in the geometric series.
Then the $z_{23}^0$ term has $\frac{1}{z_{13}}$ and
the factor $\frac{\bar{z}_{12}\, \bar{z}_{13}}{z_{12}\, z_{13}}$
becomes $\frac{\bar{z}_{13}^2}{ z_{13}^2}$.
Then we need to select the $l=m$ term in the summation over $l$
in order to have $\bar{z}_{23}^0$ term.
Therefore we are left with
\bea
&&
G^{s_1}_{\Delta_1}(z_1, \bar{z}_1) \, (G^{s_2}_{\Delta_2} \,
G^{s_3}_{\Delta_3})(z_3, \bar{z}_3)
\nonu \\
&& = \frac{\bar{z}_{13}^2}{z_{13}^2}
\sum_{m=0}^{\infty}\,
\frac{\bar{z}_{13}^m}{m!}\, B(\Delta_1-s_1+2+m,\Delta_2-s_2+1,
\Delta_3-s_3+1)\, 
\bar{\pa}^m \, G^{\rm{min}(s_1,s_2,s_3)}_{\Delta_1+\Delta_2+\Delta_3}(z_3, \bar{z}_3)
\nonu \\
&& = \frac{\bar{z}_{13}^2}{z_{13}^2}
\sum_{m=0}^{\infty}\,
\frac{\bar{z}_{13}^m}{m!}\, B(2 \bar{h}_1+2+m,2\bar{h}_2+1,
2 \bar{h}_3+1)\, 
\bar{\pa}^m \, G^{\rm{min}(s_1,s_2,s_3)}_{\Delta_1+\Delta_2+\Delta_3}(z_3, \bar{z}_3),
\label{bulkOPE}
\eea
where in the third line of (\ref{bulkOPE}) we express
the arguments inside the generalized Euler beta function
in terms of the right-weights \footnote{The (normal-ordered)
quadratic terms
on the right-hand side of the OPE
of the single-particle operators with the two-particle operators
are given by
\bea
&& \frac{\bar{z}_{13}}{z_{13}}
\sum_{m=0}^{\infty}\,
\frac{\bar{z}_{13}^m}{m!}\, B(\Delta_1-s_1+1+m,\Delta_2-s_2+1)\,
( \bar{\pa}^m \,
G^{\rm{min}(s_1,s_2)}_{\Delta_1+\Delta_2}\, G^{s_3}_{\De_3} )(z_3, \bar{z}_3)
\nonu \\
&& +  \frac{\bar{z}_{13}}{z_{13}}
\sum_{m=0}^{\infty}\,
\frac{\bar{z}_{13}^m}{m!}\, B(\Delta_1-s_1+1+m,\Delta_3-s_3+1)\, 
(G^{s_2}_{\De_2}\,
\bar{\pa}^m \, G^{\rm{min}(s_1,s_3)}_{\Delta_1+\Delta_3})(z_3, \bar{z}_3)
\nonu \\
&& = \frac{\bar{z}_{13}}{z_{13}}
\sum_{m=0}^{\infty}\,
\frac{\bar{z}_{13}^m}{m!}\, B(2 \bar{h}_1+1+m, 2 \bar{h}_2+1)\,
( \bar{\pa}^m \,
G^{\rm{min}(s_1,s_2)}_{\Delta_1+\Delta_2}\, G^{s_3}_{\De_3} )(z_3, \bar{z}_3)
\nonu \\
&& +  \frac{\bar{z}_{13}}{z_{13}}
\sum_{m=0}^{\infty}\,
\frac{\bar{z}_{13}^m}{m!}\, B(2 \bar{h}_1+1+m,2 \bar{h}_3+1)\, 
(G^{s_2}_{\De_2}\,
\bar{\pa}^m \, G^{\rm{min}(s_1,s_3)}_{\Delta_1+\Delta_3})(z_3, \bar{z}_3).
\nonu 
\eea
\nonu}.

\section{ Conclusions and outlook}

The main observation of this paper is that
i) we have found the commutator in (\ref{comm1})
for generic helicities $s_1, s_2$ and $s_3$.
ii) In Appendix A, we have obtained
the OPEs of the single-particle celestial operators
with the two-particle celestial operators relevant to
the single-particle exchange on the right-hand sides (and other terms).
iii) By putting the $SU(8)$ indices on
(\ref{comm1}), the whole ninety-five (anti)commutators
are obtained from (\ref{25comm}) and the expressions of Appendix B.
iv) From the new result of (\ref{SPLIT}),
the previous findings of (\ref{ggope}) in the boundary
are checked from the viewpoint in the  bulk theory.

In the two subsections, we briefly sketch
the single-particle contributions to the multi-particle OPEs
for the quadruple and $N$-tuple  collinear limits.

\subsection{The multi-particle OPEs of
the single-particle celestial operators with
the three-particle celestial operators}

How does one write down the single-graviton contributions
to the multi-particle graviton OPE for the quadruple-collinear limits?
The OPE is given by
\bea
&&G^{s_1}_{\Delta_1}(z_1, \bar{z}_1) \, (G^{s_2}_{\Delta_2} \,
(G^{s_3}_{\Delta_3}\, G^{s_4}_{\Delta_4}))(z_4, \bar{z}_4)\nonu \\
&& = \frac{\bar{z}_{14}^3}{z_{14}^3}
\sum_{m=0}^{\infty}\,
\frac{\bar{z}_{14}^m}{m!}\, B(\Delta_1-s_1+3+m,\Delta_2-s_2+1,
\Delta_3-s_3+1,\Delta_4-s_4+1)
\nonu \\
&& \times 
\bar{\pa}^m \, G^{\rm{min}(s_1,s_2,s_3,s_4)}_{
\Delta_1+\Delta_2+\Delta_3+\Delta_4}(z_4, \bar{z}_4)
\nonu \\
&& = \frac{\bar{z}_{14}^3}{z_{14}^3}
\sum_{m=0}^{\infty}\,
\frac{\bar{z}_{14}^m}{m!}\, B(2 \bar{h}_1+3+m,2\bar{h}_2+1,
2 \bar{h}_3+1, 2 \bar{h}_4+1)\, 
\bar{\pa}^m \, G^{\rm{min}(s_1,s_2,s_3,s_4)}_{
\Delta_1+\Delta_2+\Delta_3+\Delta_4}(z_4, \bar{z}_4).
\label{quadruple1}
\eea
We are using the convention of normal-ordered product
in \cite{BBSS} for the three-particle celestial operators
having the complex coordinates $(z_4,\bar{z}_4)$ on the left-hand side
of (\ref{quadruple1})
\footnote{One can consider the similar multiple OPE
(of two-particle operators with two-particle operators)
$(G^{s_1}_{\Delta_1} \, G^{s_2}_{\Delta_2})(z_1, \bar{z}_1)  \,
(G^{s_3}_{\Delta_3}\, G^{s_4}_{\Delta_4})(z_4, \bar{z}_4)$ for the
quadruple-collinear limits. For the first contraction,
the OPE similar to (\ref{bulkOPE}) (the ordering of OPE is reversed)
can be applied.
For the second contraction, the OPE of
the (anti)holomorphic derivatives on the single operator
located at the third complex coordinates with
the particle $4$ celestial operator should be used.
More explicitly, we have
\bea
&&
G^{s_1}_{\Delta_1}(z_1, \bar{z}_1) \, (G^{s_2}_{\Delta_2} \,
G^{s_3}_{\Delta_3})(z_3, \bar{z}_3)= \pm
(G^{s_2}_{\Delta_2} \,
G^{s_3}_{\Delta_3})(z_3, \bar{z}_3) \,
G^{s_1}_{\Delta_1}(z_1, \bar{z}_1).
\nonu
\eea
The minus sign appears when both the single-particle
and two-particle operators are fermionic.
Otherwise, the plus sign appears.
After interchanging $z_1 \leftrightarrow z_3$
and $\bar{z}_1 \leftrightarrow \bar{z}_3$ in (\ref{bulkOPE})
and applying the Taylor expansion for the two
complex variables on the linear term
on the right-hand side (See also \cite{BBSS}),
\bea
&& (G^{s_2}_{\Delta_2} \,
G^{s_3}_{\Delta_3})(z_1, \bar{z}_1) \,
G^{s_1}_{\Delta_1}(z_3, \bar{z}_3)
= \pm \frac{\bar{z}_{13}^2}{z_{13}^2}
\sum_{m=0}^{\infty}\,
\frac{(-1)^m\, \bar{z}_{13}^m}{m!}\, B(2 \bar{h}_1+2+m,2\bar{h}_2+1,
2 \bar{h}_3+1)\nonu \\
&& \times
\bar{\pa}_{\bar{z}_3}^m \,
\Bigg[\sum_{s=0}^{\infty} \, \sum_{k=0}^{s} \,
  \frac{1}{(s-k)! k!}\, z_{13}^{s-k}\, \bar{z}_{13}^k \,
\pa^{s-k}_{z_3}\, \bar{\pa}^k_{\bar{z}_3}
\,
G^{\rm{min}(s_1,s_2,s_3)}_{\Delta_1+\Delta_2+\Delta_3}(z_3, \bar{z}_3)
\Bigg].
\label{twotoone}
\eea
For the OPE $ (G^{s_1}_{\Delta_1} \,
G^{s_2}_{\Delta_2})(z_1, \bar{z}_1) \,
G^{s_3}_{\Delta_3}(z_3, \bar{z}_3)$
from the OPE
of two-particle operators with themselves,
we obtain the right-hand side by taking
$s_2 \rightarrow s_1, \De_2 \rightarrow \De_1$,
$s_3 \rightarrow s_2, \De_3 \rightarrow \De_2$ and
$s_1 \rightarrow s_3, \De_1 \rightarrow \De_3$ in (\ref{twotoone}).
Moreover we have the nontrivial relation
for the (anti)commutator of normal-ordered products
$(G^{s_1}_{\Delta_1} \, (G^{s_2}_{\Delta_2} \,
G^{s_3}_{\Delta_3}))(z, \bar{z}) \neq   \pm ((G^{s_2}_{\Delta_2} \,
G^{s_3}_{\Delta_3}) \, G^{s_1}_{\Delta_1})(z,\bar{z})$. See also \cite{BBSS}.
In the chiral or antichiral CFT \cite{TY},
the OPE of the left-hand side of
(\ref{twotoone}) is given directly and explicitly
in the context of Borcherds identity
and it is  an open problem
to generalize it in the celestial CFT.  }.
The quantities $p_{12I}$, $p_{I3J}$ and $p_{J4K}$
appearing in the second, third and fourth argument of
the above `generalized' (or multivariate) Euler beta function
are put to $1$.
The sum of these, $3$, appears
in the first argument of
the generalized Euler beta function.
The corresponding constraint for the helicity
$s_K=s_5=s_1+s_2+s_3+s_4-6$, similar to (\ref{s1234}).
The overall coupling-dependent factor is given by
$\kappa_{1,2,I}\, \kappa_{I,3,J}\, \kappa_{J,4,K}$
\footnote{In general, there are also the quadratic and cubic terms in
(\ref{quadruple1}).}.
In this case, the commutator can be written as
\bea && \bigg[
(\tilde{H}^{j})_{3-h,\bar{n}},
\, (\tilde{H}^{k_2,k_3,  k_4})_{n',\bar{n}'} \bigg]  =
\nonu \\ && \pm
N_{2}^{1-\frac{1}{2}(j-s_1),
1-\frac{1}{2}(k_2-s_2+k_3-s_3+k_4-s_4)}(\bar{n},
\bar{n}')
\,
(\tilde{H}^{j+k_2+k_3+k_4})_{(3-h)+n',\bar{n}+\bar{n}'}.
\label{quadruple2}
\eea
The degree of mode-dependent function appearing in
(\ref{quadruple2}) is three
\footnote{
In this case, the two mode-dependent functions
described in (\ref{HtHt}) 
are related to each other as follows:  
\bea
&& p_{1-\frac{1}{2}(j-s_1),1-\frac{1}{2}(k_2-s_2+k_3-s_3+k_4-s_4),
1-\frac{1}{2}(j+k_2+k_3+k_4-s_5)}(\bar{n},\bar{n}')
\nonu \\
&& =
-\Bigg[\frac{1}{6(2+j-s_1+k_2-s_2+k_3-s_3+k_4-s_4)} \nonu \\
&& \times \frac{1}
{(3+j-s_1+k_2-s_2+k_3-s_3+k_4-s_4)(4+j-s_1+k_2-s_2+k_3-s_3+k_4-s_4)}
\Bigg]
\nonu \\
&& \times
N_2^{1-\frac{1}{2}(j-s_1),1-\frac{1}{2}(k_2-s_2+k_3-s_3+k_4-s_4)}(\bar{n},
\bar{n}').
\nonu
\eea
There exists an overall minus sign, compared to the previous
case. We expect that the sign changes appear alternatively.
We can easily write down the corresponding OPE
where the highest singular term has the
fourth-order pole, similar to
(\ref{HtHt}).} and the difference between the $(1-\bar{h})$ dimension
of the left-hand side and the right-hand side in
(\ref{quadruple2}) is given by four \footnote{When we consider the
degree of $SU(8)$ indices in the ${\cal N}=8$ supergravity,
the simple counting implies the possible number of
(anti)commutators is given by $9 \times 95=855$. Of course,
we don't consider 
the redundancy from the  combinations of $s_2$, $s_3$ and $s_4$.
As we can see in Appendix A, the normal-ordered product between three
operators also appear in the OPEs where the right-hand sides
contain only quadratic terms. In general, we should consider
the OPEs of these with the nine operators. This implies that it will
be very useful to compute some OPEs via the Thielemans package.}.
See also the footnote \ref{quadrupleopepole6} in different basis.

\subsection{
The multi-particle OPEs of
the single-particle celestial operators with
the $(N-1)$-particle celestial operators}

For the $N$-tuple collinear limits,
the OPE can be described by
\bea
&& G^{s_1}_{\Delta_1}(z_1, \bar{z}_1) \, (G^{s_2}_{\Delta_2} \,
(G^{s_3}_{\Delta_3}\, \cdots )  \,G^{s_N}_{\Delta_N} ) \, \cdots)(z_N, \bar{z}_N)
\nonu \\
&& = \frac{\bar{z}_{1N}^{N-1}}{z_{1N}^{N-1}}
\sum_{m=0}^{\infty}\,
\frac{\bar{z}_{1N}^m}{m!}\, B(\Delta_1-s_1+N-1+m,\Delta_2-s_2+1,
\cdots,\Delta_N-s_N+1)\, 
\nonu \\
&& \times \bar{\pa}^m \, G^{\rm{min}(s_1,\cdots,s_N)}_{
\Delta_1+\cdots+\Delta_N}(z_N, \bar{z}_N)
\label{ntuple1} \\
&& = \frac{\bar{z}_{1N}^{N-1}}{z_{1N}^{N-1}}
\sum_{m=0}^{\infty}\,
\frac{\bar{z}_{1N}^m}{m!}\, B(2 \bar{h}_1+N-1+m,2\bar{h}_2+1,
\cdots, 2 \bar{h}_N+1)\, 
\bar{\pa}^m \, G^{\rm{min}(s_1,\cdots,s_N)}_{
\Delta_1+\cdots+\Delta_N}(z_N, \bar{z}_N).
\nonu
\eea
As anticipated in \cite{GHP},
the $(N-1)$-particle celestial operators
for this particular OPE
give rise to the higher-order poles in (\ref{ntuple1})
\footnote{
The corresponding $n$-point celestial amplitude
for the ${\cal N}=8$ supergravity  in the notation of (\ref{AMP1})
and (\ref{AMP2})
\bea
&& < \Phi_{+2}^{(h_1)}(z_1,\bar{z}_1)\,
\Phi_{+2}^{(h_2)}(z_2, \bar{z}_2)\, \Phi_{+2}^{(h_3)}(z_3,
\bar{z}_3) \, \cdots \,   \Phi_{+2}^{(h_N)}(z_N,
\bar{z}_N) \cdots >
\nonu
\eea
contains
\bea
&& z_{2 N}^0 \,
\frac{\bar{z}_{1N}^{N-1}}{z_{1N}^{N-1}} \,
\sum_{m=0}^{\infty}\, \frac{\bar{z}_{1N}^m}{m!}\,
B(2\bar{h}_1+N-1+m,2\bar{h}_2+1, \cdots, 2\bar{h}_N+1)
<
\bar{\pa}_{\bar{z}_N}^m \, \Phi_{+2}^{(h_1+h_2+h_3+\cdots + h_N)}(z_N,\bar{z}_N)
\, \cdots >
\nonu
\eea
where the $(n-N)$ celestial operators in the
amplitude can be any combinations of
Table \ref{hbar1} for $N$ positive gravitons. }.
On the
right-hand side, the single-particle (exchange) terms
appear.
We are using the fully normal-ordered product
in \cite{BBSS} for the $(N-1)$-particle celestial operators
having the complex coordinates $(z_N,\bar{z}_N)$ on the left-hand side
of (\ref{ntuple1}).
The quantities $p_{12I}$, $p_{I3J}$, $p_{J4K}$ and so on
appearing in the second, third, $\cdots$, $N$-th argument of
the above generalized Euler beta function
are put to $1$.
The sum of these, $(N-1)$,  appears
in the first argument of
the generalized Euler beta function.
The corresponding constraint for the helicity
$s_{N+1}=\sum_{i=1}^N \, s_i-2(N-1)$, similar to (\ref{s1234}).
The overall coupling-dependent factor is given by
the product of $(N-1)$ couplings
similarly. 
The generalized hypergeometric function
\bea
&&
{}_3 F_2
\Bigg[
\begin{array}{c}
\sum_{i=2}^N \,
\frac{(k_i-s_i)}{2}
+\bar{n}',\qquad 1-\frac{(j-s_1)}{2}
-\bar{n},\qquad \frac{(j-s_1)}{2}-\bar{n}
\\(2-N) -\frac{(j-s_1)}{2}-\bar{n},
\qquad
(N-1)+\frac{(j-s_1)}{2}+\sum_{i=2}^N \,
(k_i-s_i)-\bar{n}
\end{array} ; 1
\Bigg],
\label{ntuple2}
\eea
from the summation over $m$
\bea
&& \sum_{m=-\bar{n}- \frac{1}{2}(j-s_1)-(N-1)}^{-(N-1)-(j-s_1)}\, \Bigg[
\frac{1}{m!}\, \,  \frac{ (s_1+\sum_{i=2}^N \,
(k_i-s_i)-2(N-1)-j-m)!}{(-m-j+s_1-(N-1))!}  \nonu \\
&& \times \frac{(m+N-1)! \,
(-1)^{m}}{(m+N-1+\bar{n}+ \frac{1}{2}(j-s_1))!}  \,
\frac{1}{(-\bar{n}-\bar{n}'-\frac{1}{2}( j+\sum_{i=2}^{N}\,
k_i-s_{N+1})-m)!}  \Bigg],
\label{ntuple3}
\eea
contains 
the mode dependent function (\ref{Ndef}) with a degree $(N-1)$.
The $k_i$ with $i =2,3,4, \cdots, N$
stands for the conformal dimension
of the celestial operator in the $(N-1)$ normal-ordered product
on the left-hand side of (\ref{ntuple1}).
Note that the $N$ dependence in (\ref{ntuple2})
appears at several places.
Among them, there exist the $(N-2)$ and $(N-1)$ dependences
appearing on the lower arguments in the
generalized hypergeometric function.
Moreover, the power $(N-1)$ in the complex coordinates
in (\ref{ntuple1}) gives rise to the upper and lower bounds
for the dummy variable $m$ in (\ref{ntuple3}).
For $N=3,4,5,6$ cases, we can check this property explicitly
\footnote{It is an open problem how the generalized hypergeometric
function (\ref{ntuple2})
or (\ref{ntuple3}) is
related to the mode-dependent function (\ref{Ndef}) for generic
$N$ analytically.
For $N=3,4,5,6$, the number of terms in (\ref{Ndef}), when we expand
them fully, is given by $96$, $716$, $8303$ and 
$66174$ respectively.}.
Recall 
that the helicity $s_{N+1}$ is for the operator appearing on the
right-hand side.

Then the final commutator by absorbing the
normalization factors for the
$(N-1)$-tuple operators on the left-hand side can be summarized by
\bea && \bigg[
(\tilde{H}^{j})_{(N-1)-h,\bar{n}},
\, (\tilde{H}^{k_2,k_3, \cdots, k_N})_{n',\bar{n}'} \bigg]  =
\nonu \\ && \pm
{N}_{N-2}^{1-\frac{1}{2}(j-s_1),
1-\frac{1}{2}(k_2-s_2+k_3-s_3+\cdots +k_N-s_N)}(\bar{n},
\bar{n}')
\,
(\tilde{H}^{j+k_2+k_3+\cdots +k_N})_{(N-1-h)+n',\bar{n}+\bar{n}'}.
\label{ntuple4}
\eea
Note that the subscript $(N-2)$ in the mode-dependent function
(\ref{ntuple4}) has nothing to do with (\ref{Ndef}).
For $N=4$, this reproduces the previous result for the quadruple-collinear
limits above.
The difference between the $(1-\bar{h})$ dimension
of the left-hand side and the right-hand side in
(\ref{ntuple4}) is given by $N$.
It is rather nontrivial to put the $SU(8)$ indices
to the $(N-1)$-particle celestial operators for generic $N$  correctly
\footnote{For pure Einstein gravity, the result (\ref{ntuple4})
is complete.}.
Due to the $N$-dependence of the holomorphic mode for the
particle $1$ on the left-hand side, there are no nonlinear terms
on the right-hand side of (\ref{ntuple4}).
For each nonlinear term on the right-hand side, the 
holomorphic mode for the
particle $1$ on the left-hand side selects its own distinct term.
For example, for $N=3$, the quadratic terms have $(N-2)-h=1-h$
in (\ref{NONLINEAR}). This is because in (\ref{ntuple1}),
the holomorphic mode dependence appears differently and when we calculate
the contour integral, the nonzero contributions select
the particular holomorphic mode for the particle $1$.
This implies that although there are $(N-1)$ singular terms in
(\ref{ntuple1}), the number of corresponding commutators is
given by $(N-1)$ not a single one
\footnote{The OPE corresponding to
(\ref{ntuple4}), as done in (\ref{HtHt})
by absorbing the overall factor, is given by 
\bea
&& \tilde{H}^{j}(\bar{z})
\,
\tilde{H}^{k_2, k_3, \cdots, k_N}(\bar{w})
\nonu \\
&& = 
\pm
p_{1-\frac{1}{2}(j-s_1),1-\frac{1}{2}\sum_{i=2}^N\, (k_i-s_i),
1-\frac{1}{2}(j+\sum_{i=2}^N \, k_i-s_{N+1})}(\bar{\pa}_{\bar{z}},\bar{\pa}_{\bar{w}})
\,
\Bigg[ \frac{\tilde{H}^{j+\sum_{i=2}^N \, k_i}(\bar{w})}{(\bar{z}-\bar{w})}
\Bigg] +\cdots,
\label{finalOPE}
\eea
which can be expanded further by using the definition of
mode-dependent function  $p_{i,j,k}$ \cite{CFT} and
the description in the footnote \ref{pandother}.
The notations for three operators are the same as the ones in
(\ref{HtHt}) by generalizing the $N=3$ to the generic $N$.
The sum of the powers of the differential operators
$\bar{\pa}_{\bar{z}}$ and
$\bar{\pa}_{\bar{w}}$ is given by $(N-1)$.
The highest singular term of (\ref{finalOPE})
where $\bar{\pa}_{\bar{z}}^{N-1}$ acts on the
factor $\frac{1}{(\bar{z}-\bar{w})}$ is
given by the $N$-th order.
All the singular terms are fixed explicitly.
The structure constants appearing on the right-hand sides of
(\ref{finalOPE})
also appear in the construction of $W_{\infty}$ algebra
\cite{PRS,Ahn2202,Ahn2203,AK2309,AK2407,AK2501}.}.

\subsection{
The open problems}

The immediate question is how one can write down the splitting
functions for quadruple (and $N$-tuple) collinear limits
relating to the amplitudes in the (pure gravity or)
${\cal N}=8$ supergravity
by generalizing the work of \cite{BHP}?
These splitting functions should appear as
the generalized Euler beta functions after the Mellin transform.

%
%
It is an open problem to determine the twenty-five OPEs
found in \cite{AK2509} from the ${\cal N}=8$ supersymmetric soft theorem
\cite{Tropper1,Tropper}
by following the work of \cite{PRSY,HP}.
The soft theorems for the gravitinos, graviphotons, graviphotinos and
scalars will appear nontrivially.

Moreover, the OPEs between the various nine currents living in the
${\cal N}=8$ supergravity and matter fields can be determined
by applying the description of \cite{HPS}. 
These OPEs can be written in terms of the various (anti)commutators
by using the contour integrals. By acting with
the additional currents on these
twenty-five
(anti)commutators, the twenty-five (anti)commutators \cite{AK2509}
between the currents
should appear
and we expect that the ${\cal N}=8$ supersymmetric BMS algebra
\cite{2212-1,2212-2} appears
by considering the particular modes on the currents.
See also \cite{FSTZ,Ahn2111} for the ${\cal N}=1$ supersymmetric Einstein
Yang-Mills theory.

In the context of
Carrollian
holography, the results of \cite{HPS} are reproduced in \cite{MRY}. Then
it is an open problem to obtain
the OPEs between the various nine currents living in the
${\cal N}=8$ supergravity and the Carrollian operators.

In particular, the amplitudes in the ${\cal N}=8$ supergravity
depend on the singlets of the multiple tensor product of
$SU(8)$ representations \cite{BEF,KLR,Liu}. It is rather nontrivial
to write down these singlets for the $n$-point amplitudes explicitly. 

Because the MHV amplitudes at tree level in the pure Yang-Mills
theory and the ${\cal N}=4$ super Yang-Mills theory
are the same and according to the KLT relation \cite{KLT},
the MHV amplitudes at tree level in the pure gravity
theory and the ${\cal N}=8$ supergravity 
are the same. This implies that the relation of (\ref{AMP2})
for the MHV graviton amplitudes at tree level in the
${\cal N}=8$ supergravity is also valid from the work of \cite{CP}.

It is an open problem to obtain the bulk descriptions on
(\ref{AMP2}) for other amplitudes having other particles
in the ${\cal N}=8$ supergravity (totally ninety-five
celestial amplitudes) and to check
whether the Mellin transform on these results will reproduce
the celestial amplitudes or not. The relevant work is given by
\cite{CS} dealing with the recursion relations in the ${\cal N}=8$
supergravity.

\vspace{.7cm}


\centerline{\bf Acknowledgments}

CA thanks M. Pate for intensive discussions,
A. Tropper and M.H. Kim for the discussion.
This work was supported by the National
Research Foundation of Korea(NRF) grant funded by the
Korea government(MSIT) 
(No. 2023R1A2C1003750).
CA
acknowledges warm hospitality from
Institute for Convergent Fundamental Studies,
Seoul National University of Science and Technology.


\newpage

\appendix

\renewcommand{\theequation}{\Alph{section}\mbox{.}\arabic{equation}}

\section{The multi-particle OPEs}

We can calculate the celestial OPEs by following the
work of \cite{CP} and using the fundamental OPEs in
\cite{AK2509}
\footnote{We skip all the dependence of the couplings
in this Appendix and they can be read off from the
description in Appendices B and C where there are twenty-five
couplings. See also Appendix E for other twenty couplings.}.

\subsection{The OPEs of
the gravitons of helicity $+2$ with the quadratic
operators}

The twenty-five OPEs \footnote{In this case,
the conditions $s_1+s_2+s_3 \geq 0$ (more precisely
$s_1+s_2+s_3 \geq 2$), $s_1+s_2 \geq 0$ and $s_1+s_3
\geq 0$ are satisfied and three kinds of infinite sums are present.

As in the gluon amplitude \cite{GHP},
let us consider the $n$-point amplitude in the ${\cal N}=8$
supergravity $< \Phi_{+2}^{(h_1)}(z_1,\bar{z}_1)\,
\Phi_{+2}^{(h_2)}(z_2, \bar{z}_2)\, \Phi_{+2}^{(h_3)}(z_3,
\bar{z}_3) \, \cdots >$ where the first three celestial operators
have $s_1=s_2=s_3=+2$ and the remaining $(n-3)$ operators can be
any combinations among nine operators in Table \ref{hbar1}.
Of course, there are ninety-four different combinations of the
three helicities having the $SU(8)$ indicies.
By using the OPE between the particle $2$ and particle $3$, we obtain
\bea
&& < \Phi_{+2}^{(h_1)}(z_1,\bar{z}_1)\,
\Phi_{+2}^{(h_2)}(z_2, \bar{z}_2)\, \Phi_{+2}^{(h_3)}(z_3,
\bar{z}_3) \, \cdots >\nonu \\
&& =\frac{\bar{z}_{23}}{z_{23}}\,
\sum_{m=0}^{\infty}\, \frac{\bar{z}_{23}^m}{m!}\,
B(2\bar{h}_2+1+m,2\bar{h}_3+1) \,
< \Phi_{+2}^{(h_1)}(z_1,\bar{z}_1)\,
\bar{\pa}_{\bar{z}_3}^{m}\, \Phi_{+2}^{(h_2+h_3)}(z_3,
\bar{z}_3) \, \cdots >
\nonu \\
&&+ z_{23}^0 \, < \Phi_{+2}^{(h_1)}(z_1,\bar{z}_1)\,
(\Phi_{+2}^{(h_2)}\, \Phi_{+2}^{(h_3)})(z_3,
\bar{z}_3) \, \cdots >
+ O(z_{23}).
\label{AMP1}
\eea
Now we substitute the first OPE of (\ref{Aone}) into the second term of
(\ref{AMP1}).
Then we obtain the following amplitude
\bea
&& < \Phi_{+2}^{(h_1)}(z_1,\bar{z}_1)\,
\Phi_{+2}^{(h_2)}(z_2, \bar{z}_2)\, \Phi_{+2}^{(h_3)}(z_3,
\bar{z}_3) \, \cdots >\nonu \\
&& =\frac{\bar{z}_{23}}{z_{23}}\,
\sum_{m=0}^{\infty}\, \frac{\bar{z}_{23}^m}{m!}\,
B(2\bar{h}_2+1+m,2\bar{h}_3+1) \,
< \Phi_{+2}^{(h_1)}(z_1,\bar{z}_1)\,
\bar{\pa}_{\bar{z}_3}^{m}\, \Phi_{+2}^{(h_2+h_3)}(z_3,
\bar{z}_3) \, \cdots >
\nonu \\
&&+ z_{23}^0 \,
\frac{\bar{z}_{13}^2}{z_{13}^2} \,
\sum_{m=0}^{\infty}\, \frac{\bar{z}_{13}^m}{m!}\,
B(2\bar{h}_1+2+m,2\bar{h}_2+1, 2\bar{h}_3+1)
<
\bar{\pa}^m \, \Phi_{+2}^{(h_1+h_2+h_3)}(z_3,\bar{z}_3)
\, \cdots > \nonu \\
&&+ z_{23}^0 \, 
\frac{\bar{z}_{13}}{z_{13}} \,
\sum_{m=0}^{\infty}\, \frac{\bar{z}_{13}^m}{m!}\,
B(2\bar{h}_1+1+m,2\bar{h}_2+1) \,
< (\bar{\pa}^m \,
\Phi_{+2}^{(h_1+h_2)}\,\Phi_{+2}^{(h_3)})
(z_3,\bar{z}_3)
\, \cdots > \nonu \\
&&+ z_{23}^0 \,
\frac{\bar{z}_{13}}{z_{13}} \,
\sum_{m=0}^{\infty}\, \frac{\bar{z}_{13}^m}{m!}\,
B(2\bar{h}_1+1+m,2\bar{h}_3+1) \,
< (
\Phi_{+2}^{(h_2)}\,\bar{\pa}^m \, \Phi_{+2}^{(h_1+h_3)})
(z_3,\bar{z}_3) \, \cdots > + O(z_{13})
+ O(z_{23}).
\label{AMP2}
\eea
The leading term $\frac{1}{z_{23}}$
on the right-hand side of (\ref{AMP2})
is $(n-1)$-point amplitude while
the first subleading term $z_{23}^0$
on the right-hand side of (\ref{AMP2})
is $(n-2)$-point amplitude.
Other remaining ninety-four amplitudes having the $SU(8)$ indicies
can be written similarly.
The corresponding analysis in the bulk point of view for
the amplitude (\ref{AMP2}) is done in \cite{CP}.
}
are described as 
\bea
&& \Phi_{+2}^{(h_1)}(z_1,\bar{z}_1)\, (\Phi_{+2}^{(h_2)}\, \Phi_{+2}^{(h_3)})(z_3,
\bar{z}_3) = \frac{\bar{z}_{13}^2}{z_{13}^2} \,
\sum_{m=0}^{\infty}\, \frac{\bar{z}_{13}^m}{m!}\,
B(2\bar{h}_1+2+m,2\bar{h}_2+1, 2\bar{h}_3+1)
\nonu \\
&& \times
\bar{\pa}^m \, \Phi_{+2}^{(h_1+h_2+h_3)}(z_3,\bar{z}_3)
+
\frac{\bar{z}_{13}}{z_{13}} \,
\sum_{m=0}^{\infty}\, \frac{\bar{z}_{13}^m}{m!}\,
B(2\bar{h}_1+1+m,2\bar{h}_2+1) \, (\bar{\pa}^m \,
\Phi_{+2}^{(h_1+h_2)}\,\Phi_{+2}^{(h_3)})
(z_3,\bar{z}_3)
\nonu \\
&& +
\frac{\bar{z}_{13}}{z_{13}} \,
\sum_{m=0}^{\infty}\, \frac{\bar{z}_{13}^m}{m!}\,
B(2\bar{h}_1+1+m,2\bar{h}_3+1) \, (
\Phi_{+2}^{(h_2)}\,\bar{\pa}^m \, \Phi_{+2}^{(h_1+h_3)})
(z_3,\bar{z}_3),
\nonu \\
&& \Phi_{+2}^{(h_1)}(z_1,\bar{z}_1)\, (\Phi_{+2}^{(h_2)}\,
\Phi_{+\frac{3}{2}}^{(h_3),A})(z_3,
\bar{z}_3) = \frac{\bar{z}_{13}^2}{z_{13}^2} \,
\sum_{m=0}^{\infty}\, \frac{\bar{z}_{13}^m}{m!}\,
B(2\bar{h}_1+2+m,2\bar{h}_2+1, 2\bar{h}_3+1)
\nonu \\
&& \times
\bar{\pa}^m \, \Phi_{+\frac{3}{2}}^{(h_1+h_2+h_3),A}(z_3,\bar{z}_3)
+
\frac{\bar{z}_{13}}{z_{13}} \,
\sum_{m=0}^{\infty}\, \frac{\bar{z}_{13}^m}{m!}\,
B(2\bar{h}_1+1+m,2\bar{h}_2+1) \, (\bar{\pa}^m \,
\Phi_{+2}^{(h_1+h_2)}\,\Phi_{+\frac{3}{2}}^{(h_3),A})
(z_3,\bar{z}_3)
\nonu \\
&& +
\frac{\bar{z}_{13}}{z_{13}} \,
\sum_{m=0}^{\infty}\, \frac{\bar{z}_{13}^m}{m!}\,
B(2\bar{h}_1+1+m,2\bar{h}_3+1) \, (
\Phi_{+2}^{(h_2)}\,\bar{\pa}^m \, \Phi_{+\frac{3}{2}}^{(h_1+h_3),A})
(z_3,\bar{z}_3),
\nonu \\
&& \Phi_{+2}^{(h_1)}(z_1,\bar{z}_1)\, (\Phi_{+2}^{(h_2)}\, \Phi_{+1}^{(h_3),AB})(z_3,
\bar{z}_3) = \frac{\bar{z}_{13}^2}{z_{13}^2} \,
\sum_{m=0}^{\infty}\, \frac{\bar{z}_{13}^m}{m!}\,
B(2\bar{h}_1+2+m,2\bar{h}_2+1, 2\bar{h}_3+1)
\nonu \\
&& \times
\bar{\pa}^m \, \Phi_{+1}^{(h_1+h_2+h_3),AB}(z_3,\bar{z}_3)
\nonu \\
&&+
\frac{\bar{z}_{13}}{z_{13}} \,
\sum_{m=0}^{\infty}\, \frac{\bar{z}_{13}^m}{m!}\,
B(2\bar{h}_1+1+m,2\bar{h}_2+1) \, (\bar{\pa}^m \,
\Phi_{+2}^{(h_1+h_2)}\,\Phi_{+1}^{(h_3),AB})
(z_3,\bar{z}_3)
\nonu \\
&& +
\frac{\bar{z}_{13}}{z_{13}} \,
\sum_{m=0}^{\infty}\, \frac{\bar{z}_{13}^m}{m!}\,
B(2\bar{h}_1+1+m,2\bar{h}_3+1) \, (
\Phi_{+2}^{(h_2)}\,\bar{\pa}^m \, \Phi_{+1}^{(h_1+h_3),AB})
(z_3,\bar{z}_3),
\nonu \\
&& \Phi_{+2}^{(h_1)}(z_1,\bar{z}_1)\, (\Phi_{+2}^{(h_2)}\,
\Phi_{+\frac{1}{2}}^{(h_3),ABC})(z_3,
\bar{z}_3) = \frac{\bar{z}_{13}^2}{z_{13}^2} \,
\sum_{m=0}^{\infty}\, \frac{\bar{z}_{13}^m}{m!}\,
B(2\bar{h}_1+2+m,2\bar{h}_2+1, 2\bar{h}_3+1)
\nonu \\
&& \times
\bar{\pa}^m \, \Phi_{+\frac{1}{2}}^{(h_1+h_2+h_3),ABC}(z_3,\bar{z}_3)
\nonu \\
&&+
\frac{\bar{z}_{13}}{z_{13}} \,
\sum_{m=0}^{\infty}\, \frac{\bar{z}_{13}^m}{m!}\,
B(2\bar{h}_1+1+m,2\bar{h}_2+1) \, (\bar{\pa}^m \,
\Phi_{+2}^{(h_1+h_2)}\,\Phi_{+\frac{1}{2}}^{(h_3),ABC})
(z_3,\bar{z}_3)
\nonu \\
&& +
\frac{\bar{z}_{13}}{z_{13}} \,
\sum_{m=0}^{\infty}\, \frac{\bar{z}_{13}^m}{m!}\,
B(2\bar{h}_1+1+m,2\bar{h}_3+1) \, (
\Phi_{+2}^{(h_2)}\,\bar{\pa}^m \, \Phi_{+\frac{1}{2}}^{(h_1+h_3),ABC})
(z_3,\bar{z}_3),
\nonu \\
&& \Phi_{+2}^{(h_1)}(z_1,\bar{z}_1)\, (\Phi_{+2}^{(h_2)}\, \Phi_{0}^{(h_3),ABCD})(z_3,
\bar{z}_3) = \frac{\bar{z}_{13}^2}{z_{13}^2} \,
\sum_{m=0}^{\infty}\, \frac{\bar{z}_{13}^m}{m!}\,
B(2\bar{h}_1+2+m,2\bar{h}_2+1, 2\bar{h}_3+1)
\nonu \\
&& \times
\bar{\pa}^m \, \Phi_{0}^{(h_1+h_2+h_3),ABCD}(z_3,\bar{z}_3)
\nonu \\
&&+
\frac{\bar{z}_{13}}{z_{13}} \,
\sum_{m=0}^{\infty}\, \frac{\bar{z}_{13}^m}{m!}\,
B(2\bar{h}_1+1+m,2\bar{h}_2+1) \, (\bar{\pa}^m \,
\Phi_{+2}^{(h_1+h_2)}\,\Phi_{0}^{(h_3),ABCD})
(z_3,\bar{z}_3)
\nonu \\
&& +
\frac{\bar{z}_{13}}{z_{13}} \,
\sum_{m=0}^{\infty}\, \frac{\bar{z}_{13}^m}{m!}\,
B(2\bar{h}_1+1+m,2\bar{h}_3+1) \, (
\Phi_{+2}^{(h_2)}\,\bar{\pa}^m \, \Phi_{0}^{(h_1+h_3),ABCD})
(z_3,\bar{z}_3),
\nonu \\
&& \Phi_{+2}^{(h_1)}(z_1,\bar{z}_1)\, (\Phi_{+2}^{(h_2)}\,
\Phi_{ABC,-\frac{1}{2}}^{(h_3)})(z_3,
\bar{z}_3) = \frac{\bar{z}_{13}^2}{z_{13}^2} \,
\sum_{m=0}^{\infty}\, \frac{\bar{z}_{13}^m}{m!}\,
B(2\bar{h}_1+2+m,2\bar{h}_2+1, 2\bar{h}_3+1)
\nonu \\
&& \times
\bar{\pa}^m \, \Phi_{ABC,-\frac{1}{2}}^{(h_1+h_2+h_3)}(z_3,\bar{z}_3)
\nonu \\
&&+
\frac{\bar{z}_{13}}{z_{13}} \,
\sum_{m=0}^{\infty}\, \frac{\bar{z}_{13}^m}{m!}\,
B(2\bar{h}_1+1+m,2\bar{h}_2+1) \, (\bar{\pa}^m \,
\Phi_{+2}^{(h_1+h_2)}\,\Phi_{ABC,-\frac{1}{2}}^{(h_3)})
(z_3,\bar{z}_3)
\nonu \\
&& +
\frac{\bar{z}_{13}}{z_{13}} \,
\sum_{m=0}^{\infty}\, \frac{\bar{z}_{13}^m}{m!}\,
B(2\bar{h}_1+1+m,2\bar{h}_3+1) \, (
\Phi_{+2}^{(h_2)}\,\bar{\pa}^m \, \Phi_{ABC,-\frac{1}{2}}^{(h_1+h_3)})
(z_3,\bar{z}_3),
\nonu \\
&& \Phi_{+2}^{(h_1)}(z_1,\bar{z}_1)\,
(\Phi_{+2}^{(h_2)}\, \Phi_{AB,-1}^{(h_3)})(z_3,
\bar{z}_3) = \frac{\bar{z}_{13}^2}{z_{13}^2} \,
\sum_{m=0}^{\infty}\, \frac{\bar{z}_{13}^m}{m!}\,
B(2\bar{h}_1+2+m,2\bar{h}_2+1, 2\bar{h}_3+1)
\nonu \\
&& \times
\bar{\pa}^m \, \Phi_{AB,-1}^{(h_1+h_2+h_3)}(z_3,\bar{z}_3)
+
\frac{\bar{z}_{13}}{z_{13}} \,
\sum_{m=0}^{\infty}\, \frac{\bar{z}_{13}^m}{m!}\,
B(2\bar{h}_1+1+m,2\bar{h}_2+1) \, (\bar{\pa}^m \,
\Phi_{+2}^{(h_1+h_2)}\,\Phi_{AB,-1}^{(h_3)})
(z_3,\bar{z}_3)
\nonu \\
&& +
\frac{\bar{z}_{13}}{z_{13}} \,
\sum_{m=0}^{\infty}\, \frac{\bar{z}_{13}^m}{m!}\,
B(2\bar{h}_1+1+m,2\bar{h}_3+1) \, (
\Phi_{+2}^{(h_2)}\,\bar{\pa}^m \, \Phi_{AB,-1}^{(h_1+h_3)})
(z_3,\bar{z}_3),
\nonu \\
&& \Phi_{+2}^{(h_1)}(z_1,\bar{z}_1)\,
(\Phi_{+2}^{(h_2)}\, \Phi_{A,-\frac{3}{2}}^{(h_3)})(z_3,
\bar{z}_3) = \frac{\bar{z}_{13}^2}{z_{13}^2} \,
\sum_{m=0}^{\infty}\, \frac{\bar{z}_{13}^m}{m!}\,
B(2\bar{h}_1+2+m,2\bar{h}_2+1, 2\bar{h}_3+1)
\nonu \\
&& \times
\bar{\pa}^m \, \Phi_{A,-\frac{3}{2}}^{(h_1+h_2+h_3)}(z_3,\bar{z}_3)
+
\frac{\bar{z}_{13}}{z_{13}} \,
\sum_{m=0}^{\infty}\, \frac{\bar{z}_{13}^m}{m!}\,
B(2\bar{h}_1+1+m,2\bar{h}_2+1) \, (\bar{\pa}^m \,
\Phi_{+2}^{(h_1+h_2)}\,\Phi_{A,-\frac{3}{2}}^{(h_3)})
(z_3,\bar{z}_3)
\nonu \\
&& +
\frac{\bar{z}_{13}}{z_{13}} \,
\sum_{m=0}^{\infty}\, \frac{\bar{z}_{13}^m}{m!}\,
B(2\bar{h}_1+1+m,2\bar{h}_3+1) \, (
\Phi_{+2}^{(h_2)}\,\bar{\pa}^m \, \Phi_{A,-\frac{3}{2}}^{(h_1+h_3)})
(z_3,\bar{z}_3),
\nonu \\
&& \Phi_{+2}^{(h_1)}(z_1,\bar{z}_1)\, (\Phi_{+2}^{(h_2)}\, \Phi_{-2}^{(h_3)})(z_3,
\bar{z}_3) = \frac{\bar{z}_{13}^2}{z_{13}^2} \,
\sum_{m=0}^{\infty}\, \frac{\bar{z}_{13}^m}{m!}\,
B(2\bar{h}_1+2+m,2\bar{h}_2+1, 2\bar{h}_3+1)
\nonu \\
&& \times
\bar{\pa}^m \, \Phi_{-2}^{(h_1+h_2+h_3)}(z_3,\bar{z}_3)
+
\frac{\bar{z}_{13}}{z_{13}} \,
\sum_{m=0}^{\infty}\, \frac{\bar{z}_{13}^m}{m!}\,
B(2\bar{h}_1+1+m,2\bar{h}_2+1) \, (\bar{\pa}^m \,
\Phi_{+2}^{(h_1+h_2)}\,\Phi_{-2}^{(h_3)})
(z_3,\bar{z}_3)
\nonu \\
&& +
\frac{\bar{z}_{13}}{z_{13}} \,
\sum_{m=0}^{\infty}\, \frac{\bar{z}_{13}^m}{m!}\,
B(2\bar{h}_1+1+m,2\bar{h}_3+1) \, (
\Phi_{+2}^{(h_2)}\,\bar{\pa}^m \, \Phi_{-2}^{(h_1+h_3)})
(z_3,\bar{z}_3),
\nonu \\
&& \Phi_{+2}^{(h_1)}(z_1,\bar{z}_1)\,
(\Phi_{+\frac{3}{2}}^{(h_2),A}\, \Phi_{+\frac{3}{2}}^{(h_3),B})(z_3,
\bar{z}_3) = \frac{\bar{z}_{13}^2}{z_{13}^2} \,
\sum_{m=0}^{\infty}\, \frac{\bar{z}_{13}^m}{m!}\,
B(2\bar{h}_1+2+m,2\bar{h}_2+1, 2\bar{h}_3+1)
\nonu \\
&& \times
\bar{\pa}^m \, \Phi_{+1}^{(h_1+h_2+h_3),AB}(z_3,\bar{z}_3)
\nonu \\
&& +
\frac{\bar{z}_{13}}{z_{13}} \,
\sum_{m=0}^{\infty}\, \frac{\bar{z}_{13}^m}{m!}\,
B(2\bar{h}_1+1+m,2\bar{h}_2+1) \, (\bar{\pa}^m \,
\Phi_{+\frac{3}{2}}^{(h_1+h_2),A}\,\Phi_{+\frac{3}{2}}^{(h_3),B})
(z_3,\bar{z}_3)
\nonu \\
&& +
\frac{\bar{z}_{13}}{z_{13}} \,
\sum_{m=0}^{\infty}\, \frac{\bar{z}_{13}^m}{m!}\,
B(2\bar{h}_1+1+m,2\bar{h}_3+1) \, (
\Phi_{+\frac{3}{2}}^{(h_2),A}\,\bar{\pa}^m \, \Phi_{+\frac{3}{2}}^{(h_1+h_3),B})
(z_3,\bar{z}_3),
\nonu \\
&& \Phi_{+2}^{(h_1)}(z_1,\bar{z}_1)\,
(\Phi_{+\frac{3}{2}}^{(h_2),A}\, \Phi_{+1}^{(h_3),BC})(z_3,
\bar{z}_3) = \frac{\bar{z}_{13}^2}{z_{13}^2} \,
\sum_{m=0}^{\infty}\, \frac{\bar{z}_{13}^m}{m!}\,
B(2\bar{h}_1+2+m,2\bar{h}_2+1, 2\bar{h}_3+1)
\nonu \\
&& \times
\bar{\pa}^m \, \Phi_{+\frac{1}{2}}^{(h_1+h_2+h_3),ABC}(z_3,\bar{z}_3)
\nonu \\
&& +
\frac{\bar{z}_{13}}{z_{13}} \,
\sum_{m=0}^{\infty}\, \frac{\bar{z}_{13}^m}{m!}\,
B(2\bar{h}_1+1+m,2\bar{h}_2+1) \, (\bar{\pa}^m \,
\Phi_{+\frac{3}{2}}^{(h_1+h_2),A}\,\Phi_{+1}^{(h_3),BC})
(z_3,\bar{z}_3)
\nonu \\
&& +
\frac{\bar{z}_{13}}{z_{13}} \,
\sum_{m=0}^{\infty}\, \frac{\bar{z}_{13}^m}{m!}\,
B(2\bar{h}_1+1+m,2\bar{h}_3+1) \, (
\Phi_{+\frac{3}{2}}^{(h_2),A}\,\bar{\pa}^m \, \Phi_{+1}^{(h_1+h_3),BC})
(z_3,\bar{z}_3),
\nonu \\
&& \Phi_{+2}^{(h_1)}(z_1,\bar{z}_1)\,
(\Phi_{+\frac{3}{2}}^{(h_2),A}\, \Phi_{+\frac{1}{2}}^{(h_3),BCD})(z_3,
\bar{z}_3) = \frac{\bar{z}_{13}^2}{z_{13}^2} \,
\sum_{m=0}^{\infty}\, \frac{\bar{z}_{13}^m}{m!}\,
B(2\bar{h}_1+2+m,2\bar{h}_2+1, 2\bar{h}_3+1)
\nonu \\
&& \times
\bar{\pa}^m \, \Phi_{0}^{(h_1+h_2+h_3),ABCD}(z_3,\bar{z}_3)
\nonu \\
&& +
\frac{\bar{z}_{13}}{z_{13}} \,
\sum_{m=0}^{\infty}\, \frac{\bar{z}_{13}^m}{m!}\,
B(2\bar{h}_1+1+m,2\bar{h}_2+1) \, (\bar{\pa}^m \,
\Phi_{+\frac{3}{2}}^{(h_1+h_2),A}\,\Phi_{+\frac{1}{2}}^{(h_3),BCD})
(z_3,\bar{z}_3)
\nonu \\
&& +
\frac{\bar{z}_{13}}{z_{13}} \,
\sum_{m=0}^{\infty}\, \frac{\bar{z}_{13}^m}{m!}\,
B(2\bar{h}_1+1+m,2\bar{h}_3+1) \, (
\Phi_{+\frac{3}{2}}^{(h_2),A}\,\bar{\pa}^m \, \Phi_{+\frac{1}{2}}^{(h_1+h_3),BCD})
(z_3,\bar{z}_3),
\nonu \\
&& \Phi_{+2}^{(h_1)}(z_1,\bar{z}_1)\,
(\Phi_{+\frac{3}{2}}^{(h_2),A}\, \Phi_{0}^{(h_3),BCDE})(z_3,
\bar{z}_3) = \frac{\bar{z}_{13}^2}{z_{13}^2} \,
\sum_{m=0}^{\infty}\, \frac{\bar{z}_{13}^m}{m!}\,
B(2\bar{h}_1+2+m,2\bar{h}_2+1, 2\bar{h}_3+1)
\nonu \\
&& \times
\frac{1}{3!}\,
\epsilon^{ABCDEFGH}\,
\bar{\pa}^m \, \Phi_{FGH,-\frac{1}{2}}^{(h_1+h_2+h_3)}(z_3,\bar{z}_3)
\nonu \\
&& +
\frac{\bar{z}_{13}}{z_{13}} \,
\sum_{m=0}^{\infty}\, \frac{\bar{z}_{13}^m}{m!}\,
B(2\bar{h}_1+1+m,2\bar{h}_2+1) \, (\bar{\pa}^m \,
\Phi_{+\frac{3}{2}}^{(h_1+h_2),A}\,\Phi_{0}^{(h_3),BCDE})
(z_3,\bar{z}_3)
\nonu \\
&& +
\frac{\bar{z}_{13}}{z_{13}} \,
\sum_{m=0}^{\infty}\, \frac{\bar{z}_{13}^m}{m!}\,
B(2\bar{h}_1+1+m,2\bar{h}_3+1) \, (
\Phi_{+\frac{3}{2}}^{(h_2),A}\,\bar{\pa}^m \, \Phi_{0}^{(h_1+h_3),BCDE})
(z_3,\bar{z}_3),
\nonu \\
&& \Phi_{+2}^{(h_1)}(z_1,\bar{z}_1)\,
(\Phi_{+\frac{3}{2}}^{(h_2),A}\, \Phi_{BCD,-\frac{1}{2}}^{(h_3)})(z_3,
\bar{z}_3) = \frac{\bar{z}_{13}^2}{z_{13}^2} \,
\sum_{m=0}^{\infty}\, \frac{\bar{z}_{13}^m}{m!}\,
B(2\bar{h}_1+2+m,2\bar{h}_2+1, 2\bar{h}_3+1)
\nonu \\
&& \times
3 \, \delta^A_{[B} \, \bar{\pa}^m \, \Phi_{CD]-1}^{(h_1+h_2+h_3)}(z_3,\bar{z}_3)
\nonu \\
&& +
\frac{\bar{z}_{13}}{z_{13}} \,
\sum_{m=0}^{\infty}\, \frac{\bar{z}_{13}^m}{m!}\,
B(2\bar{h}_1+1+m,2\bar{h}_2+1) \, (\bar{\pa}^m \,
\Phi_{+\frac{3}{2}}^{(h_1+h_2),A}\,\Phi_{BCD,-\frac{1}{2}}^{(h_3)})
(z_3,\bar{z}_3)
\nonu \\
&& +
\frac{\bar{z}_{13}}{z_{13}} \,
\sum_{m=0}^{\infty}\, \frac{\bar{z}_{13}^m}{m!}\,
B(2\bar{h}_1+1+m,2\bar{h}_3+1) \, (
\Phi_{+\frac{3}{2}}^{(h_2),A}\,\bar{\pa}^m \, \Phi_{BCD,-\frac{1}{2}}^{(h_1+h_3)})
(z_3,\bar{z}_3),
\nonu \\
&& \Phi_{+2}^{(h_1)}(z_1,\bar{z}_1)\,
(\Phi_{+\frac{3}{2}}^{(h_2),A}\, \Phi_{BC,-1}^{(h_3)})(z_3,
\bar{z}_3) = \frac{\bar{z}_{13}^2}{z_{13}^2} \,
\sum_{m=0}^{\infty}\, \frac{\bar{z}_{13}^m}{m!}\,
B(2\bar{h}_1+2+m,2\bar{h}_2+1, 2\bar{h}_3+1)
\nonu \\
&& \times
2! \, \delta^A_{[B}\,  
  \bar{\pa}^m \, \Phi_{C],-\frac{3}{2}}^{(h_1+h_2+h_3)}(z_3,\bar{z}_3)
\nonu \\
&& +
\frac{\bar{z}_{13}}{z_{13}} \,
\sum_{m=0}^{\infty}\, \frac{\bar{z}_{13}^m}{m!}\,
B(2\bar{h}_1+1+m,2\bar{h}_2+1) \, (\bar{\pa}^m \,
\Phi_{+\frac{3}{2}}^{(h_1+h_2),A}\,\Phi_{BC,-1}^{(h_3)})
(z_3,\bar{z}_3)
\nonu \\
&& +
\frac{\bar{z}_{13}}{z_{13}} \,
\sum_{m=0}^{\infty}\, \frac{\bar{z}_{13}^m}{m!}\,
B(2\bar{h}_1+1+m,2\bar{h}_3+1) \, (
\Phi_{+\frac{3}{2}}^{(h_2),A}\,\bar{\pa}^m \, \Phi_{BC,-1}^{(h_1+h_3)})
(z_3,\bar{z}_3),
\nonu \\
&& \Phi_{+2}^{(h_1)}(z_1,\bar{z}_1)\,
(\Phi_{+\frac{3}{2}}^{(h_2),A}\, \Phi_{B,-\frac{3}{2}}^{(h_3)})(z_3,
\bar{z}_3) = \frac{\bar{z}_{13}^2}{z_{13}^2} \,
\sum_{m=0}^{\infty}\, \frac{\bar{z}_{13}^m}{m!}\,
B(2\bar{h}_1+2+m,2\bar{h}_2+1, 2\bar{h}_3+1)
\nonu \\
&& \times
\delta^A_B\,
\bar{\pa}^m \, \Phi_{-2}^{(h_1+h_2+h_3)}(z_3,\bar{z}_3)
\nonu \\
&& +
\frac{\bar{z}_{13}}{z_{13}} \,
\sum_{m=0}^{\infty}\, \frac{\bar{z}_{13}^m}{m!}\,
B(2\bar{h}_1+1+m,2\bar{h}_2+1) \, (\bar{\pa}^m \,
\Phi_{+\frac{3}{2}}^{(h_1+h_2),A}\,\Phi_{B,-\frac{3}{2}}^{(h_3)})
(z_3,\bar{z}_3)
\nonu \\
&& +
\frac{\bar{z}_{13}}{z_{13}} \,
\sum_{m=0}^{\infty}\, \frac{\bar{z}_{13}^m}{m!}\,
B(2\bar{h}_1+1+m,2\bar{h}_3+1) \, (
\Phi_{+\frac{3}{2}}^{(h_2),A}\,\bar{\pa}^m \, \Phi_{B,-\frac{3}{2}}^{(h_1+h_3)})
(z_3,\bar{z}_3),
\nonu \\
&& \Phi_{+2}^{(h_1)}(z_1,\bar{z}_1)\,
(\Phi_{+1}^{(h_2),AB}\, \Phi_{+1}^{(h_3),CD})(z_3,
\bar{z}_3) = \frac{\bar{z}_{13}^2}{z_{13}^2} \,
\sum_{m=0}^{\infty}\, \frac{\bar{z}_{13}^m}{m!}\,
B(2\bar{h}_1+2+m,2\bar{h}_2+1, 2\bar{h}_3+1)
\nonu \\
&& \times
\bar{\pa}^m \, \Phi_{0}^{(h_1+h_2+h_3),ABCD}(z_3,\bar{z}_3)
\nonu \\
&& +
\frac{\bar{z}_{13}}{z_{13}} \,
\sum_{m=0}^{\infty}\, \frac{\bar{z}_{13}^m}{m!}\,
B(2\bar{h}_1+1+m,2\bar{h}_2+1) \, (\bar{\pa}^m \,
\Phi_{+1}^{(h_1+h_2),AB}\,\Phi_{+1}^{(h_3),CD})
(z_3,\bar{z}_3)
\nonu \\
&& +
\frac{\bar{z}_{13}}{z_{13}} \,
\sum_{m=0}^{\infty}\, \frac{\bar{z}_{13}^m}{m!}\,
B(2\bar{h}_1+1+m,2\bar{h}_3+1) \, (
\Phi_{+1}^{(h_2),AB}\,\bar{\pa}^m \, \Phi_{+1}^{(h_1+h_3),CD})
(z_3,\bar{z}_3),
\nonu \\
&& \Phi_{+2}^{(h_1)}(z_1,\bar{z}_1)\,
(\Phi_{+1}^{(h_2),AB}\, \Phi_{+\frac{1}{2}}^{(h_3),CDE})(z_3,
\bar{z}_3) = \frac{\bar{z}_{13}^2}{z_{13}^2} \,
\sum_{m=0}^{\infty}\, \frac{\bar{z}_{13}^m}{m!}\,
B(2\bar{h}_1+2+m,2\bar{h}_2+1, 2\bar{h}_3+1)
\nonu \\
&& \times
\frac{1}{3!}\,
\ep^{ABCDEFGH} \,
\bar{\pa}^m \, \Phi_{FGH,-\frac{1}{2}}^{(h_1+h_2+h_3)}(z_3,\bar{z}_3)
\nonu \\
&& +
\frac{\bar{z}_{13}}{z_{13}} \,
\sum_{m=0}^{\infty}\, \frac{\bar{z}_{13}^m}{m!}\,
B(2\bar{h}_1+1+m,2\bar{h}_2+1) \, (\bar{\pa}^m \,
\Phi_{+1}^{(h_1+h_2),AB}\,\Phi_{+\frac{1}{2}}^{(h_3),CDE})
(z_3,\bar{z}_3)
\nonu \\
&& +
\frac{\bar{z}_{13}}{z_{13}} \,
\sum_{m=0}^{\infty}\, \frac{\bar{z}_{13}^m}{m!}\,
B(2\bar{h}_1+1+m,2\bar{h}_3+1) \, (
\Phi_{+1}^{(h_2),AB}\,\bar{\pa}^m \, \Phi_{+\frac{1}{2}}^{(h_1+h_3),CDE})
(z_3,\bar{z}_3),
\nonu \\
&& \Phi_{+2}^{(h_1)}(z_1,\bar{z}_1)\,
(\Phi_{+1}^{(h_2),AB}\, \Phi_{0}^{(h_3),CDEF})(z_3,
\bar{z}_3) = \frac{\bar{z}_{13}^2}{z_{13}^2} \,
\sum_{m=0}^{\infty}\, \frac{\bar{z}_{13}^m}{m!}\,
B(2\bar{h}_1+2+m,2\bar{h}_2+1, 2\bar{h}_3+1)
\nonu \\
&& \times
\frac{1}{2!}\, \ep^{ABCDEFGH} \,
\bar{\pa}^m \, \Phi_{GH,-1}^{(h_1+h_2+h_3)}(z_3,\bar{z}_3)
\nonu \\
&& +
\frac{\bar{z}_{13}}{z_{13}} \,
\sum_{m=0}^{\infty}\, \frac{\bar{z}_{13}^m}{m!}\,
B(2\bar{h}_1+1+m,2\bar{h}_2+1) \, (\bar{\pa}^m \,
\Phi_{+1}^{(h_1+h_2),AB}\,\Phi_{0}^{(h_3),CDEF})
(z_3,\bar{z}_3)
\nonu \\
&& +
\frac{\bar{z}_{13}}{z_{13}} \,
\sum_{m=0}^{\infty}\, \frac{\bar{z}_{13}^m}{m!}\,
B(2\bar{h}_1+1+m,2\bar{h}_3+1) \, (
\Phi_{+1}^{(h_2),AB}\,\bar{\pa}^m \, \Phi_{0}^{(h_1+h_3),CDEF})
(z_3,\bar{z}_3),
\nonu \\
&& \Phi_{+2}^{(h_1)}(z_1,\bar{z}_1)\,
(\Phi_{+1}^{(h_2),AB}\, \Phi_{CDE,-\frac{1}{2}}^{(h_3)})(z_3,
\bar{z}_3) = \frac{\bar{z}_{13}^2}{z_{13}^2} \,
\sum_{m=0}^{\infty}\, \frac{\bar{z}_{13}^m}{m!}\,
B(2\bar{h}_1+2+m,2\bar{h}_2+1, 2\bar{h}_3+1)
\nonu \\
&& \times
3!\, \de^A_{[C}\,
\bar{\pa}^m \, \Phi_{D,-\frac{3}{2}}^{(h_1+h_2+h_3)}(z_3,\bar{z}_3)
\, \delta^B_{E]}
\nonu \\
&& +
\frac{\bar{z}_{13}}{z_{13}} \,
\sum_{m=0}^{\infty}\, \frac{\bar{z}_{13}^m}{m!}\,
B(2\bar{h}_1+1+m,2\bar{h}_2+1) \, (\bar{\pa}^m \,
\Phi_{+1}^{(h_1+h_2),AB}\,\Phi_{CDE,-\frac{1}{2}}^{(h_3)})
(z_3,\bar{z}_3)
\nonu \\
&& +
\frac{\bar{z}_{13}}{z_{13}} \,
\sum_{m=0}^{\infty}\, \frac{\bar{z}_{13}^m}{m!}\,
B(2\bar{h}_1+1+m,2\bar{h}_3+1) \, (
\Phi_{+1}^{(h_2),AB}\,\bar{\pa}^m \, \Phi_{CDE,-\frac{1}{2}}^{(h_1+h_3)})
(z_3,\bar{z}_3),
\nonu \\
&& \Phi_{+2}^{(h_1)}(z_1,\bar{z}_1)\,
(\Phi_{+1}^{(h_2),AB}\, \Phi_{CD,-1}^{(h_3)})(z_3,
\bar{z}_3) = \frac{\bar{z}_{13}^2}{z_{13}^2} \,
\sum_{m=0}^{\infty}\, \frac{\bar{z}_{13}^m}{m!}\,
B(2\bar{h}_1+2+m,2\bar{h}_2+1, 2\bar{h}_3+1)
\nonu \\
&& \times
\delta^{AB}_{CD}\,
\bar{\pa}^m \, \Phi_{-2}^{(h_1+h_2+h_3)}(z_3,\bar{z}_3)
\nonu \\
&& +
\frac{\bar{z}_{13}}{z_{13}} \,
\sum_{m=0}^{\infty}\, \frac{\bar{z}_{13}^m}{m!}\,
B(2\bar{h}_1+1+m,2\bar{h}_2+1) \, (\bar{\pa}^m \,
\Phi_{+1}^{(h_1+h_2),AB}\,\Phi_{CD,-1}^{(h_3)})
(z_3,\bar{z}_3)
\nonu \\
&& +
\frac{\bar{z}_{13}}{z_{13}} \,
\sum_{m=0}^{\infty}\, \frac{\bar{z}_{13}^m}{m!}\,
B(2\bar{h}_1+1+m,2\bar{h}_3+1) \, (
\Phi_{+1}^{(h_2),AB}\,\bar{\pa}^m \, \Phi_{CD,-1}^{(h_1+h_3)})
(z_3,\bar{z}_3),
\nonu \\
&&
\Phi_{+2}^{(h_1)}(z_1,\bar{z}_1)\,
(\Phi_{+\frac{1}{2}}^{(h_2),ABC}\, \Phi_{+\frac{1}{2}}^{(h_3),DEF})(z_3,
\bar{z}_3) = \frac{\bar{z}_{13}^2}{z_{13}^2} \,
\sum_{m=0}^{\infty}\, \frac{\bar{z}_{13}^m}{m!}\,
B(2\bar{h}_1+2+m,2\bar{h}_2+1, 2\bar{h}_3+1)
\nonu \\
&& \times
\frac{1}{2!}\, \ep^{ABCDEFGH}
\bar{\pa}^m \, \Phi_{GH,-1}^{(h_1+h_2+h_3)}(z_3,\bar{z}_3)
\nonu \\
&& +
\frac{\bar{z}_{13}}{z_{13}} \,
\sum_{m=0}^{\infty}\, \frac{\bar{z}_{13}^m}{m!}\,
B(2\bar{h}_1+1+m,2\bar{h}_2+1) \, (\bar{\pa}^m \,
\Phi_{+\frac{1}{2}}^{(h_1+h_2),ABC}\,\Phi_{+\frac{1}{2}}^{(h_3),DEF})
(z_3,\bar{z}_3)
\nonu \\
&& +
\frac{\bar{z}_{13}}{z_{13}} \,
\sum_{m=0}^{\infty}\, \frac{\bar{z}_{13}^m}{m!}\,
B(2\bar{h}_1+1+m,2\bar{h}_3+1) \, (
\Phi_{+\frac{1}{2}}^{(h_2),ABC}\,\bar{\pa}^m \, \Phi_{+\frac{1}{2}}^{(h_1+h_3),DEF})
(z_3,\bar{z}_3),
\nonu \\
&&
\Phi_{+2}^{(h_1)}(z_1,\bar{z}_1)\,
(\Phi_{+\frac{1}{2}}^{(h_2),ABC}\, \Phi_{0}^{(h_3),DEFG})(z_3,
\bar{z}_3) = \frac{\bar{z}_{13}^2}{z_{13}^2} \,
\sum_{m=0}^{\infty}\, \frac{\bar{z}_{13}^m}{m!}\,
B(2\bar{h}_1+2+m,2\bar{h}_2+1, 2\bar{h}_3+1)
\nonu \\
&& \times
 \ep^{ABCDEFGH}
 \bar{\pa}^m \, \Phi_{H,-\frac{3}{2}}^{(h_1+h_2+h_3)}(z_3,\bar{z}_3)
 \nonu \\
&& +
\frac{\bar{z}_{13}}{z_{13}} \,
\sum_{m=0}^{\infty}\, \frac{\bar{z}_{13}^m}{m!}\,
B(2\bar{h}_1+1+m,2\bar{h}_2+1) \, (\bar{\pa}^m \,
\Phi_{+\frac{1}{2}}^{(h_1+h_2),ABC}\,\Phi_{0}^{(h_3),DEFG})
(z_3,\bar{z}_3)
\nonu \\
&& +
\frac{\bar{z}_{13}}{z_{13}} \,
\sum_{m=0}^{\infty}\, \frac{\bar{z}_{13}^m}{m!}\,
B(2\bar{h}_1+1+m,2\bar{h}_3+1) \, (
\Phi_{+\frac{1}{2}}^{(h_2),ABC}\,\bar{\pa}^m \, \Phi_{0}^{(h_1+h_3),DEFG})
(z_3,\bar{z}_3),
\nonu \\
&&
\Phi_{+2}^{(h_1)}(z_1,\bar{z}_1)\,
(\Phi_{+\frac{1}{2}}^{(h_2),ABC}\, \Phi_{DEF,-\frac{1}{2}}^{(h_3)})(z_3,
\bar{z}_3) = \frac{\bar{z}_{13}^2}{z_{13}^2} \,
\sum_{m=0}^{\infty}\, \frac{\bar{z}_{13}^m}{m!}\,
B(2\bar{h}_1+2+m,2\bar{h}_2+1, 2\bar{h}_3+1)
\nonu \\
&& \times
 \de^{ABC}_{DEF}\, 
 \bar{\pa}^m \, \Phi_{-2}^{(h_1+h_2+h_3)}(z_3,\bar{z}_3)
  \nonu \\
&& +
\frac{\bar{z}_{13}}{z_{13}} \,
\sum_{m=0}^{\infty}\, \frac{\bar{z}_{13}^m}{m!}\,
B(2\bar{h}_1+1+m,2\bar{h}_2+1) \, (\bar{\pa}^m \,
\Phi_{+\frac{1}{2}}^{(h_1+h_2),ABC}\,\Phi_{DEF,-\frac{1}{2}}^{(h_3)})
(z_3,\bar{z}_3)
\nonu \\
&& +
\frac{\bar{z}_{13}}{z_{13}} \,
\sum_{m=0}^{\infty}\, \frac{\bar{z}_{13}^m}{m!}\,
B(2\bar{h}_1+1+m,2\bar{h}_3+1) \, (
\Phi_{+\frac{1}{2}}^{(h_2),ABC}\,\bar{\pa}^m \, \Phi_{DEF,-\frac{1}{2}}^{(h_1+h_3)})
(z_3,\bar{z}_3),
\nonu \\
&&
\Phi_{+2}^{(h_1)}(z_1,\bar{z}_1)\,
(\Phi_{0}^{(h_2),ABCD}\, \Phi_{0}^{(h_3),EFGH})(z_3,
\bar{z}_3) = \frac{\bar{z}_{13}^2}{z_{13}^2} \,
\sum_{m=0}^{\infty}\, \frac{\bar{z}_{13}^m}{m!}\,
B(2\bar{h}_1+2+m,2\bar{h}_2+1, 2\bar{h}_3+1)
\nonu \\
&& \times
\ep^{ABCDEFGH}
\bar{\pa}^m \, \Phi_{-2}^{(h_1+h_2+h_3)}(z_3,\bar{z}_3)
  \nonu \\
&& +
\frac{\bar{z}_{13}}{z_{13}} \,
\sum_{m=0}^{\infty}\, \frac{\bar{z}_{13}^m}{m!}\,
B(2\bar{h}_1+1+m,2\bar{h}_2+1) \, (\bar{\pa}^m \,
\Phi_{0}^{(h_1+h_2),ABCD}\,\Phi_{0}^{(h_3),EFGH})
(z_3,\bar{z}_3)
\nonu \\
&& +
\frac{\bar{z}_{13}}{z_{13}} \,
\sum_{m=0}^{\infty}\, \frac{\bar{z}_{13}^m}{m!}\,
B(2\bar{h}_1+1+m,2\bar{h}_3+1) \, (
\Phi_{0}^{(h_2),ABCD}\,\bar{\pa}^m \, \Phi_{0}^{(h_1+h_3),EFGH})
(z_3,\bar{z}_3).
\label{Aone}
\eea
Note that there are the (holomorphic)
second and the first-order poles
in (\ref{Aone}).

\enlargethispage{\baselineskip}

\subsection{The OPEs of
the gravitinos  of helicity $+\frac{3}{2}$ with the quadratic
operators}

The twenty OPEs
\footnote{ The five OPEs we do not write down explicitly do not satisfy
the condition $s_1+s_2+s_3 \geq 2$.
Recall that from the defining OPEs of
the particle $1$ and particle $2$ in \cite{AK2509},
the twenty-five OPEs have the helicity
property of $s_1 \geq s_2$.
Sometimes we need to compute the fundamental OPEs
where $s_1 < s_2$. For example, for the first OPE of (\ref{full3half}),
we should compute the OPE of the gravitinos
with helicity $+\frac{3}{2}$ with the graviton
with helicity $+2$.
Let us start with the OPE
of the graviton
with helicity $+2$ with  the gravitinos
with helicity $+\frac{3}{2}$ in \cite{AK2509}.
This OPE is equal to
$\Phi_{+\frac{3}{2}}^{(h_1),P}(z_2,\bar{z}_2)\,
\Phi_{+2}^{(h_2)}(z_1,\bar{z}_1)$.
Now we interchange the complex coordinates
$z_1 \leftrightarrow z_2$ and $\bar{z}_1 \leftrightarrow
\bar{z}_2$. Then the previous complex coordinate
dependence $\frac{\bar{z}_{12}}{z_{12}}$ becomes
$\frac{\bar{z}_{21}}{z_{21}}$ which is equal to
$\frac{\bar{z}_{12}}{z_{12}}$. Moreover by interchanging
the conformal dimension
$\De_1 \leftrightarrow \De_2$, we have the Euler
beta function $B(\De_2-1,\De_1-\frac{1}{2})$
which is equal to $B(\De_1-\frac{1}{2},\De_2-1)=
B(2\bar{h}_1+1,2 \bar{h}_2+1)$.
Also the complex coordinate dependence  for the
gravitinos appearing on the right-hand side
can be expanded in terms of Taylor series.
A further expansion in $\bar{z}_{12}$ provides the
final result for the OPE $\Phi_{+\frac{3}{2}}^{(h_1),P}(z_1,\bar{z}_1)\,
\Phi_{+2}^{(h_2)}(z_2,\bar{z}_2)$. In this case,
the couplings satisfy
$\kappa_{+\frac{3}{2},+2, -\frac{3}{2}}=\kappa_{+2,+\frac{3}{2}, -\frac{3}{2}}$.
See also the second term of
the first OPE in (\ref{full3half}).
For the two fermionic operators, there exists an extra minus sign.
Moreover, for the two operators having several $SU(8)$ indices
respectively, we should be careful about the various signs
by moving these indices properly. See also Appendix E.}
are given by
\bea
&&\Phi_{+\frac{3}{2}}^{(h_1),P}(z_1,\bar{z}_1)\,
(\Phi_{+2}^{(h_2)}\, \Phi_{+2}^{(h_3)})(z_3,
\bar{z}_3) = \frac{\bar{z}_{13}^2}{z_{13}^2} \,
\sum_{m=0}^{\infty}\, \frac{\bar{z}_{13}^m}{m!}\,
B(2\bar{h}_1+2+m,2\bar{h}_2+1, 2\bar{h}_3+1)
\nonu \\
&& \times
\bar{\pa}^m \, \Phi_{+\frac{3}{2}}^{(h_1+h_2+h_3),P}(z_3,\bar{z}_3)
 \nonu \\
&& +
\frac{\bar{z}_{13}}{z_{13}} \,
\sum_{m=0}^{\infty}\, \frac{\bar{z}_{13}^m}{m!}\,
B(2\bar{h}_1+1+m,2\bar{h}_2+1) \, (\bar{\pa}^m \,
\Phi_{+\frac{3}{2}}^{(h_1+h_2),P}\,\Phi_{+2}^{(h_3)})
(z_3,\bar{z}_3)
\nonu \\
&& +
\frac{\bar{z}_{13}}{z_{13}} \,
\sum_{m=0}^{\infty}\, \frac{\bar{z}_{13}^m}{m!}\,
B(2\bar{h}_1+1+m,2\bar{h}_3+1) \, (
\Phi_{+2}^{(h_2)}\,\bar{\pa}^m \, \Phi_{+\frac{3}{2}}^{(h_1+h_3),P})
(z_3,\bar{z}_3),
\nonu \\
&&
\Phi_{+\frac{3}{2}}^{(h_1),P}(z_1,\bar{z}_1)\,
(\Phi_{+2}^{(h_2)}\, \Phi_{+\frac{3}{2}}^{(h_3),A})(z_3,
\bar{z}_3) = \frac{\bar{z}_{13}^2}{z_{13}^2} \,
\sum_{m=0}^{\infty}\, \frac{\bar{z}_{13}^m}{m!}\,
B(2\bar{h}_1+2+m,2\bar{h}_2+1, 2\bar{h}_3+1)
\nonu \\
&& \times
\bar{\pa}^m \, \Phi_{+1}^{(h_1+h_2+h_3),PA}(z_3,\bar{z}_3)
 \nonu \\
&& +
\frac{\bar{z}_{13}}{z_{13}} \,
\sum_{m=0}^{\infty}\, \frac{\bar{z}_{13}^m}{m!}\,
B(2\bar{h}_1+1+m,2\bar{h}_2+1) \, (\bar{\pa}^m \,
\Phi_{+\frac{3}{2}}^{(h_1+h_2),P}\,\Phi_{+\frac{3}{2}}^{(h_3),A})
(z_3,\bar{z}_3)
\nonu \\
&& +
\frac{\bar{z}_{13}}{z_{13}} \,
\sum_{m=0}^{\infty}\, \frac{\bar{z}_{13}^m}{m!}\,
B(2\bar{h}_1+1+m,2\bar{h}_3+1) \, (
\Phi_{+2}^{(h_2)}\,\bar{\pa}^m \, \Phi_{+1}^{(h_1+h_3),PA})
(z_3,\bar{z}_3),
\nonu \\
&&
\Phi_{+\frac{3}{2}}^{(h_1),P}(z_1,\bar{z}_1)\,
(\Phi_{+2}^{(h_2)}\, \Phi_{+1}^{(h_3),AB})(z_3,
\bar{z}_3) = \frac{\bar{z}_{13}^2}{z_{13}^2} \,
\sum_{m=0}^{\infty}\, \frac{\bar{z}_{13}^m}{m!}\,
B(2\bar{h}_1+2+m,2\bar{h}_2+1, 2\bar{h}_3+1)
\nonu \\
&& \times
\bar{\pa}^m \, \Phi_{+\frac{1}{2}}^{(h_1+h_2+h_3),PAB}(z_3,\bar{z}_3)
 \nonu \\
&& +
\frac{\bar{z}_{13}}{z_{13}} \,
\sum_{m=0}^{\infty}\, \frac{\bar{z}_{13}^m}{m!}\,
B(2\bar{h}_1+1+m,2\bar{h}_2+1) \, (\bar{\pa}^m \,
\Phi_{+\frac{3}{2}}^{(h_1+h_2),P}\,\Phi_{+1}^{(h_3),AB})
(z_3,\bar{z}_3)
\nonu \\
&& +
\frac{\bar{z}_{13}}{z_{13}} \,
\sum_{m=0}^{\infty}\, \frac{\bar{z}_{13}^m}{m!}\,
B(2\bar{h}_1+1+m,2\bar{h}_3+1) \, (
\Phi_{+2}^{(h_2)}\,\bar{\pa}^m \, \Phi_{+\frac{1}{2}}^{(h_1+h_3),PAB})
(z_3,\bar{z}_3),
\nonu \\
&&
\Phi_{+\frac{3}{2}}^{(h_1),P}(z_1,\bar{z}_1)\,
(\Phi_{+2}^{(h_2)}\, \Phi_{+\frac{1}{2}}^{(h_3),ABC})(z_3,
\bar{z}_3) = \frac{\bar{z}_{13}^2}{z_{13}^2} \,
\sum_{m=0}^{\infty}\, \frac{\bar{z}_{13}^m}{m!}\,
B(2\bar{h}_1+2+m,2\bar{h}_2+1, 2\bar{h}_3+1)
\nonu \\
&& \times
\bar{\pa}^m \, \Phi_{0}^{(h_1+h_2+h_3),PABC}(z_3,\bar{z}_3)
 \nonu \\
&& +
\frac{\bar{z}_{13}}{z_{13}} \,
\sum_{m=0}^{\infty}\, \frac{\bar{z}_{13}^m}{m!}\,
B(2\bar{h}_1+1+m,2\bar{h}_2+1) \, (\bar{\pa}^m \,
\Phi_{+\frac{3}{2}}^{(h_1+h_2),P}\,\Phi_{+\frac{1}{2}}^{(h_3),ABC})
(z_3,\bar{z}_3)
\nonu \\
&& +
\frac{\bar{z}_{13}}{z_{13}} \,
\sum_{m=0}^{\infty}\, \frac{\bar{z}_{13}^m}{m!}\,
B(2\bar{h}_1+1+m,2\bar{h}_3+1) \, (
\Phi_{+2}^{(h_2)}\,\bar{\pa}^m \, \Phi_{0}^{(h_1+h_3),PABC})
(z_3,\bar{z}_3),
\nonu \\
&&
\Phi_{+\frac{3}{2}}^{(h_1),P}(z_1,\bar{z}_1)\,
(\Phi_{+2}^{(h_2)}\, \Phi_{0}^{(h_3),ABCD})(z_3,
\bar{z}_3) = \frac{\bar{z}_{13}^2}{z_{13}^2} \,
\sum_{m=0}^{\infty}\, \frac{\bar{z}_{13}^m}{m!}\,
B(2\bar{h}_1+2+m,2\bar{h}_2+1, 2\bar{h}_3+1)
\nonu \\
&& \times
\frac{1}{3!}\,
\ep^{PABCDEFG}\,
\bar{\pa}^m \, \Phi_{EFG,-\frac{1}{2}}^{(h_1+h_2+h_3)}(z_3,\bar{z}_3)
 \nonu \\
&& +
\frac{\bar{z}_{13}}{z_{13}} \,
\sum_{m=0}^{\infty}\, \frac{\bar{z}_{13}^m}{m!}\,
B(2\bar{h}_1+1+m,2\bar{h}_2+1) \, (\bar{\pa}^m \,
\Phi_{+\frac{3}{2}}^{(h_1+h_2),P}\,\Phi_{0}^{(h_3),ABCD})
(z_3,\bar{z}_3)
\nonu \\
&& +
\frac{\bar{z}_{13}}{z_{13}} \,
\sum_{m=0}^{\infty}\, \frac{\bar{z}_{13}^m}{m!}\,
B(2\bar{h}_1+1+m,2\bar{h}_3+1) \, \frac{1}{3!}\,
\ep^{PABCDFGH}\, (
\Phi_{+2}^{(h_2)}\,\bar{\pa}^m \, \Phi_{FGH,-\frac{1}{2}}^{(h_1+h_3)})
(z_3,\bar{z}_3),
\nonu \\
&&
\Phi_{+\frac{3}{2}}^{(h_1),P}(z_1,\bar{z}_1)\,
(\Phi_{+2}^{(h_2)}\, \Phi_{ABC,-\frac{1}{2}}^{(h_3)})(z_3,
\bar{z}_3) = \frac{\bar{z}_{13}^2}{z_{13}^2} \,
\sum_{m=0}^{\infty}\, \frac{\bar{z}_{13}^m}{m!}\,
B(2\bar{h}_1+2+m,2\bar{h}_2+1, 2\bar{h}_3+1)
\nonu \\
&& \times
3 \, \de^{P}_{[A}
  \bar{\pa}^m \, \Phi_{BC],-1}^{(h_1+h_2+h_3)}(z_3,\bar{z}_3)
 \nonu \\
&& +
\frac{\bar{z}_{13}}{z_{13}} \,
\sum_{m=0}^{\infty}\, \frac{\bar{z}_{13}^m}{m!}\,
B(2\bar{h}_1+1+m,2\bar{h}_2+1) \, (\bar{\pa}^m \,
\Phi_{+\frac{3}{2}}^{(h_1+h_2),P}\,\Phi_{ABC,-\frac{1}{2}}^{(h_3)})
(z_3,\bar{z}_3)
\nonu \\
&& +
\frac{\bar{z}_{13}}{z_{13}} \,
\sum_{m=0}^{\infty}\, \frac{\bar{z}_{13}^m}{m!}\,
B(2\bar{h}_1+1+m,2\bar{h}_3+1) \, 3 \, \de^P_{[A} (
\Phi_{+2}^{(h_2)}\,\bar{\pa}^m \, \Phi_{BC],-1}^{(h_1+h_3)})
(z_3,\bar{z}_3),
\nonu \\
&&
\Phi_{+\frac{3}{2}}^{(h_1),P}(z_1,\bar{z}_1)\,
(\Phi_{+2}^{(h_2)}\, \Phi_{AB,-1}^{(h_3)})(z_3,
\bar{z}_3) = \frac{\bar{z}_{13}^2}{z_{13}^2} \,
\sum_{m=0}^{\infty}\, \frac{\bar{z}_{13}^m}{m!}\,
B(2\bar{h}_1+2+m,2\bar{h}_2+1, 2\bar{h}_3+1)
\nonu \\
&& \times
2! \, \de^{P}_{[A}
  \bar{\pa}^m \, \Phi_{B], -\frac{3}{2}}^{(h_1+h_2+h_3)}(z_3,\bar{z}_3)
 \nonu \\
&& +
\frac{\bar{z}_{13}}{z_{13}} \,
\sum_{m=0}^{\infty}\, \frac{\bar{z}_{13}^m}{m!}\,
B(2\bar{h}_1+1+m,2\bar{h}_2+1) \, (\bar{\pa}^m \,
\Phi_{+\frac{3}{2}}^{(h_1+h_2),P}\,\Phi_{AB,-1}^{(h_3)})
(z_3,\bar{z}_3)
\nonu \\
&& +
\frac{\bar{z}_{13}}{z_{13}} \,
\sum_{m=0}^{\infty}\, \frac{\bar{z}_{13}^m}{m!}\,
B(2\bar{h}_1+1+m,2\bar{h}_3+1) \, 2! \, \de^P_{[A} (
\Phi_{+2}^{(h_2)}\,\bar{\pa}^m \, \Phi_{B],-\frac{3}{2}}^{(h_1+h_3)})
(z_3,\bar{z}_3),
\nonu \\
&&
\Phi_{+\frac{3}{2}}^{(h_1),P}(z_1,\bar{z}_1)\,
(\Phi_{+2}^{(h_2)}\, \Phi_{A,-\frac{3}{2}}^{(h_3)})(z_3,
\bar{z}_3) = \frac{\bar{z}_{13}^2}{z_{13}^2} \,
\sum_{m=0}^{\infty}\, \frac{\bar{z}_{13}^m}{m!}\,
B(2\bar{h}_1+2+m,2\bar{h}_2+1, 2\bar{h}_3+1)
\nonu \\
&& \times
\de^{P}_{A}\,
\bar{\pa}^m \, \Phi_{-2}^{(h_1+h_2+h_3)}(z_3,\bar{z}_3)
 \nonu \\
&& +
\frac{\bar{z}_{13}}{z_{13}} \,
\sum_{m=0}^{\infty}\, \frac{\bar{z}_{13}^m}{m!}\,
B(2\bar{h}_1+1+m,2\bar{h}_2+1) \, (\bar{\pa}^m \,
\Phi_{+\frac{3}{2}}^{(h_1+h_2),P}\,\Phi_{A,-\frac{3}{2}}^{(h_3)})
(z_3,\bar{z}_3)
\nonu \\
&& +
\frac{\bar{z}_{13}}{z_{13}} \,
\sum_{m=0}^{\infty}\, \frac{\bar{z}_{13}^m}{m!}\,
B(2\bar{h}_1+1+m,2\bar{h}_3+1) \,  \de^{P}_{A}\, (
\Phi_{+2}^{(h_2)}\,\bar{\pa}^m \, \Phi_{-2}^{(h_1+h_3)})
(z_3,\bar{z}_3),
\nonu \\
&&
\Phi_{+\frac{3}{2}}^{(h_1),P}(z_1,\bar{z}_1)\,
(\Phi_{+\frac{3}{2}}^{(h_2),A}\, \Phi_{+\frac{3}{2}}^{(h_3),B})(z_3,
\bar{z}_3) = \frac{\bar{z}_{13}^2}{z_{13}^2} \,
\sum_{m=0}^{\infty}\, \frac{\bar{z}_{13}^m}{m!}\,
B(2\bar{h}_1+2+m,2\bar{h}_2+1, 2\bar{h}_3+1)
\nonu \\
&& \times
\bar{\pa}^m \, \Phi_{+\frac{1}{2}}^{(h_1+h_2+h_3),PAB}(z_3,\bar{z}_3)
 \nonu \\
&& +
\frac{\bar{z}_{13}}{z_{13}} \,
\sum_{m=0}^{\infty}\, \frac{\bar{z}_{13}^m}{m!}\,
B(2\bar{h}_1+1+m,2\bar{h}_2+1) \, (\bar{\pa}^m \,
\Phi_{+1}^{(h_1+h_2),PA}\,\Phi_{+\frac{3}{2}}^{(h_3),B})
(z_3,\bar{z}_3)
\nonu \\
&& +
\frac{\bar{z}_{13}}{z_{13}} \,
\sum_{m=0}^{\infty}\, \frac{\bar{z}_{13}^m}{m!}\,
B(2\bar{h}_1+1+m,2\bar{h}_3+1) \, (-1)\,   (
\Phi_{+\frac{3}{2}}^{(h_2),A}\,\bar{\pa}^m \, \Phi_{+1}^{(h_1+h_3),PB})
(z_3,\bar{z}_3),
\nonu \\
&&
\Phi_{+\frac{3}{2}}^{(h_1),P}(z_1,\bar{z}_1)\,
(\Phi_{+\frac{3}{2}}^{(h_2),A}\, \Phi_{+1}^{(h_3),BC})(z_3,
\bar{z}_3) = \frac{\bar{z}_{13}^2}{z_{13}^2} \,
\sum_{m=0}^{\infty}\, \frac{\bar{z}_{13}^m}{m!}\,
B(2\bar{h}_1+2+m,2\bar{h}_2+1, 2\bar{h}_3+1)
\nonu \\
&& \times
\bar{\pa}^m \, \Phi_{0}^{(h_1+h_2+h_3),PABC}(z_3,\bar{z}_3)
 \nonu \\
&& +
\frac{\bar{z}_{13}}{z_{13}} \,
\sum_{m=0}^{\infty}\, \frac{\bar{z}_{13}^m}{m!}\,
B(2\bar{h}_1+1+m,2\bar{h}_2+1) \, (\bar{\pa}^m \,
\Phi_{+1}^{(h_1+h_2),PA}\,\Phi_{+1}^{(h_3),BC})
(z_3,\bar{z}_3)
\nonu \\
&& +
\frac{\bar{z}_{13}}{z_{13}} \,
\sum_{m=0}^{\infty}\, \frac{\bar{z}_{13}^m}{m!}\,
B(2\bar{h}_1+1+m,2\bar{h}_3+1) \, (-1)\,   (
\Phi_{+\frac{3}{2}}^{(h_2),A}\,\bar{\pa}^m \, \Phi_{+\frac{1}{2}}^{(h_1+h_3),PBC})
(z_3,\bar{z}_3),
\nonu \\
&&
\Phi_{+\frac{3}{2}}^{(h_1),P}(z_1,\bar{z}_1)\,
(\Phi_{+\frac{3}{2}}^{(h_2),A}\, \Phi_{+\frac{1}{2}}^{(h_3),BCD})(z_3,
\bar{z}_3) = \frac{\bar{z}_{13}^2}{z_{13}^2} \,
\sum_{m=0}^{\infty}\, \frac{\bar{z}_{13}^m}{m!}\,
B(2\bar{h}_1+2+m,2\bar{h}_2+1, 2\bar{h}_3+1)
\nonu \\
&& \times \frac{1}{3!}\,
\ep^{PABCDEFG}\,
\bar{\pa}^m \, \Phi_{EFG,-\frac{1}{2}}^{(h_1+h_2+h_3)}(z_3,\bar{z}_3)
 \nonu \\
&& +
\frac{\bar{z}_{13}}{z_{13}} \,
\sum_{m=0}^{\infty}\, \frac{\bar{z}_{13}^m}{m!}\,
B(2\bar{h}_1+1+m,2\bar{h}_2+1) \, (\bar{\pa}^m \,
\Phi_{+1}^{(h_1+h_2),PA}\,\Phi_{+\frac{1}{2}}^{(h_3),BCD})
(z_3,\bar{z}_3)
\nonu \\
&& +
\frac{\bar{z}_{13}}{z_{13}} \,
\sum_{m=0}^{\infty}\, \frac{\bar{z}_{13}^m}{m!}\,
B(2\bar{h}_1+1+m,2\bar{h}_3+1) \, (-1)\,   (
\Phi_{+\frac{3}{2}}^{(h_2),A}\,\bar{\pa}^m \, \Phi_{0}^{(h_1+h_3),PBCD})
(z_3,\bar{z}_3),
\nonu \\
&&
\Phi_{+\frac{3}{2}}^{(h_1),P}(z_1,\bar{z}_1)\,
(\Phi_{+\frac{3}{2}}^{(h_2),A}\, \Phi_{0}^{(h_3),BCDE})(z_3,
\bar{z}_3) = \frac{\bar{z}_{13}^2}{z_{13}^2} \,
\sum_{m=0}^{\infty}\, \frac{\bar{z}_{13}^m}{m!}\,
B(2\bar{h}_1+2+m,2\bar{h}_2+1, 2\bar{h}_3+1)
\nonu \\
&& \times
\frac{1}{2!}\,
\ep^{PABCDEFG}\,
\bar{\pa}^m \, \Phi_{FG,-1}^{(h_1+h_2+h_3)}(z_3,\bar{z}_3)
 \nonu \\
&& +
\frac{\bar{z}_{13}}{z_{13}} \,
\sum_{m=0}^{\infty}\, \frac{\bar{z}_{13}^m}{m!}\,
B(2\bar{h}_1+1+m,2\bar{h}_2+1) \, (\bar{\pa}^m \,
\Phi_{+1}^{(h_1+h_2),PA}\,\Phi_{0}^{(h_3),BCDE})
(z_3,\bar{z}_3)
\nonu \\
&& +
\frac{\bar{z}_{13}}{z_{13}} \,
\sum_{m=0}^{\infty}\, \frac{\bar{z}_{13}^m}{m!}\,
B(2\bar{h}_1+1+m,2\bar{h}_3+1) \, (-1)\,\frac{1}{3!}\, \ep^{PBCDEFGH}\,   (
\Phi_{+\frac{3}{2}}^{(h_2),A}\,\bar{\pa}^m \, \Phi_{FGH,-\frac{1}{2}}^{(h_1+h_3)})
(z_3,\bar{z}_3),
\nonu \\
&&
\Phi_{+\frac{3}{2}}^{(h_1),P}(z_1,\bar{z}_1)\,
(\Phi_{+\frac{3}{2}}^{(h_2),A}\, \Phi_{BCD,-\frac{1}{2}}^{(h_3)})(z_3,
\bar{z}_3) = \frac{\bar{z}_{13}^2}{z_{13}^2} \,
\sum_{m=0}^{\infty}\, \frac{\bar{z}_{13}^m}{m!}\,
B(2\bar{h}_1+2+m,2\bar{h}_2+1, 2\bar{h}_3+1)
\nonu \\
&& \times
3! \, \de^{P}_{[B}
\bar{\pa}^m \, \Phi_{C, -\frac{3}{2}}^{(h_1+h_2+h_3)}(z_3,\bar{z}_3)
\, \de^{A}_{D]}
 \nonu \\
&& +
\frac{\bar{z}_{13}}{z_{13}} \,
\sum_{m=0}^{\infty}\, \frac{\bar{z}_{13}^m}{m!}\,
B(2\bar{h}_1+1+m,2\bar{h}_2+1) \, (\bar{\pa}^m \,
\Phi_{+1}^{(h_1+h_2),PA}\,\Phi_{BCD,-\frac{1}{2}}^{(h_3)})
(z_3,\bar{z}_3)
\nonu \\
&& +
\frac{\bar{z}_{13}}{z_{13}} \,
\sum_{m=0}^{\infty}\, \frac{\bar{z}_{13}^m}{m!}\,
B(2\bar{h}_1+1+m,2\bar{h}_3+1) \, (-1)\, 3 \, \de^P_{[B}\,   (
\Phi_{+\frac{3}{2}}^{(h_2),A}\,\bar{\pa}^m \, \Phi_{CD],-1}^{(h_1+h_3)})
(z_3,\bar{z}_3),
\nonu \\
&&
\Phi_{+\frac{3}{2}}^{(h_1),P}(z_1,\bar{z}_1)\,
(\Phi_{+\frac{3}{2}}^{(h_2),A}\, \Phi_{BC,-1}^{(h_3)})(z_3,
\bar{z}_3) = \frac{\bar{z}_{13}^2}{z_{13}^2} \,
\sum_{m=0}^{\infty}\, \frac{\bar{z}_{13}^m}{m!}\,
B(2\bar{h}_1+2+m,2\bar{h}_2+1, 2\bar{h}_3+1)
\nonu \\
&& \times
\de^{PA}_{BC}\,
\bar{\pa}^m \, \Phi_{-2}^{(h_1+h_2+h_3)}(z_3,\bar{z}_3)
 \nonu \\
&& +
\frac{\bar{z}_{13}}{z_{13}} \,
\sum_{m=0}^{\infty}\, \frac{\bar{z}_{13}^m}{m!}\,
B(2\bar{h}_1+1+m,2\bar{h}_2+1) \, (\bar{\pa}^m \,
\Phi_{+1}^{(h_1+h_2),PA}\,\Phi_{BC,-1}^{(h_3)})
(z_3,\bar{z}_3)
\nonu \\
&& +
\frac{\bar{z}_{13}}{z_{13}} \,
\sum_{m=0}^{\infty}\, \frac{\bar{z}_{13}^m}{m!}\,
B(2\bar{h}_1+1+m,2\bar{h}_3+1) \, (-1)\, 2! \, \de^P_{[B}\,   (
\Phi_{+\frac{3}{2}}^{(h_2),A}\,\bar{\pa}^m \, \Phi_{C],-\frac{3}{2}}^{(h_1+h_3)})
(z_3,\bar{z}_3),
\nonu \\
&&
\Phi_{+\frac{3}{2}}^{(h_1),P}(z_1,\bar{z}_1)\,
(\Phi_{+1}^{(h_2),AB}\, \Phi_{+1}^{(h_3),CD})(z_3,
\bar{z}_3) = \frac{\bar{z}_{13}^2}{z_{13}^2} \,
\sum_{m=0}^{\infty}\, \frac{\bar{z}_{13}^m}{m!}\,
B(2\bar{h}_1+2+m,2\bar{h}_2+1, 2\bar{h}_3+1)
\nonu \\
&& \times
\frac{1}{3!}\,
\ep^{CDPABEFG}\,
\bar{\pa}^m \, \Phi_{EFG, -\frac{1}{2}}^{(h_1+h_2+h_3)}(z_3,\bar{z}_3)
 \nonu \\
&& +
\frac{\bar{z}_{13}}{z_{13}} \,
\sum_{m=0}^{\infty}\, \frac{\bar{z}_{13}^m}{m!}\,
B(2\bar{h}_1+1+m,2\bar{h}_2+1) \, (\bar{\pa}^m \,
\Phi_{+\frac{1}{2}}^{(h_1+h_2),PAB}\,\Phi_{+1}^{(h_3),CD})
(z_3,\bar{z}_3)
\nonu \\
&& +
\frac{\bar{z}_{13}}{z_{13}} \,
\sum_{m=0}^{\infty}\, \frac{\bar{z}_{13}^m}{m!}\,
B(2\bar{h}_1+1+m,2\bar{h}_3+1) \,    (
\Phi_{+1}^{(h_2),AB}\,\bar{\pa}^m \, \Phi_{+\frac{1}{2}}^{(h_1+h_3),PCD})
(z_3,\bar{z}_3),
\nonu \\
&&
\Phi_{+\frac{3}{2}}^{(h_1),P}(z_1,\bar{z}_1)\,
(\Phi_{+1}^{(h_2),AB}\, \Phi_{+\frac{1}{2}}^{(h_3),CDE})(z_3,
\bar{z}_3) = \frac{\bar{z}_{13}^2}{z_{13}^2} \,
\sum_{m=0}^{\infty}\, \frac{\bar{z}_{13}^m}{m!}\,
B(2\bar{h}_1+2+m,2\bar{h}_2+1, 2\bar{h}_3+1)
\nonu \\
&& \times
\frac{1}{2!}\,
\ep^{PABCDEFG}\,
\bar{\pa}^m \, \Phi_{FG,-1}^{(h_1+h_2+h_3)}(z_3,\bar{z}_3)
 \nonu \\
&& +
\frac{\bar{z}_{13}}{z_{13}} \,
\sum_{m=0}^{\infty}\, \frac{\bar{z}_{13}^m}{m!}\,
B(2\bar{h}_1+1+m,2\bar{h}_2+1) \, (\bar{\pa}^m \,
\Phi_{+\frac{1}{2}}^{(h_1+h_2),PAB}\,\Phi_{+\frac{1}{2}}^{(h_3),CDE})
(z_3,\bar{z}_3)
\nonu \\
&& +
\frac{\bar{z}_{13}}{z_{13}} \,
\sum_{m=0}^{\infty}\, \frac{\bar{z}_{13}^m}{m!}\,
B(2\bar{h}_1+1+m,2\bar{h}_3+1) \,    (
\Phi_{+1}^{(h_2),AB}\,\bar{\pa}^m \, \Phi_{0}^{(h_1+h_3),PCDE})
(z_3,\bar{z}_3),
\nonu \\
&&
\Phi_{+\frac{3}{2}}^{(h_1),P}(z_1,\bar{z}_1)\,
(\Phi_{+1}^{(h_2),AB}\, \Phi_{0}^{(h_3),CDEF})(z_3,
\bar{z}_3) = \frac{\bar{z}_{13}^2}{z_{13}^2} \,
\sum_{m=0}^{\infty}\, \frac{\bar{z}_{13}^m}{m!}\,
B(2\bar{h}_1+2+m,2\bar{h}_2+1, 2\bar{h}_3+1)
\nonu \\
&& \times\
\ep^{PABCDEFG}\, 
\bar{\pa}^m \, \Phi_{G, -\frac{3}{2}}^{(h_1+h_2+h_3)}(z_3,\bar{z}_3)
 \nonu \\
&& +
\frac{\bar{z}_{13}}{z_{13}} \,
\sum_{m=0}^{\infty}\, \frac{\bar{z}_{13}^m}{m!}\,
B(2\bar{h}_1+1+m,2\bar{h}_2+1) \, (\bar{\pa}^m \,
\Phi_{+\frac{1}{2}}^{(h_1+h_2),PAB}\,\Phi_{0}^{(h_3),CDEF})
(z_3,\bar{z}_3)
\nonu \\
&& +
\frac{\bar{z}_{13}}{z_{13}} \,
\sum_{m=0}^{\infty}\, \frac{\bar{z}_{13}^m}{m!}\,
B(2\bar{h}_1+1+m,2\bar{h}_3+1) \,  \frac{1}{3!}\, \ep^{PCDEFGHI}\,  (
\Phi_{+1}^{(h_2),AB}\,\bar{\pa}^m \, \Phi_{GHI,-\frac{1}{2}}^{(h_1+h_3)})
(z_3,\bar{z}_3),
\nonu \\
&&
\Phi_{+\frac{3}{2}}^{(h_1),P}(z_1,\bar{z}_1)\,
(\Phi_{+1}^{(h_2),AB}\, \Phi_{CDE,-\frac{1}{2}}^{(h_3)})(z_3,
\bar{z}_3) = \frac{\bar{z}_{13}^2}{z_{13}^2} \,
\sum_{m=0}^{\infty}\, \frac{\bar{z}_{13}^m}{m!}\,
B(2\bar{h}_1+2+m,2\bar{h}_2+1, 2\bar{h}_3+1)
\nonu \\
&& \times
\de^{PAB}_{CDE}\,
\bar{\pa}^m \, \Phi_{-2}^{(h_1+h_2+h_3)}(z_3,\bar{z}_3)
 \nonu \\
&& +
\frac{\bar{z}_{13}}{z_{13}} \,
\sum_{m=0}^{\infty}\, \frac{\bar{z}_{13}^m}{m!}\,
B(2\bar{h}_1+1+m,2\bar{h}_2+1) \, (\bar{\pa}^m \,
\Phi_{+\frac{1}{2}}^{(h_1+h_2),PAB}\,\Phi_{CDE,-\frac{1}{2}}^{(h_3)})
(z_3,\bar{z}_3)
\nonu \\
&& +
\frac{\bar{z}_{13}}{z_{13}} \,
\sum_{m=0}^{\infty}\, \frac{\bar{z}_{13}^m}{m!}\,
B(2\bar{h}_1+1+m,2\bar{h}_3+1) \, 3 \, \de^{P}_{[C} \,  (
\Phi_{+1}^{(h_2),AB}\,\bar{\pa}^m \, \Phi_{DE],-1}^{(h_1+h_3)})
(z_3,\bar{z}_3),
\nonu \\
&&
\Phi_{+\frac{3}{2}}^{(h_1),P}(z_1,\bar{z}_1)\,
(\Phi_{+\frac{1}{2}}^{(h_2),ABC}\, \Phi_{+\frac{1}{2}}^{(h_3),DEF})(z_3,
\bar{z}_3) = \frac{\bar{z}_{13}^2}{z_{13}^2} \,
\sum_{m=0}^{\infty}\, \frac{\bar{z}_{13}^m}{m!}\,
B(2\bar{h}_1+2+m,2\bar{h}_2+1, 2\bar{h}_3+1)
\nonu \\
&& \times
\ep^{DEFPABCG}\, 
\bar{\pa}^m \, \Phi_{G, -\frac{3}{2}}^{(h_1+h_2+h_3)}(z_3,\bar{z}_3)
 \nonu \\
&& +
\frac{\bar{z}_{13}}{z_{13}} \,
\sum_{m=0}^{\infty}\, \frac{\bar{z}_{13}^m}{m!}\,
B(2\bar{h}_1+1+m,2\bar{h}_2+1) \, (\bar{\pa}^m \,
\Phi_{0}^{(h_1+h_2),PABC}\,\Phi_{+\frac{1}{2}}^{(h_3),DEF})
(z_3,\bar{z}_3)
\nonu \\
&& +
\frac{\bar{z}_{13}}{z_{13}} \,
\sum_{m=0}^{\infty}\, \frac{\bar{z}_{13}^m}{m!}\,
B(2\bar{h}_1+1+m,2\bar{h}_3+1)  \, (-1)\, (
\Phi_{+\frac{1}{2}}^{(h_2),ABC}\,\bar{\pa}^m \, \Phi_{0}^{(h_1+h_3),PDEF})
(z_3,\bar{z}_3),
\nonu \\
&&
\Phi_{+\frac{3}{2}}^{(h_1),P}(z_1,\bar{z}_1)\,
(\Phi_{+\frac{1}{2}}^{(h_2),ABC}\, \Phi_{0}^{(h_3),DEFG})(z_3,
\bar{z}_3) = \frac{\bar{z}_{13}^2}{z_{13}^2} \,
\sum_{m=0}^{\infty}\, \frac{\bar{z}_{13}^m}{m!}\,
B(2\bar{h}_1+2+m,2\bar{h}_2+1, 2\bar{h}_3+1)
\nonu \\
&& \times
\ep^{PABCDEFG}\, 
\bar{\pa}^m \, \Phi_{-2}^{(h_1+h_2+h_3)}(z_3,\bar{z}_3)
 \nonu \\
&& +
\frac{\bar{z}_{13}}{z_{13}} \,
\sum_{m=0}^{\infty}\, \frac{\bar{z}_{13}^m}{m!}\,
B(2\bar{h}_1+1+m,2\bar{h}_2+1) \, (\bar{\pa}^m \,
\Phi_{0}^{(h_1+h_2),PABC}\,\Phi_{0}^{(h_3),DEFG})
(z_3,\bar{z}_3)
\label{full3half}
\\
&& +
\frac{\bar{z}_{13}}{z_{13}} \,
\sum_{m=0}^{\infty}\, \frac{\bar{z}_{13}^m}{m!}\,
B(2\bar{h}_1+1+m,2\bar{h}_3+1) \, (-1)\,
\frac{1}{3!} \, \ep^{PDEFGHIJ}  \, (
\Phi_{+\frac{1}{2}}^{(h_2),ABC}\,\bar{\pa}^m \,
\Phi_{HIJ,-\frac{1}{2}}^{(h_1+h_3)})
(z_3,\bar{z}_3).
\nonu
\eea
Note that for the femionic operators of particle $1$ and particle
$2$, the contraction terms of the particle $1$ and particle $3$
have the extra minus signs.

There exist five nontrivial OPEs having the quadratic terms only
on the right-hand sides \footnote{The reason why
these OPEs appear is that the nonlinear (quadratic) terms arise
when the OPEs of the single-particle operators with
one of the two factors in the two-particle operators contain
the singular terms, at the level of a single contraction between the
operators. In general,
there are two kinds of infinite summation.
If one of the OPEs is zero, we are left with
a single infinite summation.
Note that the linear terms appear on the right-hand sides
when the double contractions
between the operators of particle $1$ and particle $2$ are nonzero.

For the condition $s_1+s_3 \geq 0$,
the first relation of (\ref{3halflinear}) doesn't satisfy.
Therefore, there exists only a single summation, compared to
other OPEs in (\ref{3halflinear}).
From the tensor products \cite{FKS} of
${\bf 8} \otimes [{\bf 1} \otimes \overline{\bf 1}]$,
${\bf 8} \otimes [{\bf 8} \otimes \overline{\bf 8}]$,
${\bf 8} \otimes [{\bf 28} \otimes \overline{\bf 28}]$,
${\bf 8} \otimes [{\bf 56} \otimes \overline{\bf 56}]$,
and
${\bf 8} \otimes [{\bf 70} \otimes {\bf 70}]$,
the representation ${\bf 8}$ for the gravitinos can act on
either first representation or second representation of
two-particle operators. We can associate with
the representations ${\bf 8} \otimes \overline{\bf 1}$
for the right-hand side of the first
OPE in (\ref{3halflinear}). For the second OPE,
we have the representations of ${\bf 28} \otimes \overline{ \bf 8}$
where ${\bf 28}$ comes from the tensor product
${\bf 8} \otimes {\bf 8}$ and ${\bf 8} \otimes \overline{\bf 1}$
where $\overline{\bf 1}$ originates from the tensor product ${\bf 8}
\otimes \overline{\bf 8}$. Similarly,
for the third OPE,
we have the representations of ${\bf 56} \otimes \overline{ \bf 28}$
where ${\bf 56}$ is obtained from the tensor product
${\bf 8} \otimes {\bf 28}$ and ${\bf 28} \otimes \overline{\bf 8}$
where $\overline{\bf 8}$ is obtained  from the tensor product ${\bf 8}
\otimes \overline{\bf 28}$.
For the fourth OPE,
we have the representations of ${\bf 70} \otimes \overline{ \bf 56}$
where ${\bf 70}$ can be obtained from the tensor product
${\bf 8} \otimes {\bf 56}$ and ${\bf 56} \otimes \overline{\bf 28}$
where $\overline{\bf 28}$ is obtained  from the tensor product ${\bf 8}
\otimes \overline{\bf 56}$.
Finally,
for the fifth OPE,
we have the representations of $\overline{\bf 56}
\otimes { \bf 70}$ and  ${\bf 70} \otimes \overline{\bf 56}$
where $\overline{\bf 56}$ comes from the tensor product ${\bf 8}
\otimes {\bf 70}$. For the previous case in the subsection $A.1$,
we can analyze similarly
and all of them can be done trivially due to the singlet
graviton ${\bf 1}$.}
\bea
&&
\Phi_{+\frac{3}{2}}^{(h_1),P}(z_1,\bar{z}_1)\,
(\Phi_{+2}^{(h_2)}\, \Phi_{-2}^{(h_3)})(z_3,
\bar{z}_3) =
\nonu \\
&&
\frac{\bar{z}_{13}}{z_{13}} \,
\sum_{m=0}^{\infty}\, \frac{\bar{z}_{13}^m}{m!}\,
B(2\bar{h}_1+1+m,2\bar{h}_2+1) \, (\bar{\pa}^m \,
\Phi_{\frac{3}{2}}^{(h_1+h_2),P}\,\Phi_{-2}^{(h_3)})
(z_3,\bar{z}_3),
\nonu \\
&&
\Phi_{+\frac{3}{2}}^{(h_1),P}(z_1,\bar{z}_1)\,
(\Phi_{+\frac{3}{2}}^{(h_2),A}\, \Phi_{B,-\frac{3}{2}}^{(h_3)})(z_3,
\bar{z}_3) =
\nonu \\
&& 
\frac{\bar{z}_{13}}{z_{13}} \,
\sum_{m=0}^{\infty}\, \frac{\bar{z}_{13}^m}{m!}\,
B(2\bar{h}_1+1+m,2\bar{h}_2+1) \, (\bar{\pa}^m \,
\Phi_{+1}^{(h_1+h_2),PA}\,\Phi_{B,-\frac{3}{2}}^{(h_3)})
(z_3,\bar{z}_3)
\nonu \\
&& +
\frac{\bar{z}_{13}}{z_{13}} \,
\sum_{m=0}^{\infty}\, \frac{\bar{z}_{13}^m}{m!}\,
B(2\bar{h}_1+1+m,2\bar{h}_3+1) \, (-1)\,
\de^{P}_{B}  \, (
\Phi_{+\frac{3}{2}}^{(h_2),A}\,\bar{\pa}^m \,
\Phi_{-2}^{(h_1+h_3)})
(z_3,\bar{z}_3),
\nonu \\
&&
\Phi_{+\frac{3}{2}}^{(h_1),P}(z_1,\bar{z}_1)\,
(\Phi_{+1}^{(h_2),AB}\, \Phi_{CD,-1}^{(h_3)})(z_3,
\bar{z}_3) =
\nonu \\
&& 
\frac{\bar{z}_{13}}{z_{13}} \,
\sum_{m=0}^{\infty}\, \frac{\bar{z}_{13}^m}{m!}\,
B(2\bar{h}_1+1+m,2\bar{h}_2+1) \, (\bar{\pa}^m \,
\Phi_{+\frac{1}{2}}^{(h_1+h_2),PAB}\,\Phi_{CD,-1}^{(h_3)})
(z_3,\bar{z}_3)
\nonu \\
&& +
\frac{\bar{z}_{13}}{z_{13}} \,
\sum_{m=0}^{\infty}\, \frac{\bar{z}_{13}^m}{m!}\,
B(2\bar{h}_1+1+m,2\bar{h}_3+1) \, 2! \,
\de^{P}_{[C}  \, (
\Phi_{+1}^{(h_2),AB}\,\bar{\pa}^m \,
\Phi_{D]-\frac{3}{2}}^{(h_1+h_3)})
(z_3,\bar{z}_3),
\nonu \\
&&
\Phi_{+\frac{3}{2}}^{(h_1),P}(z_1,\bar{z}_1)\,
(\Phi_{+\frac{1}{2}}^{(h_2),ABC}\, \Phi_{DEF,-\frac{1}{2}}^{(h_3)})(z_3,
\bar{z}_3) =
\nonu \\
&& 
\frac{\bar{z}_{13}}{z_{13}} \,
\sum_{m=0}^{\infty}\, \frac{\bar{z}_{13}^m}{m!}\,
B(2\bar{h}_1+1+m,2\bar{h}_2+1) \, (\bar{\pa}^m \,
\Phi_{0}^{(h_1+h_2),PABC}\,\Phi_{DEF,-\frac{1}{2}}^{(h_3)})
(z_3,\bar{z}_3)
\nonu \\
&& +
\frac{\bar{z}_{13}}{z_{13}} \,
\sum_{m=0}^{\infty}\, \frac{\bar{z}_{13}^m}{m!}\,
B(2\bar{h}_1+1+m,2\bar{h}_3+1) \, (-1) \, 3 \,
\de^{P}_{[D}  \, (
\Phi_{+\frac{1}{2}}^{(h_2),ABC}\,\bar{\pa}^m \,
\Phi_{EF],-1}^{(h_1+h_3)})
(z_3,\bar{z}_3),
\nonu \\
&&
\Phi_{+\frac{3}{2}}^{(h_1),P}(z_1,\bar{z}_1)\,
(\Phi_{0}^{(h_2),ABCD}\, \Phi_{0}^{(h_3),EFGH})(z_3,
\bar{z}_3) =
\nonu \\
&& 
\frac{\bar{z}_{13}}{z_{13}} \,
\sum_{m=0}^{\infty}\, \frac{\bar{z}_{13}^m}{m!}\,
B(2\bar{h}_1+1+m,2\bar{h}_2+1) \, \frac{1}{3!}\, \ep^{PABCDIJK}\, (\bar{\pa}^m \,
\Phi_{IJK,-\frac{1}{2}}^{(h_1+h_2)}\,\Phi_{0}^{(h_3),EFGH})
(z_3,\bar{z}_3)
\label{3halflinear}
\\
&& +
\frac{\bar{z}_{13}}{z_{13}} \,
\sum_{m=0}^{\infty}\, \frac{\bar{z}_{13}^m}{m!}\,
B(2\bar{h}_1+1+m,2\bar{h}_3+1) \, \frac{1}{3!} \,
\ep^{PEFGHIJK}\, (
\Phi_{0}^{(h_2),ABCD}\,\bar{\pa}^m \,
\Phi_{IJK,-\frac{1}{2}}^{(h_1+h_3)})
(z_3,\bar{z}_3).
\nonu
\eea
The OPEs (\ref{3halflinear}) also appear in the amplitude
relations in (\ref{AMP2}) at the subleading order in $\frac{1}{z_{13}}$.

\subsection{The OPEs of
the graviphotons  of helicity $+1$  with the quadratic
operators}

The sixteen OPEs
\footnote{Among twenty OPEs we have considered in previous example,
the four OPEs do not satisfy
the condition $s_1+s_2+s_3 \geq 2$.}
are summarized by
\bea
&& \Phi_{+1}^{(h_1),PQ}(z_1,\bar{z}_1)\,
(\Phi_{+2}^{(h_2)}\, \Phi_{+2}^{(h_3)})(z_3,
\bar{z}_3) = \frac{\bar{z}_{13}^2}{z_{13}^2} \,
\sum_{m=0}^{\infty}\, \frac{\bar{z}_{13}^m}{m!}\,
B(2\bar{h}_1+2+m,2\bar{h}_2+1, 2\bar{h}_3+1)
\nonu \\
&& \times
\bar{\pa}^m \, \Phi_{+1}^{(h_1+h_2+h_3),PQ}(z_3,\bar{z}_3)
\nonu \\
&&+ 
\frac{\bar{z}_{13}}{z_{13}} \,
\sum_{m=0}^{\infty}\, \frac{\bar{z}_{13}^m}{m!}\,
B(2\bar{h}_1+1+m,2\bar{h}_2+1) \,  (\bar{\pa}^m \,
\Phi_{+1}^{(h_1+h_2),PQ}\,\Phi_{+2}^{(h_3)})
(z_3,\bar{z}_3)
\nonu \\
&& +
\frac{\bar{z}_{13}}{z_{13}} \,
\sum_{m=0}^{\infty}\, \frac{\bar{z}_{13}^m}{m!}\,
B(2\bar{h}_1+1+m,2\bar{h}_3+1) \,  (
\Phi_{+2}^{(h_2)}\,\bar{\pa}^m \,
\Phi_{+1}^{(h_1+h_3),PQ})
(z_3,\bar{z}_3),
\nonu \\
&&
\Phi_{+1}^{(h_1),PQ}(z_1,\bar{z}_1)\,
(\Phi_{+2}^{(h_2)}\, \Phi_{+\frac{3}{2}}^{(h_3),A})(z_3,
\bar{z}_3) = \frac{\bar{z}_{13}^2}{z_{13}^2} \,
\sum_{m=0}^{\infty}\, \frac{\bar{z}_{13}^m}{m!}\,
B(2\bar{h}_1+2+m,2\bar{h}_2+1, 2\bar{h}_3+1)
\nonu \\
&& \times
\bar{\pa}^m \, \Phi_{+\frac{1}{2}}^{(h_1+h_2+h_3),APQ}(z_3,\bar{z}_3)
\nonu \\
&&+ 
\frac{\bar{z}_{13}}{z_{13}} \,
\sum_{m=0}^{\infty}\, \frac{\bar{z}_{13}^m}{m!}\,
B(2\bar{h}_1+1+m,2\bar{h}_2+1) \,  (\bar{\pa}^m \,
\Phi_{+1}^{(h_1+h_2),PQ}\,\Phi_{+\frac{3}{2}}^{(h_3),A})
(z_3,\bar{z}_3)
\nonu \\
&& +
\frac{\bar{z}_{13}}{z_{13}} \,
\sum_{m=0}^{\infty}\, \frac{\bar{z}_{13}^m}{m!}\,
B(2\bar{h}_1+1+m,2\bar{h}_3+1) \,  (
\Phi_{+2}^{(h_2)}\,\bar{\pa}^m \,
\Phi_{+\frac{1}{2}}^{(h_1+h_3),PQA})
(z_3,\bar{z}_3),
\nonu \\
&&
\Phi_{+1}^{(h_1),PQ}(z_1,\bar{z}_1)\,
(\Phi_{+2}^{(h_2)}\, \Phi_{+1}^{(h_3),AB})(z_3,
\bar{z}_3) = \frac{\bar{z}_{13}^2}{z_{13}^2} \,
\sum_{m=0}^{\infty}\, \frac{\bar{z}_{13}^m}{m!}\,
B(2\bar{h}_1+2+m,2\bar{h}_2+1, 2\bar{h}_3+1)
\nonu \\
&& \times
\bar{\pa}^m \, \Phi_{0}^{(h_1+h_2+h_3),PQAB}(z_3,\bar{z}_3)
\nonu \\
&&+ 
\frac{\bar{z}_{13}}{z_{13}} \,
\sum_{m=0}^{\infty}\, \frac{\bar{z}_{13}^m}{m!}\,
B(2\bar{h}_1+1+m,2\bar{h}_2+1) \,  (\bar{\pa}^m \,
\Phi_{+1}^{(h_1+h_2),PQ}\,\Phi_{+1}^{(h_3),AB})
(z_3,\bar{z}_3)
\nonu \\
&& +
\frac{\bar{z}_{13}}{z_{13}} \,
\sum_{m=0}^{\infty}\, \frac{\bar{z}_{13}^m}{m!}\,
B(2\bar{h}_1+1+m,2\bar{h}_3+1) \,  (
\Phi_{+2}^{(h_2)}\,\bar{\pa}^m \,
\Phi_{0}^{(h_1+h_3),PQAB})
(z_3,\bar{z}_3),
\nonu \\
&&
\Phi_{+1}^{(h_1),PQ}(z_1,\bar{z}_1)\,
(\Phi_{+2}^{(h_2)}\, \Phi_{+\frac{1}{2}}^{(h_3),ABC})(z_3,
\bar{z}_3) = \frac{\bar{z}_{13}^2}{z_{13}^2} \,
\sum_{m=0}^{\infty}\, \frac{\bar{z}_{13}^m}{m!}\,
B(2\bar{h}_1+2+m,2\bar{h}_2+1, 2\bar{h}_3+1)
\nonu \\
&& \times
\frac{1}{3!}\,
\ep^{PQABCDEF}
\bar{\pa}^m \, \Phi_{DEF,-\frac{1}{2}}^{(h_1+h_2+h_3)}(z_3,\bar{z}_3)
\nonu \\
&&+ 
\frac{\bar{z}_{13}}{z_{13}} \,
\sum_{m=0}^{\infty}\, \frac{\bar{z}_{13}^m}{m!}\,
B(2\bar{h}_1+1+m,2\bar{h}_2+1) \,  (\bar{\pa}^m \,
\Phi_{+1}^{(h_1+h_2),PQ}\,\Phi_{+\frac{1}{2}}^{(h_3),ABC})
(z_3,\bar{z}_3)
\nonu \\
&& +
\frac{\bar{z}_{13}}{z_{13}} \,
\sum_{m=0}^{\infty}\, \frac{\bar{z}_{13}^m}{m!}\,
B(2\bar{h}_1+1+m,2\bar{h}_3+1) \,  \frac{1}{3!}\, \ep^{PQABCDEF}\, (
\Phi_{+2}^{(h_2)}\,\bar{\pa}^m \,
\Phi_{DEF,-\frac{1}{2}}^{(h_1+h_3)})
(z_3,\bar{z}_3),
\nonu \\
&&
\Phi_{+1}^{(h_1),PQ}(z_1,\bar{z}_1)\,
(\Phi_{+2}^{(h_2)}\, \Phi_{0}^{(h_3),ABCD})(z_3,
\bar{z}_3) = \frac{\bar{z}_{13}^2}{z_{13}^2} \,
\sum_{m=0}^{\infty}\, \frac{\bar{z}_{13}^m}{m!}\,
B(2\bar{h}_1+2+m,2\bar{h}_2+1, 2\bar{h}_3+1)
\nonu \\
&& \times
\frac{1}{2!}\,
\ep^{PQABCDEF}\,
\bar{\pa}^m \, \Phi_{EF,-1}^{(h_1+h_2+h_3)}(z_3,\bar{z}_3)
\nonu \\
&&+ 
\frac{\bar{z}_{13}}{z_{13}} \,
\sum_{m=0}^{\infty}\, \frac{\bar{z}_{13}^m}{m!}\,
B(2\bar{h}_1+1+m,2\bar{h}_2+1) \,  (\bar{\pa}^m \,
\Phi_{+1}^{(h_1+h_2),PQ}\,\Phi_{0}^{(h_3),ABCD})
(z_3,\bar{z}_3)
\nonu \\
&& +
\frac{\bar{z}_{13}}{z_{13}} \,
\sum_{m=0}^{\infty}\, \frac{\bar{z}_{13}^m}{m!}\,
B(2\bar{h}_1+1+m,2\bar{h}_3+1) \,  \frac{1}{2!}\, \ep^{PQABCDEF}\, (
\Phi_{+2}^{(h_2)}\,\bar{\pa}^m \,
\Phi_{EF,-1}^{(h_1+h_3)})
(z_3,\bar{z}_3),
\nonu \\
&&
\Phi_{+1}^{(h_1),PQ}(z_1,\bar{z}_1)\,
(\Phi_{+2}^{(h_2)}\, \Phi_{ABC,-\frac{1}{2}}^{(h_3)})(z_3,
\bar{z}_3) = \frac{\bar{z}_{13}^2}{z_{13}^2} \,
\sum_{m=0}^{\infty}\, \frac{\bar{z}_{13}^m}{m!}\,
B(2\bar{h}_1+2+m,2\bar{h}_2+1, 2\bar{h}_3+1)
\nonu \\
&& \times
3!\,
\de^{P}_{[A}
\bar{\pa}^m \, \Phi_{B,-\frac{3}{2}}^{(h_1+h_2+h_3)}(z_3,\bar{z}_3)
\, \de^{Q}_{C]}
\nonu \\
&&+ 
\frac{\bar{z}_{13}}{z_{13}} \,
\sum_{m=0}^{\infty}\, \frac{\bar{z}_{13}^m}{m!}\,
B(2\bar{h}_1+1+m,2\bar{h}_2+1) \,  (\bar{\pa}^m \,
\Phi_{+1}^{(h_1+h_2),PQ}\,\Phi_{ABC,-\frac{1}{2}}^{(h_3)})
(z_3,\bar{z}_3)
\nonu \\
&& +
\frac{\bar{z}_{13}}{z_{13}} \,
\sum_{m=0}^{\infty}\, \frac{\bar{z}_{13}^m}{m!}\,
B(2\bar{h}_1+1+m,2\bar{h}_3+1) \, 3!\, \de^{P}_{[A} (
\Phi_{+2}^{(h_2)}\,\bar{\pa}^m \,
\Phi_{B,-\frac{3}{2}}^{(h_1+h_3)})
(z_3,\bar{z}_3)\, \de^{Q}_{C]},
\nonu \\
&& \Phi_{+1}^{(h_1),PQ}(z_1,\bar{z}_1)\,
(\Phi_{+2}^{(h_2)}\, \Phi_{AB,-1}^{(h_3)})(z_3,
\bar{z}_3) = \frac{\bar{z}_{13}^2}{z_{13}^2} \,
\sum_{m=0}^{\infty}\, \frac{\bar{z}_{13}^m}{m!}\,
B(2\bar{h}_1+2+m,2\bar{h}_2+1, 2\bar{h}_3+1)
\nonu \\
&& \times
\de^{PQ}_{AB}\,
\bar{\pa}^m \, \Phi_{-2}^{(h_1+h_2+h_3)}(z_3,\bar{z}_3)
\nonu \\
&&+ 
\frac{\bar{z}_{13}}{z_{13}} \,
\sum_{m=0}^{\infty}\, \frac{\bar{z}_{13}^m}{m!}\,
B(2\bar{h}_1+1+m,2\bar{h}_2+1) \,  (\bar{\pa}^m \,
\Phi_{+1}^{(h_1+h_2),PQ}\,\Phi_{AB,-1}^{(h_3)})
(z_3,\bar{z}_3)
\nonu \\
&& +
\frac{\bar{z}_{13}}{z_{13}} \,
\sum_{m=0}^{\infty}\, \frac{\bar{z}_{13}^m}{m!}\,
B(2\bar{h}_1+1+m,2\bar{h}_3+1) \,  \de^{PQ}_{AB} \, (
\Phi_{+2}^{(h_2)}\,\bar{\pa}^m \,
\Phi_{-2}^{(h_1+h_3)})
(z_3,\bar{z}_3),
\nonu \\
&&
\Phi_{+1}^{(h_1),PQ}(z_1,\bar{z}_1)\,
(\Phi_{+\frac{3}{2}}^{(h_2),A}\, \Phi_{+\frac{3}{2}}^{(h_3),B})(z_3,
\bar{z}_3) = \frac{\bar{z}_{13}^2}{z_{13}^2} \,
\sum_{m=0}^{\infty}\, \frac{\bar{z}_{13}^m}{m!}\,
B(2\bar{h}_1+2+m,2\bar{h}_2+1, 2\bar{h}_3+1)
\nonu \\
&& \times
\bar{\pa}^m \, \Phi_{0}^{(h_1+h_2+h_3),PQAB}(z_3,\bar{z}_3)
\nonu \\
&&+ 
\frac{\bar{z}_{13}}{z_{13}} \,
\sum_{m=0}^{\infty}\, \frac{\bar{z}_{13}^m}{m!}\,
B(2\bar{h}_1+1+m,2\bar{h}_2+1) \,  (\bar{\pa}^m \,
\Phi_{+\frac{1}{2}}^{(h_1+h_2),PQA}\,\Phi_{+\frac{3}{2}}^{(h_3),B})
(z_3,\bar{z}_3)
\nonu \\
&& +
\frac{\bar{z}_{13}}{z_{13}} \,
\sum_{m=0}^{\infty}\, \frac{\bar{z}_{13}^m}{m!}\,
B(2\bar{h}_1+1+m,2\bar{h}_3+1)  \, (
\Phi_{+\frac{3}{2}}^{(h_2),A}\,\bar{\pa}^m \,
\Phi_{+\frac{1}{2}}^{(h_1+h_3),PQB})
(z_3,\bar{z}_3),
\nonu \\
&&
\Phi_{+1}^{(h_1),PQ}(z_1,\bar{z}_1)\,
(\Phi_{+\frac{3}{2}}^{(h_2),A}\, \Phi_{+1}^{(h_3),BC})(z_3,
\bar{z}_3) = \frac{\bar{z}_{13}^2}{z_{13}^2} \,
\sum_{m=0}^{\infty}\, \frac{\bar{z}_{13}^m}{m!}\,
B(2\bar{h}_1+2+m,2\bar{h}_2+1, 2\bar{h}_3+1)
\nonu \\
&& \times
\frac{1}{3!}\,
\ep^{BCAPQDEF}\,
\bar{\pa}^m \, \Phi_{DEF,-\frac{1}{2}}^{(h_1+h_2+h_3)}(z_3,\bar{z}_3)
\nonu \\
&&+ 
\frac{\bar{z}_{13}}{z_{13}} \,
\sum_{m=0}^{\infty}\, \frac{\bar{z}_{13}^m}{m!}\,
B(2\bar{h}_1+1+m,2\bar{h}_2+1) \,  (\bar{\pa}^m \,
\Phi_{+\frac{1}{2}}^{(h_1+h_2),PQA}\,\Phi_{+1}^{(h_3),BC})
(z_3,\bar{z}_3)
\nonu \\
&& +
\frac{\bar{z}_{13}}{z_{13}} \,
\sum_{m=0}^{\infty}\, \frac{\bar{z}_{13}^m}{m!}\,
B(2\bar{h}_1+1+m,2\bar{h}_3+1)  \, (
\Phi_{+\frac{3}{2}}^{(h_2),A}\,\bar{\pa}^m \,
\Phi_{0}^{(h_1+h_3),PQBC})
(z_3,\bar{z}_3),
\nonu \\
&&
\Phi_{+1}^{(h_1),PQ}(z_1,\bar{z}_1)\,
(\Phi_{+\frac{3}{2}}^{(h_2),A}\, \Phi_{+\frac{1}{2}}^{(h_3),BCD})(z_3,
\bar{z}_3) = \frac{\bar{z}_{13}^2}{z_{13}^2} \,
\sum_{m=0}^{\infty}\, \frac{\bar{z}_{13}^m}{m!}\,
B(2\bar{h}_1+2+m,2\bar{h}_2+1, 2\bar{h}_3+1)
\nonu \\
&& \times
\frac{1}{2!}\,
\ep^{APQBCDEF}\,
\bar{\pa}^m \, \Phi_{EF,-1}^{(h_1+h_2+h_3)}(z_3,\bar{z}_3)
\nonu \\
&&+ 
\frac{\bar{z}_{13}}{z_{13}} \,
\sum_{m=0}^{\infty}\, \frac{\bar{z}_{13}^m}{m!}\,
B(2\bar{h}_1+1+m,2\bar{h}_2+1) \,  (\bar{\pa}^m \,
\Phi_{+\frac{1}{2}}^{(h_1+h_2),PQA}\,\Phi_{+\frac{1}{2}}^{(h_3),BCD})
(z_3,\bar{z}_3)
\nonu \\
&& +
\frac{\bar{z}_{13}}{z_{13}} \,
\sum_{m=0}^{\infty}\, \frac{\bar{z}_{13}^m}{m!}\,
B(2\bar{h}_1+1+m,2\bar{h}_3+1) \, \frac{1}{3!}  \, \ep^{PQBCDFGH}\, (
\Phi_{+\frac{3}{2}}^{(h_2),A}\,\bar{\pa}^m \,
\Phi_{FGH,-\frac{1}{2}}^{(h_1+h_3)})
(z_3,\bar{z}_3),
\nonu \\
&&
\Phi_{+1}^{(h_1),PQ}(z_1,\bar{z}_1)\,
(\Phi_{+\frac{3}{2}}^{(h_2),A}\, \Phi_{0}^{(h_3),BCDE})(z_3,
\bar{z}_3) = \frac{\bar{z}_{13}^2}{z_{13}^2} \,
\sum_{m=0}^{\infty}\, \frac{\bar{z}_{13}^m}{m!}\,
B(2\bar{h}_1+2+m,2\bar{h}_2+1, 2\bar{h}_3+1)
\nonu \\
&& \times
\ep^{APQBCDEF}\,
\bar{\pa}^m \, \Phi_{F,-\frac{3}{2}}^{(h_1+h_2+h_3)}(z_3,\bar{z}_3)
\nonu \\
&&+ 
\frac{\bar{z}_{13}}{z_{13}} \,
\sum_{m=0}^{\infty}\, \frac{\bar{z}_{13}^m}{m!}\,
B(2\bar{h}_1+1+m,2\bar{h}_2+1) \,  (\bar{\pa}^m \,
\Phi_{+\frac{1}{2}}^{(h_1+h_2),PQA}\,\Phi_{0}^{(h_3),BCDE})
(z_3,\bar{z}_3)
\nonu \\
&& +
\frac{\bar{z}_{13}}{z_{13}} \,
\sum_{m=0}^{\infty}\, \frac{\bar{z}_{13}^m}{m!}\,
B(2\bar{h}_1+1+m,2\bar{h}_3+1) \, \frac{1}{2!}  \, \ep^{PQBCDEFG}\, (
\Phi_{+\frac{3}{2}}^{(h_2),A}\,\bar{\pa}^m \,
\Phi_{FG,-1}^{(h_1+h_3)})
(z_3,\bar{z}_3),
\nonu \\
&&
\Phi_{+1}^{(h_1),PQ}(z_1,\bar{z}_1)\,
(\Phi_{+\frac{3}{2}}^{(h_2),A}\, \Phi_{BCD,-\frac{1}{2}}^{(h_3)})(z_3,
\bar{z}_3) = \frac{\bar{z}_{13}^2}{z_{13}^2} \,
\sum_{m=0}^{\infty}\, \frac{\bar{z}_{13}^m}{m!}\,
B(2\bar{h}_1+2+m,2\bar{h}_2+1, 2\bar{h}_3+1)
\nonu \\
&& \times
\de^{APQ}_{BCD}\,
\bar{\pa}^m \, \Phi_{-2}^{(h_1+h_2+h_3)}(z_3,\bar{z}_3)
\nonu \\
&&+ 
\frac{\bar{z}_{13}}{z_{13}} \,
\sum_{m=0}^{\infty}\, \frac{\bar{z}_{13}^m}{m!}\,
B(2\bar{h}_1+1+m,2\bar{h}_2+1) \,  (\bar{\pa}^m \,
\Phi_{+\frac{1}{2}}^{(h_1+h_2),PQA}\,\Phi_{BCD,-\frac{1}{2}}^{(h_3)})
(z_3,\bar{z}_3)
\nonu \\
&& +
\frac{\bar{z}_{13}}{z_{13}} \,
\sum_{m=0}^{\infty}\, \frac{\bar{z}_{13}^m}{m!}\,
B(2\bar{h}_1+1+m,2\bar{h}_3+1) \, 3!  \, \de^{P}_{[B}\, (
\Phi_{+\frac{3}{2}}^{(h_2),A}\,\bar{\pa}^m \,
\Phi_{C,-\frac{3}{2}}^{(h_1+h_3)})
(z_3,\bar{z}_3) \de^{Q}_{D]},
\nonu \\
&&
\Phi_{+1}^{(h_1),PQ}(z_1,\bar{z}_1)\,
(\Phi_{+1}^{(h_2),AB}\, \Phi_{+1}^{(h_3),CD})(z_3,
\bar{z}_3) = \frac{\bar{z}_{13}^2}{z_{13}^2} \,
\sum_{m=0}^{\infty}\, \frac{\bar{z}_{13}^m}{m!}\,
B(2\bar{h}_1+2+m,2\bar{h}_2+1, 2\bar{h}_3+1)
\nonu \\
&& \times
\frac{1}{2!}\,
\ep^{CDPQABEF}\,
\bar{\pa}^m \, \Phi_{EF,-1}^{(h_1+h_2+h_3)}(z_3,\bar{z}_3)
\nonu \\
&&+ 
\frac{\bar{z}_{13}}{z_{13}} \,
\sum_{m=0}^{\infty}\, \frac{\bar{z}_{13}^m}{m!}\,
B(2\bar{h}_1+1+m,2\bar{h}_2+1) \,  (\bar{\pa}^m \,
\Phi_{0}^{(h_1+h_2),PQAB}\,\Phi_{+1}^{(h_3),CD})
(z_3,\bar{z}_3)
\nonu \\
&& +
\frac{\bar{z}_{13}}{z_{13}} \,
\sum_{m=0}^{\infty}\, \frac{\bar{z}_{13}^m}{m!}\,
B(2\bar{h}_1+1+m,2\bar{h}_3+1) \, (
\Phi_{+1}^{(h_2),AB}\,\bar{\pa}^m \,
\Phi_{0}^{(h_1+h_3),PQCD})
(z_3,\bar{z}_3),
\nonu \\
&&
\Phi_{+1}^{(h_1),PQ}(z_1,\bar{z}_1)\,
(\Phi_{+1}^{(h_2),AB}\, \Phi_{+\frac{1}{2}}^{(h_3),CDE})(z_3,
\bar{z}_3) = \frac{\bar{z}_{13}^2}{z_{13}^2} \,
\sum_{m=0}^{\infty}\, \frac{\bar{z}_{13}^m}{m!}\,
B(2\bar{h}_1+2+m,2\bar{h}_2+1, 2\bar{h}_3+1)
\nonu \\
&& \times
\ep^{CDEPQABF}\,
\bar{\pa}^m \, \Phi_{F,-\frac{3}{2}}^{(h_1+h_2+h_3)}(z_3,\bar{z}_3)
\nonu \\
&&+ 
\frac{\bar{z}_{13}}{z_{13}} \,
\sum_{m=0}^{\infty}\, \frac{\bar{z}_{13}^m}{m!}\,
B(2\bar{h}_1+1+m,2\bar{h}_2+1) \,  (\bar{\pa}^m \,
\Phi_{0}^{(h_1+h_2),PQAB}\,\Phi_{+\frac{1}{2}}^{(h_3),CDE})
(z_3,\bar{z}_3)
\nonu \\
&& +
\frac{\bar{z}_{13}}{z_{13}} \,
\sum_{m=0}^{\infty}\, \frac{\bar{z}_{13}^m}{m!}\,
B(2\bar{h}_1+1+m,2\bar{h}_3+1) \, \frac{1}{3!} \, \ep^{PQCDEFGH} \, (
\Phi_{+1}^{(h_2),AB}\,\bar{\pa}^m \,
\Phi_{FGH,-\frac{1}{2}}^{(h_1+h_3)})
(z_3,\bar{z}_3),
\nonu \\
&&
\Phi_{+1}^{(h_1),PQ}(z_1,\bar{z}_1)\,
(\Phi_{+1}^{(h_2),AB}\, \Phi_{0}^{(h_3),CDEF})(z_3,
\bar{z}_3) = \frac{\bar{z}_{13}^2}{z_{13}^2} \,
\sum_{m=0}^{\infty}\, \frac{\bar{z}_{13}^m}{m!}\,
B(2\bar{h}_1+2+m,2\bar{h}_2+1, 2\bar{h}_3+1)
\nonu \\
&& \times
\ep^{PQABCDEF}\,
\bar{\pa}^m \, \Phi_{-2}^{(h_1+h_2+h_3)}(z_3,\bar{z}_3)
\nonu \\
&&+ 
\frac{\bar{z}_{13}}{z_{13}} \,
\sum_{m=0}^{\infty}\, \frac{\bar{z}_{13}^m}{m!}\,
B(2\bar{h}_1+1+m,2\bar{h}_2+1) \,  (\bar{\pa}^m \,
\Phi_{0}^{(h_1+h_2),PQAB}\,\Phi_{0}^{(h_3),CDEF})
(z_3,\bar{z}_3)
\nonu \\
&& +
\frac{\bar{z}_{13}}{z_{13}} \,
\sum_{m=0}^{\infty}\, \frac{\bar{z}_{13}^m}{m!}\,
B(2\bar{h}_1+1+m,2\bar{h}_3+1) \, \frac{1}{2!} \, \ep^{PQCDEFGH} \, (
\Phi_{+1}^{(h_2),AB}\,\bar{\pa}^m \,
\Phi_{GH,-1}^{(h_1+h_3)})
(z_3,\bar{z}_3),
\nonu \\
&&
\Phi_{+1}^{(h_1),PQ}(z_1,\bar{z}_1)\,
(\Phi_{+\frac{1}{2}}^{(h_2),ABC}\, \Phi_{+\frac{1}{2}}^{(h_3),DEF})(z_3,
\bar{z}_3) = \frac{\bar{z}_{13}^2}{z_{13}^2} \,
\sum_{m=0}^{\infty}\, \frac{\bar{z}_{13}^m}{m!}\,
B(2\bar{h}_1+2+m,2\bar{h}_2+1, 2\bar{h}_3+1)
\nonu \\
&& \times
\frac{1}{3!}\,
\ep^{PQABCGHI}\, \de^{DEF}_{GHI}\,
\bar{\pa}^m \, \Phi_{-2}^{(h_1+h_2+h_3)}(z_3,\bar{z}_3)
\nonu \\
&&+ 
\frac{\bar{z}_{13}}{z_{13}} \,
\sum_{m=0}^{\infty}\, \frac{\bar{z}_{13}^m}{m!}\,
B(2\bar{h}_1+1+m,2\bar{h}_2+1) \, \frac{1}{3!}\,  \ep^{PQABCGHI}\, (\bar{\pa}^m \,
\Phi_{GHI,-\frac{1}{2}}^{(h_1+h_2)}\,\Phi_{+\frac{1}{2}}^{(h_3),DEF})
(z_3,\bar{z}_3)
\nonu \\
&& +
\frac{\bar{z}_{13}}{z_{13}} \,
\sum_{m=0}^{\infty}\, \frac{\bar{z}_{13}^m}{m!}\,
B(2\bar{h}_1+1+m,2\bar{h}_3+1) \, \frac{1}{3!} \, \ep^{PQDEFGHI} \, (
\Phi_{+\frac{1}{2}}^{(h_2),ABC}\,\bar{\pa}^m \,
\Phi_{GHI,-\frac{1}{2}}^{(h_1+h_3)})
(z_3,\bar{z}_3).
\label{fullone}
\eea
There are nine nontrivial OPEs
\footnote{Among them, three OPEs in (\ref{onelinear})
do not satisfy $s_1+s_3 \geq 0$.}
where the quadratic terms appear on the right-hand sides
\bea
&&
\Phi_{+1}^{(h_1),PQ}(z_1,\bar{z}_1)\,
(\Phi_{+2}^{(h_2)}\, \Phi_{A,-\frac{3}{2}}^{(h_3)})(z_3,
\bar{z}_3) =
\nonu \\
&& 
\frac{\bar{z}_{13}}{z_{13}} \,
\sum_{m=0}^{\infty}\, \frac{\bar{z}_{13}^m}{m!}\,
B(2\bar{h}_1+1+m,2\bar{h}_2+1) \, (\bar{\pa}^m \,
\Phi_{+1}^{(h_1+h_2),PQ}\,\Phi_{A,-\frac{3}{2}}^{(h_3)})
(z_3,\bar{z}_3),
\nonu \\
&&
\Phi_{+1}^{(h_1),PQ}(z_1,\bar{z}_1)\,
(\Phi_{+2}^{(h_2)}\, \Phi_{-2}^{(h_3)})(z_3,
\bar{z}_3) =
\nonu \\
&& 
\frac{\bar{z}_{13}}{z_{13}} \,
\sum_{m=0}^{\infty}\, \frac{\bar{z}_{13}^m}{m!}\,
B(2\bar{h}_1+1+m,2\bar{h}_2+1) \, (\bar{\pa}^m \,
\Phi_{+1}^{(h_1+h_2),PQ}\,\Phi_{-2}^{(h_3)})
(z_3,\bar{z}_3),
\nonu \\
&&
\Phi_{+1}^{(h_1),PQ}(z_1,\bar{z}_1)\,
(\Phi_{+\frac{3}{2}}^{(h_2),A}\, \Phi_{BC,-1}^{(h_3)})(z_3,
\bar{z}_3) =
\nonu \\
&& 
\frac{\bar{z}_{13}}{z_{13}} \,
\sum_{m=0}^{\infty}\, \frac{\bar{z}_{13}^m}{m!}\,
B(2\bar{h}_1+1+m,2\bar{h}_2+1) \, (\bar{\pa}^m \,
\Phi_{+\frac{1}{2}}^{(h_1+h_2),PQA}\,\Phi_{BC,-1}^{(h_3)})
(z_3,\bar{z}_3)
\nonu \\
&& +
\frac{\bar{z}_{13}}{z_{13}} \,
\sum_{m=0}^{\infty}\, \frac{\bar{z}_{13}^m}{m!}\,
B(2\bar{h}_1+1+m,2\bar{h}_3+1)  \,
\de^{PQ}_{BC}\, (
\Phi_{+\frac{3}{2}}^{(h_2),A}\,\bar{\pa}^m \,
\Phi_{-2}^{(h_1+h_3)})
(z_3,\bar{z}_3),
\nonu \\
&&
\Phi_{+1}^{(h_1),PQ}(z_1,\bar{z}_1)\,
(\Phi_{+\frac{3}{2}}^{(h_2),A}\, \Phi_{B,-\frac{3}{2}}^{(h_3)})(z_3,
\bar{z}_3) =
\nonu \\
&& 
\frac{\bar{z}_{13}}{z_{13}} \,
\sum_{m=0}^{\infty}\, \frac{\bar{z}_{13}^m}{m!}\,
B(2\bar{h}_1+1+m,2\bar{h}_2+1) \, (\bar{\pa}^m \,
\Phi_{+\frac{1}{2}}^{(h_1+h_2),PQA}\,\Phi_{B,-\frac{3}{2}}^{(h_3)})
(z_3,\bar{z}_3),
\nonu \\
&&
\Phi_{+1}^{(h_1),PQ}(z_1,\bar{z}_1)\,
(\Phi_{+1}^{(h_2),AB}\, \Phi_{CDE,-\frac{1}{2}}^{(h_3)})(z_3,
\bar{z}_3) =
\nonu \\
&& 
\frac{\bar{z}_{13}}{z_{13}} \,
\sum_{m=0}^{\infty}\, \frac{\bar{z}_{13}^m}{m!}\,
B(2\bar{h}_1+1+m,2\bar{h}_2+1) \,
(\bar{\pa}^m \,
\Phi_{0}^{(h_1+h_2),PQAB}\,\Phi_{CDE,-\frac{1}{2}}^{(h_3)})
(z_3,\bar{z}_3)
\nonu \\
&& +
\frac{\bar{z}_{13}}{z_{13}} \,
\sum_{m=0}^{\infty}\, \frac{\bar{z}_{13}^m}{m!}\,
B(2\bar{h}_1+1+m,2\bar{h}_3+1)  \,
3!\, \de^{P}_{[C}\, (
\Phi_{+1}^{(h_2),AB}\,\bar{\pa}^m \,
\Phi_{D,-\frac{3}{2}}^{(h_1+h_3)})
(z_3,\bar{z}_3) \, \de^{Q}_{E]},
\nonu \\
&&
\Phi_{+1}^{(h_1),PQ}(z_1,\bar{z}_1)\,
(\Phi_{+1}^{(h_2),AB}\, \Phi_{CD,-1}^{(h_3)})(z_3,
\bar{z}_3) =
\nonu \\
&& 
\frac{\bar{z}_{13}}{z_{13}} \,
\sum_{m=0}^{\infty}\, \frac{\bar{z}_{13}^m}{m!}\,
B(2\bar{h}_1+1+m,2\bar{h}_2+1) \,
(\bar{\pa}^m \,
\Phi_{0}^{(h_1+h_2),PQAB}\,\Phi_{CD,-1}^{(h_3)})
(z_3,\bar{z}_3)
\nonu \\
&& +
\frac{\bar{z}_{13}}{z_{13}} \,
\sum_{m=0}^{\infty}\, \frac{\bar{z}_{13}^m}{m!}\,
B(2\bar{h}_1+1+m,2\bar{h}_3+1)  \,
\de^{PQ}_{CD}\, (
\Phi_{+1}^{(h_2),AB}\,\bar{\pa}^m \,
\Phi_{-2}^{(h_1+h_3)})
(z_3,\bar{z}_3),
\nonu \\
&&
\Phi_{+1}^{(h_1),PQ}(z_1,\bar{z}_1)\,
(\Phi_{+\frac{1}{2}}^{(h_2),ABC}\, \Phi_{0}^{(h_3),DEFG})(z_3,
\bar{z}_3) =
\nonu \\
&& 
\frac{\bar{z}_{13}}{z_{13}} \,
\sum_{m=0}^{\infty}\, \frac{\bar{z}_{13}^m}{m!}\,
B(2\bar{h}_1+1+m,2\bar{h}_2+1) \,
\frac{1}{3!}\,
(\bar{\pa}^m \,\ep^{PQABCHIJ}
\Phi_{HIJ,-\frac{1}{2}}^{(h_1+h_2)}\,\Phi_{0}^{(h_3),DEFG})
(z_3,\bar{z}_3)
\nonu \\
&& +
\frac{\bar{z}_{13}}{z_{13}} \,
\sum_{m=0}^{\infty}\, \frac{\bar{z}_{13}^m}{m!}\,
B(2\bar{h}_1+1+m,2\bar{h}_3+1)  \,
\frac{1}{2!}\, \ep^{PQDEFGHI}_{}\, (
\Phi_{+\frac{1}{2}}^{(h_2),ABC}\,\bar{\pa}^m \,
\Phi_{HI,-1}^{(h_1+h_3)})
(z_3,\bar{z}_3),
\nonu \\
&&
\Phi_{+1}^{(h_1),PQ}(z_1,\bar{z}_1)\,
(\Phi_{+\frac{1}{2}}^{(h_2),ABC}\, \Phi_{DEF,-\frac{1}{2}}^{(h_3)})(z_3,
\bar{z}_3) =
\nonu \\
&& 
\frac{\bar{z}_{13}}{z_{13}} \,
\sum_{m=0}^{\infty}\, \frac{\bar{z}_{13}^m}{m!}\,
B(2\bar{h}_1+1+m,2\bar{h}_2+1) \,
\frac{1}{3!}\,
(\bar{\pa}^m \,\ep^{PQABCHIJ}
\Phi_{HIJ,-\frac{1}{2}}^{(h_1+h_2)}\,\Phi_{DEF,-\frac{1}{2}}^{(h_3)})
(z_3,\bar{z}_3)
\nonu \\
&& +
\frac{\bar{z}_{13}}{z_{13}} \,
\sum_{m=0}^{\infty}\, \frac{\bar{z}_{13}^m}{m!}\,
B(2\bar{h}_1+1+m,2\bar{h}_3+1)  \,
3!\, \de^{P}_{[D} \, (
\Phi_{+\frac{1}{2}}^{(h_2),ABC}\,\bar{\pa}^m \,
\Phi_{E,-\frac{3}{2}}^{(h_1+h_3)})
(z_3,\bar{z}_3) \, \de^{Q}_{F]},
\nonu \\
&&
\Phi_{+1}^{(h_1),PQ}(z_1,\bar{z}_1)\,
(\Phi_{0}^{(h_2),ABCD}\, \Phi_{0}^{(h_3),EFGH})(z_3,
\bar{z}_3) =
\nonu \\
&& 
\frac{\bar{z}_{13}}{z_{13}} \,
\sum_{m=0}^{\infty}\, \frac{\bar{z}_{13}^m}{m!}\,
B(2\bar{h}_1+1+m,2\bar{h}_2+1) \,
\frac{1}{2!}\,
(\bar{\pa}^m \,\ep^{PQABCDIJ}
\Phi_{IJ,-1}^{(h_1+h_2)}\,\Phi_{0}^{(h_3),EFGH})
(z_3,\bar{z}_3)
\label{onelinear}
\\
&& +
\frac{\bar{z}_{13}}{z_{13}} \,
\sum_{m=0}^{\infty}\, \frac{\bar{z}_{13}^m}{m!}\,
B(2\bar{h}_1+1+m,2\bar{h}_3+1)  \,\frac{1}{2!}\, \ep^{PQEFGHIJ}
(
\Phi_{0}^{(h_2),ABCD}\,\bar{\pa}^m \,
\Phi_{IJ,-1}^{(h_1+h_3)})
(z_3,\bar{z}_3).
\nonu
\eea

\subsection{The operator product expansions of
the graviphotinos  of helicity $+\frac{1}{2}$  with the quadratic
operators}

The twelve OPEs
\footnote{After removing the four OPEs which do not satisfy
the condition $s_1+s_2+s_3 \geq 2$, we are left with these
twelve OPEs.}
are written as 
\bea
&&
\Phi_{+\frac{1}{2}}^{(h_1),PQR}(z_1,\bar{z}_1)\,
(\Phi_{+2}^{(h_2)}\, \Phi_{+2}^{(h_3)})(z_3,
\bar{z}_3) = \frac{\bar{z}_{13}^2}{z_{13}^2} \,
\sum_{m=0}^{\infty}\, \frac{\bar{z}_{13}^m}{m!}\,
B(2\bar{h}_1+2+m,2\bar{h}_2+1, 2\bar{h}_3+1)
\nonu \\
&& \times
\bar{\pa}^m \, \Phi_{+\frac{1}{2}}^{(h_1+h_2+h_3),PQR}(z_3,\bar{z}_3)
\nonu \\
&& + 
\frac{\bar{z}_{13}}{z_{13}} \,
\sum_{m=0}^{\infty}\, \frac{\bar{z}_{13}^m}{m!}\,
B(2\bar{h}_1+1+m,2\bar{h}_2+1) \,
(\bar{\pa}^m \,
\Phi_{+\frac{1}{2}}^{(h_1+h_2),PQR}\,\Phi_{+2}^{(h_3)})
(z_3,\bar{z}_3)
\nonu \\
&& +
\frac{\bar{z}_{13}}{z_{13}} \,
\sum_{m=0}^{\infty}\, \frac{\bar{z}_{13}^m}{m!}\,
B(2\bar{h}_1+1+m,2\bar{h}_3+1)  \,
(
\Phi_{+2}^{(h_2)}\,\bar{\pa}^m \,
\Phi_{+\frac{1}{2}}^{(h_1+h_3),PQR})
(z_3,\bar{z}_3),
\nonu \\
&&
\Phi_{+\frac{1}{2}}^{(h_1),PQR}(z_1,\bar{z}_1)\,
(\Phi_{+2}^{(h_2)}\, \Phi_{+\frac{3}{2}}^{(h_3),A})(z_3,
\bar{z}_3) = \frac{\bar{z}_{13}^2}{z_{13}^2} \,
\sum_{m=0}^{\infty}\, \frac{\bar{z}_{13}^m}{m!}\,
B(2\bar{h}_1+2+m,2\bar{h}_2+1, 2\bar{h}_3+1)
\nonu \\
&& \times
\bar{\pa}^m \, \Phi_{0}^{(h_1+h_2+h_3),PQRA}(z_3,\bar{z}_3)
\nonu \\
&& + 
\frac{\bar{z}_{13}}{z_{13}} \,
\sum_{m=0}^{\infty}\, \frac{\bar{z}_{13}^m}{m!}\,
B(2\bar{h}_1+1+m,2\bar{h}_2+1) \,
(\bar{\pa}^m \,
\Phi_{+\frac{1}{2}}^{(h_1+h_2),PQR}\,\Phi_{+\frac{3}{2}}^{(h_3),A})
(z_3,\bar{z}_3)
\nonu \\
&& +
\frac{\bar{z}_{13}}{z_{13}} \,
\sum_{m=0}^{\infty}\, \frac{\bar{z}_{13}^m}{m!}\,
B(2\bar{h}_1+1+m,2\bar{h}_3+1)  \,
(
\Phi_{+2}^{(h_2)}\,\bar{\pa}^m \,
\Phi_{0}^{(h_1+h_3),PQRA})
(z_3,\bar{z}_3),
\nonu \\
&&
\Phi_{+\frac{1}{2}}^{(h_1),PQR}(z_1,\bar{z}_1)\,
(\Phi_{+2}^{(h_2)}\, \Phi_{+1}^{(h_3),AB})(z_3,
\bar{z}_3) = \frac{\bar{z}_{13}^2}{z_{13}^2} \,
\sum_{m=0}^{\infty}\, \frac{\bar{z}_{13}^m}{m!}\,
B(2\bar{h}_1+2+m,2\bar{h}_2+1, 2\bar{h}_3+1)
\nonu \\
&& \times
\frac{1}{3!}\,
\ep^{ABPQRCDE}
\bar{\pa}^m \, \Phi_{CDE,-\frac{1}{2}}^{(h_1+h_2+h_3)}(z_3,\bar{z}_3)
\nonu \\
&& + 
\frac{\bar{z}_{13}}{z_{13}} \,
\sum_{m=0}^{\infty}\, \frac{\bar{z}_{13}^m}{m!}\,
B(2\bar{h}_1+1+m,2\bar{h}_2+1) \,
(\bar{\pa}^m \,
\Phi_{+\frac{1}{2}}^{(h_1+h_2),PQR}\,\Phi_{+1}^{(h_3),AB})
(z_3,\bar{z}_3)
\nonu \\
&& +
\frac{\bar{z}_{13}}{z_{13}} \,
\sum_{m=0}^{\infty}\, \frac{\bar{z}_{13}^m}{m!}\,
B(2\bar{h}_1+1+m,2\bar{h}_3+1)  \, \frac{1}{3!}\,
\ep^{PQRABFGH}\, (
\Phi_{+2}^{(h_2)}\,\bar{\pa}^m \,
\Phi_{FGH,-\frac{1}{2}}^{(h_1+h_3)})
(z_3,\bar{z}_3),
\nonu \\
&&
\Phi_{+\frac{1}{2}}^{(h_1),PQR}(z_1,\bar{z}_1)\,
(\Phi_{+2}^{(h_2)}\, \Phi_{+\frac{1}{2}}^{(h_3),ABC})(z_3,
\bar{z}_3) = \frac{\bar{z}_{13}^2}{z_{13}^2} \,
\sum_{m=0}^{\infty}\, \frac{\bar{z}_{13}^m}{m!}\,
B(2\bar{h}_1+2+m,2\bar{h}_2+1, 2\bar{h}_3+1)
\nonu \\
&& \times
\frac{1}{2!}\,
\ep^{PQRABCDE}
\bar{\pa}^m \, \Phi_{DE,-1}^{(h_1+h_2+h_3)}(z_3,\bar{z}_3)
\nonu \\
&& + 
\frac{\bar{z}_{13}}{z_{13}} \,
\sum_{m=0}^{\infty}\, \frac{\bar{z}_{13}^m}{m!}\,
B(2\bar{h}_1+1+m,2\bar{h}_2+1) \,
(\bar{\pa}^m \,
\Phi_{+\frac{1}{2}}^{(h_1+h_2),PQR}\,\Phi_{+\frac{1}{2}}^{(h_3),ABC})
(z_3,\bar{z}_3)
\nonu \\
&& +
\frac{\bar{z}_{13}}{z_{13}} \,
\sum_{m=0}^{\infty}\, \frac{\bar{z}_{13}^m}{m!}\,
B(2\bar{h}_1+1+m,2\bar{h}_3+1)  \, \frac{1}{2!}\,
\ep^{PQRABCDE}\, (
\Phi_{+2}^{(h_2)}\,\bar{\pa}^m \,
\Phi_{DE,-1}^{(h_1+h_3)})
(z_3,\bar{z}_3),
\nonu \\
&&
\Phi_{+\frac{1}{2}}^{(h_1),PQR}(z_1,\bar{z}_1)\,
(\Phi_{+2}^{(h_2)}\, \Phi_{0}^{(h_3),ABCD})(z_3,
\bar{z}_3) = \frac{\bar{z}_{13}^2}{z_{13}^2} \,
\sum_{m=0}^{\infty}\, \frac{\bar{z}_{13}^m}{m!}\,
B(2\bar{h}_1+2+m,2\bar{h}_2+1, 2\bar{h}_3+1)
\nonu \\
&& \times
\ep^{PQRABCDE}\,
\bar{\pa}^m \, \Phi_{E,-\frac{3}{2}}^{(h_1+h_2+h_3)}(z_3,\bar{z}_3)
\nonu \\
&& + 
\frac{\bar{z}_{13}}{z_{13}} \,
\sum_{m=0}^{\infty}\, \frac{\bar{z}_{13}^m}{m!}\,
B(2\bar{h}_1+1+m,2\bar{h}_2+1) \,
(\bar{\pa}^m \,
\Phi_{+\frac{1}{2}}^{(h_1+h_2),PQR}\,\Phi_{0}^{(h_3),ABCD})
(z_3,\bar{z}_3)
\nonu \\
&& +
\frac{\bar{z}_{13}}{z_{13}} \,
\sum_{m=0}^{\infty}\, \frac{\bar{z}_{13}^m}{m!}\,
B(2\bar{h}_1+1+m,2\bar{h}_3+1)  \, 
\ep^{PQRABCDE}\, (
\Phi_{+2}^{(h_2)}\,\bar{\pa}^m \,
\Phi_{E,-\frac{3}{2}}^{(h_1+h_3)})
(z_3,\bar{z}_3),
\nonu \\
&&
\Phi_{+\frac{1}{2}}^{(h_1),PQR}(z_1,\bar{z}_1)\,
(\Phi_{+2}^{(h_2)}\, \Phi_{ABC,-\frac{1}{2}}^{(h_3)})(z_3,
\bar{z}_3) = \frac{\bar{z}_{13}^2}{z_{13}^2} \,
\sum_{m=0}^{\infty}\, \frac{\bar{z}_{13}^m}{m!}\,
B(2\bar{h}_1+2+m,2\bar{h}_2+1, 2\bar{h}_3+1)
\nonu \\
&& \times
\de^{PQR}_{ABC}\,
\bar{\pa}^m \, \Phi_{-2}^{(h_1+h_2+h_3)}(z_3,\bar{z}_3)
\nonu \\
&& + 
\frac{\bar{z}_{13}}{z_{13}} \,
\sum_{m=0}^{\infty}\, \frac{\bar{z}_{13}^m}{m!}\,
B(2\bar{h}_1+1+m,2\bar{h}_2+1) \,
(\bar{\pa}^m \,
\Phi_{+\frac{1}{2}}^{(h_1+h_2),PQR}\,\Phi_{ABC,-\frac{1}{2}}^{(h_3)})
(z_3,\bar{z}_3)
\nonu \\
&& +
\frac{\bar{z}_{13}}{z_{13}} \,
\sum_{m=0}^{\infty}\, \frac{\bar{z}_{13}^m}{m!}\,
B(2\bar{h}_1+1+m,2\bar{h}_3+1)  \, 
\de^{PQR}_{ABC}\, (
\Phi_{+2}^{(h_2)}\,\bar{\pa}^m \,
\Phi_{-2}^{(h_1+h_3)})
(z_3,\bar{z}_3),
\nonu \\
&&
\Phi_{+\frac{1}{2}}^{(h_1),PQR}(z_1,\bar{z}_1)\,
(\Phi_{+\frac{3}{2}}^{(h_2),A}\, \Phi_{+\frac{3}{2}}^{(h_3),B})(z_3,
\bar{z}_3) = \frac{\bar{z}_{13}^2}{z_{13}^2} \,
\sum_{m=0}^{\infty}\, \frac{\bar{z}_{13}^m}{m!}\,
B(2\bar{h}_1+2+m,2\bar{h}_2+1, 2\bar{h}_3+1)
\nonu \\
&& \times
\frac{1}{3!}\,
\ep^{ABPQRCDE}\,
\bar{\pa}^m \, \Phi_{CDE,-\frac{1}{2}}^{(h_1+h_2+h_3)}(z_3,\bar{z}_3)
\nonu \\
&& + 
\frac{\bar{z}_{13}}{z_{13}} \,
\sum_{m=0}^{\infty}\, \frac{\bar{z}_{13}^m}{m!}\,
B(2\bar{h}_1+1+m,2\bar{h}_2+1) \,
(\bar{\pa}^m \,
\Phi_{0}^{(h_1+h_2),PQRA}\,\Phi_{+\frac{3}{2}}^{(h_3),B})
(z_3,\bar{z}_3)
\nonu \\
&& +
\frac{\bar{z}_{13}}{z_{13}} \,
\sum_{m=0}^{\infty}\, \frac{\bar{z}_{13}^m}{m!}\,
B(2\bar{h}_1+1+m,2\bar{h}_3+1)  \, 
(-1)\,(
\Phi_{+\frac{3}{2}}^{(h_2),A}\,\bar{\pa}^m \,
\Phi_{0}^{(h_1+h_3),PQRB})
(z_3,\bar{z}_3),
\nonu \\
&&
\Phi_{+\frac{1}{2}}^{(h_1),PQR}(z_1,\bar{z}_1)\,
(\Phi_{+\frac{3}{2}}^{(h_2),A}\, \Phi_{+1}^{(h_3),BC})(z_3,
\bar{z}_3) = \frac{\bar{z}_{13}^2}{z_{13}^2} \,
\sum_{m=0}^{\infty}\, \frac{\bar{z}_{13}^m}{m!}\,
B(2\bar{h}_1+2+m,2\bar{h}_2+1, 2\bar{h}_3+1)
\nonu \\
&& \times
\frac{1}{2!}\,
\ep^{PQRBCADE}\,
\bar{\pa}^m \, \Phi_{DE,-1}^{(h_1+h_2+h_3)}(z_3,\bar{z}_3)
\nonu \\
&& + 
\frac{\bar{z}_{13}}{z_{13}} \,
\sum_{m=0}^{\infty}\, \frac{\bar{z}_{13}^m}{m!}\,
B(2\bar{h}_1+1+m,2\bar{h}_2+1) \,
(\bar{\pa}^m \,
\Phi_{0}^{(h_1+h_2),PQRA}\,\Phi_{+1}^{(h_3),BC})
(z_3,\bar{z}_3)
\nonu \\
&& +
\frac{\bar{z}_{13}}{z_{13}} \,
\sum_{m=0}^{\infty}\, \frac{\bar{z}_{13}^m}{m!}\,
B(2\bar{h}_1+1+m,2\bar{h}_3+1)  \, (-1)\,
\frac{1}{3!}\, \ep^{PQRBCFGH}\, (
\Phi_{+\frac{3}{2}}^{(h_2),A}\,\bar{\pa}^m \,
\Phi_{FGH,-\frac{1}{2}}^{(h_1+h_3)})
(z_3,\bar{z}_3),
\nonu \\
&&
\Phi_{+\frac{1}{2}}^{(h_1),PQR}(z_1,\bar{z}_1)\,
(\Phi_{+\frac{3}{2}}^{(h_2),A}\, \Phi_{+\frac{1}{2}}^{(h_3),BCD})(z_3,
\bar{z}_3) = \frac{\bar{z}_{13}^2}{z_{13}^2} \,
\sum_{m=0}^{\infty}\, \frac{\bar{z}_{13}^m}{m!}\,
B(2\bar{h}_1+2+m,2\bar{h}_2+1, 2\bar{h}_3+1)
\nonu \\
&& \times
\ep^{ABCDPQRE}\,
\bar{\pa}^m \, \Phi_{E,-\frac{3}{2}}^{(h_1+h_2+h_3)}(z_3,\bar{z}_3)
\nonu \\
&& + 
\frac{\bar{z}_{13}}{z_{13}} \,
\sum_{m=0}^{\infty}\, \frac{\bar{z}_{13}^m}{m!}\,
B(2\bar{h}_1+1+m,2\bar{h}_2+1) \,
(\bar{\pa}^m \,
\Phi_{0}^{(h_1+h_2),PQRA}\,\Phi_{+\frac{1}{2}}^{(h_3),BCD})
(z_3,\bar{z}_3)
\nonu \\
&& +
\frac{\bar{z}_{13}}{z_{13}} \,
\sum_{m=0}^{\infty}\, \frac{\bar{z}_{13}^m}{m!}\,
B(2\bar{h}_1+1+m,2\bar{h}_3+1)  \, (-1)\,
\frac{1}{2!}\, \ep^{PQRBCDFG}\, (
\Phi_{+\frac{3}{2}}^{(h_2),A}\,\bar{\pa}^m \,
\Phi_{FG,-1}^{(h_1+h_3)})
(z_3,\bar{z}_3),
\nonu \\
&&
\Phi_{+\frac{1}{2}}^{(h_1),PQR}(z_1,\bar{z}_1)\,
(\Phi_{+\frac{3}{2}}^{(h_2),A}\, \Phi_{0}^{(h_3),BCDE})(z_3,
\bar{z}_3) = \frac{\bar{z}_{13}^2}{z_{13}^2} \,
\sum_{m=0}^{\infty}\, \frac{\bar{z}_{13}^m}{m!}\,
B(2\bar{h}_1+2+m,2\bar{h}_2+1, 2\bar{h}_3+1)
\nonu \\
&& \times
\ep^{PQRABCDE}\,
\bar{\pa}^m \, \Phi_{-2}^{(h_1+h_2+h_3)}(z_3,\bar{z}_3)
\nonu \\
&& + 
\frac{\bar{z}_{13}}{z_{13}} \,
\sum_{m=0}^{\infty}\, \frac{\bar{z}_{13}^m}{m!}\,
B(2\bar{h}_1+1+m,2\bar{h}_2+1) \,
(\bar{\pa}^m \,
\Phi_{0}^{(h_1+h_2),PQRA}\,\Phi_{0}^{(h_3),BCDE})
(z_3,\bar{z}_3)
\nonu \\
&& +
\frac{\bar{z}_{13}}{z_{13}} \,
\sum_{m=0}^{\infty}\, \frac{\bar{z}_{13}^m}{m!}\,
B(2\bar{h}_1+1+m,2\bar{h}_3+1)  \, (-1)\,
\ep^{PQRBCDEF}\, (
\Phi_{+\frac{3}{2}}^{(h_2),A}\,\bar{\pa}^m \,
\Phi_{F,-\frac{3}{2}}^{(h_1+h_3)})
(z_3,\bar{z}_3),
\nonu \\
&&
\Phi_{+\frac{1}{2}}^{(h_1),PQR}(z_1,\bar{z}_1)\,
(\Phi_{+1}^{(h_2),AB}\, \Phi_{+1}^{(h_3),CD})(z_3,
\bar{z}_3) = \frac{\bar{z}_{13}^2}{z_{13}^2} \,
\sum_{m=0}^{\infty}\, \frac{\bar{z}_{13}^m}{m!}\,
B(2\bar{h}_1+2+m,2\bar{h}_2+1, 2\bar{h}_3+1)
\nonu \\
&& \times
\ep^{ABPQREFG}
\, \de^{C}_{[E}\,
\bar{\pa}^m \, \Phi_{F,-\frac{3}{2}}^{(h_1+h_2+h_3)}(z_3,\bar{z}_3)
\, \de^{D}_{G]}
\nonu \\
&& + 
\frac{\bar{z}_{13}}{z_{13}} \,
\sum_{m=0}^{\infty}\, \frac{\bar{z}_{13}^m}{m!}\,
B(2\bar{h}_1+1+m,2\bar{h}_2+1) \, \frac{1}{3!}\,
\ep^{PQRABFGH}\,
(\bar{\pa}^m \,
\Phi_{FGH,-\frac{1}{2}}^{(h_1+h_2)}\,\Phi_{+1}^{(h_3),CD})
(z_3,\bar{z}_3)
\nonu \\
&& +
\frac{\bar{z}_{13}}{z_{13}} \,
\sum_{m=0}^{\infty}\, \frac{\bar{z}_{13}^m}{m!}\,
B(2\bar{h}_1+1+m,2\bar{h}_3+1)  \, \frac{1}{3!}\,
\ep^{PQRCDFGH}\, (
\Phi_{+1}^{(h_2),AB}\,\bar{\pa}^m \,
\Phi_{FGH,-\frac{1}{2}}^{(h_1+h_3)})
(z_3,\bar{z}_3),
\nonu \\
&&
\Phi_{+\frac{1}{2}}^{(h_1),PQR}(z_1,\bar{z}_1)\,
(\Phi_{+1}^{(h_2),AB}\, \Phi_{+\frac{1}{2}}^{(h_3),CDE})(z_3,
\bar{z}_3) = \frac{\bar{z}_{13}^2}{z_{13}^2} \,
\sum_{m=0}^{\infty}\, \frac{\bar{z}_{13}^m}{m!}\,
B(2\bar{h}_1+2+m,2\bar{h}_2+1, 2\bar{h}_3+1)
\nonu \\
&& \times
\frac{-1}{3!}\,
\ep^{ABPQRFGH}\,
\de^{CDE}_{FGH}\,
\bar{\pa}^m \, \Phi_{-2}^{(h_1+h_2+h_3)}(z_3,\bar{z}_3)
\nonu \\
&& + 
\frac{\bar{z}_{13}}{z_{13}} \,
\sum_{m=0}^{\infty}\, \frac{\bar{z}_{13}^m}{m!}\,
B(2\bar{h}_1+1+m,2\bar{h}_2+1) \, \frac{1}{3!}\,
\ep^{PQRABFGH}\,
(\bar{\pa}^m \,
\Phi_{FGH,-\frac{1}{2}}^{(h_1+h_2)}\,\Phi_{+\frac{1}{2}}^{(h_3),CDE})
(z_3,\bar{z}_3)
\nonu \\
&& +
\frac{\bar{z}_{13}}{z_{13}} \,
\sum_{m=0}^{\infty}\, \frac{\bar{z}_{13}^m}{m!}\,
B(2\bar{h}_1+1+m,2\bar{h}_3+1)  \, \frac{1}{2!}\,
\ep^{PQRCDEGH}\, (
\Phi_{+1}^{(h_2),AB}\,\bar{\pa}^m \,
\Phi_{GH,-1}^{(h_1+h_3)})
(z_3,\bar{z}_3).
\label{fullonehalf}
\eea
Similarly, there are
thirteen nontrivial OPEs
\footnote{Six of these OPEs in (\ref{onehalflinear})
do not satisfy $s_1+s_3 \geq 0$.}
with quadratic terms only
\bea
&&
\Phi_{+\frac{1}{2}}^{(h_1),PQR}(z_1,\bar{z}_1)\,
(\Phi_{+2}^{(h_2)}\, \Phi_{AB,-1}^{(h_3)})(z_3,
\bar{z}_3) =
\nonu \\
&&  
\frac{\bar{z}_{13}}{z_{13}} \,
\sum_{m=0}^{\infty}\, \frac{\bar{z}_{13}^m}{m!}\,
B(2\bar{h}_1+1+m,2\bar{h}_2+1) \, 
(\bar{\pa}^m \,
\Phi_{+\frac{1}{2}}^{(h_1+h_2),PQR}\,\Phi_{AB,-1}^{(h_3)})
(z_3,\bar{z}_3),
\nonu \\
&&
\Phi_{+\frac{1}{2}}^{(h_1),PQR}(z_1,\bar{z}_1)\,
(\Phi_{+2}^{(h_2)}\, \Phi_{A,-\frac{3}{2}}^{(h_3)})(z_3,
\bar{z}_3) =
\nonu \\
&&  
\frac{\bar{z}_{13}}{z_{13}} \,
\sum_{m=0}^{\infty}\, \frac{\bar{z}_{13}^m}{m!}\,
B(2\bar{h}_1+1+m,2\bar{h}_2+1) \, 
(\bar{\pa}^m \,
\Phi_{+\frac{1}{2}}^{(h_1+h_2),PQR}\,\Phi_{A,-\frac{3}{2}}^{(h_3)})
(z_3,\bar{z}_3),
\nonu \\
&&
\Phi_{+\frac{1}{2}}^{(h_1),PQR}(z_1,\bar{z}_1)\,
(\Phi_{+2}^{(h_2)}\, \Phi_{-2}^{(h_3)})(z_3,
\bar{z}_3) =
\nonu \\
&&  
\frac{\bar{z}_{13}}{z_{13}} \,
\sum_{m=0}^{\infty}\, \frac{\bar{z}_{13}^m}{m!}\,
B(2\bar{h}_1+1+m,2\bar{h}_2+1) \, 
(\bar{\pa}^m \,
\Phi_{+\frac{1}{2}}^{(h_1+h_2),PQR}\,\Phi_{-2}^{(h_3)})
(z_3,\bar{z}_3),
\nonu \\
&&
\Phi_{+\frac{1}{2}}^{(h_1),PQR}(z_1,\bar{z}_1)\,
(\Phi_{+\frac{3}{2}}^{(h_2),A}\, \Phi_{BCD,-\frac{1}{2}}^{(h_3)})(z_3,
\bar{z}_3) =
\nonu \\
&&  
\frac{\bar{z}_{13}}{z_{13}} \,
\sum_{m=0}^{\infty}\, \frac{\bar{z}_{13}^m}{m!}\,
B(2\bar{h}_1+1+m,2\bar{h}_2+1) \, 
(\bar{\pa}^m \,
\Phi_{0}^{(h_1+h_2),PQRA}\,\Phi_{BCD,-\frac{1}{2}}^{(h_3)})
(z_3,\bar{z}_3)
\nonu \\
&& +
\frac{\bar{z}_{13}}{z_{13}} \,
\sum_{m=0}^{\infty}\, \frac{\bar{z}_{13}^m}{m!}\,
B(2\bar{h}_1+1+m,2\bar{h}_3+1)  \, 
(-1)\, \de^{PQR}_{BCD}\, (
\Phi_{+\frac{3}{2}}^{(h_2),A}\,\bar{\pa}^m \,
\Phi_{-2}^{(h_1+h_3)})
(z_3,\bar{z}_3),
\nonu \\
&&
\Phi_{+\frac{1}{2}}^{(h_1),PQR}(z_1,\bar{z}_1)\,
(\Phi_{+\frac{3}{2}}^{(h_2),A}\, \Phi_{BC,-1}^{(h_3)})(z_3,
\bar{z}_3) =
\nonu \\
&&  
\frac{\bar{z}_{13}}{z_{13}} \,
\sum_{m=0}^{\infty}\, \frac{\bar{z}_{13}^m}{m!}\,
B(2\bar{h}_1+1+m,2\bar{h}_2+1) \, 
(\bar{\pa}^m \,
\Phi_{0}^{(h_1+h_2),PQRA}\,\Phi_{BC,-1}^{(h_3)})
(z_3,\bar{z}_3),
\nonu \\
&&
\Phi_{+\frac{1}{2}}^{(h_1),PQR}(z_1,\bar{z}_1)\,
(\Phi_{+\frac{3}{2}}^{(h_2),A}\, \Phi_{B,-\frac{3}{2}}^{(h_3)})(z_3,
\bar{z}_3) =
\nonu \\
&&  
\frac{\bar{z}_{13}}{z_{13}} \,
\sum_{m=0}^{\infty}\, \frac{\bar{z}_{13}^m}{m!}\,
B(2\bar{h}_1+1+m,2\bar{h}_2+1) \, 
(\bar{\pa}^m \,
\Phi_{0}^{(h_1+h_2),PQRA}\,\Phi_{B,-\frac{3}{2}}^{(h_3)})
(z_3,\bar{z}_3),
\nonu \\
&&
\Phi_{+\frac{1}{2}}^{(h_1),PQR}(z_1,\bar{z}_1)\,
(\Phi_{+1}^{(h_2),AB}\, \Phi_{0}^{(h_3),CDEF})(z_3,
\bar{z}_3) =
\nonu \\
&&  
\frac{\bar{z}_{13}}{z_{13}} \,
\sum_{m=0}^{\infty}\, \frac{\bar{z}_{13}^m}{m!}\,
B(2\bar{h}_1+1+m,2\bar{h}_2+1) \, \frac{1}{3!}\,
\ep^{PQRABFGH}\,
(\bar{\pa}^m \,
\Phi_{FGH,-\frac{1}{2}}^{(h_1+h_2)}\,\Phi_{0}^{(h_3),CDEF})
(z_3,\bar{z}_3)
\nonu \\
&& +
\frac{\bar{z}_{13}}{z_{13}} \,
\sum_{m=0}^{\infty}\, \frac{\bar{z}_{13}^m}{m!}\,
B(2\bar{h}_1+1+m,2\bar{h}_3+1)  \, 
\ep^{PQRCDEFG}\, (
\Phi_{+1}^{(h_2),AB}\,\bar{\pa}^m \,
\Phi_{G,-\frac{3}{2}}^{(h_1+h_3)})
(z_3,\bar{z}_3),
\nonu \\
&&
\Phi_{+\frac{1}{2}}^{(h_1),PQR}(z_1,\bar{z}_1)\,
(\Phi_{+1}^{(h_2),AB}\, \Phi_{CDE,-\frac{1}{2}}^{(h_3)})(z_3,
\bar{z}_3) =
\nonu \\
&&  
\frac{\bar{z}_{13}}{z_{13}} \,
\sum_{m=0}^{\infty}\, \frac{\bar{z}_{13}^m}{m!}\,
B(2\bar{h}_1+1+m,2\bar{h}_2+1) \, \frac{1}{3!}\,
\ep^{PQRABFGH}\,
(\bar{\pa}^m \,
\Phi_{FGH,-\frac{1}{2}}^{(h_1+h_2)}\,\Phi_{CDE,-\frac{1}{2}}^{(h_3)})
(z_3,\bar{z}_3)
\nonu \\
&& +
\frac{\bar{z}_{13}}{z_{13}} \,
\sum_{m=0}^{\infty}\, \frac{\bar{z}_{13}^m}{m!}\,
B(2\bar{h}_1+1+m,2\bar{h}_3+1)  \, 
\de^{PQR}_{CDE}\,  (
\Phi_{+1}^{(h_2),AB}\,\bar{\pa}^m \,
\Phi_{-2}^{(h_1+h_3)})
(z_3,\bar{z}_3),
\nonu \\
&&
\Phi_{+\frac{1}{2}}^{(h_1),PQR}(z_1,\bar{z}_1)\,
(\Phi_{+1}^{(h_2),AB}\, \Phi_{CD,-1}^{(h_3)})(z_3,
\bar{z}_3) =
\nonu \\
&&  
\frac{\bar{z}_{13}}{z_{13}} \,
\sum_{m=0}^{\infty}\, \frac{\bar{z}_{13}^m}{m!}\,
B(2\bar{h}_1+1+m,2\bar{h}_2+1) \, \frac{1}{3!}\,
\ep^{PQRABFGH}\,
(\bar{\pa}^m \,
\Phi_{FGH,-\frac{1}{2}}^{(h_1+h_2)}\,\Phi_{CD,-1}^{(h_3)})
(z_3,\bar{z}_3),
\nonu \\
&&
\Phi_{+\frac{1}{2}}^{(h_1),PQR}(z_1,\bar{z}_1)\,
(\Phi_{+\frac{1}{2}}^{(h_2),ABC}\, \Phi_{+\frac{1}{2}}^{(h_3),DEF})(z_3,
\bar{z}_3) =
\nonu \\
&&  
\frac{\bar{z}_{13}}{z_{13}} \,
\sum_{m=0}^{\infty}\, \frac{\bar{z}_{13}^m}{m!}\,
B(2\bar{h}_1+1+m,2\bar{h}_2+1) \, \frac{1}{2!}\,
\ep^{PQRABCGH}\,
(\bar{\pa}^m \,
\Phi_{GH,-1}^{(h_1+h_2)}\,\Phi_{+\frac{1}{2}}^{(h_3),DEF})
(z_3,\bar{z}_3)
\nonu \\
&& +
\frac{\bar{z}_{13}}{z_{13}} \,
\sum_{m=0}^{\infty}\, \frac{\bar{z}_{13}^m}{m!}\,
B(2\bar{h}_1+1+m,2\bar{h}_3+1)  \, (-1)\, \frac{1}{2!}\,
\ep^{PQRDEFGH} \,  (
\Phi_{+\frac{1}{2}}^{(h_2),ABC}\,\bar{\pa}^m \,
\Phi_{GH,-1}^{(h_1+h_3)})
(z_3,\bar{z}_3),
\nonu \\
&&
\Phi_{+\frac{1}{2}}^{(h_1),PQR}(z_1,\bar{z}_1)\,
(\Phi_{+\frac{1}{2}}^{(h_2),ABC}\, \Phi_{0}^{(h_3),DEFG})(z_3,
\bar{z}_3) =
\nonu \\
&&  
\frac{\bar{z}_{13}}{z_{13}} \,
\sum_{m=0}^{\infty}\, \frac{\bar{z}_{13}^m}{m!}\,
B(2\bar{h}_1+1+m,2\bar{h}_2+1) \, \frac{1}{2!}\,
\ep^{PQRABCGH}\,
(\bar{\pa}^m \,
\Phi_{GH,-1}^{(h_1+h_2)}\,\Phi_{0}^{(h_3),DEFG})
(z_3,\bar{z}_3)
\nonu \\
&& +
\frac{\bar{z}_{13}}{z_{13}} \,
\sum_{m=0}^{\infty}\, \frac{\bar{z}_{13}^m}{m!}\,
B(2\bar{h}_1+1+m,2\bar{h}_3+1)  \, (-1)\, 
\ep^{PQRDEFGH} \,  (
\Phi_{+\frac{1}{2}}^{(h_2),ABC}\,\bar{\pa}^m \,
\Phi_{H,-\frac{3}{2}}^{(h_1+h_3)})
(z_3,\bar{z}_3),
\nonu \\
&&
\Phi_{+\frac{1}{2}}^{(h_1),PQR}(z_1,\bar{z}_1)\,
(\Phi_{+\frac{1}{2}}^{(h_2),ABC}\, \Phi_{DEF,-\frac{1}{2}}^{(h_3)})(z_3,
\bar{z}_3) =
\nonu \\
&&  
\frac{\bar{z}_{13}}{z_{13}} \,
\sum_{m=0}^{\infty}\, \frac{\bar{z}_{13}^m}{m!}\,
B(2\bar{h}_1+1+m,2\bar{h}_2+1) \, \frac{1}{2!}\,
\ep^{PQRABCGH}\,
(\bar{\pa}^m \,
\Phi_{GH,-1}^{(h_1+h_2)}\,\Phi_{DEF,-\frac{1}{2}}^{(h_3)})
(z_3,\bar{z}_3)
\nonu \\
&& +
\frac{\bar{z}_{13}}{z_{13}} \,
\sum_{m=0}^{\infty}\, \frac{\bar{z}_{13}^m}{m!}\,
B(2\bar{h}_1+1+m,2\bar{h}_3+1)  \, (-1)\, 
\de^{PQR}_{DEF} \,  (
\Phi_{+\frac{1}{2}}^{(h_2),ABC}\,\bar{\pa}^m \,
\Phi_{-2}^{(h_1+h_3)})
(z_3,\bar{z}_3),
\nonu \\
&&
\Phi_{+\frac{1}{2}}^{(h_1),PQR}(z_1,\bar{z}_1)\,
(\Phi_{0}^{(h_2),ABCD}\, \Phi_{0}^{(h_3),EFGH})(z_3,
\bar{z}_3) =
\nonu \\
&&  
\frac{\bar{z}_{13}}{z_{13}} \,
\sum_{m=0}^{\infty}\, \frac{\bar{z}_{13}^m}{m!}\,
B(2\bar{h}_1+1+m,2\bar{h}_2+1) \, 
\ep^{PQRABCDI}\,
(\bar{\pa}^m \,
\Phi_{I,-\frac{3}{2}}^{(h_1+h_2)}\,\Phi_{0}^{(h_3),EFGH})
(z_3,\bar{z}_3)
\nonu \\
&& +
\frac{\bar{z}_{13}}{z_{13}} \,
\sum_{m=0}^{\infty}\, \frac{\bar{z}_{13}^m}{m!}\,
B(2\bar{h}_1+1+m,2\bar{h}_3+1)  \, 
\ep^{PQREFGHI} \,  (
\Phi_{0}^{(h_2),ABCD}\,\bar{\pa}^m \,
\Phi_{I,-\frac{3}{2}}^{(h_1+h_3)})
(z_3,\bar{z}_3).
\label{onehalflinear}
\eea

\subsection{The OPEs of
the scalars  of helicity $0$  with the quadratic
operators}

The nine OPEs
\footnote{After removing the three OPEs which do not satisfy
the condition $s_1+s_2+s_3 \geq 2$, these
nine OPEs remain.}
are
\bea
&&
\Phi_{0}^{(h_1),PQRS}(z_1,\bar{z}_1)\,
(\Phi_{+2}^{(h_2)}\, \Phi_{+2}^{(h_3)})(z_3,
\bar{z}_3) = \frac{\bar{z}_{13}^2}{z_{13}^2} \,
\sum_{m=0}^{\infty}\, \frac{\bar{z}_{13}^m}{m!}\,
B(2\bar{h}_1+2+m,2\bar{h}_2+1, 2\bar{h}_3+1)
\nonu \\
&& \times
\bar{\pa}^m \, \Phi_{0}^{(h_1+h_2+h_3),PQRS}(z_3,\bar{z}_3)
\nonu \\
&&+  
\frac{\bar{z}_{13}}{z_{13}} \,
\sum_{m=0}^{\infty}\, \frac{\bar{z}_{13}^m}{m!}\,
B(2\bar{h}_1+1+m,2\bar{h}_2+1) \, 
(\bar{\pa}^m \,
\Phi_{0}^{(h_1+h_2),PQRS}\,\Phi_{+2}^{(h_3)})
(z_3,\bar{z}_3)
\nonu \\
&& +
\frac{\bar{z}_{13}}{z_{13}} \,
\sum_{m=0}^{\infty}\, \frac{\bar{z}_{13}^m}{m!}\,
B(2\bar{h}_1+1+m,2\bar{h}_3+1)  \, 
(
\Phi_{+2}^{(h_2)}\,\bar{\pa}^m \,
\Phi_{0}^{(h_1+h_3),PQRS})
(z_3,\bar{z}_3),
\nonu \\
&&
\Phi_{0}^{(h_1),PQRS}(z_1,\bar{z}_1)\,
(\Phi_{+2}^{(h_2)}\, \Phi_{+\frac{3}{2}}^{(h_3),A})(z_3,
\bar{z}_3) = \frac{\bar{z}_{13}^2}{z_{13}^2} \,
\sum_{m=0}^{\infty}\, \frac{\bar{z}_{13}^m}{m!}\,
B(2\bar{h}_1+2+m,2\bar{h}_2+1, 2\bar{h}_3+1)
\nonu \\
&& \times
\frac{1}{3!}\,
\ep^{APQRSBCD}\,
\bar{\pa}^m \, \Phi_{BCD,-\frac{1}{2}}^{(h_1+h_2+h_3)}(z_3,\bar{z}_3)
\nonu \\
&&+  
\frac{\bar{z}_{13}}{z_{13}} \,
\sum_{m=0}^{\infty}\, \frac{\bar{z}_{13}^m}{m!}\,
B(2\bar{h}_1+1+m,2\bar{h}_2+1) \, 
(\bar{\pa}^m \,
\Phi_{0}^{(h_1+h_2),PQRS}\,\Phi_{+\frac{3}{2}}^{(h_3),A})
(z_3,\bar{z}_3)
\nonu \\
&& +
\frac{\bar{z}_{13}}{z_{13}} \,
\sum_{m=0}^{\infty}\, \frac{\bar{z}_{13}^m}{m!}\,
B(2\bar{h}_1+1+m,2\bar{h}_3+1)  \, 
\ep^{PQRSABCD}\, \frac{1}{3!}\, (
\Phi_{+2}^{(h_2)}\,\bar{\pa}^m \,
\Phi_{BCD,-\frac{1}{2}}^{(h_1+h_3)})
(z_3,\bar{z}_3),
\nonu \\
&&
\Phi_{0}^{(h_1),PQRS}(z_1,\bar{z}_1)\,
(\Phi_{+2}^{(h_2)}\, \Phi_{+1}^{(h_3),AB})(z_3,
\bar{z}_3) = \frac{\bar{z}_{13}^2}{z_{13}^2} \,
\sum_{m=0}^{\infty}\, \frac{\bar{z}_{13}^m}{m!}\,
B(2\bar{h}_1+2+m,2\bar{h}_2+1, 2\bar{h}_3+1)
\nonu \\
&& \times
\frac{1}{2!}\,
\ep^{ABPQRSCD}\,
\bar{\pa}^m \, \Phi_{CD,-1}^{(h_1+h_2+h_3)}(z_3,\bar{z}_3)
\nonu \\
&&+  
\frac{\bar{z}_{13}}{z_{13}} \,
\sum_{m=0}^{\infty}\, \frac{\bar{z}_{13}^m}{m!}\,
B(2\bar{h}_1+1+m,2\bar{h}_2+1) \, 
(\bar{\pa}^m \,
\Phi_{0}^{(h_1+h_2),PQRS}\,\Phi_{+1}^{(h_3),AB})
(z_3,\bar{z}_3)
\nonu \\
&& +
\frac{\bar{z}_{13}}{z_{13}} \,
\sum_{m=0}^{\infty}\, \frac{\bar{z}_{13}^m}{m!}\,
B(2\bar{h}_1+1+m,2\bar{h}_3+1)  \, 
\ep^{PQRSABCD}\, \frac{1}{2!}\, (
\Phi_{+2}^{(h_2)}\,\bar{\pa}^m \,
\Phi_{CD,-1}^{(h_1+h_3)})
(z_3,\bar{z}_3),
\nonu \\
&&
\Phi_{0}^{(h_1),PQRS}(z_1,\bar{z}_1)\,
(\Phi_{+2}^{(h_2)}\, \Phi_{+\frac{1}{2}}^{(h_3),ABC})(z_3,
\bar{z}_3) = \frac{\bar{z}_{13}^2}{z_{13}^2} \,
\sum_{m=0}^{\infty}\, \frac{\bar{z}_{13}^m}{m!}\,
B(2\bar{h}_1+2+m,2\bar{h}_2+1, 2\bar{h}_3+1)
\nonu \\
&& \times
\ep^{ABCPQRSD}\,
\bar{\pa}^m \, \Phi_{D,-\frac{3}{2}}^{(h_1+h_2+h_3)}(z_3,\bar{z}_3)
\nonu \\
&&+  
\frac{\bar{z}_{13}}{z_{13}} \,
\sum_{m=0}^{\infty}\, \frac{\bar{z}_{13}^m}{m!}\,
B(2\bar{h}_1+1+m,2\bar{h}_2+1) \, 
(\bar{\pa}^m \,
\Phi_{0}^{(h_1+h_2),PQRS}\,\Phi_{+\frac{1}{2}}^{(h_3),ABC})
(z_3,\bar{z}_3)
\nonu \\
&& +
\frac{\bar{z}_{13}}{z_{13}} \,
\sum_{m=0}^{\infty}\, \frac{\bar{z}_{13}^m}{m!}\,
B(2\bar{h}_1+1+m,2\bar{h}_3+1)  \, 
\ep^{PQRSABCD}\,  (
\Phi_{+2}^{(h_2)}\,\bar{\pa}^m \,
\Phi_{D,-\frac{3}{2}}^{(h_1+h_3)})
(z_3,\bar{z}_3),
\nonu \\
&&
\Phi_{0}^{(h_1),PQRS}(z_1,\bar{z}_1)\,
(\Phi_{+2}^{(h_2)}\, \Phi_{0}^{(h_3),ABCD})(z_3,
\bar{z}_3) = \frac{\bar{z}_{13}^2}{z_{13}^2} \,
\sum_{m=0}^{\infty}\, \frac{\bar{z}_{13}^m}{m!}\,
B(2\bar{h}_1+2+m,2\bar{h}_2+1, 2\bar{h}_3+1)
\nonu \\
&& \times
\ep^{PQRSABCD}\,
\bar{\pa}^m \, \Phi_{-2}^{(h_1+h_2+h_3)}(z_3,\bar{z}_3)
\nonu \\
&&+  
\frac{\bar{z}_{13}}{z_{13}} \,
\sum_{m=0}^{\infty}\, \frac{\bar{z}_{13}^m}{m!}\,
B(2\bar{h}_1+1+m,2\bar{h}_2+1) \, 
(\bar{\pa}^m \,
\Phi_{0}^{(h_1+h_2),PQRS}\,\Phi_{0}^{(h_3),ABCD})
(z_3,\bar{z}_3)
\nonu \\
&& +
\frac{\bar{z}_{13}}{z_{13}} \,
\sum_{m=0}^{\infty}\, \frac{\bar{z}_{13}^m}{m!}\,
B(2\bar{h}_1+1+m,2\bar{h}_3+1)  \, 
\ep^{PQRSABCD}\,  (
\Phi_{+2}^{(h_2)}\,\bar{\pa}^m \,
\Phi_{-2}^{(h_1+h_3)})
(z_3,\bar{z}_3),
\nonu \\
&&
\Phi_{0}^{(h_1),PQRS}(z_1,\bar{z}_1)\,
(\Phi_{+\frac{3}{2}}^{(h_2),A}\, \Phi_{+\frac{3}{2}}^{(h_3),B})(z_3,
\bar{z}_3) = \frac{\bar{z}_{13}^2}{z_{13}^2} \,
\sum_{m=0}^{\infty}\, \frac{\bar{z}_{13}^m}{m!}\,
B(2\bar{h}_1+2+m,2\bar{h}_2+1, 2\bar{h}_3+1)
\nonu \\
&& \times
\frac{-1}{2!}\,
\ep^{APQRSCDE}\,
\de^{B}_{[C}
  \bar{\pa}^m \, \Phi_{DE],-1}^{(h_1+h_2+h_3)}(z_3,\bar{z}_3)
\nonu \\
&&+  
\frac{\bar{z}_{13}}{z_{13}} \,
\sum_{m=0}^{\infty}\, \frac{\bar{z}_{13}^m}{m!}\,
B(2\bar{h}_1+1+m,2\bar{h}_2+1) \, 
\ep^{PQRSACDE}\, \frac{1}{3!}\, (\bar{\pa}^m \,
\Phi_{CDE,-\frac{1}{2}}^{(h_1+h_2)}\,\Phi_{+\frac{3}{2}}^{(h_3),B})
(z_3,\bar{z}_3)
\nonu \\
&& +
\frac{\bar{z}_{13}}{z_{13}} \,
\sum_{m=0}^{\infty}\, \frac{\bar{z}_{13}^m}{m!}\,
B(2\bar{h}_1+1+m,2\bar{h}_3+1)  \, 
\ep^{PQRSBCDE}\,  \frac{1}{3!} \, (
\Phi_{+\frac{3}{2}}^{(h_2),A}\,\bar{\pa}^m \,
\Phi_{CDE,-\frac{1}{2}}^{(h_1+h_3)})
(z_3,\bar{z}_3),
\nonu \\
&&
\Phi_{0}^{(h_1),PQRS}(z_1,\bar{z}_1)\,
(\Phi_{+\frac{3}{2}}^{(h_2),A}\, \Phi_{+1}^{(h_3),BC})(z_3,
\bar{z}_3) = \frac{\bar{z}_{13}^2}{z_{13}^2} \,
\sum_{m=0}^{\infty}\, \frac{\bar{z}_{13}^m}{m!}\,
B(2\bar{h}_1+2+m,2\bar{h}_2+1, 2\bar{h}_3+1)
\nonu \\
&& \times
\ep^{APQRSDEF}\,
\de^{B}_{[D}
\bar{\pa}^m \, \Phi_{E, -\frac{3}{2}}^{(h_1+h_2+h_3)}(z_3,\bar{z}_3)
\, \de^{C}_{F]}
\nonu \\
&&+  
\frac{\bar{z}_{13}}{z_{13}} \,
\sum_{m=0}^{\infty}\, \frac{\bar{z}_{13}^m}{m!}\,
B(2\bar{h}_1+1+m,2\bar{h}_2+1) \, 
\ep^{PQRSADEF}\, \frac{1}{3!}\, (\bar{\pa}^m \,
\Phi_{DEF,-\frac{1}{2}}^{(h_1+h_2)}\,\Phi_{+1}^{(h_3),BC})
(z_3,\bar{z}_3)
\nonu \\
&& +
\frac{\bar{z}_{13}}{z_{13}} \,
\sum_{m=0}^{\infty}\, \frac{\bar{z}_{13}^m}{m!}\,
B(2\bar{h}_1+1+m,2\bar{h}_3+1)  \, 
\ep^{PQRSBCDE}\,  \frac{1}{2!} \, (
\Phi_{+\frac{3}{2}}^{(h_2),A}\,\bar{\pa}^m \,
\Phi_{DE,-1}^{(h_1+h_3)})
(z_3,\bar{z}_3),
\nonu \\
&&
\Phi_{0}^{(h_1),PQRS}(z_1,\bar{z}_1)\,
(\Phi_{+\frac{3}{2}}^{(h_2),A}\, \Phi_{+\frac{1}{2}}^{(h_3),BCD})(z_3,
\bar{z}_3) = \frac{\bar{z}_{13}^2}{z_{13}^2} \,
\sum_{m=0}^{\infty}\, \frac{\bar{z}_{13}^m}{m!}\,
B(2\bar{h}_1+2+m,2\bar{h}_2+1, 2\bar{h}_3+1)
\nonu \\
&& \times
\frac{-1}{3!}\,
\ep^{APQRSDEF}\,
\de^{BCD}_{DEF}\,
\bar{\pa}^m \, \Phi_{-2}^{(h_1+h_2+h_3)}(z_3,\bar{z}_3)
\nonu \\
&&+  
\frac{\bar{z}_{13}}{z_{13}} \,
\sum_{m=0}^{\infty}\, \frac{\bar{z}_{13}^m}{m!}\,
B(2\bar{h}_1+1+m,2\bar{h}_2+1) \, 
\ep^{PQRSAEFG}\, \frac{1}{3!}\, (\bar{\pa}^m \,
\Phi_{EFG,-\frac{1}{2}}^{(h_1+h_2)}\,\Phi_{+\frac{1}{2}}^{(h_3),BCD})
(z_3,\bar{z}_3)
\nonu \\
&& +
\frac{\bar{z}_{13}}{z_{13}} \,
\sum_{m=0}^{\infty}\, \frac{\bar{z}_{13}^m}{m!}\,
B(2\bar{h}_1+1+m,2\bar{h}_3+1)  \, 
\ep^{PQRSBCDE} \, (
\Phi_{+\frac{3}{2}}^{(h_2),A}\,\bar{\pa}^m \,
\Phi_{E,-\frac{3}{2}}^{(h_1+h_3)})
(z_3,\bar{z}_3),
\nonu \\
&&
\Phi_{0}^{(h_1),PQRS}(z_1,\bar{z}_1)\,
(\Phi_{+1}^{(h_2),AB}\, \Phi_{+1}^{(h_3),CD})(z_3,
\bar{z}_3) = \frac{\bar{z}_{13}^2}{z_{13}^2} \,
\sum_{m=0}^{\infty}\, \frac{\bar{z}_{13}^m}{m!}\,
B(2\bar{h}_1+2+m,2\bar{h}_2+1, 2\bar{h}_3+1)
\nonu \\
&& \times
\frac{1}{2!}\,
\ep^{ABPQRSEF}\,
\de^{CD}_{EF}\,
\bar{\pa}^m \, \Phi_{-2}^{(h_1+h_2+h_3)}(z_3,\bar{z}_3)
\nonu \\
&&+  
\frac{\bar{z}_{13}}{z_{13}} \,
\sum_{m=0}^{\infty}\, \frac{\bar{z}_{13}^m}{m!}\,
B(2\bar{h}_1+1+m,2\bar{h}_2+1) \, 
\ep^{PQRSABEF}\, \frac{1}{2!}\, (\bar{\pa}^m \,
\Phi_{EF,-1}^{(h_1+h_2)}\,\Phi_{+1}^{(h_3),CD})
(z_3,\bar{z}_3)
\nonu \\
&& +
\frac{\bar{z}_{13}}{z_{13}} \,
\sum_{m=0}^{\infty}\, \frac{\bar{z}_{13}^m}{m!}\,
B(2\bar{h}_1+1+m,2\bar{h}_3+1)  \, 
\ep^{PQRSCDEF} \, \frac{1}{2!} \, (
\Phi_{+1}^{(h_2),AB}\,\bar{\pa}^m \,
\Phi_{EF,-1}^{(h_1+h_3)})
(z_3,\bar{z}_3).
\label{fullzero}
\eea
The additional sixteen OPEs where the
quadratic terms appear
\footnote{Ten of these OPEs in (\ref{zerolinear})
do not satisfy $s_1+s_3 \geq 0$, compared to other OPEs.}
are given by
\bea
&&
\Phi_{0}^{(h_1),PQRS}(z_1,\bar{z}_1)\,
(\Phi_{+2}^{(h_2)}\, \Phi_{ABC,-\frac{1}{2}}^{(h_3)})(z_3,
\bar{z}_3) =
\nonu \\
&&  
\frac{\bar{z}_{13}}{z_{13}} \,
\sum_{m=0}^{\infty}\, \frac{\bar{z}_{13}^m}{m!}\,
B(2\bar{h}_1+1+m,2\bar{h}_2+1) \, (\bar{\pa}^m \,
\Phi_{0}^{(h_1+h_2),PQRS}\,\Phi_{ABC,-\frac{1}{2}}^{(h_3)})
(z_3,\bar{z}_3),
\nonu \\
&&
\Phi_{0}^{(h_1),PQRS}(z_1,\bar{z}_1)\,
(\Phi_{+2}^{(h_2)}\, \Phi_{AB,-1}^{(h_3)})(z_3,
\bar{z}_3) =
\nonu \\
&&  
\frac{\bar{z}_{13}}{z_{13}} \,
\sum_{m=0}^{\infty}\, \frac{\bar{z}_{13}^m}{m!}\,
B(2\bar{h}_1+1+m,2\bar{h}_2+1) \, (\bar{\pa}^m \,
\Phi_{0}^{(h_1+h_2),PQRS}\,\Phi_{AB,-1}^{(h_3)})
(z_3,\bar{z}_3),
\nonu \\
&&
\Phi_{0}^{(h_1),PQRS}(z_1,\bar{z}_1)\,
(\Phi_{+2}^{(h_2)}\, \Phi_{A,-\frac{3}{2}}^{(h_3)})(z_3,
\bar{z}_3) =
\nonu \\
&&  
\frac{\bar{z}_{13}}{z_{13}} \,
\sum_{m=0}^{\infty}\, \frac{\bar{z}_{13}^m}{m!}\,
B(2\bar{h}_1+1+m,2\bar{h}_2+1) \, (\bar{\pa}^m \,
\Phi_{0}^{(h_1+h_2),PQRS}\,\Phi_{A,-\frac{3}{2}}^{(h_3)})
(z_3,\bar{z}_3),
\nonu \\
&&
\Phi_{0}^{(h_1),PQRS}(z_1,\bar{z}_1)\,
(\Phi_{+2}^{(h_2)}\, \Phi_{-2}^{(h_3)})(z_3,
\bar{z}_3) =
\nonu \\
&&  
\frac{\bar{z}_{13}}{z_{13}} \,
\sum_{m=0}^{\infty}\, \frac{\bar{z}_{13}^m}{m!}\,
B(2\bar{h}_1+1+m,2\bar{h}_2+1) \, (\bar{\pa}^m \,
\Phi_{0}^{(h_1+h_2),PQRS}\,\Phi_{-2}^{(h_3)})
(z_3,\bar{z}_3),
\nonu \\
&&
\Phi_{0}^{(h_1),PQRS}(z_1,\bar{z}_1)\,
(\Phi_{+\frac{3}{2}}^{(h_2),A}\, \Phi_{0}^{(h_3),BCDE})(z_3,
\bar{z}_3) =
\nonu \\
&&  
\frac{\bar{z}_{13}}{z_{13}} \,
\sum_{m=0}^{\infty}\, \frac{\bar{z}_{13}^m}{m!}\,
B(2\bar{h}_1+1+m,2\bar{h}_2+1) \,
\ep^{PQRSAFGH}\, \frac{1}{3!}\,
(\bar{\pa}^m \,
\Phi_{FGH,-\frac{1}{2}}^{(h_1+h_2)}\,\Phi_{0}^{(h_3),BCDE})
(z_3,\bar{z}_3)
\nonu \\
&& +
\frac{\bar{z}_{13}}{z_{13}} \,
\sum_{m=0}^{\infty}\, \frac{\bar{z}_{13}^m}{m!}\,
B(2\bar{h}_1+1+m,2\bar{h}_3+1)  \, 
\ep^{PQRSBCDE}  \, (
\Phi_{+\frac{3}{2}}^{(h_2),A}\,\bar{\pa}^m \,
\Phi_{-2}^{(h_1+h_3)})
(z_3,\bar{z}_3),
\nonu \\
&&
\Phi_{0}^{(h_1),PQRS}(z_1,\bar{z}_1)\,
(\Phi_{+\frac{3}{2}}^{(h_2),A}\, \Phi_{BCD,-\frac{1}{2}}^{(h_3)})(z_3,
\bar{z}_3) =
\nonu \\
&&  
\frac{\bar{z}_{13}}{z_{13}} \,
\sum_{m=0}^{\infty}\, \frac{\bar{z}_{13}^m}{m!}\,
B(2\bar{h}_1+1+m,2\bar{h}_2+1) \, \ep^{PQRSAEFG}\, \frac{1}{3!}\,
(\bar{\pa}^m \,
\Phi_{EFG,-\frac{1}{2}}^{(h_1+h_2)}\,\Phi_{BCD,-\frac{1}{2}}^{(h_3)})
(z_3,\bar{z}_3),
\nonu \\
&&
\Phi_{0}^{(h_1),PQRS}(z_1,\bar{z}_1)\,
(\Phi_{+\frac{3}{2}}^{(h_2),A}\, \Phi_{BC,-1}^{(h_3)})(z_3,
\bar{z}_3) =
\nonu \\
&&  
\frac{\bar{z}_{13}}{z_{13}} \,
\sum_{m=0}^{\infty}\, \frac{\bar{z}_{13}^m}{m!}\,
B(2\bar{h}_1+1+m,2\bar{h}_2+1) \, \ep^{PQRSAEFG}\, \frac{1}{3!}\,
(\bar{\pa}^m \,
\Phi_{EFG,-\frac{1}{2}}^{(h_1+h_2)}\,\Phi_{BC,-1}^{(h_3)})
(z_3,\bar{z}_3),
\nonu \\
&&
\Phi_{0}^{(h_1),PQRS}(z_1,\bar{z}_1)\,
(\Phi_{+\frac{3}{2}}^{(h_2),A}\, \Phi_{B,-\frac{3}{2}}^{(h_3)})(z_3,
\bar{z}_3) =
\nonu \\
&&  
\frac{\bar{z}_{13}}{z_{13}} \,
\sum_{m=0}^{\infty}\, \frac{\bar{z}_{13}^m}{m!}\,
B(2\bar{h}_1+1+m,2\bar{h}_2+1) \, \ep^{PQRSAEFG}\, \frac{1}{3!}\,
(\bar{\pa}^m \,
\Phi_{EFG,-\frac{1}{2}}^{(h_1+h_2)}\,\Phi_{B,-\frac{3}{2}}^{(h_3)})
(z_3,\bar{z}_3),
\nonu \\
&&
\Phi_{0}^{(h_1),PQRS}(z_1,\bar{z}_1)\,
(\Phi_{+1}^{(h_2),AB}\, \Phi_{+\frac{1}{2}}^{(h_3),CDE})(z_3,
\bar{z}_3) =
\nonu \\
&&  
\frac{\bar{z}_{13}}{z_{13}} \,
\sum_{m=0}^{\infty}\, \frac{\bar{z}_{13}^m}{m!}\,
B(2\bar{h}_1+1+m,2\bar{h}_2+1) \,
\ep^{PQRSABFG}\, \frac{1}{2!}\,
(\bar{\pa}^m \,
\Phi_{FG,-1}^{(h_1+h_2)}\,\Phi_{+\frac{1}{2}}^{(h_3),CDE})
(z_3,\bar{z}_3)
\nonu \\
&& +
\frac{\bar{z}_{13}}{z_{13}} \,
\sum_{m=0}^{\infty}\, \frac{\bar{z}_{13}^m}{m!}\,
B(2\bar{h}_1+1+m,2\bar{h}_3+1)  \, 
\ep^{PQRSCDEF}  \, (
\Phi_{+1}^{(h_2),AB}\,\bar{\pa}^m \,
\Phi_{F,-\frac{3}{2}}^{(h_1+h_3)})
(z_3,\bar{z}_3),
\nonu \\
&&
\Phi_{0}^{(h_1),PQRS}(z_1,\bar{z}_1)\,
(\Phi_{+1}^{(h_2),AB}\, \Phi_{0}^{(h_3),CDEF})(z_3,
\bar{z}_3) =
\nonu \\
&&  
\frac{\bar{z}_{13}}{z_{13}} \,
\sum_{m=0}^{\infty}\, \frac{\bar{z}_{13}^m}{m!}\,
B(2\bar{h}_1+1+m,2\bar{h}_2+1) \,
\ep^{PQRSABFG}\, \frac{1}{2!}\,
(\bar{\pa}^m \,
\Phi_{FG,-1}^{(h_1+h_2)}\,\Phi_{0}^{(h_3),CDEF})
(z_3,\bar{z}_3)
\nonu \\
&& +
\frac{\bar{z}_{13}}{z_{13}} \,
\sum_{m=0}^{\infty}\, \frac{\bar{z}_{13}^m}{m!}\,
B(2\bar{h}_1+1+m,2\bar{h}_3+1)  \, 
\ep^{PQRSCDEF}  \, (
\Phi_{+1}^{(h_2),AB}\,\bar{\pa}^m \,
\Phi_{-2}^{(h_1+h_3)})
(z_3,\bar{z}_3),
\nonu \\
&&
\Phi_{0}^{(h_1),PQRS}(z_1,\bar{z}_1)\,
(\Phi_{+1}^{(h_2),AB}\, \Phi_{CDE,-\frac{1}{2}}^{(h_3)})(z_3,
\bar{z}_3) =
\nonu \\
&&  
\frac{\bar{z}_{13}}{z_{13}} \,
\sum_{m=0}^{\infty}\, \frac{\bar{z}_{13}^m}{m!}\,
B(2\bar{h}_1+1+m,2\bar{h}_2+1) \,
\ep^{PQRSABFG}\, \frac{1}{2!}\,
(\bar{\pa}^m \,
\Phi_{FG,-1}^{(h_1+h_2)}\,\Phi_{CDE,-\frac{1}{2}}^{(h_3)})
(z_3,\bar{z}_3),
\nonu \\
&&
\Phi_{0}^{(h_1),PQRS}(z_1,\bar{z}_1)\,
(\Phi_{+1}^{(h_2),AB}\, \Phi_{CD,-1}^{(h_3)})(z_3,
\bar{z}_3) =
\nonu \\
&&  
\frac{\bar{z}_{13}}{z_{13}} \,
\sum_{m=0}^{\infty}\, \frac{\bar{z}_{13}^m}{m!}\,
B(2\bar{h}_1+1+m,2\bar{h}_2+1) \,
\ep^{PQRSABFG}\, \frac{1}{2!}\,
(\bar{\pa}^m \,
\Phi_{FG,-1}^{(h_1+h_2)}\,\Phi_{CD,-1}^{(h_3)})
(z_3,\bar{z}_3),
\nonu \\
&&
\Phi_{0}^{(h_1),PQRS}(z_1,\bar{z}_1)\,
(\Phi_{+\frac{1}{2}}^{(h_2),ABC}\, \Phi_{+\frac{1}{2}}^{(h_3),DEF})(z_3,
\bar{z}_3) =
\nonu \\
&&  
\frac{\bar{z}_{13}}{z_{13}} \,
\sum_{m=0}^{\infty}\, \frac{\bar{z}_{13}^m}{m!}\,
B(2\bar{h}_1+1+m,2\bar{h}_2+1) \,
\ep^{PQRSABCG}\, 
(\bar{\pa}^m \,
\Phi_{G,-\frac{3}{2}}^{(h_1+h_2)}\,\Phi_{\frac{1}{2}}^{(h_3),DEF})
(z_3,\bar{z}_3)
\nonu \\
&& +
\frac{\bar{z}_{13}}{z_{13}} \,
\sum_{m=0}^{\infty}\, \frac{\bar{z}_{13}^m}{m!}\,
B(2\bar{h}_1+1+m,2\bar{h}_3+1)  \, 
\ep^{PQRSDEFG}  \, (
\Phi_{+\frac{1}{2}}^{(h_2),ABC}\,\bar{\pa}^m \,
\Phi_{G,-\frac{3}{2}}^{(h_1+h_3)})
(z_3,\bar{z}_3),
\nonu \\
&&
\Phi_{0}^{(h_1),PQRS}(z_1,\bar{z}_1)\,
(\Phi_{+\frac{1}{2}}^{(h_2),ABC}\, \Phi_{0}^{(h_3),DEFG})(z_3,
\bar{z}_3) =
\nonu \\
&&  
\frac{\bar{z}_{13}}{z_{13}} \,
\sum_{m=0}^{\infty}\, \frac{\bar{z}_{13}^m}{m!}\,
B(2\bar{h}_1+1+m,2\bar{h}_2+1) \,
\ep^{PQRSABCH}\, 
(\bar{\pa}^m \,
\Phi_{H,-\frac{3}{2}}^{(h_1+h_2)}\,\Phi_{0}^{(h_3),DEFG})
(z_3,\bar{z}_3)
\nonu \\
&& +
\frac{\bar{z}_{13}}{z_{13}} \,
\sum_{m=0}^{\infty}\, \frac{\bar{z}_{13}^m}{m!}\,
B(2\bar{h}_1+1+m,2\bar{h}_3+1)  \, 
\ep^{PQRSDEFG}  \, (
\Phi_{+\frac{1}{2}}^{(h_2),ABC}\,\bar{\pa}^m \,
\Phi_{-2}^{(h_1+h_3)})
(z_3,\bar{z}_3),
\nonu \\
&&
\Phi_{0}^{(h_1),PQRS}(z_1,\bar{z}_1)\,
(\Phi_{+\frac{1}{2}}^{(h_2),ABC}\, \Phi_{DEF,-\frac{1}{2}}^{(h_3)})(z_3,
\bar{z}_3) =
\nonu \\
&&  
\frac{\bar{z}_{13}}{z_{13}} \,
\sum_{m=0}^{\infty}\, \frac{\bar{z}_{13}^m}{m!}\,
B(2\bar{h}_1+1+m,2\bar{h}_2+1) \,
\ep^{PQRSABCH}\, 
(\bar{\pa}^m \,
\Phi_{H,-\frac{3}{2}}^{(h_1+h_2)}\,\Phi_{DEF,-\frac{1}{2}}^{(h_3)})
(z_3,\bar{z}_3),
\nonu \\
&&
\Phi_{0}^{(h_1),PQRS}(z_1,\bar{z}_1)\,
(\Phi_{0}^{(h_2),ABCD}\, \Phi_{0}^{(h_3),EFGH})(z_3,
\bar{z}_3) =
\nonu \\
&&  
\frac{\bar{z}_{13}}{z_{13}} \,
\sum_{m=0}^{\infty}\, \frac{\bar{z}_{13}^m}{m!}\,
B(2\bar{h}_1+1+m,2\bar{h}_2+1) \,
\ep^{PQRSABCD}\, 
(\bar{\pa}^m \,
\Phi_{-2}^{(h_1+h_2)}\,\Phi_{0}^{(h_3),EFGH})
(z_3,\bar{z}_3)
\label{zerolinear}
\\
&& +
\frac{\bar{z}_{13}}{z_{13}} \,
\sum_{m=0}^{\infty}\, \frac{\bar{z}_{13}^m}{m!}\,
B(2\bar{h}_1+1+m,2\bar{h}_3+1)  \, 
\ep^{PQRSEFGH}  \, (
\Phi_{0}^{(h_2),ABCD}\,\bar{\pa}^m \,
\Phi_{-2}^{(h_1+h_3)})
(z_3,\bar{z}_3).
\nonu
\eea

\subsection{The operator product expansions of
the graviphotinos  of helicity $-\frac{1}{2}$   with the quadratic
operators}

The six OPEs
\footnote{In this case, the three OPEs which do not satisfy
the condition $s_1+s_2+s_3 \geq 2$ are removed and we are left with
these OPEs.}
are
\bea
&&
\Phi_{PQR,-\frac{1}{2}}^{(h_1)}(z_1,\bar{z}_1)\,
(\Phi_{+2}^{(h_2)}\, \Phi_{+2}^{(h_3)})(z_3,
\bar{z}_3) = \frac{\bar{z}_{13}^2}{z_{13}^2} \,
\sum_{m=0}^{\infty}\, \frac{\bar{z}_{13}^m}{m!}\,
B(2\bar{h}_1+2+m,2\bar{h}_2+1, 2\bar{h}_3+1)
\nonu \\
&& \times
\bar{\pa}^m \, \Phi_{PQR,-\frac{1}{2}}^{(h_1+h_2+h_3)}(z_3,\bar{z}_3)
\nonu \\
&&+  
\frac{\bar{z}_{13}}{z_{13}} \,
\sum_{m=0}^{\infty}\, \frac{\bar{z}_{13}^m}{m!}\,
B(2\bar{h}_1+1+m,2\bar{h}_2+1) \, 
(\bar{\pa}^m \,
\Phi_{PQR,-\frac{1}{2}}^{(h_1+h_2)}\,\Phi_{+2}^{(h_3)})
(z_3,\bar{z}_3)
\nonu \\
&& +
\frac{\bar{z}_{13}}{z_{13}} \,
\sum_{m=0}^{\infty}\, \frac{\bar{z}_{13}^m}{m!}\,
B(2\bar{h}_1+1+m,2\bar{h}_3+1)  \, 
(
\Phi_{+2}^{(h_2)}\,\bar{\pa}^m \,
\Phi_{PQR,-\frac{1}{2}}^{(h_1+h_3)})
(z_3,\bar{z}_3),
\nonu \\
&&
\Phi_{PQR,-\frac{1}{2}}^{(h_1)}(z_1,\bar{z}_1)\,
(\Phi_{+2}^{(h_2)}\, \Phi_{+\frac{3}{2}}^{(h_3),A})(z_3,
\bar{z}_3) = \frac{\bar{z}_{13}^2}{z_{13}^2} \,
\sum_{m=0}^{\infty}\, \frac{\bar{z}_{13}^m}{m!}\,
B(2\bar{h}_1+2+m,2\bar{h}_2+1, 2\bar{h}_3+1)
\nonu \\
&& \times
(-3) \,
\de^{A}_{[P}\,
  \bar{\pa}^m \, \Phi_{QR],-1}^{(h_1+h_2+h_3)}(z_3,\bar{z}_3)
\nonu \\
&&+  
\frac{\bar{z}_{13}}{z_{13}} \,
\sum_{m=0}^{\infty}\, \frac{\bar{z}_{13}^m}{m!}\,
B(2\bar{h}_1+1+m,2\bar{h}_2+1) \, 
(\bar{\pa}^m \,
\Phi_{PQR,-\frac{1}{2}}^{(h_1+h_2)}\,\Phi_{+\frac{3}{2}}^{(h_3),A})
(z_3,\bar{z}_3)
\nonu \\
&& +
\frac{\bar{z}_{13}}{z_{13}} \,
\sum_{m=0}^{\infty}\, \frac{\bar{z}_{13}^m}{m!}\,
B(2\bar{h}_1+1+m,2\bar{h}_3+1)  \,
(-3)\, \de^{A}_{[P}\, 
(
\Phi_{+2}^{(h_2)}\,\bar{\pa}^m \,
\Phi_{QR],-1}^{(h_1+h_3)})
(z_3,\bar{z}_3),
\nonu \\
&&
\Phi_{PQR,-\frac{1}{2}}^{(h_1)}(z_1,\bar{z}_1)\,
(\Phi_{+2}^{(h_2)}\, \Phi_{+1}^{(h_3),AB})(z_3,
\bar{z}_3) = \frac{\bar{z}_{13}^2}{z_{13}^2} \,
\sum_{m=0}^{\infty}\, \frac{\bar{z}_{13}^m}{m!}\,
B(2\bar{h}_1+2+m,2\bar{h}_2+1, 2\bar{h}_3+1)
\nonu \\
&& \times
3!\,
\de^{A}_{[P}\,
\bar{\pa}^m \, \Phi_{Q,-\frac{3}{2}}^{(h_1+h_2+h_3)}(z_3,\bar{z}_3)
\,  \de^{B}_{R]}
\nonu \\
&&+  
\frac{\bar{z}_{13}}{z_{13}} \,
\sum_{m=0}^{\infty}\, \frac{\bar{z}_{13}^m}{m!}\,
B(2\bar{h}_1+1+m,2\bar{h}_2+1) \, 
(\bar{\pa}^m \,
\Phi_{PQR,-\frac{1}{2}}^{(h_1+h_2)}\,\Phi_{+1}^{(h_3),AB})
(z_3,\bar{z}_3)
\nonu \\
&& +
\frac{\bar{z}_{13}}{z_{13}} \,
\sum_{m=0}^{\infty}\, \frac{\bar{z}_{13}^m}{m!}\,
B(2\bar{h}_1+1+m,2\bar{h}_3+1)  \,
3! \,
\de^{A}_{[P} \,
(
\Phi_{+2}^{(h_2)}\,\bar{\pa}^m \,
\Phi_{Q,-\frac{3}{2}}^{(h_1+h_3)})
(z_3,\bar{z}_3) \, \de^{B}_{R]},
\nonu \\
&&
\Phi_{PQR,-\frac{1}{2}}^{(h_1)}(z_1,\bar{z}_1)\,
(\Phi_{+2}^{(h_2)}\, \Phi_{+\frac{1}{2}}^{(h_3),ABC})(z_3,
\bar{z}_3) = \frac{\bar{z}_{13}^2}{z_{13}^2} \,
\sum_{m=0}^{\infty}\, \frac{\bar{z}_{13}^m}{m!}\,
B(2\bar{h}_1+2+m,2\bar{h}_2+1, 2\bar{h}_3+1)
\nonu \\
&& \times
(-1) \,
\de^{ABC}_{PQR}\, 
\bar{\pa}^m \, \Phi_{-2}^{(h_1+h_2+h_3)}(z_3,\bar{z}_3)
\nonu \\
&&+  
\frac{\bar{z}_{13}}{z_{13}} \,
\sum_{m=0}^{\infty}\, \frac{\bar{z}_{13}^m}{m!}\,
B(2\bar{h}_1+1+m,2\bar{h}_2+1) \, 
(\bar{\pa}^m \,
\Phi_{PQR,-\frac{1}{2}}^{(h_1+h_2)}\,\Phi_{+\frac{1}{2}}^{(h_3),ABC})
(z_3,\bar{z}_3)
\nonu \\
&& +
\frac{\bar{z}_{13}}{z_{13}} \,
\sum_{m=0}^{\infty}\, \frac{\bar{z}_{13}^m}{m!}\,
B(2\bar{h}_1+1+m,2\bar{h}_3+1)  \,
(-1)\, \de^{ABC}_{PQR} \,
(
\Phi_{+2}^{(h_2)}\,\bar{\pa}^m \,
\Phi_{-2}^{(h_1+h_3)})
(z_3,\bar{z}_3),
\nonu \\
&&
\Phi_{PQR,-\frac{1}{2}}^{(h_1)}(z_1,\bar{z}_1)\,
(\Phi_{+\frac{3}{2}}^{(h_2),A}\, \Phi_{+\frac{3}{2}}^{(h_3),B})(z_3,
\bar{z}_3) = \frac{\bar{z}_{13}^2}{z_{13}^2} \,
\sum_{m=0}^{\infty}\, \frac{\bar{z}_{13}^m}{m!}\,
B(2\bar{h}_1+2+m,2\bar{h}_2+1, 2\bar{h}_3+1)
\nonu \\
&& \times
3!\,
\de^{A}_{R}\,
\bar{\pa}^m \, \Phi_{P,-\frac{3}{2}}^{(h_1+h_2+h_3)}(z_3,\bar{z}_3)
\, \de^{B}_{Q]}
\nonu \\
&&+  
\frac{\bar{z}_{13}}{z_{13}} \,
\sum_{m=0}^{\infty}\, \frac{\bar{z}_{13}^m}{m!}\,
B(2\bar{h}_1+1+m,2\bar{h}_2+1) \,
(-1) \, 3 \, \de^{A}_{[P}\, 
(\bar{\pa}^m \,
\Phi_{QR],-1}^{(h_1+h_2)}\,\Phi_{+\frac{3}{2}}^{(h_3),B})
(z_3,\bar{z}_3)
\nonu \\
&& +
\frac{\bar{z}_{13}}{z_{13}} \,
\sum_{m=0}^{\infty}\, \frac{\bar{z}_{13}^m}{m!}\,
B(2\bar{h}_1+1+m,2\bar{h}_3+1)  \,
 3 \, \de^{B}_{[P}\,  \,
(
\Phi_{+\frac{3}{2}}^{(h_2),A}\,\bar{\pa}^m \,
\Phi_{QR],-1}^{(h_1+h_3)})
(z_3,\bar{z}_3),
\nonu \\
&&
\Phi_{PQR,-\frac{1}{2}}^{(h_1)}(z_1,\bar{z}_1)\,
(\Phi_{+\frac{3}{2}}^{(h_2),A}\, \Phi_{+1}^{(h_3),BC})(z_3,
\bar{z}_3) = \frac{\bar{z}_{13}^2}{z_{13}^2} \,
\sum_{m=0}^{\infty}\, \frac{\bar{z}_{13}^m}{m!}\,
B(2\bar{h}_1+2+m,2\bar{h}_2+1, 2\bar{h}_3+1)
\nonu \\
&& \times
3!\,
\de^{A}_{R}\, \de^{B}_{Q}\, \de^{C}_{P}\, 
\bar{\pa}^m \, \Phi_{-2}^{(h_1+h_2+h_3)}(z_3,\bar{z}_3)
\nonu \\
&&+  
\frac{\bar{z}_{13}}{z_{13}} \,
\sum_{m=0}^{\infty}\, \frac{\bar{z}_{13}^m}{m!}\,
B(2\bar{h}_1+1+m,2\bar{h}_2+1) \,
(-1) \, 3 \, \de^{A}_{[P}\, 
(\bar{\pa}^m \,
\Phi_{QR],-1}^{(h_1+h_2)}\,\Phi_{+1}^{(h_3),BC})
(z_3,\bar{z}_3)
\nonu \\
&& +
\frac{\bar{z}_{13}}{z_{13}} \,
\sum_{m=0}^{\infty}\, \frac{\bar{z}_{13}^m}{m!}\,
B(2\bar{h}_1+1+m,2\bar{h}_3+1)  \,
(-1)\, 3! \, \de^{B}_{[P}\,  \,
(
\Phi_{+\frac{3}{2}}^{(h_2),A}\,\bar{\pa}^m \,
\Phi_{Q,-\frac{3}{2}}^{(h_1+h_3)})
(z_3,\bar{z}_3) \, \de^{C}_{R]}.
\label{fullminusonehalf}
\eea
We obtain the additional eighteen OPEs
\footnote{The fourteen OPEs in (\ref{minusonehalflinear})
do not satisfy the
previous condition $s_1+s_3 \geq 0$. The reason why
the OPE of the single-particle operators having the
helicity $-\frac{1}{2}$ with the two-particle operators
associated with the last element in (\ref{REDEFINITION})
doesn't appear
is that the condition $s_1+s_2 \geq 0$ is not satisfied.}
where the quadratic terms
appear only
\bea
&&
\Phi_{PQR,-\frac{1}{2}}^{(h_1)}(z_1,\bar{z}_1)\,
(\Phi_{+2}^{(h_2)}\, \Phi_{0}^{(h_3),ABCD})(z_3,
\bar{z}_3) =
\nonu \\
&&  
\frac{\bar{z}_{13}}{z_{13}} \,
\sum_{m=0}^{\infty}\, \frac{\bar{z}_{13}^m}{m!}\,
B(2\bar{h}_1+1+m,2\bar{h}_2+1) \, (\bar{\pa}^m \,
\Phi_{PQR,-\frac{1}{2}}^{(h_1+h_2)}\,\Phi_{0}^{(h_3),ABCD})
(z_3,\bar{z}_3),
\nonu \\
&&
\Phi_{PQR,-\frac{1}{2}}^{(h_1)}(z_1,\bar{z}_1)\,
(\Phi_{+2}^{(h_2)}\, \Phi_{ABC,-\frac{1}{2}}^{(h_3)})(z_3,
\bar{z}_3) =
\nonu \\
&&  
\frac{\bar{z}_{13}}{z_{13}} \,
\sum_{m=0}^{\infty}\, \frac{\bar{z}_{13}^m}{m!}\,
B(2\bar{h}_1+1+m,2\bar{h}_2+1) \, (\bar{\pa}^m \,
\Phi_{PQR,-\frac{1}{2}}^{(h_1+h_2)}\,\Phi_{ABC,-\frac{1}{2}}^{(h_3)})
(z_3,\bar{z}_3),
\nonu \\
&&
\Phi_{PQR,-\frac{1}{2}}^{(h_1)}(z_1,\bar{z}_1)\,
(\Phi_{+2}^{(h_2)}\, \Phi_{AB,-1}^{(h_3)})(z_3,
\bar{z}_3) =
\nonu \\
&&  
\frac{\bar{z}_{13}}{z_{13}} \,
\sum_{m=0}^{\infty}\, \frac{\bar{z}_{13}^m}{m!}\,
B(2\bar{h}_1+1+m,2\bar{h}_2+1) \, (\bar{\pa}^m \,
\Phi_{PQR,-\frac{1}{2}}^{(h_1+h_2)}\,\Phi_{AB,-1}^{(h_3)})
(z_3,\bar{z}_3),
\nonu \\
&&
\Phi_{PQR,-\frac{1}{2}}^{(h_1)}(z_1,\bar{z}_1)\,
(\Phi_{+2}^{(h_2)}\, \Phi_{A,-\frac{3}{2}}^{(h_3)})(z_3,
\bar{z}_3) =
\nonu \\
&&  
\frac{\bar{z}_{13}}{z_{13}} \,
\sum_{m=0}^{\infty}\, \frac{\bar{z}_{13}^m}{m!}\,
B(2\bar{h}_1+1+m,2\bar{h}_2+1) \, (\bar{\pa}^m \,
\Phi_{PQR,-\frac{1}{2}}^{(h_1+h_2)}\,\Phi_{A,-\frac{3}{2}}^{(h_3)})
(z_3,\bar{z}_3),
\nonu \\
&&
\Phi_{PQR,-\frac{1}{2}}^{(h_1)}(z_1,\bar{z}_1)\,
(\Phi_{+2}^{(h_2)}\, \Phi_{-2}^{(h_3)})(z_3,
\bar{z}_3) =
\nonu \\
&&  
\frac{\bar{z}_{13}}{z_{13}} \,
\sum_{m=0}^{\infty}\, \frac{\bar{z}_{13}^m}{m!}\,
B(2\bar{h}_1+1+m,2\bar{h}_2+1) \, (\bar{\pa}^m \,
\Phi_{PQR,-\frac{1}{2}}^{(h_1+h_2)}\,\Phi_{-2}^{(h_3)})
(z_3,\bar{z}_3),
\nonu \\
&&
\Phi_{PQR,-\frac{1}{2}}^{(h_1)}(z_1,\bar{z}_1)\,
(\Phi_{+\frac{3}{2}}^{(h_2),A}\, \Phi_{+\frac{1}{2}}^{(h_3),BCD})(z_3,
\bar{z}_3) =
\nonu \\
&&  
\frac{\bar{z}_{13}}{z_{13}} \,
\sum_{m=0}^{\infty}\, \frac{\bar{z}_{13}^m}{m!}\,
B(2\bar{h}_1+1+m,2\bar{h}_2+1) \,
(-1) \, 3 \, \de^{A}_{[P} \,
(\bar{\pa}^m \,
\Phi_{QR],-1}^{(h_1+h_2)}\,\Phi_{+\frac{1}{2}}^{(h_3),BCD})
(z_3,\bar{z}_3)
\nonu \\
&& +
\frac{\bar{z}_{13}}{z_{13}} \,
\sum_{m=0}^{\infty}\, \frac{\bar{z}_{13}^m}{m!}\,
B(2\bar{h}_1+1+m,2\bar{h}_3+1)  \, 
\de^{BCD}_{PQR} \, (
\Phi_{+\frac{3}{2}}^{(h_2),A}\,\bar{\pa}^m \,
\Phi_{-2}^{(h_1+h_3)})
(z_3,\bar{z}_3),
\nonu \\
&&
\Phi_{PQR,-\frac{1}{2}}^{(h_1)}(z_1,\bar{z}_1)\,
(\Phi_{+\frac{3}{2}}^{(h_2),A}\, \Phi_{0}^{(h_3),BCDE})(z_3,
\bar{z}_3) =
\nonu \\
&&  
\frac{\bar{z}_{13}}{z_{13}} \,
\sum_{m=0}^{\infty}\, \frac{\bar{z}_{13}^m}{m!}\,
B(2\bar{h}_1+1+m,2\bar{h}_2+1) \,
(-1)\, 3\, \de^{A}_{[P}
(\bar{\pa}^m \,
\Phi_{QR],-1}^{(h_1+h_2)}\,\Phi_{0}^{(h_3),BCDE})
(z_3,\bar{z}_3),
\nonu \\
&&
\Phi_{PQR,-\frac{1}{2}}^{(h_1)}(z_1,\bar{z}_1)\,
(\Phi_{+\frac{3}{2}}^{(h_2),A}\, \Phi_{BCD,-\frac{1}{2}}^{(h_3)})(z_3,
\bar{z}_3) =
\nonu \\
&&  
\frac{\bar{z}_{13}}{z_{13}} \,
\sum_{m=0}^{\infty}\, \frac{\bar{z}_{13}^m}{m!}\,
B(2\bar{h}_1+1+m,2\bar{h}_2+1) \, (-1) \, 3\,
\de^{A}_{[P} \, (\bar{\pa}^m \,
\Phi_{QR],-1}^{(h_1+h_2)}\,\Phi_{BCD,-\frac{1}{2}}^{(h_3)})
(z_3,\bar{z}_3),
\nonu \\
&&
\Phi_{PQR,-\frac{1}{2}}^{(h_1)}(z_1,\bar{z}_1)\,
(\Phi_{+\frac{3}{2}}^{(h_2),A}\, \Phi_{BC,-1}^{(h_3)})(z_3,
\bar{z}_3) =
\nonu \\
&&  
\frac{\bar{z}_{13}}{z_{13}} \,
\sum_{m=0}^{\infty}\, \frac{\bar{z}_{13}^m}{m!}\,
B(2\bar{h}_1+1+m,2\bar{h}_2+1) \, (-1) \, 3 \,
\de^{A}_{[P}\, (\bar{\pa}^m \,
\Phi_{QR],-1}^{(h_1+h_2)}\,\Phi_{BC,-1}^{(h_3)})
(z_3,\bar{z}_3),
\nonu \\
&&
\Phi_{PQR,-\frac{1}{2}}^{(h_1)}(z_1,\bar{z}_1)\,
(\Phi_{+\frac{3}{2}}^{(h_2),A}\, \Phi_{B,-\frac{3}{2}}^{(h_3)})(z_3,
\bar{z}_3) =
\nonu \\
&&  
\frac{\bar{z}_{13}}{z_{13}} \,
\sum_{m=0}^{\infty}\, \frac{\bar{z}_{13}^m}{m!}\,
B(2\bar{h}_1+1+m,2\bar{h}_2+1) \, (-1) \, 3 \,
\de^{A}_{[P}\, (\bar{\pa}^m \,
\Phi_{QR],-1}^{(h_1+h_2)}\,\Phi_{B,-\frac{3}{2}}^{(h_3)})
(z_3,\bar{z}_3),
\nonu \\
&&
\Phi_{PQR,-\frac{1}{2}}^{(h_1)}(z_1,\bar{z}_1)\,
(\Phi_{+1}^{(h_2),AB}\, \Phi_{+1}^{(h_3),CD})(z_3,
\bar{z}_3) =
\nonu \\
&&  
\frac{\bar{z}_{13}}{z_{13}} \,
\sum_{m=0}^{\infty}\, \frac{\bar{z}_{13}^m}{m!}\,
B(2\bar{h}_1+1+m,2\bar{h}_2+1) \,
3! \, \de^{A}_{[P} \,
(\bar{\pa}^m \,
\Phi_{Q,-\frac{3}{2}}^{(h_1+h_2)}\,\Phi_{+1}^{(h_3),CD})
(z_3,\bar{z}_3) \, \de^{B}_{R]}
\nonu \\
&& +
\frac{\bar{z}_{13}}{z_{13}} \,
\sum_{m=0}^{\infty}\, \frac{\bar{z}_{13}^m}{m!}\,
B(2\bar{h}_1+1+m,2\bar{h}_3+1)  \, 
3! \, \de^{C}_{[P} \, (
\Phi_{+1}^{(h_2),AB}\,\bar{\pa}^m \,
\Phi_{Q,-\frac{3}{2}}^{(h_1+h_3)})
(z_3,\bar{z}_3) \, \de^{D}_{R]},
\nonu \\
&&
\Phi_{PQR,-\frac{1}{2}}^{(h_1)}(z_1,\bar{z}_1)\,
(\Phi_{+1}^{(h_2),AB}\, \Phi_{+\frac{1}{2}}^{(h_3),CDE})(z_3,
\bar{z}_3) =
\nonu \\
&&  
\frac{\bar{z}_{13}}{z_{13}} \,
\sum_{m=0}^{\infty}\, \frac{\bar{z}_{13}^m}{m!}\,
B(2\bar{h}_1+1+m,2\bar{h}_2+1) \,
3! \, \de^{A}_{[P}\, 
(\bar{\pa}^m \,
\Phi_{Q,-\frac{3}{2}}^{(h_1+h_2)}\,\Phi_{+\frac{1}{2}}^{(h_3),CDE})
(z_3,\bar{z}_3) \, \de^{B}_{R]}
\nonu \\
&& +
\frac{\bar{z}_{13}}{z_{13}} \,
\sum_{m=0}^{\infty}\, \frac{\bar{z}_{13}^m}{m!}\,
B(2\bar{h}_1+1+m,2\bar{h}_3+1)  \, 
(-1) \, \de^{CDE}_{PQR}  \, (
\Phi_{+1}^{(h_2),AB}\,\bar{\pa}^m \,
\Phi_{-2}^{(h_1+h_3)})
(z_3,\bar{z}_3),
\nonu \\
&&
\Phi_{PQR,-\frac{1}{2}}^{(h_1)}(z_1,\bar{z}_1)\,
(\Phi_{+1}^{(h_2),AB}\, \Phi_{0}^{(h_3),CDEF})(z_3,
\bar{z}_3) =
\nonu \\
&&  
\frac{\bar{z}_{13}}{z_{13}} \,
\sum_{m=0}^{\infty}\, \frac{\bar{z}_{13}^m}{m!}\,
B(2\bar{h}_1+1+m,2\bar{h}_2+1) \,
3! \, \de^{A}_{[P}\, 
(\bar{\pa}^m \,
\Phi_{Q,-\frac{3}{2}}^{(h_1+h_2)}\,\Phi_{0}^{(h_3),CDEF})
(z_3,\bar{z}_3) \, \de^{B}_{R]},
\nonu \\
&&
\Phi_{PQR,-\frac{1}{2}}^{(h_1)}(z_1,\bar{z}_1)\,
(\Phi_{+1}^{(h_2),AB}\, \Phi_{CDE,-\frac{1}{2}}^{(h_3)})(z_3,
\bar{z}_3) =
\nonu \\
&&  
\frac{\bar{z}_{13}}{z_{13}} \,
\sum_{m=0}^{\infty}\, \frac{\bar{z}_{13}^m}{m!}\,
B(2\bar{h}_1+1+m,2\bar{h}_2+1) \,
3! \, \de^{A}_{[P}\, 
(\bar{\pa}^m \,
\Phi_{Q,-\frac{3}{2}}^{(h_1+h_2)}\,\Phi_{CDE,-\frac{1}{2}}^{(h_3)})
(z_3,\bar{z}_3) \, \de^{B}_{R]},
\nonu \\
&&
\Phi_{PQR,-\frac{1}{2}}^{(h_1)}(z_1,\bar{z}_1)\,
(\Phi_{+1}^{(h_2),AB}\, \Phi_{CD,-1}^{(h_3)})(z_3,
\bar{z}_3) =
\nonu \\
&&  
\frac{\bar{z}_{13}}{z_{13}} \,
\sum_{m=0}^{\infty}\, \frac{\bar{z}_{13}^m}{m!}\,
B(2\bar{h}_1+1+m,2\bar{h}_2+1) \,
3! \, \de^{A}_{[P}\,
(\bar{\pa}^m \,
\Phi_{Q,-\frac{3}{2}}^{(h_1+h_2)}\,\Phi_{CD,-1}^{(h_3)})
(z_3,\bar{z}_3) \, \de^{B}_{R]},
\nonu \\
&&
\Phi_{PQR,-\frac{1}{2}}^{(h_1)}(z_1,\bar{z}_1)\,
(\Phi_{+\frac{1}{2}}^{(h_2),ABC}\, \Phi_{+\frac{1}{2}}^{(h_3),DEF})(z_3,
\bar{z}_3) =
\nonu \\
&&  
\frac{\bar{z}_{13}}{z_{13}} \,
\sum_{m=0}^{\infty}\, \frac{\bar{z}_{13}^m}{m!}\,
B(2\bar{h}_1+1+m,2\bar{h}_2+1) \,
(-1) \, \de^{ABC}_{PQR}\, 
(\bar{\pa}^m \,
\Phi_{-2}^{(h_1+h_2)}\,\Phi_{\frac{1}{2}}^{(h_3),DEF})
(z_3,\bar{z}_3)
\nonu \\
&& +
\frac{\bar{z}_{13}}{z_{13}} \,
\sum_{m=0}^{\infty}\, \frac{\bar{z}_{13}^m}{m!}\,
B(2\bar{h}_1+1+m,2\bar{h}_3+1)  \, 
\de^{DEF}_{PQR}  \, (
\Phi_{+\frac{1}{2}}^{(h_2),ABC}\,\bar{\pa}^m \,
\Phi_{-2}^{(h_1+h_3)})
(z_3,\bar{z}_3),
\nonu \\
&&
\Phi_{PQR,-\frac{1}{2}}^{(h_1)}(z_1,\bar{z}_1)\,
(\Phi_{+\frac{1}{2}}^{(h_2),ABC}\, \Phi_{0}^{(h_3),DEFG})(z_3,
\bar{z}_3) =
\nonu \\
&&  
\frac{\bar{z}_{13}}{z_{13}} \,
\sum_{m=0}^{\infty}\, \frac{\bar{z}_{13}^m}{m!}\,
B(2\bar{h}_1+1+m,2\bar{h}_2+1) \,
(-1) \, \de^{ABC}_{PQR}\,  
(\bar{\pa}^m \,
\Phi_{-2}^{(h_1+h_2)}\,\Phi_{0}^{(h_3),DEFG})
(z_3,\bar{z}_3),
\nonu \\
&&
\Phi_{PQR,-\frac{1}{2}}^{(h_1)}(z_1,\bar{z}_1)\,
(\Phi_{+\frac{1}{2}}^{(h_2),ABC}\, \Phi_{DEF,-\frac{1}{2}}^{(h_3)})(z_3,
\bar{z}_3) =
\nonu \\
&&  
\frac{\bar{z}_{13}}{z_{13}} \,
\sum_{m=0}^{\infty}\, \frac{\bar{z}_{13}^m}{m!}\,
B(2\bar{h}_1+1+m,2\bar{h}_2+1) \,
(-1)\, \de^{ABC}_{PQR}\, 
(\bar{\pa}^m \,
\Phi_{-2}^{(h_1+h_2)}\,\Phi_{DEF,-\frac{1}{2}}^{(h_3)})
(z_3,\bar{z}_3).
\label{minusonehalflinear}
\eea

\subsection{The operator product expansions of
the graviphotons  of helicity $-1$  with the quadratic
operators}

The four OPEs
\footnote{After the two OPEs which do not satisfy
the condition $s_1+s_2+s_3 \geq 2$ are removed,
these OPEs remain.}
are given by
\bea
&&
\Phi_{PQ,-1}^{(h_1)}(z_1,\bar{z}_1)\,
(\Phi_{+2}^{(h_2)}\, \Phi_{+2}^{(h_3)})(z_3,
\bar{z}_3) = \frac{\bar{z}_{13}^2}{z_{13}^2} \,
\sum_{m=0}^{\infty}\, \frac{\bar{z}_{13}^m}{m!}\,
B(2\bar{h}_1+2+m,2\bar{h}_2+1, 2\bar{h}_3+1)
\nonu \\
&& \times
\bar{\pa}^m \, \Phi_{PQ,-1}^{(h_1+h_2+h_3)}(z_3,\bar{z}_3)
\nonu \\
&&+  
\frac{\bar{z}_{13}}{z_{13}} \,
\sum_{m=0}^{\infty}\, \frac{\bar{z}_{13}^m}{m!}\,
B(2\bar{h}_1+1+m,2\bar{h}_2+1) \, 
(\bar{\pa}^m \,
\Phi_{PQ,-1}^{(h_1+h_2)}\,\Phi_{+2}^{(h_3)})
(z_3,\bar{z}_3)
\nonu \\
&& +
\frac{\bar{z}_{13}}{z_{13}} \,
\sum_{m=0}^{\infty}\, \frac{\bar{z}_{13}^m}{m!}\,
B(2\bar{h}_1+1+m,2\bar{h}_3+1)  \,
(
\Phi_{+2}^{(h_2)}\,\bar{\pa}^m \,
\Phi_{PQ,-1}^{(h_1+h_3)})
(z_3,\bar{z}_3),
\nonu \\
&&
\Phi_{PQ,-1}^{(h_1)}(z_1,\bar{z}_1)\,
(\Phi_{+2}^{(h_2)}\, \Phi_{+\frac{3}{2}}^{(h_3),A})(z_3,
\bar{z}_3) = \frac{\bar{z}_{13}^2}{z_{13}^2} \,
\sum_{m=0}^{\infty}\, \frac{\bar{z}_{13}^m}{m!}\,
B(2\bar{h}_1+2+m,2\bar{h}_2+1, 2\bar{h}_3+1)
\nonu \\
&& \times
2!\,
\de^{A}_{[P}\,
  \bar{\pa}^m \, \Phi_{Q],-\frac{3}{2}}^{(h_1+h_2+h_3)}(z_3,\bar{z}_3)
\nonu \\
&&+  
\frac{\bar{z}_{13}}{z_{13}} \,
\sum_{m=0}^{\infty}\, \frac{\bar{z}_{13}^m}{m!}\,
B(2\bar{h}_1+1+m,2\bar{h}_2+1) \, 
(\bar{\pa}^m \,
\Phi_{PQ,-1}^{(h_1+h_2)}\,\Phi_{+\frac{3}{2}}^{(h_3),A})
(z_3,\bar{z}_3)
\nonu \\
&& +
\frac{\bar{z}_{13}}{z_{13}} \,
\sum_{m=0}^{\infty}\, \frac{\bar{z}_{13}^m}{m!}\,
B(2\bar{h}_1+1+m,2\bar{h}_3+1)  \,
2! \, \de^{A}_{[P}\, 
(
\Phi_{+2}^{(h_2)}\,\bar{\pa}^m \,
\Phi_{Q],-\frac{3}{2}}^{(h_1+h_3)})
(z_3,\bar{z}_3),
\nonu \\
&&
\Phi_{PQ,-1}^{(h_1)}(z_1,\bar{z}_1)\,
(\Phi_{+2}^{(h_2)}\, \Phi_{+1}^{(h_3),AB})(z_3,
\bar{z}_3) = \frac{\bar{z}_{13}^2}{z_{13}^2} \,
\sum_{m=0}^{\infty}\, \frac{\bar{z}_{13}^m}{m!}\,
B(2\bar{h}_1+2+m,2\bar{h}_2+1, 2\bar{h}_3+1)
\nonu \\
&& \times
\de^{AB}_{PQ}\,
\bar{\pa}^m \, \Phi_{-2}^{(h_1+h_2+h_3)}(z_3,\bar{z}_3)
\nonu \\
&&+  
\frac{\bar{z}_{13}}{z_{13}} \,
\sum_{m=0}^{\infty}\, \frac{\bar{z}_{13}^m}{m!}\,
B(2\bar{h}_1+1+m,2\bar{h}_2+1) \, 
(\bar{\pa}^m \,
\Phi_{PQ,-1}^{(h_1+h_2)}\,\Phi_{+1}^{(h_3),AB})
(z_3,\bar{z}_3)
\nonu \\
&& +
\frac{\bar{z}_{13}}{z_{13}} \,
\sum_{m=0}^{\infty}\, \frac{\bar{z}_{13}^m}{m!}\,
B(2\bar{h}_1+1+m,2\bar{h}_3+1)  \,
\de^{AB}_{PQ}\, 
(
\Phi_{+2}^{(h_2)}\,\bar{\pa}^m \,
\Phi_{-2}^{(h_1+h_3)})
(z_3,\bar{z}_3),
\nonu \\
&&
\Phi_{PQ,-1}^{(h_1)}(z_1,\bar{z}_1)\,
(\Phi_{+\frac{3}{2}}^{(h_2),A}\, \Phi_{+\frac{3}{2}}^{(h_3),B})(z_3,
\bar{z}_3) = \frac{\bar{z}_{13}^2}{z_{13}^2} \,
\sum_{m=0}^{\infty}\, \frac{\bar{z}_{13}^m}{m!}\,
B(2\bar{h}_1+2+m,2\bar{h}_2+1, 2\bar{h}_3+1)
\nonu \\
&& \times \de^{AB}_{QP}\,
\bar{\pa}^m \, \Phi_{-2}^{(h_1+h_2+h_3)}(z_3,\bar{z}_3)
\nonu \\
&&+  
\frac{\bar{z}_{13}}{z_{13}} \,
\sum_{m=0}^{\infty}\, \frac{\bar{z}_{13}^m}{m!}\,
B(2\bar{h}_1+1+m,2\bar{h}_2+1) \,
2! \, \de^{A}_{[P}
(\bar{\pa}^m \,
\Phi_{Q],-\frac{3}{2}}^{(h_1+h_2)}\,\Phi_{+\frac{3}{2}}^{(h_3),B})
(z_3,\bar{z}_3)
\nonu \\
&& +
\frac{\bar{z}_{13}}{z_{13}} \,
\sum_{m=0}^{\infty}\, \frac{\bar{z}_{13}^m}{m!}\,
B(2\bar{h}_1+1+m,2\bar{h}_3+1)  \,
2! \, \de^{B}_{[P}
(
\Phi_{+\frac{3}{2}}^{(h_2),A}\,\bar{\pa}^m \,
\Phi_{Q],-\frac{3}{2}}^{(h_1+h_3)})
(z_3,\bar{z}_3).
\label{fullminusone}
\eea
The following twenty-one OPEs
\footnote{In this case, only two OPEs
in (\ref{minusonelinear}) satisfy the condition $s_1+s_3 \geq 0$.
Moreover, for the last four two-particle operators in
(\ref{REDEFINITION}), the corresponding OPEs of the single-particle
operators with helicity $-1$ are not present due to the condition
$s_1+s_2 < 0$.}
contain
the quadratic terms on the right-hand sides
\bea
&&
\Phi_{PQ,-1}^{(h_1)}(z_1,\bar{z}_1)\,
(\Phi_{+2}^{(h_2)}\, \Phi_{\frac{1}{2}}^{(h_3),ABC})(z_3,
\bar{z}_3) =
\nonu \\
&&  
\frac{\bar{z}_{13}}{z_{13}} \,
\sum_{m=0}^{\infty}\, \frac{\bar{z}_{13}^m}{m!}\,
B(2\bar{h}_1+1+m,2\bar{h}_2+1) \, (\bar{\pa}^m \,
\Phi_{PQ,-1}^{(h_1+h_2)}\,\Phi_{+\frac{1}{2}}^{(h_3),ABC})
(z_3,\bar{z}_3),
\nonu \\
&&
\Phi_{PQ,-1}^{(h_1)}(z_1,\bar{z}_1)\,
(\Phi_{+2}^{(h_2)}\, \Phi_{0}^{(h_3),ABCD})(z_3,
\bar{z}_3) =
\nonu \\
&&  
\frac{\bar{z}_{13}}{z_{13}} \,
\sum_{m=0}^{\infty}\, \frac{\bar{z}_{13}^m}{m!}\,
B(2\bar{h}_1+1+m,2\bar{h}_2+1) \, (\bar{\pa}^m \,
\Phi_{PQ,-1}^{(h_1+h_2)}\,\Phi_{0}^{(h_3),ABCD})
(z_3,\bar{z}_3),
\nonu \\
&&
\Phi_{PQ,-1}^{(h_1)}(z_1,\bar{z}_1)\,
(\Phi_{+2}^{(h_2)}\, \Phi_{ABC,-\frac{1}{2}}^{(h_3)})(z_3,
\bar{z}_3) =
\nonu \\
&&  
\frac{\bar{z}_{13}}{z_{13}} \,
\sum_{m=0}^{\infty}\, \frac{\bar{z}_{13}^m}{m!}\,
B(2\bar{h}_1+1+m,2\bar{h}_2+1) \, (\bar{\pa}^m \,
\Phi_{PQ,-1}^{(h_1+h_2)}\,\Phi_{ABC,-\frac{1}{2}}^{(h_3)})
(z_3,\bar{z}_3),
\nonu \\
&&
\Phi_{PQ,-1}^{(h_1)}(z_1,\bar{z}_1)\,
(\Phi_{+2}^{(h_2)}\, \Phi_{AB,-1}^{(h_3)})(z_3,
\bar{z}_3) =
\nonu \\
&&  
\frac{\bar{z}_{13}}{z_{13}} \,
\sum_{m=0}^{\infty}\, \frac{\bar{z}_{13}^m}{m!}\,
B(2\bar{h}_1+1+m,2\bar{h}_2+1) \, (\bar{\pa}^m \,
\Phi_{PQ,-1}^{(h_1+h_2)}\,\Phi_{AB,-1}^{(h_3)})
(z_3,\bar{z}_3),
\nonu \\
&&
\Phi_{PQ,-1}^{(h_1)}(z_1,\bar{z}_1)\,
(\Phi_{+2}^{(h_2)}\, \Phi_{A,-\frac{3}{2}}^{(h_3)})(z_3,
\bar{z}_3) =
\nonu \\
&&  
\frac{\bar{z}_{13}}{z_{13}} \,
\sum_{m=0}^{\infty}\, \frac{\bar{z}_{13}^m}{m!}\,
B(2\bar{h}_1+1+m,2\bar{h}_2+1) \, (\bar{\pa}^m \,
\Phi_{PQ,-1}^{(h_1+h_2)}\,\Phi_{A,-\frac{3}{2}}^{(h_3)})
(z_3,\bar{z}_3),
\nonu \\
&&
\Phi_{PQ,-1}^{(h_1)}(z_1,\bar{z}_1)\,
(\Phi_{+2}^{(h_2)}\, \Phi_{-2}^{(h_3)})(z_3,
\bar{z}_3) =
\nonu \\
&&  
\frac{\bar{z}_{13}}{z_{13}} \,
\sum_{m=0}^{\infty}\, \frac{\bar{z}_{13}^m}{m!}\,
B(2\bar{h}_1+1+m,2\bar{h}_2+1) \, (\bar{\pa}^m \,
\Phi_{PQ,-1}^{(h_1+h_2)}\,\Phi_{-2}^{(h_3)})
(z_3,\bar{z}_3),
\nonu \\
&&
\Phi_{PQ,-1}^{(h_1)}(z_1,\bar{z}_1)\,
(\Phi_{+\frac{3}{2}}^{(h_2),A}\, \Phi_{+1}^{(h_3),BC})(z_3,
\bar{z}_3) =
\nonu \\
&&  
\frac{\bar{z}_{13}}{z_{13}} \,
\sum_{m=0}^{\infty}\, \frac{\bar{z}_{13}^m}{m!}\,
B(2\bar{h}_1+1+m,2\bar{h}_2+1) \,
2! \, \de^{A}_{[P} \,
(\bar{\pa}^m \,
\Phi_{Q],-\frac{3}{2}}^{(h_1+h_2)}\,\Phi_{+1}^{(h_3),BC})
(z_3,\bar{z}_3)
\nonu \\
&& +
\frac{\bar{z}_{13}}{z_{13}} \,
\sum_{m=0}^{\infty}\, \frac{\bar{z}_{13}^m}{m!}\,
B(2\bar{h}_1+1+m,2\bar{h}_3+1)  \, 
\de^{BC}_{PQ} \, (
\Phi_{+\frac{3}{2}}^{(h_2),A}\,\bar{\pa}^m \,
\Phi_{-2}^{(h_1+h_3)})
(z_3,\bar{z}_3),
\nonu \\
&&
\Phi_{PQ,-1}^{(h_1)}(z_1,\bar{z}_1)\,
(\Phi_{+\frac{3}{2}}^{(h_2),A}\, \Phi_{+\frac{1}{2}}^{(h_3),BCD})(z_3,
\bar{z}_3) =
\nonu \\
&&  
\frac{\bar{z}_{13}}{z_{13}} \,
\sum_{m=0}^{\infty}\, \frac{\bar{z}_{13}^m}{m!}\,
B(2\bar{h}_1+1+m,2\bar{h}_2+1) \,
2! \, \de^{A}_{[P} \,
(\bar{\pa}^m \,
\Phi_{Q],-\frac{3}{2}}^{(h_1+h_2)}\,\Phi_{+\frac{1}{2}}^{(h_3),BCD})
(z_3,\bar{z}_3),
\nonu \\
&&
\Phi_{PQ,-1}^{(h_1)}(z_1,\bar{z}_1)\,
(\Phi_{+\frac{3}{2}}^{(h_2),A}\, \Phi_{0}^{(h_3),BCDE})(z_3,
\bar{z}_3) =
\nonu \\
&&  
\frac{\bar{z}_{13}}{z_{13}} \,
\sum_{m=0}^{\infty}\, \frac{\bar{z}_{13}^m}{m!}\,
B(2\bar{h}_1+1+m,2\bar{h}_2+1) \,
2! \, \de^{A}_{[P}
(\bar{\pa}^m \,
\Phi_{Q],-\frac{3}{2}}^{(h_1+h_2)}\,\Phi_{0}^{(h_3),BCDE})
(z_3,\bar{z}_3),
\nonu \\
&&
\Phi_{PQ,-1}^{(h_1)}(z_1,\bar{z}_1)\,
(\Phi_{+\frac{3}{2}}^{(h_2),A}\, \Phi_{BCD,-\frac{1}{2}}^{(h_3)})(z_3,
\bar{z}_3) =
\nonu \\
&&  
\frac{\bar{z}_{13}}{z_{13}} \,
\sum_{m=0}^{\infty}\, \frac{\bar{z}_{13}^m}{m!}\,
B(2\bar{h}_1+1+m,2\bar{h}_2+1) \, 2!\,
\de^{A}_{[P} \, (\bar{\pa}^m \,
\Phi_{Q],-\frac{3}{2}}^{(h_1+h_2)}\,\Phi_{BCD,-\frac{1}{2}}^{(h_3)})
(z_3,\bar{z}_3),
\nonu \\
&&
\Phi_{PQ,-1}^{(h_1)}(z_1,\bar{z}_1)\,
(\Phi_{+\frac{3}{2}}^{(h_2),A}\, \Phi_{BC,-1}^{(h_3)})(z_3,
\bar{z}_3) =\nonu \\
&&  
\frac{\bar{z}_{13}}{z_{13}} \,
\sum_{m=0}^{\infty}\, \frac{\bar{z}_{13}^m}{m!}\,
B(2\bar{h}_1+1+m,2\bar{h}_2+1) \, 2! \,
\de^{A}_{[P}\, (\bar{\pa}^m \,
\Phi_{Q],-\frac{3}{2}}^{(h_1+h_2)}\,\Phi_{BC,-1}^{(h_3)})
(z_3,\bar{z}_3),
\nonu \\
&&
\Phi_{PQ,-1}^{(h_1)}(z_1,\bar{z}_1)\,
(\Phi_{+\frac{3}{2}}^{(h_2),A}\, \Phi_{B,-\frac{3}{2}}^{(h_3)})(z_3,
\bar{z}_3) =
\nonu \\
&&  
\frac{\bar{z}_{13}}{z_{13}} \,
\sum_{m=0}^{\infty}\, \frac{\bar{z}_{13}^m}{m!}\,
B(2\bar{h}_1+1+m,2\bar{h}_2+1) \, 2! \,
\de^{A}_{[P}\, (\bar{\pa}^m \,
\Phi_{Q],-\frac{3}{2}}^{(h_1+h_2)}\,\Phi_{B,-\frac{3}{2}}^{(h_3)})
(z_3,\bar{z}_3),
\nonu \\
&&
\Phi_{PQ,-1}^{(h_1)}(z_1,\bar{z}_1)\,
(\Phi_{+1}^{(h_2),AB}\, \Phi_{+1}^{(h_3),CD})(z_3,
\bar{z}_3) =
\nonu \\
&&  
\frac{\bar{z}_{13}}{z_{13}} \,
\sum_{m=0}^{\infty}\, \frac{\bar{z}_{13}^m}{m!}\,
B(2\bar{h}_1+1+m,2\bar{h}_2+1) \,
\de^{AB}_{PQ} \,
(\bar{\pa}^m \,
\Phi_{-2}^{(h_1+h_2)}\,\Phi_{+1}^{(h_3),CD})
(z_3,\bar{z}_3)
\nonu \\
&& +
\frac{\bar{z}_{13}}{z_{13}} \,
\sum_{m=0}^{\infty}\, \frac{\bar{z}_{13}^m}{m!}\,
B(2\bar{h}_1+1+m,2\bar{h}_3+1)  \, 
\de^{CD}_{PQ} \, (
\Phi_{+1}^{(h_2),AB}\,\bar{\pa}^m \,
\Phi_{-2}^{(h_1+h_3)})
(z_3,\bar{z}_3),
\nonu \\
&&
\Phi_{PQ,-1}^{(h_1)}(z_1,\bar{z}_1)\,
(\Phi_{+1}^{(h_2),AB}\, \Phi_{+\frac{1}{2}}^{(h_3),CDE})(z_3,
\bar{z}_3) =
\nonu \\
&&  
\frac{\bar{z}_{13}}{z_{13}} \,
\sum_{m=0}^{\infty}\, \frac{\bar{z}_{13}^m}{m!}\,
B(2\bar{h}_1+1+m,2\bar{h}_2+1) \,
\de^{AB}_{PQ}\, 
(\bar{\pa}^m \,
\Phi_{-2}^{(h_1+h_2)}\,\Phi_{+\frac{1}{2}}^{(h_3),CDE})
(z_3,\bar{z}_3),
\nonu \\
&&
\Phi_{PQ,-1}^{(h_1)}(z_1,\bar{z}_1)\,
(\Phi_{+1}^{(h_2),AB}\, \Phi_{0}^{(h_3),CDEF})(z_3,
\bar{z}_3) =
\nonu \\
&&  
\frac{\bar{z}_{13}}{z_{13}} \,
\sum_{m=0}^{\infty}\, \frac{\bar{z}_{13}^m}{m!}\,
B(2\bar{h}_1+1+m,2\bar{h}_2+1) \,
\de^{AB}_{PQ}\, 
(\bar{\pa}^m \,
\Phi_{-2}^{(h_1+h_2)}\,\Phi_{0}^{(h_3),CDEF})
(z_3,\bar{z}_3),
\nonu \\
&&
\Phi_{PQ,-1}^{(h_1)}(z_1,\bar{z}_1)\,
(\Phi_{+1}^{(h_2),AB}\, \Phi_{CDE,-\frac{1}{2}}^{(h_3)})(z_3,
\bar{z}_3) =
\nonu \\
&&  
\frac{\bar{z}_{13}}{z_{13}} \,
\sum_{m=0}^{\infty}\, \frac{\bar{z}_{13}^m}{m!}\,
B(2\bar{h}_1+1+m,2\bar{h}_2+1) \,
\de^{AB}_{PQ}\, 
(\bar{\pa}^m \,
\Phi_{-2}^{(h_1+h_2)}\,\Phi_{CDE,-\frac{1}{2}}^{(h_3)})
(z_3,\bar{z}_3),
\nonu \\
&&
\Phi_{PQ,-1}^{(h_1)}(z_1,\bar{z}_1)\,
(\Phi_{+1}^{(h_2),AB}\, \Phi_{CD,-1}^{(h_3)})(z_3,
\bar{z}_3) =
\nonu \\
&&  
\frac{\bar{z}_{13}}{z_{13}} \,
\sum_{m=0}^{\infty}\, \frac{\bar{z}_{13}^m}{m!}\,
B(2\bar{h}_1+1+m,2\bar{h}_2+1) \,
\de^{AB}_{PQ}\,
(\bar{\pa}^m \,
\Phi_{-2}^{(h_1+h_2)}\,\Phi_{CD,-1}^{(h_3)})
(z_3,\bar{z}_3).
\label{minusonelinear}
\eea

\subsection{The OPEs of
the gravitinos   of helicity $-\frac{3}{2}$  with the quadratic
operators}

The two OPEs
\footnote{Similarly, after the two OPEs which do not satisfy
the condition $s_1+s_2+s_3 \geq 2$ are removed,
we are left with these OPEs.}
are
\bea
&&
\Phi_{P,-\frac{3}{2}}^{(h_1)}(z_1,\bar{z}_1)\,
(\Phi_{+2}^{(h_2)}\, \Phi_{+2}^{(h_3)})(z_3,
\bar{z}_3) = \frac{\bar{z}_{13}^2}{z_{13}^2} \,
\sum_{m=0}^{\infty}\, \frac{\bar{z}_{13}^m}{m!}\,
B(2\bar{h}_1+2+m,2\bar{h}_2+1, 2\bar{h}_3+1)
\nonu \\
&& \times 
\bar{\pa}^m \, \Phi_{P,-\frac{3}{2}}^{(h_1+h_2+h_3)}(z_3,\bar{z}_3)
\nonu \\
&& +  
\frac{\bar{z}_{13}}{z_{13}} \,
\sum_{m=0}^{\infty}\, \frac{\bar{z}_{13}^m}{m!}\,
B(2\bar{h}_1+1+m,2\bar{h}_2+1) \,
(\bar{\pa}^m \,
\Phi_{P,-\frac{3}{2}}^{(h_1+h_2)}\,\Phi_{+2}^{(h_3)})
(z_3,\bar{z}_3)
\nonu \\
&& +
\frac{\bar{z}_{13}}{z_{13}} \,
\sum_{m=0}^{\infty}\, \frac{\bar{z}_{13}^m}{m!}\,
B(2\bar{h}_1+1+m,2\bar{h}_3+1)  \, 
(
\Phi_{+2}^{(h_2)}\,\bar{\pa}^m \,
\Phi_{P,-\frac{3}{2}}^{(h_1+h_3)})
(z_3,\bar{z}_3),
\nonu \\
&&
\Phi_{P,-\frac{3}{2}}^{(h_1)}(z_1,\bar{z}_1)\,
(\Phi_{+2}^{(h_2)}\, \Phi_{+\frac{3}{2}}^{(h_3),A})(z_3,
\bar{z}_3) = \frac{\bar{z}_{13}^2}{z_{13}^2} \,
\sum_{m=0}^{\infty}\, \frac{\bar{z}_{13}^m}{m!}\,
B(2\bar{h}_1+2+m,2\bar{h}_2+1, 2\bar{h}_3+1)
\nonu \\
&& \times
(-1) \, \de^{A}_{P}\,
\bar{\pa}^m \, \Phi_{-2}^{(h_1+h_2+h_3)}(z_3,\bar{z}_3)
\nonu \\
&& +  
\frac{\bar{z}_{13}}{z_{13}} \,
\sum_{m=0}^{\infty}\, \frac{\bar{z}_{13}^m}{m!}\,
B(2\bar{h}_1+1+m,2\bar{h}_2+1) \,
(\bar{\pa}^m \,
\Phi_{P,-\frac{3}{2}}^{(h_1+h_2)}\,\Phi_{+\frac{3}{2}}^{(h_3),A})
(z_3,\bar{z}_3)
\nonu \\
&& +
\frac{\bar{z}_{13}}{z_{13}} \,
\sum_{m=0}^{\infty}\, \frac{\bar{z}_{13}^m}{m!}\,
B(2\bar{h}_1+1+m,2\bar{h}_3+1)  \, (-1) \, \de^{A}_{P}\,
(
\Phi_{+2}^{(h_2)}\,\bar{\pa}^m \,
\Phi_{-2}^{(h_1+h_3)})
(z_3,\bar{z}_3).
\label{fullminusthreehalf}
\eea
There are fourteen OPEs
\footnote{In this case, only one OPE
in (\ref{minusthreehalflinear}) satisfies the condition $s_1+s_3 \geq 0$.
Moreover, for the last nine two-particle operators in
(\ref{REDEFINITION}), the corresponding OPEs of the single-particle
operators with helicity $-\frac{3}{2}$
are not present due to the condition
$s_1+s_2 < 0$.}
with quadratic terms 
\bea
&&
\Phi_{P,-\frac{3}{2}}^{(h_1)}(z_1,\bar{z}_1)\,
(\Phi_{+2}^{(h_2)}\, \Phi_{+1}^{(h_3),AB})(z_3,
\bar{z}_3) =
\nonu \\
&&  
\frac{\bar{z}_{13}}{z_{13}} \,
\sum_{m=0}^{\infty}\, \frac{\bar{z}_{13}^m}{m!}\,
B(2\bar{h}_1+1+m,2\bar{h}_2+1) \, (\bar{\pa}^m \,
\Phi_{P,-\frac{3}{2}}^{(h_1+h_2)}\,\Phi_{+1}^{(h_3),AB})
(z_3,\bar{z}_3),
\nonu \\
&&
\Phi_{P,-\frac{3}{2}}^{(h_1)}(z_1,\bar{z}_1)\,
(\Phi_{+2}^{(h_2)}\, \Phi_{\frac{1}{2}}^{(h_3),ABC})(z_3,
\bar{z}_3) =
\nonu \\
&&  
\frac{\bar{z}_{13}}{z_{13}} \,
\sum_{m=0}^{\infty}\, \frac{\bar{z}_{13}^m}{m!}\,
B(2\bar{h}_1+1+m,2\bar{h}_2+1) \, (\bar{\pa}^m \,
\Phi_{P,-\frac{3}{2}}^{(h_1+h_2)}\,\Phi_{+\frac{1}{2}}^{(h_3),ABC})
(z_3,\bar{z}_3),
\nonu \\
&&
\Phi_{P,-\frac{3}{2}}^{(h_1)}(z_1,\bar{z}_1)\,
(\Phi_{+2}^{(h_2)}\, \Phi_{0}^{(h_3),ABCD})(z_3,
\bar{z}_3) =
\nonu \\
&&  
\frac{\bar{z}_{13}}{z_{13}} \,
\sum_{m=0}^{\infty}\, \frac{\bar{z}_{13}^m}{m!}\,
B(2\bar{h}_1+1+m,2\bar{h}_2+1) \, (\bar{\pa}^m \,
\Phi_{P,-\frac{3}{2}}^{(h_1+h_2)}\,\Phi_{0}^{(h_3),ABCD})
(z_3,\bar{z}_3),
\nonu \\
&&
\Phi_{P,-\frac{3}{2}}^{(h_1)}(z_1,\bar{z}_1)\,
(\Phi_{+2}^{(h_2)}\, \Phi_{ABC,-\frac{1}{2}}^{(h_3)})(z_3,
\bar{z}_3) =
\nonu \\
&&  
\frac{\bar{z}_{13}}{z_{13}} \,
\sum_{m=0}^{\infty}\, \frac{\bar{z}_{13}^m}{m!}\,
B(2\bar{h}_1+1+m,2\bar{h}_2+1) \, (\bar{\pa}^m \,
\Phi_{P,-\frac{3}{2}}^{(h_1+h_2)}\,\Phi_{ABC,-\frac{1}{2}}^{(h_3)})
(z_3,\bar{z}_3),
\nonu \\
&&
\Phi_{P,-\frac{3}{2}}^{(h_1)}(z_1,\bar{z}_1)\,
(\Phi_{+2}^{(h_2)}\, \Phi_{AB,-1}^{(h_3)})(z_3,
\bar{z}_3) =
\nonu \\
&&  
\frac{\bar{z}_{13}}{z_{13}} \,
\sum_{m=0}^{\infty}\, \frac{\bar{z}_{13}^m}{m!}\,
B(2\bar{h}_1+1+m,2\bar{h}_2+1) \, (\bar{\pa}^m \,
\Phi_{P,-\frac{3}{2}}^{(h_1+h_2)}\,\Phi_{AB,-1}^{(h_3)})
(z_3,\bar{z}_3),
\nonu \\
&&
\Phi_{P,-\frac{3}{2}}^{(h_1)}(z_1,\bar{z}_1)\,
(\Phi_{+2}^{(h_2)}\, \Phi_{A,-\frac{3}{2}}^{(h_3)})(z_3,
\bar{z}_3) =
\nonu \\
&&  
\frac{\bar{z}_{13}}{z_{13}} \,
\sum_{m=0}^{\infty}\, \frac{\bar{z}_{13}^m}{m!}\,
B(2\bar{h}_1+1+m,2\bar{h}_2+1) \, (\bar{\pa}^m \,
\Phi_{P,-\frac{3}{2}}^{(h_1+h_2)}\,\Phi_{A,-\frac{3}{2}}^{(h_3)})
(z_3,\bar{z}_3),
\nonu \\
&&
\Phi_{P,-\frac{3}{2}}^{(h_1)}(z_1,\bar{z}_1)\,
(\Phi_{+2}^{(h_2)}\, \Phi_{-2}^{(h_3)})(z_3,
\bar{z}_3) =
\nonu \\
&&  
\frac{\bar{z}_{13}}{z_{13}} \,
\sum_{m=0}^{\infty}\, \frac{\bar{z}_{13}^m}{m!}\,
B(2\bar{h}_1+1+m,2\bar{h}_2+1) \, (\bar{\pa}^m \,
\Phi_{P,-\frac{3}{2}}^{(h_1+h_2)}\,\Phi_{-2}^{(h_3)})
(z_3,\bar{z}_3),
\nonu \\
&&
\Phi_{P,-\frac{3}{2}}^{(h_1)}(z_1,\bar{z}_1)\,
(\Phi_{+\frac{3}{2}}^{(h_2),A}\, \Phi_{+\frac{3}{2}}^{(h_3),B})(z_3,
\bar{z}_3) =
\nonu \\
&&  
\frac{\bar{z}_{13}}{z_{13}} \,
\sum_{m=0}^{\infty}\, \frac{\bar{z}_{13}^m}{m!}\,
B(2\bar{h}_1+1+m,2\bar{h}_2+1) \,
(-1) \, \de^{A}_{P} \,
(\bar{\pa}^m \,
\Phi_{-2}^{(h_1+h_2)}\,\Phi_{+\frac{3}{2}}^{(h_3),B})
(z_3,\bar{z}_3)
\nonu \\
&& +
\frac{\bar{z}_{13}}{z_{13}} \,
\sum_{m=0}^{\infty}\, \frac{\bar{z}_{13}^m}{m!}\,
B(2\bar{h}_1+1+m,2\bar{h}_3+1)  \, 
\de^{B}_{P} \, (
\Phi_{+\frac{3}{2}}^{(h_2),A}\,\bar{\pa}^m \,
\Phi_{-2}^{(h_1+h_3)})
(z_3,\bar{z}_3),
\nonu \\
&&
\Phi_{P,-\frac{3}{2}}^{(h_1)}(z_1,\bar{z}_1)\,
(\Phi_{+\frac{3}{2}}^{(h_2),A}\, \Phi_{+1}^{(h_3),BC})(z_3,
\bar{z}_3) =
\nonu \\
&&  
\frac{\bar{z}_{13}}{z_{13}} \,
\sum_{m=0}^{\infty}\, \frac{\bar{z}_{13}^m}{m!}\,
B(2\bar{h}_1+1+m,2\bar{h}_2+1) \,
(-1) \, \de^{A}_{P} \,
(\bar{\pa}^m \,
\Phi_{-2}^{(h_1+h_2)}\,\Phi_{+1}^{(h_3),BC})
(z_3,\bar{z}_3),
\nonu \\
&&
\Phi_{P,-\frac{3}{2}}^{(h_1)}(z_1,\bar{z}_1)\,
(\Phi_{+\frac{3}{2}}^{(h_2),A}\, \Phi_{+\frac{1}{2}}^{(h_3),BCD})(z_3,
\bar{z}_3) =
\nonu \\
&&  
\frac{\bar{z}_{13}}{z_{13}} \,
\sum_{m=0}^{\infty}\, \frac{\bar{z}_{13}^m}{m!}\,
B(2\bar{h}_1+1+m,2\bar{h}_2+1) \,
(-1) \, \de^{A}_{P} \,
(\bar{\pa}^m \,
\Phi_{-2}^{(h_1+h_2)}\,\Phi_{+\frac{1}{2}}^{(h_3),BCD})
(z_3,\bar{z}_3),
\nonu \\
&&
\Phi_{P,-\frac{3}{2}}^{(h_1)}(z_1,\bar{z}_1)\,
(\Phi_{+\frac{3}{2}}^{(h_2),A}\, \Phi_{0}^{(h_3),BCDE})(z_3,
\bar{z}_3) =
\nonu \\
&&  
\frac{\bar{z}_{13}}{z_{13}} \,
\sum_{m=0}^{\infty}\, \frac{\bar{z}_{13}^m}{m!}\,
B(2\bar{h}_1+1+m,2\bar{h}_2+1) \,
(-1) \, \de^{A}_{P}
(\bar{\pa}^m \,
\Phi_{-2}^{(h_1+h_2)}\,\Phi_{0}^{(h_3),BCDE})
(z_3,\bar{z}_3),
\nonu \\
&&
\Phi_{P,-\frac{3}{2}}^{(h_1)}(z_1,\bar{z}_1)\,
(\Phi_{+\frac{3}{2}}^{(h_2),A}\, \Phi_{BCD,-\frac{1}{2}}^{(h_3)})(z_3,
\bar{z}_3) =
\nonu \\
&&  
\frac{\bar{z}_{13}}{z_{13}} \,
\sum_{m=0}^{\infty}\, \frac{\bar{z}_{13}^m}{m!}\,
B(2\bar{h}_1+1+m,2\bar{h}_2+1) \, (-1)\,
\de^{A}_{P} \, (\bar{\pa}^m \,
\Phi_{-2}^{(h_1+h_2)}\,\Phi_{BCD,-\frac{1}{2}}^{(h_3)})
(z_3,\bar{z}_3),
\nonu \\
&&
\Phi_{P,-\frac{3}{2}}^{(h_1)}(z_1,\bar{z}_1)\,
(\Phi_{+\frac{3}{2}}^{(h_2),A}\, \Phi_{BC,-1}^{(h_3)})(z_3,
\bar{z}_3) =\nonu \\
&&  
\frac{\bar{z}_{13}}{z_{13}} \,
\sum_{m=0}^{\infty}\, \frac{\bar{z}_{13}^m}{m!}\,
B(2\bar{h}_1+1+m,2\bar{h}_2+1) \, (-1)\,
\de^{A}_{P}\, (\bar{\pa}^m \,
\Phi_{-2}^{(h_1+h_2)}\,\Phi_{BC,-1}^{(h_3)})
(z_3,\bar{z}_3),
\nonu \\
&&
\Phi_{P,-\frac{3}{2}}^{(h_1)}(z_1,\bar{z}_1)\,
(\Phi_{+\frac{3}{2}}^{(h_2),A}\, \Phi_{B,-\frac{3}{2}}^{(h_3)})(z_3,
\bar{z}_3) =
\nonu \\
&&  
\frac{\bar{z}_{13}}{z_{13}} \,
\sum_{m=0}^{\infty}\, \frac{\bar{z}_{13}^m}{m!}\,
B(2\bar{h}_1+1+m,2\bar{h}_2+1) \, (-1) \,
\de^{A}_{P}\, (\bar{\pa}^m \,
\Phi_{-2}^{(h_1+h_2)}\,\Phi_{B,-\frac{3}{2}}^{(h_3)})
(z_3,\bar{z}_3).
\label{minusthreehalflinear}
\eea

\subsection{The OPEs of
the gravitons  of helicity $-2$  with the quadratic
operators}

The final OPE
\footnote{We see that the remaining second OPE
from the previous two OPEs doesn't satisfy
the condition $s_1+s_2+s_3 \geq 2$ and we are left with this
final OPE.}
is
\bea
&&
\Phi_{-2}^{(h_1)}(z_1,\bar{z}_1)\,
(\Phi_{+2}^{(h_2)}\, \Phi_{+2}^{(h_3)})(z_3,
\bar{z}_3) = \frac{\bar{z}_{13}^2}{z_{13}^2} \,
\sum_{m=0}^{\infty}\, \frac{\bar{z}_{13}^m}{m!}\,
B(2\bar{h}_1+2+m,2\bar{h}_2+1, 2\bar{h}_3+1)
\nonu \\
&& \times
\bar{\pa}^m \, \Phi_{-2}^{(h_1+h_2+h_3)}(z_3,\bar{z}_3)
\nonu \\
&& +  
\frac{\bar{z}_{13}}{z_{13}} \,
\sum_{m=0}^{\infty}\, \frac{\bar{z}_{13}^m}{m!}\,
B(2\bar{h}_1+1+m,2\bar{h}_2+1) \,
(\bar{\pa}^m \,
\Phi_{-2}^{(h_1+h_2)}\,\Phi_{+2}^{(h_3)})
(z_3,\bar{z}_3)
\nonu \\
&& +
\frac{\bar{z}_{13}}{z_{13}} \,
\sum_{m=0}^{\infty}\, \frac{\bar{z}_{13}^m}{m!}\,
B(2\bar{h}_1+1+m,2\bar{h}_3+1)  \, 
(
\Phi_{+2}^{(h_2)}\,\bar{\pa}^m \,
\Phi_{-2}^{(h_1+h_3)})
(z_3,\bar{z}_3).
\label{fullminustwo}
\eea
There are eight OPEs
\footnote{In this case, all of the OPEs
 in (\ref{minustwolinear}) do not satisfy
the condition $s_1+s_3 \geq 0$.
Moreover, for the last sixteen two-particle operators in
(\ref{REDEFINITION}), the corresponding OPEs of the single-particle
operators with helicity $-2$
are not present due to the condition
$s_1+s_2 < 0$.}
where the quadratic terms  appear on the right-hand sides
\bea
&&
\Phi_{-2}^{(h_1)}(z_1,\bar{z}_1)\,
(\Phi_{+2}^{(h_2)}\, \Phi_{+\frac{3}{2}}^{(h_3),A})(z_3,
\bar{z}_3) =
\nonu \\
&&  
\frac{\bar{z}_{13}}{z_{13}} \,
\sum_{m=0}^{\infty}\, \frac{\bar{z}_{13}^m}{m!}\,
B(2\bar{h}_1+1+m,2\bar{h}_2+1) \, (\bar{\pa}^m \,
\Phi_{-2}^{(h_1+h_2)}\,\Phi_{+\frac{3}{2}}^{(h_3),A})
(z_3,\bar{z}_3),
\nonu \\
&&
\Phi_{-2}^{(h_1)}(z_1,\bar{z}_1)\,
(\Phi_{+2}^{(h_2)}\, \Phi_{+1}^{(h_3),AB})(z_3,
\bar{z}_3) =
\nonu \\
&&  
\frac{\bar{z}_{13}}{z_{13}} \,
\sum_{m=0}^{\infty}\, \frac{\bar{z}_{13}^m}{m!}\,
B(2\bar{h}_1+1+m,2\bar{h}_2+1) \, (\bar{\pa}^m \,
\Phi_{-2}^{(h_1+h_2)}\,\Phi_{+1}^{(h_3),AB})
(z_3,\bar{z}_3),
\nonu \\
&&
\Phi_{-2}^{(h_1)}(z_1,\bar{z}_1)\,
(\Phi_{+2}^{(h_2)}\, \Phi_{\frac{1}{2}}^{(h_3),ABC})(z_3,
\bar{z}_3) =
\nonu \\
&&  
\frac{\bar{z}_{13}}{z_{13}} \,
\sum_{m=0}^{\infty}\, \frac{\bar{z}_{13}^m}{m!}\,
B(2\bar{h}_1+1+m,2\bar{h}_2+1) \, (\bar{\pa}^m \,
\Phi_{-2}^{(h_1+h_2)}\,\Phi_{+\frac{1}{2}}^{(h_3),ABC})
(z_3,\bar{z}_3),
\nonu \\
&&
\Phi_{-2}^{(h_1)}(z_1,\bar{z}_1)\,
(\Phi_{+2}^{(h_2)}\, \Phi_{0}^{(h_3),ABCD})(z_3,
\bar{z}_3) =
\nonu \\
&&  
\frac{\bar{z}_{13}}{z_{13}} \,
\sum_{m=0}^{\infty}\, \frac{\bar{z}_{13}^m}{m!}\,
B(2\bar{h}_1+1+m,2\bar{h}_2+1) \, (\bar{\pa}^m \,
\Phi_{-2}^{(h_1+h_2)}\,\Phi_{0}^{(h_3),ABCD})
(z_3,\bar{z}_3),
\nonu \\
&&
\Phi_{-2}^{(h_1)}(z_1,\bar{z}_1)\,
(\Phi_{+2}^{(h_2)}\, \Phi_{ABC,-\frac{1}{2}}^{(h_3)})(z_3,
\bar{z}_3) =
\nonu \\
&&  
\frac{\bar{z}_{13}}{z_{13}} \,
\sum_{m=0}^{\infty}\, \frac{\bar{z}_{13}^m}{m!}\,
B(2\bar{h}_1+1+m,2\bar{h}_2+1) \, (\bar{\pa}^m \,
\Phi_{-2}^{(h_1+h_2)}\,\Phi_{ABC,-\frac{1}{2}}^{(h_3)})
(z_3,\bar{z}_3),
\nonu \\
&&
\Phi_{-2}^{(h_1)}(z_1,\bar{z}_1)\,
(\Phi_{+2}^{(h_2)}\, \Phi_{AB,-1}^{(h_3)})(z_3,
\bar{z}_3) =
\nonu \\
&&  
\frac{\bar{z}_{13}}{z_{13}} \,
\sum_{m=0}^{\infty}\, \frac{\bar{z}_{13}^m}{m!}\,
B(2\bar{h}_1+1+m,2\bar{h}_2+1) \, (\bar{\pa}^m \,
\Phi_{-2}^{(h_1+h_2)}\,\Phi_{AB,-1}^{(h_3)})
(z_3,\bar{z}_3),
\nonu \\
&&
\Phi_{-2}^{(h_1)}(z_1,\bar{z}_1)\,
(\Phi_{+2}^{(h_2)}\, \Phi_{A,-\frac{3}{2}}^{(h_3)})(z_3,
\bar{z}_3) =
\nonu \\
&&  
\frac{\bar{z}_{13}}{z_{13}} \,
\sum_{m=0}^{\infty}\, \frac{\bar{z}_{13}^m}{m!}\,
B(2\bar{h}_1+1+m,2\bar{h}_2+1) \, (\bar{\pa}^m \,
\Phi_{-2}^{(h_1+h_2)}\,\Phi_{A,-\frac{3}{2}}^{(h_3)})
(z_3,\bar{z}_3),
\nonu \\
&&
\Phi_{-2}^{(h_1)}(z_1,\bar{z}_1)\,
(\Phi_{+2}^{(h_2)}\, \Phi_{-2}^{(h_3)})(z_3,
\bar{z}_3) =
\nonu \\
&&  
\frac{\bar{z}_{13}}{z_{13}} \,
\sum_{m=0}^{\infty}\, \frac{\bar{z}_{13}^m}{m!}\,
B(2\bar{h}_1+1+m,2\bar{h}_2+1) \, (\bar{\pa}^m \,
\Phi_{-2}^{(h_1+h_2)}\,\Phi_{-2}^{(h_3)})
(z_3,\bar{z}_3).
\label{minustwolinear}
\eea
Therefore, there are
the whole multi-particle OPEs,
(\ref{Aone}), (\ref{full3half}), (\ref{3halflinear}),
(\ref{fullone}),
(\ref{onelinear}),
(\ref{fullonehalf}),
(\ref{onehalflinear}),
(\ref{fullzero}), (\ref{zerolinear}),
(\ref{fullminusonehalf}),
(\ref{minusonehalflinear}),
(\ref{fullminusone}), (\ref{minusonelinear}),
(\ref{fullminusthreehalf}),
(\ref{minusthreehalflinear}),
(\ref{fullminustwo}) and (\ref{minustwolinear}). 
Note that although
the overall complex coordinates dependent factor,
the general term which depends on the dummy variable $m$
and the antiholomorphic derivative are common in each case,
the right-weights, $\bar{h}_1$, $\bar{h}_2$ and $\bar{h}_3$,
behave differently according to Tables \ref{hbar1},
\ref{hbar2}, and \ref{hbar2-1}. 
This is manifest when we write the above OPEs
in terms of the conformal dimension $\De_i$ and
the helicity $s_i$ rather than the right-weight $\bar{h}_i$
($i=1,2,3$). We can write down the amplitudes from all the OPEs like as
(\ref{AMP2})
\footnote{Note that the normal-ordered products
(cubic terms)
on the right-hand sides of all the OPEs in Appendix A are present.
These depend on the complex coordinates $(z_3, \bar{z}_3)$
and do not depend on $z_{13}$ and $\bar{z}_{13}$ as described
before.}.

\section{The remaining (anti)commutators for the multi particle
OPEs}

The previous construction in (\ref{25comm})
can be generalized to other cases in Appendix A.

\subsection{The (anti)commutators
of the gravitinos and the quadratic
operators}

The twenty (anti)commutators are
\bea
&& \bigg[
(\hat{\Phi}_{+\frac{3}{2}}^{(\frac{1}{4}-\frac{j}{2}),P})_{2-h,\bar{n}},
(\hat{\Phi}_{+2}^{(-\frac{k}{2})}\,
\hat{\Phi}_{+2}^{(-\frac{l}{2})}
)_{n',\bar{n}'} \bigg]  =
\pm  \frac{
\kappa_{+2,+\frac{3}{2},-\frac{3}{2}}^2}{ (1-k)! (1-l)! (-3+k+l)!}  \nonu \\ && \times
N^{\frac{7}{4}-\frac{j}{2},\frac{1}{2} (-k-l+6)}_{1} (\bar{n},\bar{n}') \,
(\hat{\Phi}_{+\frac{3}{2}}^{(-\frac{j}{2}-\frac{k}{2}
-\frac{l}{2}+\frac{1}{4}),P})_{(2-h)+n',\bar{n}+\bar{n}'},
\nonu \\
&& \bigg\{
(\hat{\Phi}_{+\frac{3}{2}}^{(\frac{1}{4}-\frac{j}{2}),P})_{2-h,\bar{n}},
(\hat{\Phi}_{+2}^{(-\frac{k}{2})}\,
\hat{\Phi}_{+\frac{3}{2}}^{(-\frac{l}{2}+\frac{1}{4}),A}
)_{n',\bar{n}'} \bigg\}  =
\pm  \frac{
\kappa_{+2,+\frac{3}{2},-\frac{3}{2}}\,
\kappa_{+\frac{3}{2},+\frac{3}{2},-1}}{ (1-k)! (\frac{1}{2}-l)! (k+l-\frac{5}{2})!}  \nonu \\ && \times
N^{\frac{7}{4}-\frac{j}{2},\frac{1}{2}
(-k-l+\frac{11}{2})}_{1} (\bar{n},\bar{n}') \,
(\hat{\Phi}_{+1}^{(-\frac{j}{2}-\frac{k}{2}
-\frac{l}{2}+\frac{1}{2}),PA})_{(2-h)+n',\bar{n}+\bar{n}'},
\nonu \\
&& \bigg[
(
\hat{\Phi}_{+\frac{3}{2}}^{(\frac{1}{4}-\frac{j}{2}),P})_{2-h,\bar{n}},
(\hat{\Phi}_{+2}^{(-\frac{k}{2})}\,
\hat{\Phi}_{+1}^{(-\frac{l}{2}+\frac{1}{2}),AB}
)_{n',\bar{n}'} \bigg]  =
\pm  \frac{
\kappa_{+2,+\frac{3}{2},-\frac{3}{2}}\,
\kappa_{+\frac{3}{2},+1,-\frac{1}{2}}}{ (1-k)! (-l)! (k+l-2)!}  \nonu \\ && \times
N^{\frac{7}{4}-\frac{j}{2},\frac{1}{2} (-k-l+5)}_{1} (\bar{n},\bar{n}') \,
(\hat{\Phi}_{+\frac{1}{2}}^{(-\frac{j}{2}-\frac{k}{2}
-\frac{l}{2}+\frac{3}{4}),PAB})_{(2-h)+n',\bar{n}+\bar{n}'},
\nonu \\
&& \bigg\{
(
\hat{\Phi}_{+\frac{3}{2}}^{(\frac{1}{4}-\frac{j}{2}),P})_{2-h,\bar{n}},
(\hat{\Phi}_{+2}^{(-\frac{k}{2})}\,
\hat{\Phi}_{+\frac{1}{2}}^{(-\frac{l}{2}+\frac{3}{4}),ABC}
)_{n',\bar{n}'} \bigg\}  =
\pm  \frac{
\kappa_{+2,+\frac{3}{2},-\frac{3}{2}}\,
\kappa_{+\frac{3}{2},+\frac{1}{2},0}}{ (1-k)! (-\frac{1}{2}-l)! (k+l-\frac{3}{2})!}  \nonu \\ && \times
N^{\frac{7}{4}-\frac{j}{2},\frac{1}{2} (-k-l+\frac{9}{2})}_{1} (\bar{n},\bar{n}') \,
(\hat{\Phi}_{0}^{(-\frac{j}{2}-\frac{k}{2}
  -\frac{l}{2}+1),PABC})_{(2-h)+n',\bar{n}+\bar{n}'},
\nonu \\
&& \bigg[
(
\hat{\Phi}_{+\frac{3}{2}}^{(\frac{1}{4}-\frac{j}{2}),P})_{2-h,\bar{n}},
(\hat{\Phi}_{+2}^{(-\frac{k}{2})}\,
\hat{\Phi}_{0}^{(-\frac{l}{2}+1),ABCD}
)_{n',\bar{n}'} \bigg]  =
\pm  \frac{
\kappa_{+2,+\frac{3}{2},-\frac{3}{2}}\,
\kappa_{+\frac{3}{2},0,+\frac{1}{2}}}{ (1-k)! (-1-l)! (k+l-1)!}  \nonu \\ && \times
N^{\frac{7}{4}-\frac{j}{2},\frac{1}{2} (-k-l+4)}_{1} (\bar{n},\bar{n}') \,
\frac{1}{3!}\, \epsilon^{PABCDEFG} \,
(\hat{\Phi}_{EFG,-\frac{1}{2}}^{(-\frac{j}{2}-\frac{k}{2}
-\frac{l}{2}+\frac{5}{4})})_{(2-h)+n',\bar{n}+\bar{n}'},
\nonu \\
&& \bigg\{
(
\hat{\Phi}_{+\frac{3}{2}}^{(\frac{1}{4}-\frac{j}{2}),P})_{2-h,\bar{n}},
(\hat{\Phi}_{+2}^{(-\frac{k}{2})}\,
\hat{\Phi}_{ABC,-\frac{1}{2}}^{(-\frac{l}{2}+\frac{5}{4})}
)_{n',\bar{n}'} \bigg\}  =
\pm  \frac{
\kappa_{+2,+\frac{3}{2},-\frac{3}{2}}\,
\kappa_{+\frac{3}{2},-\frac{1}{2},+1}}{ (1-k)! (-\frac{3}{2}-l)! (k+l-\frac{1}{2})!}  \nonu \\ && \times
N^{\frac{7}{4}-\frac{j}{2},\frac{1}{2} (-k-l+\frac{7}{2})}_{1} (\bar{n},\bar{n}') \,
3 \,
\delta^P_{[A}\, (\hat{\Phi}_{BC],-1}^{(-\frac{j}{2}-\frac{k}{2}
-\frac{l}{2}+\frac{3}{2})})_{(2-h)+n',\bar{n}+\bar{n}'},
\nonu \\
&& \bigg[
(
\hat{\Phi}_{+\frac{3}{2}}^{(\frac{1}{4}-\frac{j}{2}),P})_{2-h,\bar{n}},
(\hat{\Phi}_{+2}^{(-\frac{k}{2})}\,
\hat{\Phi}_{AB,-1}^{(-\frac{l}{2}+\frac{3}{2})}
)_{n',\bar{n}'} \bigg]  =
\pm  \frac{
\kappa_{+2,+\frac{3}{2},-\frac{3}{2}}\,
\kappa_{+\frac{3}{2},-1,+\frac{3}{2}}}{ (1-k)! (-2-l)! (k+l)!}  \nonu \\ && \times
N^{\frac{7}{4}-\frac{j}{2},\frac{1}{2} (-k-l+3)}_{1} (\bar{n},\bar{n}') \,
2\,
\delta^P_{[A} \, (\hat{\Phi}_{B],-\frac{3}{2}}^{(-\frac{j}{2}-\frac{k}{2}
-\frac{l}{2}+\frac{7}{4})})_{(2-h)+n',\bar{n}+\bar{n}'},
\nonu \\
&& \bigg\{
(
\hat{\Phi}_{+\frac{3}{2}}^{(\frac{1}{4}-\frac{j}{2}),P})_{2-h,\bar{n}},
(\hat{\Phi}_{+2}^{(-\frac{k}{2})}\,
\hat{\Phi}_{A,-\frac{3}{2}}^{(-\frac{l}{2}+\frac{7}{4})}
)_{n',\bar{n}'} \bigg\}  =
\pm  \frac{
\kappa_{+2,+\frac{3}{2},-\frac{3}{2}}\,
\kappa_{+\frac{3}{2},-\frac{3}{2},+2}}{ (1-k)! (-\frac{5}{2}-l)! (\frac{1}{2}+k+l)!}  \nonu \\ && \times
N^{\frac{7}{4}-\frac{j}{2},\frac{1}{2} (-k-l+\frac{5}{2})}_{1} (\bar{n},\bar{n}') \,
\delta^P_A\, (\hat{\Phi}_{-2}^{(-\frac{j}{2}-\frac{k}{2}
-\frac{l}{2}+2)})_{(2-h)+n',\bar{n}+\bar{n}'},
\nonu \\
&& \bigg[
(
\hat{\Phi}_{+\frac{3}{2}}^{(\frac{1}{4}-\frac{j}{2}),P})_{2-h,\bar{n}},
(\hat{\Phi}_{+\frac{3}{2}}^{(-\frac{k}{2}+\frac{1}{4}),A}\,
\hat{\Phi}_{+\frac{3}{2}}^{(-\frac{l}{2}+\frac{1}{4}),B}
)_{n',\bar{n}'} \bigg]  =
\pm  \frac{
\kappa_{+\frac{3}{2},+\frac{3}{2},-1}\,
\kappa_{+\frac{3}{2},+1,-\frac{1}{2}}}{ (\frac{1}{2}-k)! (\frac{1}{2}-l)! (-2+k+l)!}  \nonu \\ && \times
N^{\frac{7}{4}-\frac{j}{2},\frac{1}{2} (-k-l+5)}_{1} (\bar{n},\bar{n}') \,
(\hat{\Phi}_{+\frac{1}{2}}^{(-\frac{j}{2}-\frac{k}{2}
-\frac{l}{2}+\frac{3}{4}),PAB})_{(2-h)+n',\bar{n}+\bar{n}'},
\nonu \\
&& \bigg\{
(
\hat{\Phi}_{+\frac{3}{2}}^{(\frac{1}{4}-\frac{j}{2}),P})_{2-h,\bar{n}},
(\hat{\Phi}_{+\frac{3}{2}}^{(-\frac{k}{2}+\frac{1}{4}),A}\,
\hat{\Phi}_{+1}^{(-\frac{l}{2}+\frac{1}{2}),BC}
)_{n',\bar{n}'} \bigg\}  =
\pm  \frac{
\kappa_{+\frac{3}{2},+\frac{3}{2},-1}\,
\kappa_{+1,+1,0}}{ (\frac{1}{2}-k)! (-l)! (-\frac{3}{2}+k+l)!}
\nonu \\ && \times
N^{\frac{7}{4}-\frac{j}{2},\frac{1}{2} (-k-l+\frac{9}{2})}_{1} (\bar{n},\bar{n}') \,
(\hat{\Phi}_{0}^{(-\frac{j}{2}-\frac{k}{2}
-\frac{l}{2}+1),PABC})_{(2-h)+n',\bar{n}+\bar{n}'},
\nonu \\
&& \bigg[
(
\hat{\Phi}_{+\frac{3}{2}}^{(\frac{1}{4}-\frac{j}{2}),P})_{2-h,\bar{n}},
(\hat{\Phi}_{+\frac{3}{2}}^{(-\frac{k}{2}+\frac{1}{4}),A}\,
\hat{\Phi}_{+\frac{1}{2}}^{(-\frac{l}{2}+\frac{3}{4}),BCD}
)_{n',\bar{n}'} \bigg]  =
\pm  \frac{
\kappa_{+\frac{3}{2},+\frac{3}{2},-1}\,
\kappa_{+1,+\frac{1}{2},+\frac{1}{2}}}{ (\frac{1}{2}-k)! (-\frac{1}{2}-l)! (-1+k+l)!}
\nonu \\ && \times
N^{\frac{7}{4}-\frac{j}{2},\frac{1}{2} (-k-l+4)}_{1} (\bar{n},\bar{n}') \,
\frac{1}{3!}\, \epsilon^{PABCDFGH}
(\hat{\Phi}_{FGH,-\frac{1}{2}}^{(-\frac{j}{2}-\frac{k}{2}
-\frac{l}{2}+\frac{5}{4})})_{(2-h)+n',\bar{n}+\bar{n}'},
\nonu \\
&& \bigg\{
\hat{\Phi}_{+\frac{3}{2}}^{(\frac{1}{4}-\frac{j}{2}),P})_{2-h,\bar{n}},
(\hat{\Phi}_{+\frac{3}{2}}^{(-\frac{k}{2}+\frac{1}{4}),A}\,
\hat{\Phi}_{0}^{(-\frac{l}{2}+1),BCDE}
)_{n',\bar{n}'} \bigg\}  =
\pm  \frac{
\kappa_{+\frac{3}{2},+\frac{3}{2},-1}\,
\kappa_{+1,0,+1}}{ (\frac{1}{2}-k)! (-1-l)! (-\frac{1}{2}+k+l)!}
\nonu \\ && \times
N^{\frac{7}{4}-\frac{j}{2},\frac{1}{2} (-k-l+\frac{7}{2})}_{1} (\bar{n},\bar{n}') \,
\frac{1}{2!} \, \epsilon^{PABCDEFG}\, 
(\hat{\Phi}_{FG,-1}^{(-\frac{j}{2}-\frac{k}{2}
-\frac{l}{2}+\frac{3}{2})})_{(2-h)+n',\bar{n}+\bar{n}'},
\nonu \\
&& \bigg[
(
\hat{\Phi}_{+\frac{3}{2}}^{(\frac{1}{4}-\frac{j}{2}),P})_{2-h,\bar{n}},
(\hat{\Phi}_{+\frac{3}{2}}^{(-\frac{k}{2}+\frac{1}{4}),A}\,
\hat{\Phi}_{BCD,-\frac{1}{2}}^{(-\frac{l}{2}+\frac{5}{4})}
)_{n',\bar{n}'} \bigg]  =
\pm  \frac{
\kappa_{+\frac{3}{2},+\frac{3}{2},-1}\,
\kappa_{+1,-\frac{1}{2},+\frac{3}{2}}}{ (\frac{1}{2}-k)! (-\frac{3}{2}-l)! (k+l)!}
\nonu \\ && \times
N^{\frac{7}{4}-\frac{j}{2},\frac{1}{2} (-k-l+3)}_{1} (\bar{n},\bar{n}') \,
3!\, \delta^{P}_{[B}
(\hat{\Phi}_{C,-\frac{3}{2}}^{(-\frac{j}{2}-\frac{k}{2}
  -\frac{l}{2}+\frac{7}{4})})_{(2-h)+n',\bar{n}+\bar{n}'} \,
\delta^{A}_{D]},
\nonu \\
&& \bigg\{(
\hat{\Phi}_{+\frac{3}{2}}^{(\frac{1}{4}-\frac{j}{2}),P})_{2-h,\bar{n}},
(\hat{\Phi}_{+\frac{3}{2}}^{(-\frac{k}{2}+\frac{1}{4}),A}\,
\hat{\Phi}_{BC,-1}^{(-\frac{l}{2}+\frac{3}{2})}
)_{n',\bar{n}'} \bigg\}  =
\pm  \frac{
\kappa_{+\frac{3}{2},+\frac{3}{2},-1}\,
\kappa_{+1,-1,+2}}{ (\frac{1}{2}-k)! (-2-l)! (\frac{1}{2}+k+l)!}
\nonu \\ && \times
N^{\frac{7}{4}-\frac{j}{2},\frac{1}{2} (-k-l+\frac{5}{2})}_{1} (\bar{n},\bar{n}') \,
\delta^{PA}_{BC}\,
(\hat{\Phi}_{-2}^{(-\frac{j}{2}-\frac{k}{2}
-\frac{l}{2}+2)})_{(2-h)+n',\bar{n}+\bar{n}'},
\nonu \\
&& \bigg[
(
\hat{\Phi}_{+\frac{3}{2}}^{(\frac{1}{4}-\frac{j}{2}),P})_{2-h,\bar{n}},
(\hat{\Phi}_{+1}^{(-\frac{k}{2}+\frac{1}{2}),AB}\,
\hat{\Phi}_{+1}^{(-\frac{l}{2}+\frac{1}{2}),CD}
)_{n',\bar{n}'} \bigg]  =
\pm  \frac{\kappa_{+\frac{3}{2},+1,-\frac{1}{2}}\,
\kappa_{+1,+\frac{1}{2},+\frac{1}{2}}}{(-k)! (-l)! (-1+k+l)!}
\nonu \\ && \times
N^{\frac{7}{4}-\frac{j}{2},\frac{1}{2} (-k-l+4)}_{1} (\bar{n},\bar{n}') \,
\frac{1}{3!}\,
\epsilon^{PABCDEFG}
(\hat{\Phi}_{EFG,-\frac{1}{2}}^{(-\frac{j}{2}-\frac{k}{2}
-\frac{l}{2}+\frac{5}{4})})_{(2-h)+n',\bar{n}+\bar{n}'},
\nonu \\
&& \bigg\{
(
\hat{\Phi}_{+\frac{3}{2}}^{(\frac{1}{4}-\frac{j}{2}),P})_{2-h,\bar{n}},
(\hat{\Phi}_{+1}^{(-\frac{k}{2}+\frac{1}{2}),AB}\,
\hat{\Phi}_{+\frac{1}{2}}^{(-\frac{l}{2}+\frac{3}{4}),CDE}
)_{n',\bar{n}'} \bigg\}  =
\pm  \frac{
\kappa_{+\frac{3}{2},+1,-\frac{1}{2}}\,
\kappa_{+\frac{1}{2},+\frac{1}{2},+1}}{(-k)! (-\frac{1}{2}-l)! (-\frac{1}{2}+k+l)!}
\nonu \\ && \times
N^{\frac{7}{4}-\frac{j}{2},\frac{1}{2} (-k-l+\frac{7}{2})}_{1} (\bar{n},\bar{n}') \,
\frac{1}{2!}\,
\epsilon^{PABCDEFG} \,
(\hat{\Phi}_{FG,-1}^{(-\frac{j}{2}-\frac{k}{2}
-\frac{l}{2}+\frac{3}{2})})_{(2-h)+n',\bar{n}+\bar{n}'},
\nonu \\
&& \bigg[
(
\hat{\Phi}_{+\frac{3}{2}}^{(\frac{1}{4}-\frac{j}{2}),P})_{2-h,\bar{n}},
(\hat{\Phi}_{+1}^{(-\frac{k}{2}+\frac{1}{2}),AB}\,
\hat{\Phi}_{0}^{(-\frac{l}{2}+1),CDEF}
)_{n',\bar{n}'} \bigg]  =
\pm  \frac{\kappa_{+\frac{3}{2},+1,-\frac{1}{2}}\,
\kappa_{+\frac{1}{2},0,+\frac{3}{2}}}{(-k)! (-1-l)! (k+l)!}
\nonu \\ && \times
N^{\frac{7}{4}-\frac{j}{2},\frac{1}{2} (-k-l+3)}_{1} (\bar{n},\bar{n}') \,
\epsilon^{PABCDEFG} \,
(\hat{\Phi}_{G,-\frac{3}{2}}^{(-\frac{j}{2}-\frac{k}{2}
-\frac{l}{2}+\frac{7}{4})})_{(2-h)+n',\bar{n}+\bar{n}'},
\nonu \\
&& \bigg\{
(
\hat{\Phi}_{+\frac{3}{2}}^{(\frac{1}{4}-\frac{j}{2}),P})_{2-h,\bar{n}},
(\hat{\Phi}_{+1}^{(-\frac{k}{2}+\frac{1}{2}),AB}\,
\hat{\Phi}_{CDE,-\frac{1}{2}}^{(-\frac{l}{2}+\frac{5}{4})}
)_{n',\bar{n}'} \bigg\}  =
\pm  \frac{
\kappa_{+\frac{3}{2},+1,-\frac{1}{2}}\,
\kappa_{+\frac{1}{2},-\frac{1}{2},+2}}{(-k)! (-\frac{3}{2}-l)! (\frac{1}{2}+k+l)!}
\nonu \\ && \times
N^{\frac{7}{4}-\frac{j}{2},\frac{1}{2} (-k-l+\frac{5}{2})}_{1} (\bar{n},\bar{n}') \,
\frac{1}{5!}\,
\epsilon^{PABFGHIJ} \, \epsilon_{CDEFGHIJ}
(\hat{\Phi}_{-2}^{(-\frac{j}{2}-\frac{k}{2}
-\frac{l}{2}+2)})_{(2-h)+n',\bar{n}+\bar{n}'},
\nonu \\
&& \bigg[
(
\hat{\Phi}_{+\frac{3}{2}}^{(\frac{1}{4}-\frac{j}{2}),P})_{2-h,\bar{n}},
(\hat{\Phi}_{+\frac{1}{2}}^{(-\frac{k}{2}+\frac{3}{4}),ABC}\,
\hat{\Phi}_{+\frac{1}{2}}^{(-\frac{l}{2}+\frac{3}{4}),DEF}
)_{n',\bar{n}'} \bigg]  =
\pm  \frac{
\kappa_{+\frac{3}{2},+\frac{1}{2},0}\,
\kappa_{+\frac{1}{2},0,+\frac{3}{2}}}{(-\frac{1}{2}-k)! (-\frac{1}{2}-l)! (k+l)!}
\nonu \\ && \times
N^{\frac{7}{4}-\frac{j}{2},\frac{1}{2} (-k-l+3)}_{1} (\bar{n},\bar{n}') \,
\epsilon^{PABCDEFG} \,
(\hat{\Phi}_{G,-\frac{3}{2}}^{(-\frac{j}{2}-\frac{k}{2}
-\frac{l}{2}+\frac{7}{4})})_{(2-h)+n',\bar{n}+\bar{n}'},
\nonu  \\
&& \bigg\{
(
\hat{\Phi}_{+\frac{3}{2}}^{(\frac{1}{4}-\frac{j}{2}),P})_{2-h,\bar{n}},
(\hat{\Phi}_{+\frac{1}{2}}^{(-\frac{k}{2}+\frac{3}{4}),ABC}\,
\hat{\Phi}_{0}^{(-\frac{l}{2}+1),DEFG}
)_{n',\bar{n}'} \bigg\}  =
\pm  \frac{
\kappa_{+\frac{3}{2},+\frac{1}{2},0}\,
\kappa_{0,0,+2}}{(-\frac{1}{2}-k)! (-1-l)! (\frac{1}{2}+k+l)!}
\nonu \\ && \times
N^{\frac{7}{4}-\frac{j}{2},\frac{1}{2} (-k-l+\frac{5}{2})}_{1} (\bar{n},\bar{n}') \,
\epsilon^{PABCDEFG} \,
(\hat{\Phi}_{-2}^{(-\frac{j}{2}-\frac{k}{2}
-\frac{l}{2}+2)})_{(2-h)+n',\bar{n}+\bar{n}'}.
\label{20case}
\eea
We present the $SU(8)$ representations corresponding to
(\ref{20case}) as follows
\footnote{\label{20casesfoot}
From the tensor product of $SU(8)$, it is known in \cite{FKS} that
\bea
{\bf 8} \otimes {\bf 1} \otimes {\bf 1} &=& {\bf 8},
\qquad
{\bf 8} \otimes {\bf 1} \otimes {\bf 8} = {\bf 28} \oplus \cdots,
\qquad
{\bf 8} \otimes {\bf 1} \otimes {\bf 28} = {\bf 56}  \oplus \cdots,
\qquad
{\bf 8} \otimes {\bf 1} \otimes {\bf 56} = {\bf 70}  \oplus \cdots,
\nonu \\
{\bf 8} \otimes {\bf 1} \otimes {\bf 70} &=& \overline{\bf 56}
\oplus \cdots,
\qquad
{\bf 8} \otimes {\bf 1} \otimes \overline{\bf 56} =
\overline{\bf 28}  \oplus \cdots,
\qquad
{\bf 8} \otimes {\bf 1} \otimes \overline{\bf 28} = \overline{\bf 8}
\oplus \cdots,
\qquad
{\bf 8} \otimes {\bf 1} \otimes \overline{\bf 8} = \overline{\bf 1}
\oplus \cdots,
\nonu \\
{\bf 8} \otimes {\bf 8} \otimes {\bf 8} & = & {\bf 56} \oplus  \cdots,
\qquad
{\bf 8} \otimes {\bf 8} \otimes {\bf 28} = {\bf 70} \oplus  \cdots,
\qquad
{\bf 8} \otimes {\bf 8} \otimes {\bf 56}  = 
\overline{\bf 56} \oplus  \cdots,
\nonu \\
{\bf 8} \otimes {\bf 8} \otimes {\bf 70}  &=&  \overline{\bf 28} \oplus  \cdots,
\qquad
{\bf 8} \otimes {\bf 8} \otimes \overline{\bf 56} =
\overline{\bf 8} \oplus 
\cdots,
\qquad
{\bf 8} \otimes {\bf 8} \otimes \overline{\bf 28}  = 
\overline{\bf 1} \oplus  \cdots,
\nonu \\
{\bf 8} \otimes {\bf 28} \otimes {\bf 28} & =& \overline{\bf 56}
\oplus  \cdots,
\qquad
{\bf 8} \otimes {\bf 28} \otimes {\bf 56} = 
\overline{\bf 28} \oplus  \cdots,
\qquad
{\bf 8} \otimes {\bf 28} \otimes {\bf 70} =
\overline{\bf 8} \oplus  \cdots,
\nonu \\
{\bf 8} \otimes {\bf 28} \otimes \overline{\bf 56} & = &
\overline{\bf 1} \oplus  \cdots,
\qquad
{\bf 8} \otimes {\bf 56} \otimes {\bf 56} = \overline{\bf 8} \oplus  \cdots,
\qquad
{\bf 8} \otimes {\bf 56} \otimes {\bf 70} = \overline{\bf 1} \oplus  \cdots.
\nonu
\eea
Note that compared to  (\ref{25comm}),
there are five trivial cases (the (anti)commutators
between the gravitinos and the ninth, the sixteenth, the
twenty-first, the twenty-fourth and the last of
(\ref{REDEFINITION}))
leading to the twenty nontrivial
(anti)commutators in (\ref{20case}).
}.
\subsection{The commutators between the graviphotons and the quadratic
operators}

The sixteen commutators are
\bea
&& \bigg[
(\hat{\Phi}_{+1}^{(\frac{1}{2}-\frac{j}{2}),PQ})_{2-h,\bar{n}},
(\hat{\Phi}_{+2}^{(-\frac{k}{2})}\,
\hat{\Phi}_{+2}^{(-\frac{l}{2})}
)_{n',\bar{n}'} \bigg]  =
\pm  \frac{
\kappa_{+2,+1,-1}^2
}{ (1-k)! (1-l)! (-3+k+l)!}  \nonu \\ && \times
N^{\frac{3}{2}-\frac{j}{2},\frac{1}{2} (-k-l+6)}_{1} (\bar{n},\bar{n}') \,
(\hat{\Phi}_{+1}^{(-\frac{j}{2}-\frac{k}{2}
-\frac{l}{2}+\frac{1}{2}),PQ})_{(2-h)+n',\bar{n}+\bar{n}'},
\nonu \\
&& \bigg[
(\hat{\Phi}_{+1}^{(\frac{1}{2}-\frac{j}{2}),PQ})_{2-h,\bar{n}},
(\hat{\Phi}_{+2}^{(-\frac{k}{2})}\,
\hat{\Phi}_{+\frac{3}{2}}^{(-\frac{l}{2}+\frac{1}{4}),A}
)_{n',\bar{n}'} \bigg]  =
\pm  \frac{
\kappa_{+2,+1,-1} \,
\kappa_{+\frac{3}{2},+1,-\frac{1}{2}}}{ (1-k)! (\frac{1}{2}-l)! (k+l-\frac{5}{2})!}  \nonu \\ && \times
N^{\frac{3}{2}-\frac{j}{2},\frac{1}{2}
(-k-l+\frac{11}{2})}_{1} (\bar{n},\bar{n}') \,
(\hat{\Phi}_{+\frac{1}{2}}^{(-\frac{j}{2}-\frac{k}{2}
-\frac{l}{2}+\frac{3}{4}),PQA})_{(2-h)+n',\bar{n}+\bar{n}'},
\nonu \\
&& \bigg[
(\hat{\Phi}_{+1}^{(\frac{1}{2}-\frac{j}{2}),PQ})_{2-h,\bar{n}},
(\hat{\Phi}_{+2}^{(-\frac{k}{2})}\,
\hat{\Phi}_{+1}^{(-\frac{l}{2}+\frac{1}{2}),AB}
)_{n',\bar{n}'} \bigg]  =
\pm  \frac{
\kappa_{+2,+1,-1} \,
\kappa_{+1,+1,0}}{ (1-k)! (-l)! (k+l-2)!}  \nonu \\ && \times
N^{\frac{3}{2}-\frac{j}{2},\frac{1}{2} (-k-l+5)}_{1} (\bar{n},\bar{n}') \,
(\hat{\Phi}_{0}^{(-\frac{j}{2}-\frac{k}{2}
-\frac{l}{2}+1),PQAB})_{(2-h)+n',\bar{n}+\bar{n}'},
\nonu \\
&& \bigg[
(
\hat{\Phi}_{+1}^{(\frac{1}{2}-\frac{j}{2}),PQ})_{2-h,\bar{n}},
(\hat{\Phi}_{+2}^{(-\frac{k}{2})}\,
\hat{\Phi}_{+\frac{1}{2}}^{(-\frac{l}{2}+\frac{3}{4}),ABC}
)_{n',\bar{n}'} \bigg]  =
\pm  \frac{
\kappa_{+2,+1,-1} \,
\kappa_{1,+\frac{1}{2},+\frac{1}{2}}}{ (1-k)! (-\frac{1}{2}-l)! (k+l-\frac{3}{2})!}  \nonu \\ && \times
N^{\frac{3}{2}-\frac{j}{2},\frac{1}{2} (-k-l+\frac{9}{2})}_{1} (\bar{n},\bar{n}') \,
\frac{1}{3!}\,
\epsilon^{PQABCDEF} \,
(\hat{\Phi}_{DEF,-\frac{1}{2}}^{(-\frac{j}{2}-\frac{k}{2}
-\frac{l}{2}+\frac{5}{4})})_{(2-h)+n',\bar{n}+\bar{n}'},
\nonu \\
&& \bigg[
(
\hat{\Phi}_{+1}^{(\frac{1}{2}-\frac{j}{2}),PQ})_{2-h,\bar{n}},
(\hat{\Phi}_{+2}^{(-\frac{k}{2})}\,
\hat{\Phi}_{0}^{(-\frac{l}{2}+1),ABCD}
)_{n',\bar{n}'} \bigg]  =
\pm  \frac{
\kappa_{+2,+1,-1} \,
\kappa_{+1,0,+1}}{ (1-k)! (-1-l)! (k+l-1)!}  \nonu \\ && \times
N^{\frac{3}{2}-\frac{j}{2},\frac{1}{2} (-k-l+4)}_{1} (\bar{n},\bar{n}') \,
\frac{1}{2!} \,
\epsilon^{PQABCDEF} \,
(\hat{\Phi}_{EF,-1}^{(-\frac{j}{2}-\frac{k}{2}
-\frac{l}{2}+\frac{3}{2})})_{(2-h)+n',\bar{n}+\bar{n}'},
\nonu \\
&& \bigg[
(
\hat{\Phi}_{+1}^{(\frac{1}{2}-\frac{j}{2}),PQ})_{2-h,\bar{n}},
(\hat{\Phi}_{+2}^{(-\frac{k}{2})}\,
\hat{\Phi}_{ABC,-\frac{1}{2}}^{(-\frac{l}{2}+\frac{5}{4})}
)_{n',\bar{n}'} \bigg]  =
\pm  \frac{
\kappa_{+2,+1,-1} \,
\kappa_{1,-\frac{1}{2},+\frac{3}{2}}}{ (1-k)! (-\frac{3}{2}-l)! (k+l-\frac{1}{2})!}  \nonu \\ && \times
N^{\frac{3}{2}-\frac{j}{2},\frac{1}{2} (-k-l+\frac{7}{2})}_{1} (\bar{n},\bar{n}') \,
3! \,
\delta^P_{[A}\, (\hat{\Phi}_{B,-\frac{3}{2}}^{(-\frac{j}{2}-\frac{k}{2}
-\frac{l}{2}+\frac{7}{4})})_{(2-h)+n',\bar{n}+\bar{n}'} \,
\delta^Q_{C]},
\nonu \\
&& \bigg[
(
\hat{\Phi}_{+1}^{(\frac{1}{2}-\frac{j}{2}),PQ})_{2-h,\bar{n}},
(\hat{\Phi}_{+2}^{(-\frac{k}{2})}\,
\hat{\Phi}_{AB,-1}^{(-\frac{l}{2}+\frac{3}{2})}
)_{n',\bar{n}'} \bigg]  =
\pm  \frac{
\kappa_{+2,+1,-1} \,
\kappa_{+1,-1,+2}}{ (1-k)! (-2-l)! (k+l)!}  \nonu \\ && \times
N^{\frac{3}{2}-\frac{j}{2},\frac{1}{2} (-k-l+3)}_{1} (\bar{n},\bar{n}') \,
\delta^{PQ}_{AB}
\, (\hat{\Phi}_{-2}^{(-\frac{j}{2}-\frac{k}{2}
-\frac{l}{2}+2)})_{(2-h)+n',\bar{n}+\bar{n}'},
\nonu \\
&& \bigg[
(\hat{\Phi}_{+1}^{(\frac{1}{2}-\frac{j}{2}),PQ})_{2-h,\bar{n}},
(\hat{\Phi}_{+\frac{3}{2}}^{(-\frac{k}{2}+\frac{1}{4}),A}\,
\hat{\Phi}_{+\frac{3}{2}}^{(-\frac{l}{2}+\frac{1}{4}),B}
)_{n',\bar{n}'} \bigg]  =
\pm  \frac{\kappa_{+\frac{3}{2},+1,-\frac{1}{2}}\,
\kappa_{+\frac{3}{2},+\frac{1}{2},0} 
}{ (\frac{1}{2}-k)! (\frac{1}{2}-l)! (-2+k+l)!}  \nonu \\ && \times
N^{\frac{3}{2}-\frac{j}{2},\frac{1}{2} (-k-l+5)}_{1} (\bar{n},\bar{n}') \,
(\hat{\Phi}_{0}^{(-\frac{j}{2}-\frac{k}{2}
-\frac{l}{2}+1),PQAB})_{(2-h)+n',\bar{n}+\bar{n}'},
\nonu \\
&& \bigg[
(
\hat{\Phi}_{+1}^{(\frac{1}{2}-\frac{j}{2}),PQ})_{2-h,\bar{n}},
(\hat{\Phi}_{+\frac{3}{2}}^{(-\frac{k}{2}+\frac{1}{4}),A}\,
\hat{\Phi}_{+1}^{(-\frac{l}{2}+\frac{1}{2}),BC}
)_{n',\bar{n}'} \bigg]  =
\pm  \frac{
\kappa_{+\frac{3}{2},+1,-\frac{1}{2}}\,
\kappa_{+1,+\frac{1}{2},+\frac{1}{2}} }{ (\frac{1}{2}-k)! (-l)! (-\frac{3}{2}+k+l)!}
\nonu \\ && \times
N^{\frac{3}{2}-\frac{j}{2},\frac{1}{2} (-k-l+\frac{9}{2})}_{1} (\bar{n},\bar{n}') \,
\frac{1}{3!}\,
\epsilon^{PQABCDEF}\, 
(\hat{\Phi}_{DEF,-\frac{1}{2}}^{(-\frac{j}{2}-\frac{k}{2}
-\frac{l}{2}+\frac{5}{4})})_{(2-h)+n',\bar{n}+\bar{n}'},
\nonu \\
&& \bigg[
(\hat{\Phi}_{+1}^{(\frac{1}{2}-\frac{j}{2}),PQ})_{2-h,\bar{n}},
(\hat{\Phi}_{+\frac{3}{2}}^{(-\frac{k}{2}+\frac{1}{4}),A}\,
\hat{\Phi}_{+\frac{1}{2}}^{(-\frac{l}{2}+\frac{3}{4}),BCD}
)_{n',\bar{n}'} \bigg]  =
\pm  \frac{
\kappa_{+\frac{3}{2},+1,-\frac{1}{2}}\,
\kappa_{+\frac{1}{2},+\frac{1}{2},+1} }{ (\frac{1}{2}-k)! (-\frac{1}{2}-l)! (-1+k+l)!}
\nonu \\ && \times
N^{\frac{3}{2}-\frac{j}{2},\frac{1}{2} (-k-l+4)}_{1} (\bar{n},\bar{n}') \,
\frac{1}{2!}\, \epsilon^{PQABCDFG}
(\hat{\Phi}_{FG,-1}^{(-\frac{j}{2}-\frac{k}{2}
-\frac{l}{2}+\frac{3}{2})})_{(2-h)+n',\bar{n}+\bar{n}'},
\nonu \\
&& \bigg[
(\hat{\Phi}_{+1}^{(\frac{1}{2}-\frac{j}{2}),PQ})_{2-h,\bar{n}},
(\hat{\Phi}_{+\frac{3}{2}}^{(-\frac{k}{2}+\frac{1}{4}),A}\,
\hat{\Phi}_{0}^{(-\frac{l}{2}+1),BCDE}
)_{n',\bar{n}'} \bigg]  =
\pm  \frac{
\kappa_{+\frac{3}{2},+1,-\frac{1}{2}}\,
\kappa_{+\frac{1}{2},0,+\frac{3}{2}} }{ (\frac{1}{2}-k)! (-1-l)! (-\frac{1}{2}+k+l)!}
\nonu \\ && \times
N^{\frac{3}{2}-\frac{j}{2},\frac{1}{2} (-k-l+\frac{7}{2})}_{1} (\bar{n},\bar{n}') \,
\epsilon^{PQABCDEF}\, 
(\hat{\Phi}_{F,-\frac{3}{2}}^{(-\frac{j}{2}-\frac{k}{2}
-\frac{l}{2}+\frac{7}{4})})_{(2-h)+n',\bar{n}+\bar{n}'},
\nonu \\
&& \bigg[
(
\hat{\Phi}_{+1}^{(\frac{1}{2}-\frac{j}{2}),PQ})_{2-h,\bar{n}},
(\hat{\Phi}_{+\frac{3}{2}}^{(-\frac{k}{2}+\frac{1}{4}),A}\,
\hat{\Phi}_{BCD,-\frac{1}{2}}^{(-\frac{l}{2}+\frac{5}{4})}
)_{n',\bar{n}'} \bigg]  =
\pm  \frac{
\kappa_{+\frac{3}{2},+1,-\frac{1}{2}}\,
\kappa_{+\frac{1}{2},-\frac{1}{2},+2} }{ (\frac{1}{2}-k)! (-\frac{3}{2}-l)! (k+l)!}
\nonu \\ && \times
N^{\frac{3}{2}-\frac{j}{2},\frac{1}{2} (-k-l+3)}_{1} (\bar{n},\bar{n}') \,
\frac{1}{5!} \,
\epsilon^{PQAEFGHI}\, \epsilon_{BCDEFGHI}
(\hat{\Phi}_{-2}^{(-\frac{j}{2}-\frac{k}{2}
-\frac{l}{2}+2)})_{(2-h)+n',\bar{n}+\bar{n}'} \,
\delta^{A}_{D]},
\nonu \\
&& \bigg[
(
\hat{\Phi}_{+1}^{(\frac{1}{2}-\frac{j}{2}),PQ})_{2-h,\bar{n}},
(\hat{\Phi}_{+1}^{(-\frac{k}{2}+\frac{1}{2}),AB}\,
\hat{\Phi}_{+1}^{(-\frac{l}{2}+\frac{1}{2}),CD}
)_{n',\bar{n}'} \bigg]  =
\pm  \frac{
\kappa_{+1,+1,0}\,
\kappa_{+1,0,+1} }{(-k)! (-l)! (-1+k+l)!}
\nonu \\ && \times
N^{\frac{3}{2}-\frac{j}{2},\frac{1}{2} (-k-l+4)}_{1} (\bar{n},\bar{n}') \,
\frac{1}{2!} \,
\epsilon^{PQABCDEF}
(\hat{\Phi}_{EF,-1}^{(-\frac{j}{2}-\frac{k}{2}
-\frac{l}{2}+\frac{3}{2})})_{(2-h)+n',\bar{n}+\bar{n}'},
\nonu \\
&& \bigg[
(
\hat{\Phi}_{+1}^{(\frac{1}{2}-\frac{j}{2}),PQ})_{2-h,\bar{n}},
(\hat{\Phi}_{+1}^{(-\frac{k}{2}+\frac{1}{2}),AB}\,
\hat{\Phi}_{+\frac{1}{2}}^{(-\frac{l}{2}+\frac{3}{4}),CDE}
)_{n',\bar{n}'} \bigg]  =
\pm  \frac{
\kappa_{+1,+1,0}\,
\kappa_{+\frac{1}{2},0,+\frac{3}{2}} }{(-k)! (-\frac{1}{2}-l)! (-\frac{1}{2}+k+l)!}
\nonu \\ && \times
N^{\frac{3}{2}-\frac{j}{2},\frac{1}{2} (-k-l+\frac{7}{2})}_{1} (\bar{n},\bar{n}') \,
\epsilon^{PQABCDEF} \,
(\hat{\Phi}_{F,-\frac{3}{2}}^{(-\frac{j}{2}-\frac{k}{2}
-\frac{l}{2}+\frac{7}{4})})_{(2-h)+n',\bar{n}+\bar{n}'},
\nonu \\
&& \bigg[
(
\hat{\Phi}_{+1}^{(\frac{1}{2}-\frac{j}{2}),PQ})_{2-h,\bar{n}},
(\hat{\Phi}_{+1}^{(-\frac{k}{2}+\frac{1}{2}),AB}\,
\hat{\Phi}_{0}^{(-\frac{l}{2}+1),CDEF}
)_{n',\bar{n}'} \bigg]  =
\pm  \frac{
\kappa_{+1,+1,0}\,
\kappa_{0,0,+2} }{(-k)! (-1-l)! (k+l)!}
\nonu \\ && \times
N^{\frac{3}{2}-\frac{j}{2},\frac{1}{2} (-k-l+3)}_{1} (\bar{n},\bar{n}') \,
\epsilon^{PQABCDEF} \,
(\hat{\Phi}_{-2}^{(-\frac{j}{2}-\frac{k}{2}
-\frac{l}{2}+2)})_{(2-h)+n',\bar{n}+\bar{n}'},
\nonu \\
&& \bigg[
(
\hat{\Phi}_{+1}^{(\frac{1}{2}-\frac{j}{2}),PQ})_{2-h,\bar{n}},
(\hat{\Phi}_{+\frac{1}{2}}^{(-\frac{k}{2}+\frac{3}{4}),ABC}\,
\hat{\Phi}_{+\frac{1}{2}}^{(-\frac{l}{2}+\frac{3}{4}),DEF}
)_{n',\bar{n}'} \bigg]  =
\pm  \frac{
\kappa_{+1,+\frac{1}{2},+\frac{1}{2}}\,
\kappa_{+\frac{1}{2},-\frac{1}{2},+2} }{(-\frac{1}{2}-k)! (-\frac{1}{2}-l)! (k+l)!}
\label{16case} \\ && \times
N^{\frac{3}{2}-\frac{j}{2},\frac{1}{2} (-k-l+3)}_{1} (\bar{n},\bar{n}') \,
\frac{1}{6!}\,
\epsilon^{PQABCGHI} \,
\epsilon_{GHIJKLM}\, \epsilon^{DEFJKLM}
(\hat{\Phi}_{-2}^{(-\frac{j}{2}-\frac{k}{2}
-\frac{l}{2}+2)})_{(2-h)+n',\bar{n}+\bar{n}'}.
\nonu
\eea
As before, the $SU(8)$ representations for (\ref{16case})
can be analyzed similarly
\footnote{
\label{16casefoot}
The following relations satisfy
\bea
{\bf 28} \otimes {\bf 1} \otimes {\bf 1} &=& {\bf 28},
\qquad
{\bf 28} \otimes {\bf 1} \otimes {\bf 8} = {\bf 56} \oplus \cdots,
\qquad
{\bf 28} \otimes {\bf 1} \otimes {\bf 28} = {\bf 70}  \oplus \cdots,
\qquad
{\bf 28} \otimes {\bf 1} \otimes {\bf 56} = \overline{\bf 56}  \oplus \cdots,
\nonu \\
{\bf 28} \otimes {\bf 1} \otimes {\bf 70} &=& \overline{\bf 28}
\oplus \cdots,
\qquad
{\bf 28} \otimes {\bf 1} \otimes \overline{\bf 56} =
\overline{\bf 8}  \oplus \cdots,
\qquad
{\bf 28} \otimes {\bf 1} \otimes \overline{\bf 28} = \overline{\bf 1}
\oplus \cdots,
\nonu \\
{\bf 28} \otimes {\bf 8} \otimes {\bf 8} & = & {\bf 70} \oplus  \cdots,
\qquad
{\bf 28} \otimes {\bf 8} \otimes {\bf 28} = \overline{\bf 56} \oplus  \cdots,
\qquad
{\bf 28} \otimes {\bf 8} \otimes {\bf 56}  = 
\overline{\bf 28} \oplus  \cdots,
\nonu \\
{\bf 28} \otimes {\bf 8} \otimes {\bf 70}  &=&  \overline{\bf 8} \oplus  \cdots,
\qquad
{\bf 28} \otimes {\bf 8} \otimes \overline{\bf 56} =
\overline{\bf 1} \oplus 
\cdots,
\qquad
{\bf 28} \otimes {\bf 28} \otimes {\bf 28}  = \overline{\bf 28}
\oplus  \cdots,
\nonu \\
{\bf 28} \otimes {\bf 28} \otimes {\bf 56} &=& 
\overline{\bf 8} \oplus  \cdots,
\qquad
{\bf 28} \otimes {\bf 28} \otimes {\bf 70} =
\overline{\bf 1} \oplus  \cdots,
\qquad
{\bf 28} \otimes {\bf 56} \otimes {\bf 56}  = 
\overline{\bf 1} \oplus  \cdots.
\nonu
\eea
Compared to the footnote \ref{20casesfoot},
there are further four trivial  cases (the commutators
between the graviphotons and the eighth, the fifteenth, the
twentieth, and the twenty-third  of
(\ref{REDEFINITION}))
leading to the sixteen nontrivial
commutators in (\ref{16case}).
}.
\subsection{The (anti)commutators between the graviphotinos and the quadratic
operators}

The twelve (anti)commutators are
\bea
&& \bigg[
(\hat{\Phi}_{+\frac{1}{2}}^{(\frac{3}{4}-\frac{j}{2}),PQR})_{2-h,\bar{n}},
(\hat{\Phi}_{+2}^{(-\frac{k}{2})}\,
\hat{\Phi}_{+2}^{(-\frac{l}{2})}
)_{n',\bar{n}'} \bigg]  =
\pm  \frac{
\kappa_{+2,+\frac{1}{2},-\frac{1}{2}}^2 }
{ (1-k)! (1-l)! (-3+k+l)!}  \nonu \\ && \times
N^{\frac{5}{4}-\frac{j}{2},\frac{1}{2} (-k-l+6)}_{1} (\bar{n},\bar{n}') \,
(\hat{\Phi}_{+\frac{1}{2}}^{(-\frac{j}{2}-\frac{k}{2}
-\frac{l}{2}+\frac{3}{4}),PQR})_{(2-h)+n',\bar{n}+\bar{n}'},
\nonu \\
&& \bigg\{
(\hat{\Phi}_{+\frac{1}{2}}^{(\frac{3}{4}-\frac{j}{2}),PQR})_{2-h,\bar{n}},
(\hat{\Phi}_{+2}^{(-\frac{k}{2})}\,
\hat{\Phi}_{+\frac{3}{2}}^{(-\frac{l}{2}+\frac{1}{4}),A}
)_{n',\bar{n}'} \bigg\}  =
\pm  \frac{
\kappa_{+2,+\frac{1}{2},-\frac{1}{2}}\,
\kappa_{+\frac{3}{2},+\frac{1}{2},0}}{ (1-k)! (\frac{1}{2}-l)! (k+l-\frac{5}{2})!}  \nonu \\ && \times
N^{\frac{5}{4}-\frac{j}{2},\frac{1}{2}
(-k-l+\frac{11}{2})}_{1} (\bar{n},\bar{n}') \,
(\hat{\Phi}_{0}^{(-\frac{j}{2}-\frac{k}{2}
-\frac{l}{2}+1),PQRA})_{(2-h)+n',\bar{n}+\bar{n}'},
\nonu \\
&& \bigg[
(
\hat{\Phi}_{+\frac{1}{2}}^{(\frac{3}{4}-\frac{j}{2}),PQR})_{2-h,\bar{n}},
(\hat{\Phi}_{+2}^{(-\frac{k}{2})}\,
\hat{\Phi}_{+1}^{(-\frac{l}{2}+\frac{1}{2}),AB}
)_{n',\bar{n}'} \bigg]  =
\pm  \frac{
\kappa_{+2,+\frac{1}{2},-\frac{1}{2}}\,
\kappa_{+1,+\frac{1}{2},+\frac{1}{2}}}{ (1-k)! (-l)! (k+l-2)!}  \nonu \\&& \times
N^{\frac{5}{4}-\frac{j}{2},\frac{1}{2} (-k-l+5)}_{1} (\bar{n},\bar{n}') \,
\frac{1}{3!} \, \epsilon^{PQRABCDE}
(\hat{\Phi}_{CDE,-\frac{1}{2}}^{(-\frac{j}{2}-\frac{k}{2}
-\frac{l}{2}+\frac{5}{4})})_{(2-h)+n',\bar{n}+\bar{n}'},
\nonu \\
&& \bigg\{
(
\hat{\Phi}_{+\frac{1}{2}}^{(\frac{3}{4}-\frac{j}{2}),PQR})_{2-h,\bar{n}},
(\hat{\Phi}_{+2}^{(-\frac{k}{2})}\,
\hat{\Phi}_{+\frac{1}{2}}^{(-\frac{l}{2}+\frac{3}{4}),ABC}
)_{n',\bar{n}'} \bigg\}  =
\pm  \frac{
\kappa_{+2,+\frac{1}{2},-\frac{1}{2}}\,
\kappa_{+\frac{1}{2},+\frac{1}{2},+1}}{ (1-k)! (-\frac{1}{2}-l)! (k+l-\frac{3}{2})!}  \nonu \\ && \times
N^{\frac{5}{4}-\frac{j}{2},\frac{1}{2} (-k-l+\frac{9}{2})}_{1} (\bar{n},\bar{n}') \,
\frac{1}{2!}\,
\epsilon^{PQRABCDE} \,
(\hat{\Phi}_{DE,-1}^{(-\frac{j}{2}-\frac{k}{2}
  -\frac{l}{2}+\frac{3}{2})})_{(2-h)+n',\bar{n}+\bar{n}'},
\nonu \\
&& \bigg[
(
\hat{\Phi}_{+\frac{1}{2}}^{(\frac{3}{4}-\frac{j}{2}),PQR})_{2-h,\bar{n}},
(\hat{\Phi}_{+2}^{(-\frac{k}{2})}\,
\hat{\Phi}_{0}^{(-\frac{l}{2}+1),ABCD}
)_{n',\bar{n}'} \bigg]  =
\pm  \frac{
\kappa_{+2,+\frac{1}{2},-\frac{1}{2}}\,
\kappa_{+\frac{1}{2},0,+\frac{3}{2}}}{ (1-k)! (-1-l)! (k+l-1)!}  \nonu \\ && \times
N^{\frac{5}{4}-\frac{j}{2},\frac{1}{2} (-k-l+4)}_{1} (\bar{n},\bar{n}') \,
\epsilon^{PQRABCDE} \,
(\hat{\Phi}_{E,-\frac{3}{2}}^{(-\frac{j}{2}-\frac{k}{2}
-\frac{l}{2}+\frac{7}{4})})_{(2-h)+n',\bar{n}+\bar{n}'},
\nonu \\
&& \bigg\{
(\hat{\Phi}_{+\frac{1}{2}}^{(\frac{3}{4}-\frac{j}{2}),PQR})_{2-h,\bar{n}},
(\hat{\Phi}_{+2}^{(-\frac{k}{2})}\,
\hat{\Phi}_{ABC,-\frac{1}{2}}^{(-\frac{l}{2}+\frac{5}{4})}
)_{n',\bar{n}'} \bigg\}  =
\pm  \frac{
\kappa_{+2,+\frac{1}{2},-\frac{1}{2}}\,
\kappa_{+\frac{1}{2},-\frac{1}{2},+2}}{ (1-k)! (-\frac{3}{2}-l)! (k+l-\frac{1}{2})!}  \nonu \\ && \times
N^{\frac{5}{4}-\frac{j}{2},\frac{1}{2} (-k-l+\frac{7}{2})}_{1} (\bar{n},\bar{n}') \,
\frac{1}{5!} \,
\epsilon^{PQRDEFGH}\, \epsilon_{ABCDEFGH} \,
(\hat{\Phi}_{-2}^{(-\frac{j}{2}-\frac{k}{2}
-\frac{l}{2}+2)})_{(2-h)+n',\bar{n}+\bar{n}'}, \,
\nonu \\
&& \bigg[
(\hat{\Phi}_{+\frac{1}{2}}^{(\frac{3}{4}-\frac{j}{2}),PQR})_{2-h,\bar{n}},
(\hat{\Phi}_{+\frac{3}{2}}^{(-\frac{k}{2}+\frac{1}{4}),A}\,
\hat{\Phi}_{+\frac{3}{2}}^{(-\frac{l}{2}+\frac{1}{4}),B}
)_{n',\bar{n}'} \bigg]  =
\pm  \frac{
\kappa_{+\frac{3}{2},+\frac{1}{2},0}\,
\kappa_{+\frac{3}{2},0,+\frac{1}{2}}}{ (\frac{1}{2}-k)! (\frac{1}{2}-l)! (-2+k+l)!}  \nonu \\ && \times
N^{\frac{5}{4}-\frac{j}{2},\frac{1}{2} (-k-l+5)}_{1} (\bar{n},\bar{n}') \,
\frac{1}{3!} \,
\epsilon^{PQRABCDE}
(\hat{\Phi}_{CDE,-\frac{1}{2}}^{(-\frac{j}{2}-\frac{k}{2}
-\frac{l}{2}+\frac{5}{4})})_{(2-h)+n',\bar{n}+\bar{n}'},
\nonu \\
&& \bigg\{
(\hat{\Phi}_{+\frac{1}{2}}^{(\frac{3}{4}-\frac{j}{2}),PQR})_{2-h,\bar{n}},
(\hat{\Phi}_{+\frac{3}{2}}^{(-\frac{k}{2}+\frac{1}{4}),A}\,
\hat{\Phi}_{+1}^{(-\frac{l}{2}+\frac{1}{2}),BC}
)_{n',\bar{n}'} \bigg\}  =
\pm  \frac{
\kappa_{+\frac{3}{2},+\frac{1}{2},0}\,
\kappa_{+1,0,+1}}{ (\frac{1}{2}-k)! (-l)! (-\frac{3}{2}+k+l)!}
\nonu \\ && \times
N^{\frac{5}{4}-\frac{j}{2},\frac{1}{2} (-k-l+\frac{9}{2})}_{1} (\bar{n},\bar{n}') \,
\frac{1}{2!}\,
\epsilon^{PQRABCDE}\, 
(\hat{\Phi}_{DE,-1}^{(-\frac{j}{2}-\frac{k}{2}
-\frac{l}{2}+\frac{3}{2})})_{(2-h)+n',\bar{n}+\bar{n}'},
\nonu \\
&& \bigg[
(
\hat{\Phi}_{+\frac{1}{2}}^{(\frac{3}{4}-\frac{j}{2}),PQR})_{2-h,\bar{n}},
(\hat{\Phi}_{+\frac{3}{2}}^{(-\frac{k}{2}+\frac{1}{4}),A}\,
\hat{\Phi}_{+\frac{1}{2}}^{(-\frac{l}{2}+\frac{3}{4}),BCD}
)_{n',\bar{n}'} \bigg]  =
\pm  \frac{
\kappa_{+\frac{3}{2},+\frac{1}{2},0}\,
\kappa_{+\frac{1}{2},0,+\frac{3}{2}}}{ (\frac{1}{2}-k)! (-\frac{1}{2}-l)! (-1+k+l)!}
\nonu \\ && \times
N^{\frac{5}{4}-\frac{j}{2},\frac{1}{2} (-k-l+4)}_{1} (\bar{n},\bar{n}') \,
\epsilon^{PQRABCDF}
(\hat{\Phi}_{F,-\frac{3}{2}}^{(-\frac{j}{2}-\frac{k}{2}
-\frac{l}{2}+\frac{7}{4})})_{(2-h)+n',\bar{n}+\bar{n}'},
\nonu \\
&& \bigg\{
(\hat{\Phi}_{+\frac{1}{2}}^{(\frac{3}{4}-\frac{j}{2}),PQR})_{2-h,\bar{n}},
(\hat{\Phi}_{+\frac{3}{2}}^{(-\frac{k}{2}+\frac{1}{4}),A}\,
\hat{\Phi}_{0}^{(-\frac{l}{2}+1),BCDE}
)_{n',\bar{n}'} \bigg\}  =
\pm  \frac{
\kappa_{+\frac{3}{2},+\frac{1}{2},0}\,
\kappa_{0,0,+2}}{ (\frac{1}{2}-k)! (-1-l)! (-\frac{1}{2}+k+l)!}
\nonu \\ && \times
N^{\frac{5}{4}-\frac{j}{2},\frac{1}{2} (-k-l+\frac{7}{2})}_{1} (\bar{n},\bar{n}') \,
\epsilon^{PQRABCDE}\, 
(\hat{\Phi}_{-2}^{(-\frac{j}{2}-\frac{k}{2}
-\frac{l}{2}+2)})_{(2-h)+n',\bar{n}+\bar{n}'},
\nonu \\
&& \bigg[
(\hat{\Phi}_{+\frac{1}{2}}^{(\frac{3}{4}-\frac{j}{2}),PQR})_{2-h,\bar{n}},
(\hat{\Phi}_{+1}^{(-\frac{k}{2}+\frac{1}{2}),AB}\,
\hat{\Phi}_{+1}^{(-\frac{l}{2}+\frac{1}{2}),CD}
)_{n',\bar{n}'} \bigg]  =
\pm  \frac{
\kappa_{+1,+\frac{1}{2},+\frac{1}{2}}\,
\kappa_{+1,-\frac{1}{2},+\frac{3}{2}}}{(-k)! (-l)! (-1+k+l)!}
\nonu \\ && \times
N^{\frac{5}{4}-\frac{j}{2},\frac{1}{2} (-k-l+4)}_{1} (\bar{n},\bar{n}') \,
\epsilon^{PQRABEFG} \, \delta^{C}_{[F} \, \delta^{D}_{G} \,
(\hat{\Phi}_{E],-\frac{3}{2}}^{(-\frac{j}{2}-\frac{k}{2}
-\frac{l}{2}+\frac{7}{4})})_{(2-h)+n',\bar{n}+\bar{n}'},
\label{12case}
\\
&& \bigg\{
(
\hat{\Phi}_{+\frac{1}{2}}^{(\frac{3}{4}-\frac{j}{2}),PQR})_{2-h,\bar{n}},
(\hat{\Phi}_{+1}^{(-\frac{k}{2}+\frac{1}{2}),AB}\,
\hat{\Phi}_{+\frac{1}{2}}^{(-\frac{l}{2}+\frac{3}{4}),CDE}
)_{n',\bar{n}'} \bigg\}  =
\pm  \frac{
\kappa_{+1,+\frac{1}{2},+\frac{1}{2}}\,
\kappa_{+\frac{1}{2},-\frac{1}{2},+2}}{(-k)! (-\frac{1}{2}-l)! (-\frac{1}{2}+k+l)!}
\nonu \\ && \times
N^{\frac{5}{4}-\frac{j}{2},\frac{1}{2} (-k-l+\frac{7}{2})}_{1} (\bar{n},
\bar{n}') \,
\frac{1}{6!}\,
\epsilon^{QPRABFGH} \,
\epsilon_{FGHIJKLM} \,
\epsilon^{CDEIJKLM}
(\hat{\Phi}_{-2}^{(-\frac{j}{2}-\frac{k}{2}
-\frac{l}{2}+2)})_{(2-h)+n',\bar{n}+\bar{n}'}.
\nonu
\eea
In this case, the $SU(8)$ representations
for (\ref{12case}) can be described
\footnote{
\label{12casefoot} The following relations hold
\bea
{\bf 56} \otimes {\bf 1} \otimes {\bf 1} &=& {\bf 56},
\qquad
{\bf 56} \otimes {\bf 1} \otimes {\bf 8} = {\bf 70} \oplus \cdots,
\qquad
{\bf 56} \otimes {\bf 1} \otimes {\bf 28} = \overline{\bf 56}  \oplus \cdots,
\qquad
{\bf 56} \otimes {\bf 1} \otimes {\bf 56} = \overline{\bf 28}  \oplus \cdots,
\nonu \\
{\bf 56} \otimes {\bf 1} \otimes {\bf 70} &=& \overline{\bf 8}
\oplus \cdots,
\qquad
{\bf 56} \otimes {\bf 1} \otimes \overline{\bf 56} =
\overline{\bf 1}  \oplus \cdots,
\qquad
{\bf 56} \otimes {\bf 8} \otimes {\bf 8}  =  \overline{\bf 56} \oplus  \cdots,
\nonu \\
{\bf 56} \otimes {\bf 8} \otimes {\bf 28} & = & \overline{\bf 28} \oplus  \cdots,
\qquad
{\bf 56} \otimes {\bf 8} \otimes {\bf 56}  = 
\overline{\bf 8} \oplus  \cdots,
\qquad
{\bf 56} \otimes {\bf 8} \otimes {\bf 70}  = \overline{\bf 1} \oplus  \cdots,
\nonu \\
{\bf 56} \otimes {\bf 28} \otimes {\bf 28}  & = & \overline{\bf 8}
\oplus  \cdots,
\qquad
{\bf 56} \otimes {\bf 28} \otimes {\bf 56}  = 
\overline{\bf 1} \oplus  \cdots.
\nonu
\eea
Compared to the footnote \ref{16casefoot},
there exist further four trivial  cases (the (anti)commutators
between the graviphotinos and the seventh, the fourteenth, the
nineteenth and the twenty-second  of
(\ref{REDEFINITION}))
which provide the twelve nontrivial
(anti)commutators in (\ref{12case}).
}.

\subsection{The commutators between the scalars and the quadratic
operators}

The nine commutators are
\bea
&& \bigg[
(\hat{\Phi}_{0}^{(1-\frac{j}{2}),PQRS})_{2-h,\bar{n}},
(\hat{\Phi}_{+2}^{(-\frac{k}{2})}\,
\hat{\Phi}_{+2}^{(-\frac{l}{2})}
)_{n',\bar{n}'} \bigg]  =
\pm  \frac{
\kappa_{+2,0,0}^2}{ (1-k)! (1-l)! (-3+k+l)!}  \nonu \\ && \times
N^{1-\frac{j}{2},\frac{1}{2} (-k-l+6)}_{1} (\bar{n},\bar{n}') \,
(\hat{\Phi}_{0}^{(-\frac{j}{2}-\frac{k}{2}
-\frac{l}{2}+1),PQRS})_{(2-h)+n',\bar{n}+\bar{n}'},
\nonu \\
&& \bigg[
(\hat{\Phi}_{0}^{(1-\frac{j}{2}),PQRS})_{2-h,\bar{n}},
(\hat{\Phi}_{+2}^{(-\frac{k}{2})}\,
\hat{\Phi}_{+\frac{3}{2}}^{(-\frac{l}{2}+\frac{1}{4}),A}
)_{n',\bar{n}'} \bigg]  =
\pm  \frac{
\kappa_{+2,0,0}\,
\kappa_{+\frac{3}{2},0,+\frac{1}{2}}}{ (1-k)! (\frac{1}{2}-l)! (k+l-\frac{5}{2})!}  \nonu \\ && \times
N^{1-\frac{j}{2},\frac{1}{2}
(-k-l+\frac{11}{2})}_{1} (\bar{n},\bar{n}') \,
\frac{1}{3!} \,\epsilon^{PQRSABCD}
(\hat{\Phi}_{BCD,-\frac{1}{2}}^{(-\frac{j}{2}-\frac{k}{2}
-\frac{l}{2}+\frac{5}{4})})_{(2-h)+n',\bar{n}+\bar{n}'},
\nonu \\
&& \bigg[
(
\hat{\Phi}_{0}^{(1-\frac{j}{2}),PQRS})_{2-h,\bar{n}},
(\hat{\Phi}_{+2}^{(-\frac{k}{2})}\,
\hat{\Phi}_{+1}^{(-\frac{l}{2}+\frac{1}{2}),AB}
)_{n',\bar{n}'} \bigg]  =
\pm  \frac{
\kappa_{+2,0,0}\,
\kappa_{+1,0,+1}}{ (1-k)! (-l)! (k+l-2)!}  \nonu \\ && \times
N^{1-\frac{j}{2},\frac{1}{2} (-k-l+5)}_{1} (\bar{n},\bar{n}') \,
\frac{1}{2!} \,
\epsilon^{PQRSABCD}
(\hat{\Phi}_{CD,-1}^{(-\frac{j}{2}-\frac{k}{2}
-\frac{l}{2}+\frac{3}{2})})_{(2-h)+n',\bar{n}+\bar{n}'},
\nonu \\
&& \bigg[
(
\hat{\Phi}_{0}^{(1-\frac{j}{2}),PQRS})_{2-h,\bar{n}},
(\hat{\Phi}_{+2}^{(-\frac{k}{2})}\,
\hat{\Phi}_{+\frac{1}{2}}^{(-\frac{l}{2}+\frac{3}{4}),ABC}
)_{n',\bar{n}'} \bigg]  =
\pm  \frac{
\kappa_{+2,0,0}\,
\kappa_{+\frac{1}{2},0,+\frac{3}{2}}}{ (1-k)! (-\frac{1}{2}-l)! (k+l-\frac{3}{2})!}  \nonu \\ && \times
N^{1-\frac{j}{2},\frac{1}{2} (-k-l+\frac{9}{2})}_{1} (\bar{n},\bar{n}') \,
\epsilon^{PQRSABCD} \,
(\hat{\Phi}_{D,-\frac{3}{2}}^{(-\frac{j}{2}-\frac{k}{2}
-\frac{l}{2}+\frac{7}{4})})_{(2-h)+n',\bar{n}+\bar{n}'},
\nonu \\
&& \bigg[
(
\hat{\Phi}_{0}^{(1-\frac{j}{2}),PQRS})_{2-h,\bar{n}},
(\hat{\Phi}_{+2}^{(-\frac{k}{2})}\,
\hat{\Phi}_{0}^{(-\frac{l}{2}+1),ABCD}
)_{n',\bar{n}'} \bigg]  =
\pm  \frac{
\kappa_{+2,0,0}\,
\kappa_{0,0,+2}}{ (1-k)! (-1-l)! (k+l-1)!}  \nonu \\ && \times
N^{1-\frac{j}{2},\frac{1}{2} (-k-l+4)}_{1} (\bar{n},\bar{n}') \,
\epsilon^{PQRSABCD} \,
(\hat{\Phi}_{-2}^{(-\frac{j}{2}-\frac{k}{2}
-\frac{l}{2}+2)})_{(2-h)+n',\bar{n}+\bar{n}'},
\nonu \\
&& \bigg[
(\hat{\Phi}_{0}^{(1-\frac{j}{2}),PQRS})_{2-h,\bar{n}},
(\hat{\Phi}_{+\frac{3}{2}}^{(-\frac{k}{2}+\frac{1}{4}),A}\,
\hat{\Phi}_{+\frac{3}{2}}^{(-\frac{l}{2}+\frac{1}{4}),B}
)_{n',\bar{n}'} \bigg]  =
\pm  \frac{
\kappa_{+\frac{3}{2},0,+\frac{1}{2}}\,
\kappa_{+\frac{3}{2},-\frac{1}{2},+1}}{ (\frac{1}{2}-k)! (\frac{1}{2}-l)! (-2+k+l)!}  \nonu \\ && \times
N^{1-\frac{j}{2},\frac{1}{2} (-k-l+5)}_{1} (\bar{n},\bar{n}') \,
\frac{1}{2!} \,
\epsilon^{PQRSACDE} \, \delta^{B}_{[D}
(\hat{\Phi}_{CE],-1}^{(-\frac{j}{2}-\frac{k}{2}
-\frac{l}{2}+\frac{3}{2})})_{(2-h)+n',\bar{n}+\bar{n}'},
\nonu \\
&& \bigg[
(\hat{\Phi}_{0}^{(1-\frac{j}{2}),PQRS})_{2-h,\bar{n}},
(\hat{\Phi}_{+\frac{3}{2}}^{(-\frac{k}{2}+\frac{1}{4}),A}\,
\hat{\Phi}_{+1}^{(-\frac{l}{2}+\frac{1}{2}),BC}
)_{n',\bar{n}'} \bigg]  =
\pm  \frac{
\kappa_{+\frac{3}{2},0,+\frac{1}{2}}\,
\kappa_{+1,-\frac{1}{2},+\frac{3}{2}}}{ (\frac{1}{2}-k)! (-l)! (-\frac{3}{2}+k+l)!}
\nonu \\ && \times
N^{1-\frac{j}{2},\frac{1}{2} (-k-l+\frac{9}{2})}_{1} (\bar{n},\bar{n}') \,
\epsilon^{PQRSADEF}\, \delta^{B}_{[E}\, \delta^{C}_{D}\,
(\hat{\Phi}_{F],-\frac{3}{2}}^{(-\frac{j}{2}-\frac{k}{2}
-\frac{l}{2}+\frac{7}{4})})_{(2-h)+n',\bar{n}+\bar{n}'},
\nonu \\
&& \bigg[
(
\hat{\Phi}_{0}^{(1-\frac{j}{2}),PQRS})_{2-h,\bar{n}},
(\hat{\Phi}_{+\frac{3}{2}}^{(-\frac{k}{2}+\frac{1}{4}),A}\,
\hat{\Phi}_{+\frac{1}{2}}^{(-\frac{l}{2}+\frac{3}{4}),BCD}
)_{n',\bar{n}'} \bigg]  =
\pm  \frac{
\kappa_{+\frac{3}{2},0,+\frac{1}{2}}\,
\kappa_{+\frac{1}{2},-\frac{1}{2},+2}}{ (\frac{1}{2}-k)! (-\frac{1}{2}-l)! (-1+k+l)!}
\nonu \\ && \times
N^{1-\frac{j}{2},\frac{1}{2} (-k-l+4)}_{1} (\bar{n},\bar{n}') \,
\frac{-1}{6!}\,
\epsilon^{PQRSAEFG}
\epsilon_{EFGHIJKL}\,
\epsilon^{BCDHIJKL}\,
(\hat{\Phi}_{-2}^{(-\frac{j}{2}-\frac{k}{2}
-\frac{l}{2}+2)})_{(2-h)+n',\bar{n}+\bar{n}'},
\nonu \\
&& \bigg[
(\hat{\Phi}_{0}^{(1-\frac{j}{2}),PQRS})_{2-h,\bar{n}},
(\hat{\Phi}_{+1}^{(-\frac{k}{2}+\frac{1}{2}),AB}\,
\hat{\Phi}_{+1}^{(-\frac{l}{2}+\frac{1}{2}),CD}
)_{n',\bar{n}'} \bigg]  =
\pm  \frac{
\kappa_{+1,0,+1}\,
\kappa_{+1,-1,+2}}{(-k)! (-l)! (-1+k+l)!}
\nonu \\ && \times
N^{1-\frac{j}{2},\frac{1}{2} (-k-l+4)}_{1} (\bar{n},\bar{n}') \,
\frac{1}{2!} \,
\epsilon^{PQRSABEF} \,
\delta^{CD}_{EF}
\, (\hat{\Phi}_{-2}^{(-\frac{j}{2}-\frac{k}{2}
-\frac{l}{2}+2)})_{(2-h)+n',\bar{n}+\bar{n}'}.
\label{9case} 
\eea
It is easy to see that the $SU(8)$ representations can be summarized
\footnote{
\label{9casefoot}
The following relations are
satisfied:
\bea
{\bf 70} \otimes {\bf 1} \otimes {\bf 1} &=& {\bf 70},
\qquad
{\bf 70} \otimes {\bf 1} \otimes {\bf 8} = \overline{\bf 56} \oplus \cdots,
\qquad
{\bf 70} \otimes {\bf 1} \otimes {\bf 28} = \overline{\bf 28}  \oplus \cdots,
\qquad
{\bf 70} \otimes {\bf 1} \otimes {\bf 56} = \overline{\bf 8}  \oplus \cdots,
\nonu \\
{\bf 70} \otimes {\bf 1} \otimes {\bf 70} &=& \overline{\bf 1}
\oplus \cdots,
\qquad
{\bf 70} \otimes {\bf 8} \otimes {\bf 8}  =  \overline{\bf 28} \oplus  \cdots,
\qquad
{\bf 70} \otimes {\bf 8} \otimes {\bf 28}  =  \overline{\bf 8} \oplus  \cdots,
\nonu \\
{\bf 70} \otimes {\bf 8} \otimes {\bf 56}  &=& 
\overline{\bf 1} \oplus  \cdots,
\qquad
{\bf 70} \otimes {\bf 28} \otimes {\bf 28}   =  \overline{\bf 1}
\oplus  \cdots.
\nonu
\eea
When we compare to the footnote \ref{12casefoot},
there exist further three trivial  cases (the commutators
between the scalars and the sixth,  the
thirteenth, and the eighteenth  of
(\ref{REDEFINITION}))
which reduce to the nine nontrivial
commutators in (\ref{9case}).
}.
\subsection{The other
(anti)commutators between the graviphotinos and the quadratic
operators}

The six (anti)commutators are
\bea
&& \bigg[
(\hat{\Phi}_{PQR,-\frac{1}{2}}^{(\frac{5}{4}-\frac{j}{2})})_{2-h,\bar{n}},
(\hat{\Phi}_{+2}^{(-\frac{k}{2})}\,
\hat{\Phi}_{+2}^{(-\frac{l}{2})}
)_{n',\bar{n}'} \bigg]  =
\pm  \frac{
\kappa_{+2,-\frac{1}{2},+\frac{1}{2}}^2}{
(1-k)! (1-l)! (-3+k+l)!}  \nonu \\ && \times
N^{\frac{3}{4}-\frac{j}{2},\frac{1}{2} (-k-l+6)}_{1} (\bar{n},\bar{n}') \,
(\hat{\Phi}_{PQR,-\frac{1}{2}}^{(-\frac{j}{2}-\frac{k}{2}
-\frac{l}{2}+\frac{5}{4})})_{(2-h)+n',\bar{n}+\bar{n}'},
\nonu \\
&& \bigg\{
(\hat{\Phi}_{PQR,-\frac{1}{2}}^{(\frac{5}{4}-\frac{j}{2})})_{2-h,\bar{n}},
(\hat{\Phi}_{+2}^{(-\frac{k}{2})}\,
\hat{\Phi}_{+\frac{3}{2}}^{(-\frac{l}{2}+\frac{1}{4}),A}
)_{n',\bar{n}'} \bigg\}  =
\pm  \frac{
\kappa_{+2,-\frac{1}{2},+\frac{1}{2}}\,
\kappa_{+\frac{3}{2},-\frac{1}{2},+1}}{ (1-k)! (\frac{1}{2}-l)! (k+l-\frac{5}{2})!}  \nonu \\ && \times
N^{\frac{3}{4}-\frac{j}{2},\frac{1}{2}
(-k-l+\frac{11}{2})}_{1} (\bar{n},\bar{n}') \,
(-3) \,\delta^{A}_{[P}
(\hat{\Phi}_{QR],-1}^{(-\frac{j}{2}-\frac{k}{2}
-\frac{l}{2}+\frac{3}{2})})_{(2-h)+n',\bar{n}+\bar{n}'},
\nonu \\
&& \bigg[
(
\hat{\Phi}_{PQR,-\frac{1}{2}}^{(\frac{5}{4}-\frac{j}{2})})_{2-h,\bar{n}},
(\hat{\Phi}_{+2}^{(-\frac{k}{2})}\,
\hat{\Phi}_{+1}^{(-\frac{l}{2}+\frac{1}{2}),AB}
)_{n',\bar{n}'} \bigg]  =
\pm  \frac{
\kappa_{+2,-\frac{1}{2},+\frac{1}{2}}\,
\kappa_{+1,-\frac{1}{2},+\frac{3}{2}}}{ (1-k)! (-l)! (k+l-2)!}  \nonu \\ && \times
N^{\frac{3}{4}-\frac{j}{2},\frac{1}{2} (-k-l+5)}_{1} (\bar{n},\bar{n}') \,
3! \,
\delta^A_{[R} \, \delta^B_{Q} \,
(\hat{\Phi}_{P],-\frac{3}{2}}^{(-\frac{j}{2}-\frac{k}{2}
-\frac{l}{2}+\frac{7}{4})})_{(2-h)+n',\bar{n}+\bar{n}'},
\nonu \\
&& \bigg\{
(
\hat{\Phi}_{PQR,-\frac{1}{2}}^{(\frac{5}{4}-\frac{j}{2})})_{2-h,\bar{n}},
(\hat{\Phi}_{+2}^{(-\frac{k}{2})}\,
\hat{\Phi}_{+\frac{1}{2}}^{(-\frac{l}{2}+\frac{3}{4}),ABC}
)_{n',\bar{n}'} \bigg\}  =
\pm  \frac{
\kappa_{+2,-\frac{1}{2},+\frac{1}{2}}\,
\kappa_{+\frac{1}{2},-\frac{1}{2},+2}}{ (1-k)! (-\frac{1}{2}-l)! (k+l-\frac{3}{2})!}  \nonu \\ && \times
N^{\frac{3}{4}-\frac{j}{2},\frac{1}{2} (-k-l+\frac{9}{2})}_{1} (\bar{n},\bar{n}') \,
\frac{-1}{5!}\,
\epsilon^{ABCDEFGH}
\epsilon_{PQRDEFGH} \,
(\hat{\Phi}_{-2}^{(-\frac{j}{2}-\frac{k}{2}
-\frac{l}{2}+2)})_{(2-h)+n',\bar{n}+\bar{n}'},
\nonu \\
&& \bigg[
(\hat{\Phi}_{PQR,-\frac{1}{2}}^{(\frac{5}{4}-\frac{j}{2})})_{2-h,\bar{n}},
(\hat{\Phi}_{+\frac{3}{2}}^{(-\frac{k}{2}+\frac{1}{4}),A}\,
\hat{\Phi}_{+\frac{3}{2}}^{(-\frac{l}{2}+\frac{1}{4}),B}
)_{n',\bar{n}'} \bigg]  =
\pm  \frac{
\kappa_{+\frac{3}{2},-\frac{1}{2},+1}\,
\kappa_{+\frac{3}{2},-1,+\frac{3}{2}}}{ (\frac{1}{2}-k)! (\frac{1}{2}-l)! (-2+k+l)!}  \nonu \\ && \times
N^{\frac{3}{4}-\frac{j}{2},\frac{1}{2} (-k-l+5)}_{1} (\bar{n},\bar{n}') \,
3! \,
\delta^{A}_{[R} \, \delta^{B}_{Q}
(\hat{\Phi}_{P],-\frac{3}{2}}^{(-\frac{j}{2}-\frac{k}{2}
-\frac{l}{2}+\frac{7}{4})})_{(2-h)+n',\bar{n}+\bar{n}'},
\nonu \\
&& \bigg\{
(\hat{\Phi}_{PQR,-\frac{1}{2}}^{(\frac{5}{4}-\frac{j}{2})})_{2-h,\bar{n}},
(\hat{\Phi}_{+\frac{3}{2}}^{(-\frac{k}{2}+\frac{1}{4}),A}\,
\hat{\Phi}_{+1}^{(-\frac{l}{2}+\frac{1}{2}),BC}
)_{n',\bar{n}'} \bigg\}  =
\pm  \frac{
\kappa_{+\frac{3}{2},-\frac{1}{2},+1}\,
\kappa_{+1,-1,+2}}{ (\frac{1}{2}-k)! (-l)! (-\frac{3}{2}+k+l)!}
\nonu \\ && \times
N^{\frac{3}{4}-\frac{j}{2},\frac{1}{2} (-k-l+\frac{9}{2})}_{1} (\bar{n},\bar{n}') \,
3! \,
\delta^{A}_{[R}\, \delta^{B}_{Q}\, \delta^{C}_{P]}
(\hat{\Phi}_{-2}^{(-\frac{j}{2}-\frac{k}{2}
-\frac{l}{2}+2)})_{(2-h)+n',\bar{n}+\bar{n}'}.
\label{6case}
\eea
The $SU(8)$ representations can be written as 
follows \footnote{
The following relations 
are satisfied:
\bea
\overline{\bf 56} \otimes {\bf 1} \otimes {\bf 1} &=& \overline{\bf 56},
\qquad
\overline{\bf 56} \otimes {\bf 1} \otimes {\bf 8} = \overline{\bf 28} \oplus \cdots,
\qquad
\overline{\bf 56}
\otimes {\bf 1} \otimes {\bf 28} = \overline{\bf 8}  \oplus \cdots,
\qquad
\overline{\bf 8}
\otimes {\bf 1} \otimes {\bf 56} = \overline{\bf 1}  \oplus \cdots,
\nonu \\
\overline{\bf 56}
\otimes {\bf 8} \otimes {\bf 8}  &=&  \overline{\bf 8} \oplus  \cdots,
\qquad
\overline{\bf 56} \otimes {\bf 8} \otimes {\bf 28}  =  \overline{\bf 1} \oplus  \cdots.
\nonu
\eea
Compared to the footnote \ref{9casefoot},
there exist further three trivial  cases (the (anti)commutators
between the graviphotinos and the fifth, the twelfth,
and the seventeenth  of
(\ref{REDEFINITION}))
which give the six nontrivial
(anti)commutators in (\ref{6case}).
\label{6casefoot}}.

\subsection{The other commutators between the graviphotons and the quadratic
operators}

The four commutators are
\bea
&& \bigg[
(\hat{\Phi}_{PQ,-1}^{(\frac{3}{2}-\frac{j}{2})})_{2-h,\bar{n}},
(\hat{\Phi}_{+2}^{(-\frac{k}{2})}\,
\hat{\Phi}_{+2}^{(-\frac{l}{2})}
)_{n',\bar{n}'} \bigg]  =
\pm  \frac{
\kappa_{+2,-1,+1}^2 }{ (1-k)! (1-l)! (-3+k+l)!}  \nonu \\ && \times
N^{\frac{1}{2}-\frac{j}{2},\frac{1}{2} (-k-l+6)}_{1} (\bar{n},\bar{n}') \,
(\hat{\Phi}_{PQ,-1}^{(-\frac{j}{2}-\frac{k}{2}
-\frac{l}{2}+\frac{3}{2})})_{(2-h)+n',\bar{n}+\bar{n}'},
\nonu \\
&& \bigg[
(\hat{\Phi}_{PQ,-1}^{(\frac{3}{2}-\frac{j}{2})})_{2-h,\bar{n}},
(\hat{\Phi}_{+2}^{(-\frac{k}{2})}\,
\hat{\Phi}_{+\frac{3}{2}}^{(-\frac{l}{2}+\frac{1}{4}),A}
)_{n',\bar{n}'} \bigg]  =
\pm  \frac{
\kappa_{+2,-1,+1} \,
\kappa_{+\frac{3}{2},-1,+\frac{3}{2}}}{ (1-k)! (\frac{1}{2}-l)! (k+l-\frac{5}{2})!}  \nonu \\ && \times
N^{\frac{1}{2}-\frac{j}{2},\frac{1}{2}
(-k-l+\frac{11}{2})}_{1} (\bar{n},\bar{n}') \,
2! \,\delta^{A}_{[P}
(\hat{\Phi}_{Q],-\frac{3}{2}}^{(-\frac{j}{2}-\frac{k}{2}
-\frac{l}{2}+\frac{7}{4})})_{(2-h)+n',\bar{n}+\bar{n}'},
\nonu \\
&& \bigg[
(
\hat{\Phi}_{PQ,-1}^{(\frac{3}{2}-\frac{j}{2})})_{2-h,\bar{n}},
(\hat{\Phi}_{+2}^{(-\frac{k}{2})}\,
\hat{\Phi}_{+1}^{(-\frac{l}{2}+\frac{1}{2}),AB}
)_{n',\bar{n}'} \bigg]  =
\pm  \frac{
\kappa_{+2,-1,+1} \,
\kappa_{+1,-1,+2}}{ (1-k)! (-l)! (k+l-2)!}  \nonu \\ && \times
N^{\frac{1}{2}-\frac{j}{2},\frac{1}{2} (-k-l+5)}_{1} (\bar{n},\bar{n}') \,
\delta^{AB}_{PQ}
\, \hat{\Phi}_{-2}^{(-\frac{j}{2}-\frac{k}{2}
-\frac{l}{2}+2)})_{(2-h)+n',\bar{n}+\bar{n}'},
\nonu \\
&& \bigg[
(\hat{\Phi}_{PQ,-1}^{(\frac{3}{2}-\frac{j}{2})})_{2-h,\bar{n}},
(\hat{\Phi}_{+\frac{3}{2}}^{(-\frac{k}{2}+\frac{1}{4}),A}\,
\hat{\Phi}_{+\frac{3}{2}}^{(-\frac{l}{2}+\frac{1}{4}),B}
)_{n',\bar{n}'} \bigg]  =
\pm  \frac{
\kappa_{+\frac{3}{2},-1,+\frac{3}{2}}
\, \kappa_{+\frac{3}{2},-\frac{3}{2},+2} }{ (\frac{1}{2}-k)! (\frac{1}{2}-l)! (-2+k+l)!}  \nonu \\ && \times
N^{\frac{1}{2}-\frac{j}{2},\frac{1}{2} (-k-l+5)}_{1} (\bar{n},\bar{n}') \,
\delta^{AB}_{QP}
\, (\hat{\Phi}_{-2}^{(-\frac{j}{2}-\frac{k}{2}
-\frac{l}{2}+2)})_{(2-h)+n',\bar{n}+\bar{n}'}.
\label{4case}
\eea
The $SU(8)$ representations can be written  similarly 
\footnote{
\label{4casefoot}
The relations are
satisfied:
\bea
\overline{\bf 28} \otimes {\bf 1} \otimes {\bf 1} &=& \overline{\bf 28},
\qquad
\overline{\bf 28} \otimes {\bf 1} \otimes {\bf 8} = \overline{\bf 8} \oplus \cdots,
\qquad
\overline{\bf 28}
\otimes {\bf 1} \otimes {\bf 28} = \overline{\bf 1}  \oplus \cdots,
\qquad
\overline{\bf 28}
\otimes {\bf 8} \otimes {\bf 8}  =  \overline{\bf 1} \oplus  \cdots.
\nonu
\eea
When we compare to the footnote \ref{6casefoot},
there exist further two trivial  cases (the commutators
between the graviphotons and the fourth,
and the eleventh  of
(\ref{REDEFINITION}))
implying the four nontrivial
commutators in (\ref{4case}).
}.

\subsection{The other (anti)commutators between the
gravitinos and the quadratic
operators}

The two (anti)commutators
\footnote{
\label{2casefoot}
We observe that the relations are
satisfied:
\bea
\overline{\bf 8} \otimes {\bf 1} \otimes {\bf 1} &=& \overline{\bf 8},
\qquad
\overline{\bf 8} \otimes {\bf 1} \otimes {\bf 8} = \overline{\bf 1} \oplus
\cdots.
\nonu
\eea
Compared to the footnote \ref{4casefoot},
there exist further two trivial cases (the (anti)commutators
between the gravitinos and the third,
and the tenth  of
(\ref{REDEFINITION}))
leading to the two nontrivial
(anti)commutators in (\ref{2case}).
}
are
\bea
&& \bigg[
(\hat{\Phi}_{P,-\frac{3}{2}}^{(\frac{7}{4}-\frac{j}{2})})_{2-h,\bar{n}},
(\hat{\Phi}_{+2}^{(-\frac{k}{2})}\,
\hat{\Phi}_{+2}^{(-\frac{l}{2})}
)_{n',\bar{n}'} \bigg]  =
\pm  \frac{
\kappa_{+2,-\frac{3}{2},+\frac{3}{2}}^2}{ (1-k)! (1-l)! (-3+k+l)!}  \nonu \\ && \times
N^{\frac{1}{4}-\frac{j}{2},\frac{1}{2} (-k-l+6)}_{1} (\bar{n},\bar{n}') \,
(\hat{\Phi}_{P,-\frac{3}{2}}^{(-\frac{j}{2}-\frac{k}{2}
-\frac{l}{2}+\frac{7}{4})})_{(2-h)+n',\bar{n}+\bar{n}'},
\nonu \\
&& \bigg\{
(\hat{\Phi}_{P,-\frac{3}{2}}^{(\frac{7}{4}-\frac{j}{2})})_{2-h,\bar{n}},
(\hat{\Phi}_{+2}^{(-\frac{k}{2})}\,
\hat{\Phi}_{+\frac{3}{2}}^{(-\frac{l}{2}+\frac{1}{4}),A}
)_{n',\bar{n}'} \bigg\}  =
\pm  \frac{
\kappa_{+2,-\frac{3}{2},+\frac{3}{2}}\,
\kappa_{+\frac{3}{2},-\frac{3}{2},+2}}{ (1-k)! (\frac{1}{2}-l)! (k+l-\frac{5}{2})!}  \nonu \\ && \times
N^{\frac{1}{4}-\frac{j}{2},\frac{1}{2}
(-k-l+\frac{11}{2})}_{1} (\bar{n},\bar{n}') \,
(-1)\, \delta^{A}_{P}
(\hat{\Phi}_{-2}^{(-\frac{j}{2}-\frac{k}{2}
-\frac{l}{2}+2)})_{(2-h)+n',\bar{n}+\bar{n}'}.
\label{2case}
\eea

\subsection{The commutator between the graviton and the quadratic
operator}

The one commutator
\footnote{
The relation is
satisfied
\bea
\overline{\bf 1} \otimes {\bf 1} \otimes {\bf 1} &=& \overline{\bf 1}.
\nonu
\eea
When we  compare to the footnote \ref{2casefoot},
there exists further one trivial  case (the commutator
between the gravitons and the second  of
(\ref{REDEFINITION}))
which allows us to have the one nontrivial
commutator in (\ref{1case}).
}
is
\bea
&& \bigg[
(\hat{\Phi}_{-2}^{(2-\frac{j}{2})})_{2-h,\bar{n}},
(\hat{\Phi}_{+2}^{(-\frac{k}{2})}\,
\hat{\Phi}_{+2}^{(-\frac{l}{2})}
)_{n',\bar{n}'} \bigg]  =
\pm  \frac{
\kappa_{+2,-2,+2}^2 }{ (1-k)! (1-l)! (-3+k+l)!}  \nonu \\ && \times
N^{-\frac{j}{2},\frac{1}{2} (-k-l+6)}_{1} (\bar{n},\bar{n}') \,
(\hat{\Phi}_{-2}^{(-\frac{j}{2}-\frac{k}{2}
-\frac{l}{2}+2)})_{(2-h)+n',\bar{n}+\bar{n}'}.
\label{1case}
\eea
We can find the
corresponding (anti)commutators coming from the
quadratic terms on the right-hand sides in Appendix A,
by putting the $SU(8)$ indices onto the equation (\ref{nonlinearcase}). 

\section{Some computations of the multi-particle OPEs
inside the Thielemans package}

Let us calculate some OPEs inside the Thielemans package \cite{Thielemans}.
In the following OPEs, there exists the overall factor
\bea
(\text{\tt h1}+{\tt 3} \text{\tt h2}+\text{\tt s1}+{\tt 3}
\text{\tt s2}-{\tt 4}) (\text{\tt h1}+\text{\tt h2}+\text{\tt h3}+
\text{\tt s1}+\text{\tt s2}+\text{\tt s3}-{\tt 4}),
\label{NORM}
\eea
on the right-hand sides of the OPEs.
For simplicity, we ignore this factor in all the OPEs (except the ones
in the last subsection) in this Appendix.
We have the expressions starting from the page $68$ (the twenty-seventh line
from the below)
to the page $71$
(until the fourteenth line from the below)
of \cite{AK2509}. After that, we introduce the following
twenty-five operators (\ref{REDEFINITION}) as follows:
\bea
&& \text{\tt comp}=
\left\{\text{\tt NO}[\text{\tt Phi}[\text{\tt h2},{\tt +2}],
\text{\tt Phi}[\text{\tt h3},{\tt +2}]], \right.
\nonu \\
&& \text{\tt NO}\left[\text{\tt Phi}[\text{\tt h2},{\tt +2}],
\text{\tt Phi}\left[\text{\tt h3},
\{\text{\tt AA1}\},{\tt +\frac{3}{2}}\right]\right],
\nonu \\
&& \text{\tt NO}[\text{\tt Phi}[\text{\tt h2},{\tt +2}],
\text{\tt Phi}[\text{\tt h3},\{\text{\tt AA1},\text{\tt BB1}\},{\tt +1}]],
\nonu \\
&& \text{\tt NO}\left[\text{\tt Phi}[\text{\tt h2},{\tt +2}],
\text{\tt Phi}\left[\text{\tt h3},\{\text{\tt AA1},\text{\tt BB1},
\text{\tt CC1}\},{\tt +\frac{1}{2}}\right]\right],
\nonu \\
&& \text{\tt NO}[\text{\tt Phi}[\text{\tt h2},{\tt +2}],
\text{\tt Phi}[\text{\tt h3},\{\text{\tt AA1},\text{\tt BB1},
\text{\tt CC1},\text{\tt DD1}\},{\tt 0}]],\nonu \\
&& \text{\tt NO}
\left[\text{\tt Phi}[\text{\tt h2},{\tt +2}],\text{\tt Phi}
\left[\text{\tt h3},\{\text{\tt AA1},\text{\tt BB1},
\text{\tt CC1}\},{\tt -\frac{1}{2}}\right]\right],
\nonu \\
&&
\text{\tt NO}[\text{\tt Phi}[\text{\tt h2},{\tt +2}],
\text{\tt Phi}[\text{\tt h3},\{\text{\tt AA1},\text{\tt BB1}\},{\tt -1}]],
\nonu \\
&&
\text{\tt NO}\left[\text{\tt Phi}[\text{\tt h2},{\tt +2}],
\text{\tt Phi}\left[\text{\tt h3},\{\text{\tt AA1}\},{\tt -\frac{3}{2}}
\right]\right],
\nonu \\
&&
\text{\tt NO}[\text{\tt Phi}[\text{\tt h2},{\tt +2}],
\text{\tt Phi}[\text{\tt h3},{\tt -2}]],
\nonu \\
&&\text{\tt NO}\left[\text{\tt Phi}
\left[\text{\tt h2},\{\text{\tt AA1}\},{\tt +\frac{3}{2}}\right],
\text{\tt Phi}\left[\text{\tt h3},\{\text{\tt BB1}\},{\tt +\frac{3}{2}}
\right]\right],
\nonu \\
&&
\text{\tt NO}\left[\text{\tt Phi}
\left[\text{\tt h2},\{\text{\tt AA1}\},{\tt +\frac{3}{2}}\right],
\text{\tt Phi}[\text{\tt h3},\{\text{\tt BB1},\text{\tt CC1}\},
{\tt +1}]\right],
\nonu \\
&&
\text{\tt NO}\left[\text{\tt Phi}
\left[\text{\tt h2},\{\text{\tt AA1}\},{\tt +\frac{3}{2}}\right],
\text{\tt Phi}\left[\text{\tt h3},\{\text{\tt BB1},\text{\tt CC1},
\text{\tt DD1}\},{\tt +\frac{1}{2}}\right]\right],
\nonu \\
&&
\text{\tt NO}
\left[\text{\tt Phi}\left[\text{\tt h2},\{\text{\tt AA1}\},
{\tt +\frac{3}{2}}\right],
\text{\tt Phi}[\text{\tt h3},\{\text{\tt BB1},\text{\tt CC1},\text{\tt DD1},
\text{\tt EE1}\},{\tt 0}]\right],
\nonu \\
&& \text{\tt NO}\left[\text{\tt Phi}
\left[\text{\tt h2},\{\text{\tt AA1}\},{\tt +\frac{3}{2}}\right],
\text{\tt Phi}
\left[\text{\tt h3},\{\text{\tt BB1},\text{\tt CC1},
\text{\tt DD1}\},{\tt -\frac{1}{2}}\right]
\right],
\nonu \\
&& \text{\tt NO}\left[\text{\tt Phi}
\left[\text{\tt h2},\{\text{\tt AA1}\},{\tt +\frac{3}{2}}\right],
\text{\tt Phi}[\text{\tt h3},\{\text{\tt BB1},\text{\tt CC1}\},
{\tt -1}]\right],
\nonu \\
&&
\text{\tt NO}\left[\text{\tt Phi}
\left[\text{\tt h2},\{\text{\tt AA1}\},{\tt +\frac{3}{2}}\right],
\text{\tt Phi}\left[\text{\tt h3},\{\text{\tt BB1}\},
{\tt -\frac{3}{2}}\right]\right],
\nonu \\
&&
\text{\tt NO}[\text{\tt Phi}[\text{\tt h2},\{\text{\tt AA1},
\text{\tt BB1}\},{\tt +1}],\text{\tt Phi}[\text{\tt h3},\{\text{\tt CC1},
\text{\tt DD1}\},{\tt +1}]],
\nonu \\
&&
\text{\tt NO}\left[\text{\tt Phi}[\text{\tt h2},
\{\text{\tt AA1},\text{\tt BB1}\},{\tt +1}],
\text{\tt Phi}\left[\text{\tt h3},
\{\text{\tt CC1},\text{\tt DD1},\text{\tt EE1}\},
{\tt +\frac{1}{2}}\right]\right],
\nonu \\
&&
\text{\tt NO}[\text{\tt Phi}[\text{\tt h2},\{\text{\tt AA1},
\text{\tt BB1}\},{\tt +1}],
\text{\tt Phi}[\text{\tt h3},\{\text{\tt CC1},\text{\tt DD1},\text{\tt EE1},
\text{\tt FF1}\},{\tt 0}]],
\nonu \\
&&
\text{\tt NO}\left[\text{\tt Phi}[\text{\tt h2},\{\text{\tt AA1},
\text{\tt BB1}\},{\tt +1}],\text{\tt Phi}
\left[\text{\tt h3},\{\text{\tt CC1},\text{\tt DD1},
\text{\tt EE1}\},{\tt -\frac{1}{2}}\right]\right],
\nonu \\
&&
\text{\tt NO}[\text{\tt Phi}[\text{\tt h2},\{\text{\tt AA1},
\text{\tt BB1}\},{\tt +1}],\text{\tt Phi}[\text{\tt h3},\{\text{\tt CC1},
\text{\tt DD1}\},{\tt -1}]],
\nonu \\
&& \text{\tt NO}\left[\text{\tt Phi}
\left[\text{\tt h2},\{\text{\tt AA1},
\text{\tt BB1},\text{\tt CC1}\},{\tt +\frac{1}{2}}\right],
\text{\tt Phi}\left[\text{\tt h3},\{\text{\tt DD1},\text{\tt EE1},
\text{\tt FF1}\},{\tt +\frac{1}{2}}\right]\right],
\nonu \\
&&
\text{\tt NO}\left[\text{\tt Phi}\left[\text{\tt h2},
\{\text{\tt AA1},\text{\tt BB1},
\text{\tt CC1}\},{\tt +\frac{1}{2}}\right],
\text{\tt Phi}[\text{\tt h3},\{\text{\tt DD1},
\text{\tt EE1},\text{\tt FF1},\text{\tt GG1}\},{\tt 0}]\right],
\nonu \\
&&
\text{\tt NO}\left[\text{\tt Phi}\left[\text{\tt h2},
\{\text{\tt AA1},\text{\tt BB1},
\text{\tt CC1}\},{\tt +\frac{1}{2}}\right],\text{\tt Phi}
\left[\text{\tt h3},\{\text{\tt DD1},\text{\tt EE1},
\text{\tt FF1}\},{\tt -\frac{1}{2}}
\right]\right],
\nonu \\
&&
\left.
\text{\tt NO}\left[\text{\tt Phi}[\text{\tt h2},\{\text{\tt AA1},
\text{\tt BB1},\text{\tt CC1},\text{\tt DD1}\},{\tt 0}],
\text{\tt Phi}[\text{\tt h3},\{\text{\tt EE1},\text{\tt FF1},\text{\tt GG1},
\text{\tt HH1}\},{\tt 0}]\right]\right\};
\label{COMP}
\eea
We present the ninety-five OPEs together with (\ref{NORM}) and
(\ref{COMP})
as follows
\footnote{From these OPEs, we can observe the signs, the structures of
the product of two couplings, the $SU(8)$ group structures, the
overall normalizations, 
and the types of celestial operators on the right-hand sides.
This information provides a double check for the ninety-five
(anti)commutators in (\ref{25comm}) and Appendix B.
In principle, we can compute any OPE of any multiple-particle
(normal-ordered) operator with other types of
multi-particle (normal-ordered) operator inside the package.
For example, for the quadruple-collinear limit in (\ref{quadruple1})
or the multiple OPE of two-particle operators with two-particle
operators,
the highest pole is given by the sixth-order pole.
For the $N$-tuple collinear limit, the highest singular term is
the $2(N-1)$-th order pole because the number of pole in the
(anti)holomorphic coordinates is increased by $2$ as we add
a single operator to the second operator on the left-hand side.
Moreover, the descendant operators of the operator living in
the highest order pole do not appear in all the lower order poles
because these operators are not quasiprimary, contrary to (\ref{HtHt}).
Only after the correct quasiprimary operators (by considering
all the derivative terms)
are constructed, then all the descendant operators
appear in all the singular terms (until to the first-order pole).
Recall that the description of (\ref{ntuple1}) or (\ref{ntuple4})
implies that the highest singular term is $N$-th order.
When $N=2$ (the OPE of a single-particle operator with itself),
the two highest order poles are the same. For $N \neq 2$,
the highest pole of former behaves as $4,6,8, \dots, 2(N-1), \dots$
while the highest pole of
the latter does as $3,4,5, \dots, N, \dots$ with $N \geq 3$.}. 

\subsection{The fourth order pole of the OPEs
of the gravitons of helicity $+2$ and the quadratic operators }

\bea
&& \text{\tt OPESimplify}
\left[\text{\tt OPEPole}[{\tt 4}][\text{\tt Phi}[\text{\tt h1},{\tt +2}],
\text{\tt comp}[[{\tt 1}]]],
\text{\tt Factor}\right]
\nonu \\
&& = \text{\tt kappa}\left[{\tt 2,2,-2}\right]^2 \, \text{\tt Phi}[\text{\tt h1}+
\text{\tt h2}+\text{\tt h3},{\tt 2}],
\nonu \\
&& \text{\tt OPESimplify}
\left[\text{\tt OPEPole}[{\tt 4}][\text{\tt Phi}[\text{\tt h1},{\tt +2}],
\text{\tt comp}[[{\tt 2}]]],
\text{\tt Factor}\right]
\nonu \\
&& = \text{\tt kappa}\left[{\tt 2,\frac{3}{2},-\frac{3}{2}}\right]\,
\text{\tt kappa}\left[{\tt 2,2,-2}\right]
\, \text{\tt Phi}\left[\text{\tt h1}+
\text{\tt h2}+\text{\tt h3},{\tt \{AA1\}}, {\tt \frac{3}{2}}\right],
\nonu \\
&& \text{\tt OPESimplify}
\left[\text{\tt OPEPole}[{\tt 4}][\text{\tt Phi}[\text{\tt h1},{\tt +2}],
\text{\tt comp}[[{\tt 3}]]],
\text{\tt Factor}\right]
\nonu \\
&& = \text{\tt kappa}\left[{\tt 2,1,-1}\right]\,
\text{\tt kappa}\left[{\tt 2,2,-2}\right]
\, \text{\tt Phi}[\text{\tt h1}+
\text{\tt h2}+\text{\tt h3},{\tt \{ AA1,BB1 \}}, {\tt 1}],
\nonu \\
&& \text{\tt OPESimplify}
\left[\text{\tt OPEPole}[{\tt 4}][\text{\tt Phi}[\text{\tt h1},{\tt +2}],
\text{\tt comp}[[{\tt 4}]]],
\text{\tt Factor}\right]
\nonu \\
&& = \text{\tt kappa}\left[{\tt 2,\frac{1}{2},-\frac{1}{2}}\right]\,
\text{\tt kappa}\left[{\tt 2,2,-2}\right]
\, \text{\tt Phi}\left[\text{\tt h1}+
\text{\tt h2}+\text{\tt h3},{\tt \{ AA1,BB1,CC1 \}},
{\tt \frac{1}{2}} \right],
\nonu \\
&& \text{\tt OPESimplify}
\left[\text{\tt OPEPole}[{\tt 4}][\text{\tt Phi}[\text{\tt h1},{\tt +2}],
\text{\tt comp}[[{\tt 5}]]],
\text{\tt Factor}\right]
\nonu \\
&& = \text{\tt kappa}\left[{\tt 2,0,0}\right]\,
\text{\tt kappa}\left[{\tt 2,2,-2}\right]
\, \text{\tt Phi}[\text{\tt h1}+
\text{\tt h2}+\text{\tt h3},{\tt \{ AA1,BB1,CC1,DD1 \}}, {\tt 0}],
\nonu \\
&& \text{\tt OPESimplify}
\left[\text{\tt OPEPole}[{\tt 4}][\text{\tt Phi}[\text{\tt h1},{\tt +2}],
\text{\tt comp}[[{\tt 6}]]],
\text{\tt Factor}\right]
\nonu \\
&& = \text{\tt kappa}\left[{\tt 2,-\frac{1}{2},\frac{1}{2}}\right]\,
\text{\tt kappa}\left[{\tt 2,2,-2}\right]
\, \text{\tt Phi}\left[\text{\tt h1}+
\text{\tt h2}+\text{\tt h3},{\tt \{ AA1,BB1,CC1 \}},
{\tt -\frac{1}{2}}\right],
\nonu \\
&& \text{\tt OPESimplify}
\left[\text{\tt OPEPole}[{\tt 4}][\text{\tt Phi}[\text{\tt h1},{\tt +2}],
\text{\tt comp}[[{\tt 7}]]],
\text{\tt Factor}\right]
\nonu \\
&& = \text{\tt kappa}\left[{\tt 2,-1,1}\right]\,
\text{\tt kappa}\left[{\tt 2,2,-2}\right]
\, \text{\tt Phi}[\text{\tt h1}+
\text{\tt h2}+\text{\tt h3},{\tt \{ AA1,BB1 \}}, {\tt -1}],
\nonu \\
&& \text{\tt OPESimplify}
\left[\text{\tt OPEPole}[{\tt 4}][\text{\tt Phi}[\text{\tt h1},{\tt +2}],
\text{\tt comp}[[{\tt 8}]]],
\text{\tt Factor}\right]
\nonu \\
&& = \text{\tt kappa}\left[{\tt 2,-\frac{3}{2},\frac{3}{2}}\right]\,
\text{\tt kappa}\left[{\tt 2,2,-2}\right]
\, \text{\tt Phi}\left[\text{\tt h1}+
\text{\tt h2}+\text{\tt h3},{\tt \{ AA1 \}}, {\tt -\frac{3}{2}}\right],
\nonu \\
&& \text{\tt OPESimplify}
\left[\text{\tt OPEPole}[{\tt 4}][\text{\tt Phi}[\text{\tt h1},{\tt +2}],
\text{\tt comp}[[{\tt 9}]]],
\text{\tt Factor}\right]
\nonu \\
&& = \text{\tt kappa}\left[{\tt 2,-2,2}\right]\,
\text{\tt kappa}\left[{\tt 2,2,-2}\right]
\, \text{\tt Phi}[\text{\tt h1}+
\text{\tt h2}+\text{\tt h3}, {\tt -2}],
\nonu \\
&& \text{\tt OPESimplify}
\left[\text{\tt OPEPole}[{\tt 4}][\text{\tt Phi}[\text{\tt h1},{\tt +2}],
\text{\tt comp}[[{\tt 10}]]],
\text{\tt Factor}\right]
\nonu \\
&& = \text{\tt kappa}\left[{\tt \frac{3}{2},\frac{3}{2},-1}\right]\,
\text{\tt kappa}\left[{\tt 2,\frac{3}{2},-\frac{3}{2}}\right]
\, \text{\tt Phi}[\text{\tt h1}+
\text{\tt h2}+\text{\tt h3}, {\tt \{AA1, BB1\}}, {\tt 1}],
\nonu \\
&& \text{\tt OPESimplify}
\left[\text{\tt OPEPole}[{\tt 4}][\text{\tt Phi}[\text{\tt h1},{\tt +2}],
\text{\tt comp}[[{\tt 11}]]],
\text{\tt Factor}\right]
\nonu \\
&& = \text{\tt kappa}\left[{\tt \frac{3}{2},1,-\frac{1}{2}}\right]\,
\text{\tt kappa}\left[{\tt 2,\frac{3}{2},-\frac{3}{2}}\right]
\, \text{\tt Phi}\left[\text{\tt h1}+
\text{\tt h2}+\text{\tt h3}, {\tt \{AA1, BB1, CC1\}},
{\tt \frac{1}{2}}\right],
\nonu \\
&& \text{\tt OPESimplify}
\left[\text{\tt OPEPole}[{\tt 4}][\text{\tt Phi}[\text{\tt h1},{\tt +2}],
\text{\tt comp}[[{\tt 12}]]],
\text{\tt Factor}\right]
\nonu \\
&& = \text{\tt kappa}\left[{\tt \frac{3}{2},\frac{1}{2},0}\right]\,
\text{\tt kappa}\left[{\tt 2,\frac{3}{2},-\frac{3}{2}}\right]
\, \text{\tt Phi}[\text{\tt h1}+
\text{\tt h2}+\text{\tt h3}, {\tt \{AA1, BB1, CC1,DD1\}}, {\tt 0}],
\nonu \\
&& \text{\tt OPESimplify}
\left[\text{\tt OPEPole}[{\tt 4}][\text{\tt Phi}[\text{\tt h1},{\tt +2}],
\text{\tt comp}[[{\tt 13}]]],
\text{\tt Factor}\right]
\nonu \\
&& = -\text{\tt kappa}\left[{\tt \frac{3}{2},0, \frac{1}{2}}\right]\,
\text{\tt kappa}\left[{\tt 2,\frac{3}{2},-\frac{3}{2}}\right]\,
\text{\tt Epsilon}
[{\tt 1,2,3},\text{\tt AA1},\text{\tt BB1},\text{\tt CC1},\text{\tt DD1},
\text{\tt EE1}]
\nonu \\
&& \times \text{\tt Phi}\left[\text{\tt h1}+
\text{\tt h2}+\text{\tt h3}, {\tt \{1,2,3\}}, {\tt -\frac{1}{2}}\right]
+ \cdots,
\nonu \\
&& \text{\tt OPESimplify}
\left[\text{\tt OPEPole}[{\tt 4}][\text{\tt Phi}[\text{\tt h1},{\tt +2}],
\text{\tt comp}[[{\tt 14}]]],
\text{\tt Factor}\right]
\nonu \\
&& = \text{\tt kappa}\left[{\tt \frac{3}{2},-\frac{1}{2},1}\right]\,
\text{\tt kappa}\left[{\tt 2,\frac{3}{2},-\frac{3}{2}}\right]\,
\text{\tt Delta}
[\text{\tt AA1},\text{\tt DD1}]
\, \text{\tt Phi}[\text{\tt h1}+
\text{\tt h2}+\text{\tt h3}, {\tt \{BB1,CC1\}}, {\tt -1}]
\nonu \\
&& + \cdots,
\nonu \\
&& \text{\tt OPESimplify}
\left[\text{\tt OPEPole}[{\tt 4}][\text{\tt Phi}[\text{\tt h1},{\tt +2}],
    \text{\tt comp}[[{\tt 15}]]],
\text{\tt Factor}\right]
\nonu \\
&& = -\text{\tt kappa}\left[{\tt \frac{3}{2},-1, \frac{3}{2}}\right]\,
\text{\tt kappa}\left[{\tt 2,\frac{3}{2},-\frac{3}{2}}\right]\,
\text{\tt Delta}
[\text{\tt AA1},\text{\tt CC1}]
\, \text{\tt Phi}\left[\text{\tt h1}+
\text{\tt h2}+\text{\tt h3}, {\tt \{BB1\}}, {\tt -\frac{3}{2}}\right]
\nonu \\
&& + \cdots,
\nonu \\
&& \text{\tt OPESimplify}
\left[\text{\tt OPEPole}[{\tt 4}][\text{\tt Phi}[\text{\tt h1},{\tt +2}],
    \text{\tt comp}[[{\tt 16}]]],
\text{\tt Factor}\right]
\nonu \\
&& = \text{\tt kappa}\left[{\tt \frac{3}{2},-\frac{3}{2},2}\right]\,
\text{\tt kappa}\left[{\tt 2,\frac{3}{2},-\frac{3}{2}}\right]\,
\text{\tt Delta}
[\text{\tt AA1},\text{\tt BB1}]
\, \text{\tt Phi}[\text{\tt h1}+
\text{\tt h2}+\text{\tt h3}, {\tt -2}],
\nonu \\
&& \text{\tt OPESimplify}
\left[\text{\tt OPEPole}[{\tt 4}][\text{\tt Phi}[\text{\tt h1},{\tt +2}],
\text{\tt comp}[[{\tt 17}]]],
\text{\tt Factor}\right]
\nonu \\
&& = \text{\tt kappa}\left[{\tt 1,1,0}\right]\,
\text{\tt kappa}\left[{\tt 2,1,-1}\right]
\, \text{\tt Phi}[\text{\tt h1}+
\text{\tt h2}+\text{\tt h3},{\tt \{AA1,BB1,CC1,DD1\} }, {\tt 0}],
\nonu \\
&& \text{\tt OPESimplify}
\left[\text{\tt OPEPole}[{\tt 4}][\text{\tt Phi}[\text{\tt h1},{\tt +2}],
\text{\tt comp}[[{\tt 18}]]],
\text{\tt Factor}\right]
\nonu \\
&& = -\text{\tt kappa}\left[{\tt 1,\frac{1}{2},\frac{1}{2}}\right]\,
\text{\tt kappa}\left[{\tt 2,1,-1}\right]\,
\text{\tt Epsilon}
[{\tt 1,2,3},\text{\tt AA1},\text{\tt BB1},\text{\tt CC1},\text{\tt DD1},
  \text{\tt EE1}]
\nonu \\
&& \times  \text{\tt Phi}\left[\text{\tt h1}+
\text{\tt h2}+\text{\tt h3},{\tt \{1,2,3\} }, {\tt -\frac{1}{2}}\right]
+ \cdots,
\nonu \\
&& \text{\tt OPESimplify}
\left[\text{\tt OPEPole}[{\tt 4}][\text{\tt Phi}[\text{\tt h1},{\tt +2}],
\text{\tt comp}[[{\tt 19}]]],
\text{\tt Factor}\right]
\nonu \\
&& = \text{\tt kappa}\left[{\tt 1,0,1}\right]\,
\text{\tt kappa}\left[{\tt 2,1,-1}\right]\,
\text{\tt Epsilon}
[{\tt 1,2},\text{\tt AA1},\text{\tt BB1},\text{\tt CC1},\text{\tt DD1},
\text{\tt EE1},  \text{\tt FF1}]
\nonu \\
&& \times  \text{\tt Phi}[\text{\tt h1}+
\text{\tt h2}+\text{\tt h3},{\tt \{1,2\} }, {\tt -1}]
+ \cdots,
\nonu \\
&& \text{\tt OPESimplify}
\left[\text{\tt OPEPole}[{\tt 4}][\text{\tt Phi}[\text{\tt h1},{\tt +2}],
\text{\tt comp}[[{\tt 20}]]],
\text{\tt Factor}\right]
\nonu \\
&& = \text{\tt kappa}\left[{\tt 1,-\frac{1}{2},\frac{3}{2}}\right]\,
\text{\tt kappa}\left[{\tt 2,1,-1}\right]\,
\text{\tt Delta}[\text{\tt AA1},\text{\tt EE1}]
\text{\tt Delta}[\text{\tt BB1},\text{\tt DD1}]
\nonu \\
&& \times  \text{\tt Phi}\left[\text{\tt h1}+
\text{\tt h2}+\text{\tt h3},{\tt \{CC1\} }, {\tt -\frac{3}{2}}\right]
+ \cdots,
\nonu \\
&& \text{\tt OPESimplify}
\left[\text{\tt OPEPole}[{\tt 4}][\text{\tt Phi}[\text{\tt h1},{\tt +2}],
\text{\tt comp}[[{\tt 21}]]],
\text{\tt Factor}\right]
\nonu \\
&& = -\text{\tt kappa}\left[{\tt 1,-1,2}\right]\,
\text{\tt kappa}\left[{\tt 2,1,-1}\right]\,
\text{\tt Delta}[\text{\tt AA1},\text{\tt DD1}]
\text{\tt Delta}[\text{\tt BB1},\text{\tt CC1}]
\nonu \\
&& \times  \text{\tt Phi}[\text{\tt h1}+
\text{\tt h2}+\text{\tt h3}, {\tt -2}]
+ \cdots,
\nonu \\
&& \text{\tt OPESimplify}
\left[\text{\tt OPEPole}[{\tt 4}][\text{\tt Phi}[\text{\tt h1},{\tt +2}],
\text{\tt comp}[[{\tt 22}]]],
\text{\tt Factor}\right]
\nonu \\
&& = \text{\tt kappa}\left[{\tt \frac{1}{2},\frac{1}{2},1}\right]\,
\text{\tt kappa}\left[{\tt 2,\frac{1}{2},-\frac{1}{2}}\right]\,
\text{\tt
Epsilon}[{\tt 1,2},\text{\tt AA1},\text{\tt BB1},\text{\tt CC1},
\text{\tt DD1},\text{\tt EE1},\text{\tt FF1}]
\nonu \\
&& \times  \text{\tt Phi}[\text{\tt h1}+
\text{\tt h2}+\text{\tt h3}, \{ {\tt 1, 2}\}, {\tt -1}]
+ \cdots,
\nonu \\
&& \text{\tt OPESimplify}
\left[\text{\tt OPEPole}[{\tt 4}][\text{\tt Phi}[\text{\tt h1},{\tt +2}],
\text{\tt comp}[[{\tt 23}]]],
\text{\tt Factor}\right]
\nonu \\
&& = -\text{\tt kappa}\left[{\tt \frac{1}{2},0,\frac{3}{2}}\right]\,
\text{\tt kappa}\left[{\tt 2,\frac{1}{2},-\frac{1}{2}}\right]\,
\text{\tt
Epsilon}[{\tt 1},\text{\tt AA1},\text{\tt BB1},\text{\tt CC1},
\text{\tt DD1},\text{\tt EE1},\text{\tt FF1},\text{\tt GG1} ]
\nonu \\
&& \times  \text{\tt Phi}\left[\text{\tt h1}+
\text{\tt h2}+\text{\tt h3}, \{ {\tt 1}\}, {\tt -\frac{3}{2}}\right]
+ \cdots,
\nonu \\
&& \text{\tt OPESimplify}
\left[\text{\tt OPEPole}[{\tt 4}][\text{\tt Phi}[\text{\tt h1},{\tt +2}],
\text{\tt comp}[[{\tt 24}]]],
\text{\tt Factor}\right]
\nonu \\
&& = \text{\tt kappa}\left[{\tt \frac{1}{2},-\frac{1}{2},2}\right]\,
\text{\tt kappa}\left[{\tt 2,\frac{1}{2},-\frac{1}{2}}\right]\,
\text{\tt Epsilon}[{\tt 1,2,3,4,5},
\text{\tt AA1},\text{\tt BB1},\text{\tt CC1}]
\nonu \\
&& \times \text{\tt Epsilon}[{\tt 1,2,3,4,5},
\text{\tt DD1},\text{\tt EE1},\text{\tt FF1}]
\,  \text{\tt Phi}[\text{\tt h1}+
\text{\tt h2}+\text{\tt h3}, {\tt -2}]
+ \cdots,
\nonu \\
&& \text{\tt OPESimplify}
\left[\text{\tt OPEPole}[{\tt 4}][\text{\tt Phi}[\text{\tt h1},{\tt +2}],
\text{\tt comp}[[{\tt 25}]]],
\text{\tt Factor}\right]
\nonu \\
&& = \text{\tt kappa}\left[{\tt 0,0,2}\right]\,
\text{\tt kappa}\left[{\tt 2,0,0}\right]\,
\text{\tt Epsilon}[\text{\tt AA1},\text{\tt BB1},
\text{\tt CC1},\text{\tt DD1},\text{\tt EE1},\text{\tt FF1},
\text{\tt GG1},\text{\tt HH1}]
\nonu \\
&& \times   \text{\tt Phi}[\text{\tt h1}+
\text{\tt h2}+\text{\tt h3}, {\tt -2}].
\label{Cone}
\eea
Due to the
many terms in some OPEs, we present only one single term and
the remaining terms are ignored and are denoted by
$+ \cdots$.
There exist the overall minus signs at the five places
in (\ref{Cone}). These can be absorbed by changing the
$SU(8)$ indices properly according to the arrangement of Appendix A.
The locations of $SU(8)$ indices
with the overall plus signs
(which are consistent with the arrangement of Appendix A)
appearing on the right-hand sides
are distributed correctly.

\subsection{The fourth order pole of the OPEs
of the gravitinos of helicity $+\frac{3}{2}$ and the quadratic operators }

\bea
&&
\text{\tt OPESimplify}\left[\text{\tt OPEPole}[\tt 4]\left[\text{\tt
Phi}\left[\text{\tt h1},\{\tt P\},{\tt +\frac{3}{2}}\right],
\text{\tt comp}[[{\tt 1}]]],
\text{\tt Factor}\right] \right.
\nonu \\
&&  =\text{\tt kappa}\left[{\tt 2,\frac{3}{2},-\frac{3}{2}}\right]^{\tt 2}
\text{\tt Phi}\left[\text{\tt h1}+\text{\tt h2}+\text{\tt h3},\{\tt P\},
{\tt \frac{3}{2}}\right],
\nonu \\
&& \text{\tt OPESimplify}\left[\text{\tt OPEPole}[\tt 4]\left[\text{\tt
Phi}\left[\text{\tt h1},\{\tt P\},{\tt +\frac{3}{2}}\right],
\text{\tt comp}[[{\tt 2}]]],
\text{\tt Factor}\right] \right.
\nonu \\
&& =-
\text{\tt kappa}\left[{\tt \frac{3}{2},\frac{3}{2},-1}\right]\,
\text{\tt kappa}\left[{\tt 2,\frac{3}{2},-\frac{3}{2}}\right]
\text{\tt Phi}\left[\text{\tt h1}+\text{\tt h2}+\text{\tt h3},\{\tt AA1, P\},
{\tt 1}\right],
\nonu \\
&& \text{\tt OPESimplify}\left[\text{\tt OPEPole}[\tt 4]\left[\text{\tt
Phi}\left[\text{\tt h1},\{\tt P\},{\tt +\frac{3}{2}}\right],
\text{\tt comp}[[{\tt 3}]]],
\text{\tt Factor}\right] \right.
\nonu \\ && =
\text{\tt kappa}\left[{\tt \frac{3}{2},1,-\frac{1}{2}}\right]\,
\text{\tt kappa}\left[{\tt 2,\frac{3}{2},-\frac{3}{2}}\right]
\text{\tt Phi}\left[\text{\tt h1}+\text{\tt h2}+\text{\tt h3},\{\tt AA1,
BB1, P\},
{\tt \frac{1}{2}}\right],
\nonu \\
&& \text{\tt OPESimplify}\left[\text{\tt OPEPole}[\tt 4]\left[\text{\tt
Phi}\left[\text{\tt h1},\{\tt P\},{\tt +\frac{3}{2}}\right],
\text{\tt comp}[[{\tt 4}]]],
\text{\tt Factor}\right] \right.
\nonu \\ && =-
\text{\tt kappa}\left[{\tt \frac{3}{2},\frac{1}{2},0}\right]\,
\text{\tt kappa}\left[{\tt 2,\frac{3}{2},-\frac{3}{2}}\right]
\text{\tt Phi}\left[\text{\tt h1}+\text{\tt h2}+\text{\tt h3},\{\tt AA1,
BB1, CC1, P\},
{\tt 0}\right],
\nonu \\
&& \text{\tt OPESimplify}\left[\text{\tt OPEPole}[\tt 4]\left[\text{\tt
Phi}\left[\text{\tt h1},\{\tt P\},{\tt +\frac{3}{2}}\right],
\text{\tt comp}[[{\tt 5}]]],
\text{\tt Factor}\right] \right.
\nonu \\ && =-
\text{\tt kappa}\left[{\tt \frac{3}{2},0,\frac{1}{2}}\right]\,
\text{\tt kappa}\left[{\tt 2,\frac{3}{2},-\frac{3}{2}}\right]\,
\text{\tt Epsilon}[{\tt 1,2,3},\text{\tt AA1},
\text{\tt BB1},\text{\tt CC1},\text{\tt DD1},{\tt P}]\nonu \\
&& \times 
\text{\tt Phi}\left[\text{\tt h1}+\text{\tt h2}+\text{\tt h3},\{\tt 1,
2, 3 \},
{\tt -\frac{1}{2}}\right] + \cdots,
\nonu \\
&& \text{\tt OPESimplify}\left[\text{\tt OPEPole}[\tt 4]\left[\text{\tt
Phi}\left[\text{\tt h1},\{\tt P\},{\tt +\frac{3}{2}}\right],
\text{\tt comp}[[{\tt 6}]]],
\text{\tt Factor}\right] \right.
\nonu \\
&& =
\text{\tt kappa}\left[{\tt \frac{3}{2},-\frac{1}{2},1}\right]\,
\text{\tt kappa}\left[{\tt 2,\frac{3}{2},-\frac{3}{2}}\right]\,
\text{\tt Delta}[\text{\tt CC1},{\tt P}]
\nonu \\
&& \times 
\text{\tt Phi}\left[\text{\tt h1}+\text{\tt h2}+\text{\tt h3},\{\tt AA1,
BB1 \},
{\tt -1}\right] + \cdots,
\nonu \\
&& \text{\tt OPESimplify}\left[\text{\tt OPEPole}[\tt 4]\left[\text{\tt
Phi}\left[\text{\tt h1},\{\tt P\},{\tt +\frac{3}{2}}\right],
\text{\tt comp}[[{\tt 7}]]],
\text{\tt Factor}\right] \right.
\nonu \\
&& =
-\text{\tt kappa}\left[{\tt \frac{3}{2},-1,\frac{3}{2}}\right]\,
\text{\tt kappa}\left[{\tt 2,\frac{3}{2},-\frac{3}{2}}\right]\,
\text{\tt Delta}[\text{\tt BB1},{\tt P}]
\nonu \\
&& \times 
\text{\tt Phi}\left[\text{\tt h1}+\text{\tt h2}+\text{\tt h3},\{\tt AA1\},
{\tt -\frac{3}{2}}\right] + \cdots,
\nonu \\
&& \text{\tt OPESimplify}\left[\text{\tt OPEPole}[\tt 4]\left[\text{\tt
Phi}\left[\text{\tt h1},\{\tt P\},{\tt +\frac{3}{2}}\right],
\text{\tt comp}[[{\tt 8}]]],
\text{\tt Factor}\right] \right.
\nonu \\
&& =
\text{\tt kappa}\left[{\tt \frac{3}{2},-\frac{3}{2},2}\right]\,
\text{\tt kappa}\left[{\tt 2,\frac{3}{2},-\frac{3}{2}}\right]\,
\text{\tt Delta}[\text{\tt AA1},{\tt P}]
\, 
\text{\tt Phi}\left[\text{\tt h1}+\text{\tt h2}+\text{\tt h3},
{\tt -2}\right],
\nonu \\
&& \text{\tt OPESimplify}\left[\text{\tt OPEPole}[\tt 4]\left[\text{\tt
Phi}\left[\text{\tt h1},\{\tt P\},{\tt +\frac{3}{2}}\right],
\text{\tt comp}[[{\tt 10}]]],
\text{\tt Factor}\right] \right.
\nonu \\
&& =
\text{\tt kappa}\left[{\tt \frac{3}{2},1,-\frac{1}{2}}\right]\,
\text{\tt kappa}\left[{\tt \frac{3}{2},\frac{3}{2},-1}\right]\,
\text{\tt Phi}\left[\text{\tt h1}+\text{\tt h2}+\text{\tt h3},
\{{\tt AA1, BB1,P} \},
{\tt \frac{1}{2}}\right],
\nonu \\
&& \text{\tt OPESimplify}\left[\text{\tt OPEPole}[\tt 4]\left[\text{\tt
Phi}\left[\text{\tt h1},\{\tt P\},{\tt +\frac{3}{2}}\right],
\text{\tt comp}[[{\tt 11}]]],
\text{\tt Factor}\right] \right.
\nonu \\
&& =-
\text{\tt kappa}\left[{\tt 1,1,0}\right]\,
\text{\tt kappa}\left[{\tt \frac{3}{2},\frac{3}{2},-1}\right]\,
\text{\tt Phi}\left[\text{\tt h1}+\text{\tt h2}+\text{\tt h3},
\{{\tt AA1, BB1,CC1,P} \},
{\tt 0}\right],
\nonu \\
&& \text{\tt OPESimplify}\left[\text{\tt OPEPole}[\tt 4]\left[\text{\tt
Phi}\left[\text{\tt h1},\{\tt P\},{\tt +\frac{3}{2}}\right],
\text{\tt comp}[[{\tt 12}]]],
\text{\tt Factor}\right] \right.
\nonu \\
&& =-
\text{\tt kappa}\left[{\tt 1,\frac{1}{2},\frac{1}{2}}\right]\,
\text{\tt kappa}\left[{\tt \frac{3}{2},\frac{3}{2},-1}\right]\,
\text{\tt Epsilon}[{\tt 1,2,3},
\text{\tt AA1},\text{\tt BB1},\text{\tt CC1},\text{\tt DD1},
{\tt P}]\nonu \\
&& \times 
\text{\tt Phi}\left[\text{\tt h1}+\text{\tt h2}+\text{\tt h3},
\{{\tt 1,2,3} \},
{\tt -\frac{1}{2}}\right]+ \cdots,
\nonu \\
&& \text{\tt OPESimplify}\left[\text{\tt OPEPole}[\tt 4]\left[\text{\tt
Phi}\left[\text{\tt h1},\{\tt P\},{\tt +\frac{3}{2}}\right],
\text{\tt comp}[[{\tt 13}]]],
\text{\tt Factor}\right] \right.
\nonu \\
&& =-
\text{\tt kappa}\left[{\tt 1,0,1}\right]\,
\text{\tt kappa}\left[{\tt \frac{3}{2},\frac{3}{2},-1}\right]\,
\text{\tt Epsilon}[{\tt 1,2},
\text{\tt AA1},\text{\tt BB1},\text{\tt CC1},\text{\tt DD1},
\text{\tt EE1},  
{\tt P}]\nonu \\
&& \times 
\text{\tt Phi}\left[\text{\tt h1}+\text{\tt h2}+\text{\tt h3},
\{{\tt 1,2} \},
{\tt -1}\right]+ \cdots,
\nonu \\
&& \text{\tt OPESimplify}\left[\text{\tt OPEPole}[\tt 4]\left[\text{\tt
Phi}\left[\text{\tt h1},\{\tt P\},{\tt +\frac{3}{2}}\right],
\text{\tt comp}[[{\tt 14}]]],
\text{\tt Factor}\right] \right.
\nonu \\
&& =-
\text{\tt kappa}\left[{\tt 1,-\frac{1}{2},\frac{3}{2}}\right]\,
\text{\tt kappa}\left[{\tt \frac{3}{2},\frac{3}{2},-1}\right]\,
\text{\tt Delta}[\text{\tt AA1},\text{\tt DD1}]
\text{\tt Delta}[\text{
\tt CC1},{\tt P}]
\nonu \\
&& \times 
\text{\tt Phi}\left[\text{\tt h1}+\text{\tt h2}+\text{\tt h3},
\{{\tt BB1} \},
{\tt -\frac{3}{2}}\right]+ \cdots,
\nonu \\
&& \text{\tt OPESimplify}\left[\text{\tt OPEPole}[\tt 4]\left[\text{\tt
Phi}\left[\text{\tt h1},\{\tt P\},{\tt +\frac{3}{2}}\right],
\text{\tt comp}[[{\tt 15}]]],
\text{\tt Factor}\right] \right.
\nonu \\
&& =
\text{\tt kappa}\left[{\tt 1,-1,2}\right]\,
\text{\tt kappa}\left[{\tt \frac{3}{2},\frac{3}{2},-1}\right]\,
\text{\tt Delta}[\text{\tt AA1},\text{\tt CC1}]
\text{\tt Delta}[\text{
\tt BB1},{\tt P}]
\nonu \\
&& \times 
\text{\tt Phi}\left[\text{\tt h1}+\text{\tt h2}+\text{\tt h3},
{\tt -2}\right]+ \cdots,
\nonu \\
&& \text{\tt OPESimplify}\left[\text{\tt OPEPole}[\tt 4]\left[\text{\tt
Phi}\left[\text{\tt h1},\{\tt P\},{\tt +\frac{3}{2}}\right],
\text{\tt comp}[[{\tt 17}]]],
\text{\tt Factor}\right] \right.
\nonu \\
&& =-
\text{\tt kappa}\left[{\tt 1,\frac{1}{2},\frac{1}{2}}\right]\,
\text{\tt kappa}\left[{\tt \frac{3}{2},1,-\frac{1}{2}}\right]\,
\text{\tt Epsilon}[{\tt 1,2,3},\text{\tt AA1},\text{\tt BB1},\text{\tt
CC1},\text{\tt DD1},{\tt P}]
\nonu \\
&& \times 
\text{\tt Phi}\left[\text{\tt h1}+\text{\tt h2}+
\text{\tt h3},\{{\tt 1,2,3} \},
{\tt -\frac{1}{2}}\right]+ \cdots,
\nonu \\
&& \text{\tt OPESimplify}\left[\text{\tt OPEPole}[\tt 4]\left[\text{\tt
Phi}\left[\text{\tt h1},\{\tt P\},{\tt +\frac{3}{2}}\right],
\text{\tt comp}[[{\tt 18}]]],
\text{\tt Factor}\right] \right.
\nonu \\
&& =-
\text{\tt kappa}\left[{\tt \frac{1}{2},\frac{1}{2},1}\right]\,
\text{\tt kappa}\left[{\tt \frac{3}{2},1,-\frac{1}{2}}\right]\,
\text{\tt Epsilon}[{\tt 1,2},\text{\tt AA1},\text{\tt BB1},\text{\tt
CC1},\text{\tt DD1},\text{\tt EE1},{\tt P}]
\nonu \\
&& \times 
\text{\tt Phi}\left[\text{\tt h1}+\text{\tt h2}+
\text{\tt h3},\{{\tt 1,2} \},
{\tt -1}\right]+ \cdots,
\nonu \\
&& \text{\tt OPESimplify}\left[\text{\tt OPEPole}[\tt 4]\left[\text{\tt
Phi}\left[\text{\tt h1},\{\tt P\},{\tt +\frac{3}{2}}\right],
\text{\tt comp}[[{\tt 19}]]],
\text{\tt Factor}\right] \right.
\nonu \\
&&=-
\text{\tt kappa}\left[{\tt \frac{1}{2},0,\frac{3}{2}}\right]\,
\text{\tt kappa}\left[{\tt \frac{3}{2},1,-\frac{1}{2}}\right]\,
\text{\tt Epsilon}[{\tt 1},\text{\tt AA1},\text{\tt BB1},\text{\tt
CC1},\text{\tt DD1},\text{\tt EE1},\text{\tt FF1},{\tt P}]
\nonu \\
&& \times 
\text{\tt Phi}\left[\text{\tt h1}+\text{\tt h2}+
\text{\tt h3},\{{\tt 1} \},
{\tt -\frac{3}{2}}\right]+ \cdots,
\nonu \\
&& \text{\tt OPESimplify}\left[\text{\tt OPEPole}[\tt 4]\left[\text{\tt
Phi}\left[\text{\tt h1},\{\tt P\},{\tt +\frac{3}{2}}\right],
\text{\tt comp}[[{\tt 20}]]],
\text{\tt Factor}\right] \right.
\nonu \\
&&=
\text{\tt kappa}\left[{\tt \frac{1}{2},-\frac{1}{2},2}\right]\,
\text{\tt kappa}\left[{\tt \frac{3}{2},1,-\frac{1}{2}}\right]\,
\text{\tt Epsilon}[{\tt 1,2,3,4,5},\text{\tt AA1},\text{\tt BB1},{\tt P}]
\nonu \\
&& \times
\text{\tt Epsilon}[{\tt 1,2,3,4,5},
\text{\tt CC1},\text{\tt DD1},\text{\tt EE1}]
\, 
\text{\tt Phi}\left[\text{\tt h1}+\text{\tt h2}+
\text{\tt h3},
{\tt -2}\right]+ \cdots,
\nonu \\
&& \text{\tt OPESimplify}\left[\text{\tt OPEPole}[\tt 4]\left[\text{\tt
Phi}\left[\text{\tt h1},\{\tt P\},{\tt +\frac{3}{2}}\right],
\text{\tt comp}[[{\tt 22}]]],
\text{\tt Factor}\right] \right.
\nonu \\
&&=-
\text{\tt kappa}\left[{\tt \frac{1}{2},0,\frac{3}{2}}\right]\,
\text{\tt kappa}\left[{\tt \frac{3}{2},\frac{1}{2},0}\right]\,
\text{\tt Epsilon}[{\tt 1},\text{\tt AA1},\text{\tt BB1},
\text{\tt CC1}, \text{\tt DD1}, \text{\tt EE1},\text{\tt FF1},{\tt P}]
\nonu \\
&& \times
\text{\tt Phi}\left[\text{\tt h1}+\text{\tt h2}+
\text{\tt h3}, \{ {\tt 1} \},
{\tt -\frac{3}{2}}\right]+ \cdots,
\nonu \\
&& \text{\tt OPESimplify}\left[\text{\tt OPEPole}[\tt 4]\left[\text{\tt
Phi}\left[\text{\tt h1},\{\tt P\},{\tt +\frac{3}{2}}\right],
\text{\tt comp}[[{\tt 23}]]],
\text{\tt Factor}\right] \right.
\nonu \\
&& =-
\text{\tt kappa}\left[{\tt 0,0,2}\right]\,
\text{\tt kappa}\left[{\tt \frac{3}{2},\frac{1}{2},0}\right]\,
\text{\tt Epsilon}[\text{\tt AA1},\text{\tt BB1},
\text{\tt CC1}, \text{\tt DD1}, \text{\tt EE1},\text{\tt FF1},
\text{\tt GG1}, {\tt P}]
\nonu \\
&& \times
\text{\tt Phi}\left[\text{\tt h1}+\text{\tt h2}+
\text{\tt h3},
{\tt -2}\right].
\label{Ctwo}
\eea
There exist the overall minus signs at the thirteen  places
in (\ref{Ctwo}). As before in previous
subsection, these can be absorbed by changing the
locations of $SU(8)$ indices appropriately
according to  the ordering of Appendix A.

\subsection{The fourth order pole of the OPEs
of the graviphotons of helicity $+1$ and the quadratic operators }

\bea
&& \text{\tt OPESimplify}
\left[\text{\tt OPEPole}[{\tt 4}][\text{\tt Phi}[\text{\tt h1},\{{\tt
P,Q}\},{\tt +1}],
\text{\tt comp}[[{\tt 1}]]],
\text{\tt Factor}\right]
\nonu \\
&& =\text{\tt kappa}[{\tt 2,1,-1}]^{\tt 2} \,
\text{\tt Phi}[\text{\tt h1}+\text{\tt h2}+\text{\tt h3},\{{\tt P,Q}\},
{\tt 1}],
\nonu \\
&& \text{\tt OPESimplify}
\left[\text{\tt OPEPole}[{\tt 4}][\text{\tt Phi}[\text{\tt h1},\{{\tt
P,Q}\},{\tt +1}],
\text{\tt comp}[[{\tt 2}]]],
\text{\tt Factor}\right]
\nonu \\
&& =
\text{\tt kappa}\left[{\tt \frac{3}{2},1,-\frac{1}{2}}\right]\,
\text{\tt kappa}[{\tt 2,1,-1}]
\,
\text{\tt Phi}
\left[\text{\tt h1}+\text{\tt h2}+\text{\tt h3},\{{\tt AA1,P,Q}\},
{\tt \frac{1}{2}}\right],
\nonu \\
&& \text{\tt OPESimplify}
\left[\text{\tt OPEPole}[{\tt 4}][\text{\tt Phi}[\text{\tt h1},\{{\tt
P,Q}\},{\tt +1}],
\text{\tt comp}[[{\tt 3}]]],
\text{\tt Factor}\right]
\nonu \\
&& =
\text{\tt kappa}\left[{\tt 1,1,0}\right]\,
\text{\tt kappa}[{\tt 2,1,-1}]
\,
\text{\tt Phi}[\text{\tt h1}+\text{\tt h2}+
\text{\tt h3},\{{\tt AA1,BB1,P,Q}\},
{\tt 0}],
\nonu \\
&& \text{\tt OPESimplify}
\left[\text{\tt OPEPole}[{\tt 4}][\text{\tt Phi}[\text{\tt h1},\{{\tt
P,Q}\},{\tt +1}],
\text{\tt comp}[[{\tt 4}]]],
\text{\tt Factor}\right]
\nonu \\
&& =-
\text{\tt kappa}\left[{\tt 1,\frac{1}{2},\frac{1}{2}}\right]\,
\text{\tt kappa}[{\tt 2,1,-1}]
\,
\text{\tt Epsilon}[{\tt 1,2,3},\text{\tt AA1},\text{\tt BB1},
\text{\tt CC1,P,Q}]
\nonu \\
&& \times
\text{\tt Phi}\left[\text{\tt h1}+\text{\tt h2}+
\text{\tt h3},\{{\tt 1,2,3}\},
{\tt -\frac{1}{2}}\right]+\cdots,
\nonu \\
&& \text{\tt OPESimplify}
\left[\text{\tt OPEPole}[{\tt 4}][\text{\tt Phi}[\text{\tt h1},\{{\tt
P,Q}\},{\tt +1}],
\text{\tt comp}[[{\tt 5}]]],
\text{\tt Factor}\right]
\nonu \\
&& =
\text{\tt kappa}\left[{\tt 1,0,1}\right]\,
\text{\tt kappa}[{\tt 2,1,-1}]
\,
\text{\tt Epsilon}[{\tt 1,2},\text{\tt AA1},\text{\tt BB1},
\text{\tt CC1,DD1,P,Q}]
\nonu \\
&& \times
\text{\tt Phi}[\text{\tt h1}+\text{\tt h2}+
\text{\tt h3},\{{\tt 1,2}\},
{\tt -1}]+\cdots,
\nonu \\
&& \text{\tt OPESimplify}
\left[\text{\tt OPEPole}[{\tt 4}][\text{\tt Phi}[\text{\tt h1},\{{\tt
P,Q}\},{\tt +1}],
\text{\tt comp}[[{\tt 6}]]],
\text{\tt Factor}\right]
\nonu \\
&& =
\text{\tt kappa}\left[{\tt 1,-\frac{1}{2},\frac{3}{2}}\right]\,
\text{\tt kappa}[{\tt 2,1,-1}]
\,
\text{\tt Delta}[\text{\tt BB1},{\tt Q}]\,
\text{\tt Delta}[\text{\tt CC1},{\tt P}]
\nonu \\
&& \times
\text{\tt Phi}\left[\text{\tt h1}+\text{\tt h2}+
\text{\tt h3},\{{\tt AA1}\},
{\tt -\frac{3}{2}}\right]+\cdots,
\nonu \\
&& \text{\tt OPESimplify}
\left[\text{\tt OPEPole}[{\tt 4}][\text{\tt Phi}[\text{\tt h1},\{{\tt
P,Q}\},{\tt +1}],
\text{\tt comp}[[{\tt 7}]]],
\text{\tt Factor}\right]
\nonu \\
&& =-
\text{\tt kappa}\left[{\tt 1,-1,2}\right]\,
\text{\tt kappa}[{\tt 2,1,-1}]
\,
\text{\tt Delta}[\text{\tt AA1},{\tt Q}]\,
\text{\tt Delta}[\text{\tt BB1},{\tt P}]
\nonu \\
&& \times
\text{\tt Phi}[\text{\tt h1}+\text{\tt h2}+
\text{\tt h3},
{\tt -2}]+\cdots,
\nonu \\
&& \text{\tt OPESimplify}
\left[\text{\tt OPEPole}[{\tt 4}][\text{\tt Phi}[\text{\tt h1},\{{\tt
P,Q}\},{\tt +1}],
\text{\tt comp}[[{\tt 10}]]],
\text{\tt Factor}\right]
\nonu \\
&& =
\text{\tt kappa}\left[{\tt \frac{3}{2},\frac{1}{2},0}\right]\,
\text{\tt kappa}\left[{\tt \frac{3}{2},1,-\frac{1}{2}}\right]
\,
\text{\tt Phi}[\text{\tt h1}+\text{\tt h2}+
\text{\tt h3}, \{{\tt AA1, BB1, P,Q }\},
{\tt 0}],
\nonu \\
&& \text{\tt OPESimplify}
\left[\text{\tt OPEPole}[{\tt 4}][\text{\tt Phi}[\text{\tt h1},\{{\tt
P,Q}\},{\tt +1}],
\text{\tt comp}[[{\tt 11}]]],
\text{\tt Factor}\right]
\nonu \\
&& =-
\text{\tt kappa}\left[{\tt 1,\frac{1}{2},\frac{1}{2}}\right]\,
\text{\tt kappa}\left[{\tt \frac{3}{2},1,-\frac{1}{2}}\right]
\,
\text{\tt Epsilon}[{\tt 1,2,3},\text{\tt AA1},
\text{\tt BB1},\text{\tt CC1},{\tt P,Q}]
\nonu \\
&& \times
\text{\tt Phi}\left[\text{\tt h1}+\text{\tt h2}+
\text{\tt h3}, \{{\tt 1, 2, 3}\},
{\tt -\frac{1}{2}}\right] + \cdots,
\nonu \\
&& \text{\tt OPESimplify}
\left[\text{\tt OPEPole}[{\tt 4}][\text{\tt Phi}[\text{\tt h1},\{{\tt
P,Q}\},{\tt +1}],
\text{\tt comp}[[{\tt 12}]]],
\text{\tt Factor}\right]
\nonu \\
&& =
\text{\tt kappa}\left[{\tt \frac{1}{2},\frac{1}{2},1}\right]\,
\text{\tt kappa}\left[{\tt \frac{3}{2},1,-\frac{1}{2}}\right]
\,
\text{\tt Epsilon}[{\tt 1,2},\text{\tt AA1},
\text{\tt BB1},\text{\tt CC1},\text{\tt DD1},{\tt P,Q}]
\nonu \\
&& \times
\text{\tt Phi}[\text{\tt h1}+\text{\tt h2}+
\text{\tt h3}, \{{\tt 1, 2}\},
{\tt -1}] + \cdots,
\nonu \\
&& \text{\tt OPESimplify}
\left[\text{\tt OPEPole}[{\tt 4}][\text{\tt Phi}[\text{\tt h1},\{{\tt
P,Q}\},{\tt +1}],
\text{\tt comp}[[{\tt 13}]]],
\text{\tt Factor}\right]
\nonu \\
&& =-
\text{\tt kappa}\left[{\tt \frac{1}{2},0,\frac{3}{2}}\right]\,
\text{\tt kappa}\left[{\tt \frac{3}{2},1,-\frac{1}{2}}\right]
\,
\text{\tt Epsilon}[{\tt 1},\text{\tt AA1},
\text{\tt BB1},\text{\tt CC1},\text{\tt DD1},\text{\tt EE1}, {\tt P,Q}]
\nonu \\
&& \times
\text{\tt Phi}\left[\text{\tt h1}+\text{\tt h2}+
\text{\tt h3}, \{{\tt 1}\},
{\tt -\frac{3}{2}}\right] + \cdots,
\nonu \\
&& \text{\tt OPESimplify}
\left[\text{\tt OPEPole}[{\tt 4}][\text{\tt Phi}[\text{\tt h1},\{{\tt
P,Q}\},{\tt +1}],
\text{\tt comp}[[{\tt 14}]]],
\text{\tt Factor}\right]
\nonu \\
&& =
\text{\tt kappa}\left[{\tt \frac{1}{2},-\frac{1}{2},2}\right]\,
\text{\tt kappa}\left[{\tt \frac{3}{2},1,-\frac{1}{2}}\right]
\,
\text{\tt Epsilon}[{\tt 1,2,3,4,5},\text{\tt AA1},{\tt P,Q}]
\nonu \\
&& \text{\tt Epsilon}[{\tt 1,2,3,4,5},
\text{\tt BB1},\text{\tt CC1},\text{\tt DD1}]
\,
\text{\tt Phi}[\text{\tt h1}+\text{\tt h2}+
\text{\tt h3},
{\tt -2}] + \cdots,
\nonu \\
&& \text{\tt OPESimplify}
\left[\text{\tt OPEPole}[{\tt 4}][\text{\tt Phi}[\text{\tt h1},\{{\tt
P,Q}\},{\tt +1}],
\text{\tt comp}[[{\tt 17}]]],
\text{\tt Factor}\right]
\nonu \\
&& =
\text{\tt kappa}\left[{\tt 1,0,1}\right]\,
\text{\tt kappa}\left[{\tt 1,1,0}\right]
\,
\text{\tt Epsilon}[{\tt 1,2},\text{\tt AA1},
\text{\tt BB1},
\text{\tt CC1},
\text{\tt DD1},
{\tt P,Q}]
\nonu \\
&& \times
\text{\tt Phi}[\text{\tt h1}+\text{\tt h2}+
\text{\tt h3}, \{{\tt 1,2} \},
{\tt -1}] + \cdots,
\nonu \\
&& \text{\tt OPESimplify}
\left[\text{\tt OPEPole}[{\tt 4}][\text{\tt Phi}[\text{\tt h1},\{{\tt
P,Q}\},{\tt +1}],
\text{\tt comp}[[{\tt 18}]]],
\text{\tt Factor}\right]
\nonu \\
&& =-
\text{\tt kappa}\left[{\tt \frac{1}{2},0,\frac{3}{2}}\right]\,
\text{\tt kappa}\left[{\tt 1,1,0}\right]
\,
\text{\tt Epsilon}[{\tt 1},\text{\tt AA1},\text{\tt BB1},
\text{\tt CC1},\text{\tt DD1},\text{\tt EE1},{\tt P,Q}]
\nonu \\
&& \times
\text{\tt Phi}\left[\text{\tt h1}+\text{\tt h2}+
\text{\tt h3}, \{{\tt 1} \},
{\tt -\frac{3}{2}}\right] + \cdots,
\nonu \\
&& \text{\tt OPESimplify}
\left[\text{\tt OPEPole}[{\tt 4}][\text{\tt Phi}[\text{\tt h1},\{{\tt
P,Q}\},{\tt +1}],
\text{\tt comp}[[{\tt 19}]]],
\text{\tt Factor}\right]
\nonu \\
&& =
\text{\tt kappa}\left[{\tt 0,0,2}\right]\,
\text{\tt kappa}\left[{\tt 1,1,0}\right]
\,
\text{\tt Epsilon}[\text{\tt AA1},\text{\tt BB1},
\text{\tt CC1},\text{\tt DD1},\text{\tt EE1},\text{\tt FF1},{\tt P,Q}]
\nonu \\
&& \times
\text{\tt Phi}[\text{\tt h1}+\text{\tt h2}+
\text{\tt h3},
{\tt -2}],
\nonu \\
&& \text{\tt OPESimplify}
\left[\text{\tt OPEPole}[{\tt 4}][\text{\tt Phi}[\text{\tt h1},\{{\tt
P,Q}\},{\tt +1}],
\text{\tt comp}[[{\tt 22}]]],
\text{\tt Factor}\right]
\nonu \\
&& =-
\text{\tt kappa}\left[{\tt \frac{1}{2},-\frac{1}{2},2}\right]\,
\text{\tt kappa}\left[{\tt 1,\frac{1}{2},\frac{1}{2}}\right]
\,
\text{\tt Epsilon}[{\tt 2,3,4,5,6},\text{\tt DD1},
\text{\tt EE1},\text{\tt FF1}]
\nonu \\
&& \times \text{\tt Epsilon}[{\tt 1,7,8},
\text{\tt AA1},\text{\tt BB1},\text{\tt CC1},{\tt P,Q}]
\,
\text{\tt Phi}[\text{\tt h1}+\text{\tt h2}+
\text{\tt h3},
{\tt -2}] + \cdots.
\label{Cthree}
\eea
The overall minus signs at the six  places
in (\ref{Cthree}) appear and
these can be absorbed by changing the
locations of $SU(8)$ indices appropriately.
At the final relation of (\ref{Cthree}),
in addition to the rearrangement of $SU(8)$ indices,
the property of the coupling between the first two
helicity indices \cite{AK2509}
is used.
The additional epsilon tensor (from Appendix A) is implicit here.

\subsection{The fourth order pole of the OPEs
of the graviphotinos of helicity $+\frac{1}{2}$ and the quadratic
operators }

\bea
&& \text{\tt OPESimplify}\left[\text{\tt OPEPole}[{\tt 4}]
\left[\text{\tt Phi}\left[\text{\tt h1},\{{\tt P,Q,R}\},
+{\tt \frac{1}{2}}\right],
\text{\tt comp}[[{\tt 1}]]],
\text{\tt Factor}
\right] \right.
\nonu \\
&& =
\text{\tt kappa}\left[{\tt 2,\frac{1}{2},-\frac{1}{2}}\right]^{\tt 2}
\text{\tt Phi}\left[\text{\tt h1}+\text{\tt h2}+\text{\tt h3},
\{{ \tt P,Q,R}\},{\tt \frac{1}{2}}\right],
\nonu \\
&& \text{\tt OPESimplify}\left[\text{\tt OPEPole}[{\tt 4}]
\left[\text{\tt Phi}\left[\text{\tt h1},\{{\tt P,Q,R}\},
+{\tt \frac{1}{2}}\right],
\text{\tt comp}[[{\tt 2}]]],
\text{\tt Factor}
\right] \right.
\nonu \\
&& =-
\text{\tt kappa}\left[{\tt \frac{3}{2},\frac{1}{2},0}\right]\,
\text{\tt kappa}\left[{\tt 2,\frac{1}{2},-\frac{1}{2}} \right] \,
\text{\tt Phi}\left[\text{\tt h1}+\text{\tt h2}+\text{\tt h3},
\{{ \tt AA1,P,Q,R}\},{\tt 0}\right],
\nonu \\
&& \text{\tt OPESimplify}\left[\text{\tt OPEPole}[{\tt 4}]
\left[\text{\tt Phi}\left[\text{\tt h1},\{{\tt P,Q,R}\},
+{\tt \frac{1}{2}}\right],
\text{\tt comp}[[{\tt 3}]]],
\text{\tt Factor}
\right] \right.
\nonu \\
&& =-
\text{\tt kappa}\left[{\tt 1,\frac{1}{2},\frac{1}{2}}\right]\,
\text{\tt kappa}\left[{\tt 2,\frac{1}{2},-\frac{1}{2}} \right] \,
\text{\tt Epsilon}[{\tt 1,2,3},\text{\tt AA1},\text{\tt BB1},{\tt
P,Q,R}] \nonu \\
&& \times
\text{\tt Phi}\left[\text{\tt h1}+\text{\tt h2}+\text{\tt h3},
\{{ \tt 1,2,3}\},{\tt -\frac{1}{2}}\right]+\cdots,
\nonu \\
&& \text{\tt OPESimplify}\left[\text{\tt OPEPole}[{\tt 4}]
\left[\text{\tt Phi}\left[\text{\tt h1},\{{\tt P,Q,R}\},
+{\tt \frac{1}{2}}\right],
\text{\tt comp}[[{\tt 4}]]],
\text{\tt Factor}
\right] \right.
\nonu \\
&& =-
\text{\tt kappa}\left[{\tt \frac{1}{2},\frac{1}{2},1}\right]\,
\text{\tt kappa}\left[{\tt 2,\frac{1}{2},-\frac{1}{2}} \right] \,
\text{\tt Epsilon}[{\tt 1,2},
\text{\tt AA1},\text{\tt BB1},\text{\tt CC1}, {\tt
P,Q,R}] \nonu \\
&& \times
\text{\tt Phi}\left[\text{\tt h1}+\text{\tt h2}+\text{\tt h3},
\{{ \tt 1,2}\},{\tt -1}\right]+\cdots,
\nonu \\
&& \text{\tt OPESimplify}\left[\text{\tt OPEPole}[{\tt 4}]
\left[\text{\tt Phi}\left[\text{\tt h1},\{{\tt P,Q,R}\},
+{\tt \frac{1}{2}}\right],
\text{\tt comp}[[{\tt 5}]]],
\text{\tt Factor}
\right] \right.
\nonu \\
&& =-
\text{\tt kappa}\left[{\tt \frac{1}{2},0,\frac{3}{2}}\right]\,
\text{\tt kappa}\left[{\tt 2,\frac{1}{2},-\frac{1}{2}} \right] \,
\text{\tt Epsilon}[{\tt 1},
\text{\tt AA1},\text{\tt BB1},\text{\tt CC1}, \text{\tt DD1}, {\tt
P,Q,R}] \nonu \\
&& \times
\text{\tt Phi}\left[\text{\tt h1}+\text{\tt h2}+\text{\tt h3},
\{{ \tt 1}\},{\tt -\frac{3}{2}}\right]+\cdots,
\nonu \\
&& \text{\tt OPESimplify}\left[\text{\tt OPEPole}[{\tt 4}]
\left[\text{\tt Phi}\left[\text{\tt h1},\{{\tt P,Q,R}\},
+{\tt \frac{1}{2}}\right],
\text{\tt comp}[[{\tt 6}]]],
\text{\tt Factor}
\right] \right.
\nonu \\
&& =
\text{\tt kappa}\left[{\tt \frac{1}{2},-\frac{1}{2},2}\right]\,
\text{\tt kappa}\left[{\tt 2,\frac{1}{2},-\frac{1}{2}} \right] \,
\text{\tt Epsilon}[{\tt 1,2,3,4,5},\text{\tt AA1},\text{\tt BB1},
\text{\tt CC1}]
\nonu \\
&& \text{\tt Epsilon}[{\tt 1,2,3,4,5,P,Q,R}]
\,
\text{\tt Phi}\left[\text{\tt h1}+\text{\tt h2}+\text{\tt h3},
{\tt -2}\right]+\cdots,
\nonu \\
&& \text{\tt OPESimplify}\left[\text{\tt OPEPole}[{\tt 4}]
\left[\text{\tt Phi}\left[\text{\tt h1},\{{\tt P,Q,R}\},
+{\tt \frac{1}{2}}\right],
\text{\tt comp}[[{\tt 10}]]],
\text{\tt Factor}
\right] \right.
\nonu \\
&& =-
\text{\tt kappa}\left[{\tt \frac{3}{2},0,\frac{1}{2}}\right]\,
\text{\tt kappa}\left[{\tt \frac{3}{2},\frac{1}{2},0} \right] \,
\text{\tt Epsilon}[{\tt 1,2,3},\text{\tt AA1},\text{\tt BB1},
\text{\tt P,Q,R}]
\nonu \\
&& \times
\text{\tt Phi}\left[\text{\tt h1}+\text{\tt h2}+\text{\tt h3},
\{{\tt 1,2,3} \},
{\tt -\frac{1}{2}}\right]+\cdots,
\nonu \\
&& \text{\tt OPESimplify}\left[\text{\tt OPEPole}[{\tt 4}]
\left[\text{\tt Phi}\left[\text{\tt h1},\{{\tt P,Q,R}\},
+{\tt \frac{1}{2}}\right],
\text{\tt comp}[[{\tt 11}]]],
\text{\tt Factor}
\right] \right.
\nonu \\
&& =-
\text{\tt kappa}\left[{\tt 1,0,1}\right]\,
\text{\tt kappa}\left[{\tt \frac{3}{2},\frac{1}{2},0} \right] \,
\text{\tt Epsilon}[{\tt 1,2},\text{\tt AA1},\text{\tt BB1},
\text{\tt CC1},\text{\tt P,Q,R}]
\nonu \\
&& \times
\text{\tt Phi}\left[\text{\tt h1}+\text{\tt h2}+\text{\tt h3},
\{{\tt 1,2} \},
{\tt -1}\right]+\cdots,
\nonu \\
&& \text{\tt OPESimplify}\left[\text{\tt OPEPole}[{\tt 4}]
\left[\text{\tt Phi}\left[\text{\tt h1},\{{\tt P,Q,R}\},
+{\tt \frac{1}{2}}\right],
\text{\tt comp}[[{\tt 12}]]],
\text{\tt Factor}
\right] \right.
\nonu \\
&& =-
\text{\tt kappa}\left[{\tt \frac{1}{2},0,\frac{3}{2}}\right]\,
\text{\tt kappa}\left[{\tt \frac{3}{2},\frac{1}{2},0} \right] \,
\text{\tt Epsilon}[{\tt 1},\text{\tt AA1},\text{\tt BB1},
\text{\tt CC1},\text{\tt DD1},\text{\tt P,Q,R}]
\nonu \\
&& \times
\text{\tt Phi}\left[\text{\tt h1}+\text{\tt h2}+\text{\tt h3},
\{{\tt 1} \},
{\tt -\frac{3}{2}}\right]+\cdots,
\nonu \\
&& \text{\tt OPESimplify}\left[\text{\tt OPEPole}[{\tt 4}]
\left[\text{\tt Phi}\left[\text{\tt h1},\{{\tt P,Q,R}\},
+{\tt \frac{1}{2}}\right],
\text{\tt comp}[[{\tt 13}]]],
\text{\tt Factor}
\right] \right.
\nonu \\
&& =-
\text{\tt kappa}\left[{\tt 0,0,2}\right]\,
\text{\tt kappa}\left[{\tt \frac{3}{2},\frac{1}{2},0} \right] \,
\text{\tt Epsilon}[\text{\tt AA1},\text{\tt BB1},
\text{\tt CC1},\text{\tt DD1},\text{\tt EE1},\text{\tt P,Q,R}]
\nonu \\
&& \times
\text{\tt Phi}\left[\text{\tt h1}+\text{\tt h2}+\text{\tt h3},
{\tt -2}\right],
\nonu \\
&& \text{\tt OPESimplify}\left[\text{\tt OPEPole}[{\tt 4}]
\left[\text{\tt Phi}\left[\text{\tt h1},\{{\tt P,Q,R}\},
+{\tt \frac{1}{2}}\right],
\text{\tt comp}[[{\tt 17}]]],
\text{\tt Factor}
\right] \right.
\nonu \\
&& =-
\text{\tt kappa}\left[{\tt 1,-\frac{1}{2},\frac{3}{2}}\right]\,
\text{\tt kappa}\left[{\tt 1,\frac{1}{2},\frac{1}{2}} \right] \,
\text{\tt Delta}[{\tt 3},\text{\tt CC1}] \,
\text{\tt Delta}[{\tt 2},\text{\tt DD1}]
\nonu \\
&& \times \text{\tt Epsilon}[{\tt 1,2,3},\text{\tt AA1},
\text{\tt BB1},{\tt P,Q,R}]
\,
\text{\tt Phi}\left[\text{\tt h1}+\text{\tt h2}+\text{\tt h3},
\{ {\tt 1} \},
{\tt -\frac{3}{2}}\right]+\cdots,
\nonu \\
&& \text{\tt OPESimplify}\left[\text{\tt OPEPole}[{\tt 4}]
\left[\text{\tt Phi}\left[\text{\tt h1},\{{\tt P,Q,R}\},
+{\tt \frac{1}{2}}\right],
\text{\tt comp}[[{\tt 18}]]],
\text{\tt Factor}
\right] \right.
\nonu \\
&& =-
\text{\tt kappa}\left[{\tt \frac{1}{2},-\frac{1}{2},2}\right]\,
\text{\tt kappa}\left[{\tt 1,\frac{1}{2},\frac{1}{2}} \right] \,
\text{\tt Epsilon}[{\tt 2,3,4,5,6},\text{\tt CC1},
\text{\tt DD1},\text{\tt EE1}]
\nonu \\
&& \times \text{\tt Epsilon}[{\tt 1,7,8},
\text{\tt AA1},\text{\tt BB1},{\tt P,Q,R}]
\,
\text{\tt Phi}\left[\text{\tt h1}+\text{\tt h2}+\text{\tt h3},
{\tt -2}\right]+\cdots.
\label{Cfour}
\eea
The overall minus signs at the ten  places
in (\ref{Cfour}) appear and
these can be absorbed by changing the
$SU(8)$ indices appropriately.
The extra minus signs can appear
when the orders of OPEs of the two fermionic operators
are reversed.
The additional epsilon tensor (from Appendix A) is implicit 
at the final relation of (\ref{Cfour}) and 
the property of the coupling between the first two
helicity indices 
is used as before.

\subsection{The fourth order pole of the OPEs
of the scalars of helicity $0$ and the quadratic operators }

\bea
&& \text{\tt OPESimplify}\left[\text{\tt OPEPole}[\tt 4][\text{\tt
Phi}\left[\text{\tt h1},\{{\tt P,Q,R,S}\},{\tt 0}],
\text{\tt comp}[[{\tt 1}]]],
\text{\tt Factor}\right] \right.
\nonu \\
&& =\text{\tt kappa}\left[{\tt 2,0,0} \right]^{\tt 2}
\text{\tt Phi}\left[\text{\tt h1}+\text{\tt h2}+\text{\tt h3},
\{{\tt P,Q,R,S} \},{\tt 0} \right],
\nonu \\
&& \text{\tt OPESimplify}\left[\text{\tt OPEPole}[\tt 4][\text{\tt
Phi}\left[\text{\tt h1},\{{\tt P,Q,R,S}\},{\tt 0}],
\text{\tt comp}[[{\tt 2}]]],
\text{\tt Factor}\right] \right.
\nonu \\
&& =-\text{\tt kappa}\left[{\tt \frac{3}{2},0,\frac{1}{2}} \right]
\,
\text{\tt kappa}\left[{\tt 2,0,0} \right]\,
\text{\tt Epsilon}[{\tt 1,2,3},\text{\tt AA1},{\tt P,Q,R,S}]\,
\nonu \\
&& \times
\text{\tt Phi}\left[\text{\tt h1}+\text{\tt h2}+\text{\tt h3},
\{{\tt 1,2,3} \},{\tt -\frac{1}{2}} \right]+\cdots,
\nonu \\
&& \text{\tt OPESimplify}\left[\text{\tt OPEPole}[\tt 4][\text{\tt
Phi}\left[\text{\tt h1},\{{\tt P,Q,R,S}\},{\tt 0}],
\text{\tt comp}[[{\tt 3}]]],
\text{\tt Factor}\right] \right.
\nonu \\
&& =\text{\tt kappa}\left[{\tt 1,0,1} \right]
\,
\text{\tt kappa}\left[{\tt 2,0,0} \right]\,
\text{\tt Epsilon}[{\tt 1,2},\text{\tt AA1},\text{\tt BB1},{\tt P,Q,R,S}]\,
\nonu \\
&& \times
\text{\tt Phi}\left[\text{\tt h1}+\text{\tt h2}+\text{\tt h3},
\{{\tt 1,2} \},{\tt -1} \right]+\cdots,
\nonu \\
&& \text{\tt OPESimplify}\left[\text{\tt OPEPole}[\tt 4][\text{\tt
Phi}\left[\text{\tt h1},\{{\tt P,Q,R,S}\},{\tt 0}],
\text{\tt comp}[[{\tt 4}]]],
\text{\tt Factor}\right] \right.
\nonu \\
&& =-\text{\tt kappa}\left[{\tt \frac{1}{2},0,\frac{3}{2}} \right]
\,
\text{\tt kappa}\left[{\tt 2,0,0} \right]\,
\text{\tt Epsilon}[{\tt 1},\text{\tt AA1},\text{\tt BB1},
\text{\tt CC1},{\tt P,Q,R,S}]\,
\nonu \\
&& \times
\text{\tt Phi}\left[\text{\tt h1}+\text{\tt h2}+\text{\tt h3},
\{{\tt 1} \},{\tt -\frac{3}{2}} \right]+\cdots,
\nonu \\
&& \text{\tt OPESimplify}\left[\text{\tt OPEPole}[\tt 4][\text{\tt
Phi}\left[\text{\tt h1},\{{\tt P,Q,R,S}\},{\tt 0}],
\text{\tt comp}[[{\tt 5}]]],
\text{\tt Factor}\right] \right.
\nonu \\
&& =\text{\tt kappa}\left[{\tt 0,0,2} \right]
\,
\text{\tt kappa}\left[{\tt 2,0,0} \right]\,
\text{\tt Epsilon}[\text{\tt AA1},\text{\tt BB1},
\text{\tt CC1},\text{\tt DD1}, {\tt P,Q,R,S}]\,
\nonu \\
&& \times
\text{\tt Phi}\left[\text{\tt h1}+\text{\tt h2}+\text{\tt h3},
{\tt -2} \right],
\nonu \\
&& \text{\tt OPESimplify}\left[\text{\tt OPEPole}[\tt 4][\text{\tt
Phi}\left[\text{\tt h1},\{{\tt P,Q,R,S}\},{\tt 0}],
\text{\tt comp}[[{\tt 10}]]],
\text{\tt Factor}\right] \right.
\nonu \\
&& =\text{\tt kappa}\left[{\tt \frac{3}{2},-\frac{1}{2},1} \right]
\,
\text{\tt kappa}\left[{\tt \frac{3}{2},0,\frac{1}{2}} \right]\,
\text{\tt Delta}[{\tt 3},\text{\tt BB1}]
\text{\tt Epsilon}[{\tt 1,2,3},\text{\tt AA1},{\tt P,Q,R,S}]
\nonu \\
&& \times
\text{\tt Phi}\left[\text{\tt h1}+\text{\tt h2}+\text{\tt h3},\
\{{\tt 1,2} \},
{\tt -1} \right]+\cdots,
\nonu \\
&& \text{\tt OPESimplify}\left[\text{\tt OPEPole}[\tt 4][\text{\tt
Phi}\left[\text{\tt h1},\{{\tt P,Q,R,S}\},{\tt 0}],
\text{\tt comp}[[{\tt 11}]]],
\text{\tt Factor}\right] \right.
\nonu \\
&& =-
\text{\tt kappa}\left[{\tt 1,-\frac{1}{2},\frac{3}{2}} \right]
\,
\text{\tt kappa}\left[{\tt \frac{3}{2},0,\frac{1}{2}} \right]\,
\text{\tt Delta}[{\tt 3},\text{\tt BB1}]
\text{\tt Delta}[{\tt 2},\text{\tt CC1}]
\nonu \\
&& \times
\text{\tt Epsilon}[{\tt 1,2,3},\text{\tt AA1},{\tt P,Q,R,S}]
\,
\text{\tt Phi}\left[\text{\tt h1}+\text{\tt h2}+\text{\tt h3},\
\{{\tt 1} \},
{\tt -\frac{3}{2}} \right]+\cdots,
\nonu \\
&& \text{\tt OPESimplify}\left[\text{\tt OPEPole}[\tt 4][\text{\tt
Phi}\left[\text{\tt h1},\{{\tt P,Q,R,S}\},{\tt 0}],
\text{\tt comp}[[{\tt 12}]]],
\text{\tt Factor}\right] \right.
\nonu \\
&& =-
\text{\tt kappa}\left[{\tt \frac{1}{2},-\frac{1}{2},2} \right]
\,
\text{\tt kappa}\left[{\tt \frac{3}{2},0,\frac{1}{2}} \right]\,
\text{\tt Epsilon}[{\tt 2,3,4,5,6},\text{\tt BB1},
\text{\tt CC1},\text{\tt DD1}]
\nonu \\
&& \text{\tt Epsilon}[{\tt 1,7,8},\text{\tt AA1},{\tt P,Q,R,S}]
\,
\text{\tt Phi}\left[\text{\tt h1}+\text{\tt h2}+\text{\tt h3},
{\tt -2} \right]+\cdots,
\nonu \\
&& \text{\tt OPESimplify}\left[\text{\tt OPEPole}[\tt 4][\text{\tt
Phi}\left[\text{\tt h1},\{{\tt P,Q,R,S}\},{\tt 0}],
\text{\tt comp}[[{\tt 17}]]],
\text{\tt Factor}\right] \right.
\nonu \\
&& =-
\text{\tt kappa}\left[{\tt 1,-1,2} \right]
\,
\text{\tt kappa}\left[{\tt 1,0,1} \right]\,
\text{\tt Delta}[{\tt 2},\text{\tt CC1}]
\text{\tt Delta}[{\tt 1},\text{\tt DD1}]
\nonu \\
&& \times
\text{\tt Epsilon}[{\tt 1,2},\text{\tt AA1},\text{\tt BB1},
{\tt P,Q,R,S}]
\, 
\text{\tt Phi}\left[\text{\tt h1}+\text{\tt h2}+\text{\tt h3},
{\tt -2} \right]+\cdots.
\label{Cfive}
\eea
The overall minus signs at the five  places
in (\ref{Cfive}) 
can be absorbed by changing the
$SU(8)$ indices appropriately as done before.
The additional epsilon tensor (from Appendix A) is implicit 
at the eighth relation of (\ref{Cfive}) and 
the property of the coupling between the first two
helicity indices 
is used.

\subsection{The fourth order pole of the OPEs
of the gravitphotinos of helicity $-\frac{1}{2}$ and the quadratic operators }

\bea
&& \text{\tt OPESimplify}\left[\text{\tt OPEPole}[{\tt 4}]
\left[\text{\tt Phi}\left[\text{\tt h1},\{{\tt P,Q,R}\},
{\tt -\frac{1}{2}}\right],\text{\tt comp}[[{\tt 1}]]],
\text{\tt Factor}\right] \right.
\nonu \\
&& = \text{\tt kappa}\left[{\tt 2,-\frac{1}{2},\frac{1}{2}}\right]^{\tt 2}
\text{\tt Phi}\left[\text{\tt h1}+\text{\tt h2}+\text{\tt h3},
\{{\tt P,Q,R}\},{\tt -\frac{1}{2}} \right],
\nonu \\
&& \text{\tt OPESimplify}\left[\text{\tt OPEPole}[{\tt 4}]
\left[\text{\tt Phi}\left[\text{\tt h1},\{{\tt P,Q,R}\},
{\tt -\frac{1}{2}}\right],\text{\tt comp}[[{\tt 2}]]],
\text{\tt Factor}\right] \right.
\nonu \\
&& = -\text{\tt kappa}\left[{\tt \frac{3}{2},-\frac{1}{2},1}\right]\,
\text{\tt kappa}\left[{\tt 2,-\frac{1}{2},\frac{1}{2}}\right]\,
\text{\tt Delta}[\text{\tt AA1},{\tt R}]\,
\text{\tt Phi}\left[\text{\tt h1}+\text{\tt h2}+\text{\tt h3},
\{{\tt P,Q}\},{\tt -1} \right]+\cdots,
\nonu \\
&& \text{\tt OPESimplify}\left[\text{\tt OPEPole}[{\tt 4}]
\left[\text{\tt Phi}\left[\text{\tt h1},\{{\tt P,Q,R}\},
{\tt -\frac{1}{2}}\right],\text{\tt comp}[[{\tt 3}]]],
\text{\tt Factor}\right] \right.
\nonu \\
&& = \text{\tt kappa}\left[{\tt 1,-\frac{1}{2},\frac{3}{2}}\right]\,
\text{\tt kappa}\left[{\tt 2,-\frac{1}{2},\frac{1}{2}}\right]\,
\text{\tt Delta}[\text{\tt AA1},{\tt R}]
\text{\tt Delta}[\text{\tt BB1},{\tt Q}] \,
\nonu \\
&& \times \text{\tt Phi}\left[\text{\tt h1}+\text{\tt h2}+\text{\tt h3},
\{{\tt P}\},{\tt -\frac{3}{2}} \right]+\cdots,
\nonu \\
&& \text{\tt OPESimplify}\left[\text{\tt OPEPole}[{\tt 4}]
\left[\text{\tt Phi}\left[\text{\tt h1},\{{\tt P,Q,R}\},
{\tt -\frac{1}{2}}\right],\text{\tt comp}[[{\tt 4}]]],
\text{\tt Factor}\right] \right.
\nonu \\
&& = -\text{\tt kappa}\left[{\tt \frac{1}{2},-\frac{1}{2},2}\right]\,
\text{\tt kappa}\left[{\tt 2,-\frac{1}{2},\frac{1}{2}}\right]\,
\text{\tt Epsilon}[{\tt 1,2,3,4,5},\text{\tt AA1},
\text{\tt BB1},\text{\tt CC1}]
\nonu \\
&& \times \text{\tt Epsilon}[{\tt 1,2,3,4,5,P,Q,R}]
\,
\text{\tt Phi}\left[\text{\tt h1}+\text{\tt h2}+\text{\tt h3},
{\tt -2} \right]+\cdots,
\nonu \\
&& \text{\tt OPESimplify}\left[\text{\tt OPEPole}[{\tt 4}]
\left[\text{\tt Phi}\left[\text{\tt h1},\{{\tt P,Q,R}\},
{\tt -\frac{1}{2}}\right],\text{\tt comp}[[{\tt 10}]]],
\text{\tt Factor}\right] \right.
\nonu \\
&& = \text{\tt kappa}\left[{\tt \frac{3}{2},-1,\frac{3}{2}}\right]\,
\text{\tt kappa}\left[{\tt \frac{3}{2},-\frac{1}{2},1}\right]\,
\text{\tt Delta}[\text{\tt AA1},{\tt R}]
\text{\tt Delta}[\text{\tt BB1},{\tt Q}]
\nonu \\
&& \times 
\text{\tt Phi}\left[\text{\tt h1}+\text{\tt h2}+\text{\tt h3},
\{{\tt P} \},
{\tt -\frac{3}{2}} \right]+\cdots,
\nonu \\
&& \text{\tt OPESimplify}\left[\text{\tt OPEPole}[{\tt 4}]
\left[\text{\tt Phi}\left[\text{\tt h1},\{{\tt P,Q,R}\},
{\tt -\frac{1}{2}}\right],\text{\tt comp}[[{\tt 11}]]],
\text{\tt Factor}\right] \right.
\nonu \\
&& = \text{\tt kappa}\left[{\tt 1,-1,2}\right]\,
\text{\tt kappa}\left[{\tt \frac{3}{2},-\frac{1}{2},1}\right]\,
\text{\tt Delta}[\text{\tt AA1},{\tt R}]
\text{\tt Delta}[\text{\tt BB1},{\tt Q}]
\text{\tt Delta}[\text{\tt CC1},{\tt P}]
\nonu \\
&& \times 
\text{\tt Phi}\left[\text{\tt h1}+\text{\tt h2}+\text{\tt h3},
{\tt -2} \right]+\cdots.
\label{Csix}
\eea
The overall minus signs at the two  places
in (\ref{Csix}) with the symmetry of the couplings 
can be absorbed by changing the
$SU(8)$ indices appropriately as done before.

\subsection{The fourth order pole of the OPEs
of the graviphotons of helicity $-1$ and the quadratic operators }

\bea
&& \text{\tt OPESimplify}\left[\text{\tt OPEPole}[{\tt 4}]
[\text{\tt Phi}[\text{\tt h1},\{{\tt P,Q}\},{\tt -1}],
\text{\tt comp}[[{\tt 1}]]],
\text{\tt Factor}\right]
\nonu \\
&&
=
\text{\tt kappa}\left[{\tt 2,-1,1}\right]^{\tt 2}
\text{\tt Phi}\left[\text{\tt h1}+\text{\tt h2}+\text{\tt h3},
\{{\tt P,Q}\},-1\right],
\nonu  \\
&& \text{\tt OPESimplify}\left[\text{\tt OPEPole}[{\tt 4}]
[\text{\tt Phi}[\text{\tt h1},\{{\tt P,Q}\},{\tt -1}],
\text{\tt comp}[[{\tt 2}]]],
\text{\tt Factor}\right]
\nonu \\
&&
=-
\text{\tt kappa}\left[{\tt \frac{3}{2},-1,\frac{3}{2}}\right]\,
\text{\tt kappa}\left[{\tt 2,-1,1}\right]\,
\text{\tt Delta}[\text{\tt AA1},{\tt Q}]
\text{\tt Phi}\left[\text{\tt h1}+\text{\tt h2}+\text{\tt h3},
\{{\tt P}\},{\tt -\frac{3}{2}}\right]+ \cdots,
\nonu  \\
&& \text{\tt OPESimplify}\left[\text{\tt OPEPole}[{\tt 4}]
[\text{\tt Phi}[\text{\tt h1},\{{\tt P,Q}\},{\tt -1}],
\text{\tt comp}[[{\tt 3}]]],
\text{\tt Factor}\right]
\nonu \\
&&
=-
\text{\tt kappa}\left[{\tt 1,-1,2}\right]\,
\text{\tt kappa}\left[{\tt 2,-1,1}\right]\,
\text{\tt Delta}[\text{\tt AA1},{\tt Q}]\,
\text{\tt Delta}[\text{\tt BB1},{\tt P}] \nonu \\
&& \times
\text{\tt Phi}\left[\text{\tt h1}+\text{\tt h2}+\text{\tt h3},
{\tt -2}\right]+ \cdots,
\nonu \\
&& \text{\tt OPESimplify}\left[\text{\tt OPEPole}[{\tt 4}]
[\text{\tt Phi}[\text{\tt h1},\{{\tt P,Q}\},{\tt -1}],
\text{\tt comp}[[{\tt 10}]]],
\text{\tt Factor}\right]
\nonu \\
&&
=
\text{\tt kappa}\left[{\tt \frac{3}{2},-\frac{3}{2},2}\right]\,
\text{\tt kappa}\left[{\tt \frac{3}{2},-1,\frac{3}{2}}\right]\,
\text{\tt Delta}[\text{\tt AA1},{\tt Q}]\,
\text{\tt Delta}[\text{\tt BB1},{\tt P}]\,
\nonu \\
&& \times
\text{\tt Phi}\left[\text{\tt h1}+\text{\tt h2}+\text{\tt h3},
{\tt -2}\right]+ \cdots.
\label{Cseven}
\eea
We observe that the overall extra minus signs in (\ref{Cseven})
can be explained by changing the $SU(8)$ indices
according to Appendix A.

\subsection{The fourth order pole of the OPEs
of the gravitinos of helicity $-\frac{3}{2}$ and the quadratic operators }

\bea
&& \text{\tt OPESimplify}\left[\text{\tt OPEPole}[{\tt 4}]
\left[\text{\tt Phi}[\text{\tt h1},\{{\tt P}\},{\tt -\frac{3}{2}}\right],
\text{\tt comp}[[{\tt 1}]]],
\text{\tt Factor}\right]
\nonu \\
&&
=\text{\tt kappa}\left[{\tt 2,-\frac{3}{2},\frac{3}{2}}\right]^{\tt 2}
\text{\tt Phi}\left[\text{\tt h1}+\text{\tt h2}+\text{\tt h3},
\{{\tt P}\},{\tt -\frac{3}{2}}\right],
\nonu \\
&& \text{\tt OPESimplify}\left[\text{\tt OPEPole}[{\tt 4}]
\left[\text{\tt Phi}[\text{\tt h1},\{{\tt P}\},{\tt -\frac{3}{2}}\right],
\text{\tt comp}[[{\tt 2}]]],
\text{\tt Factor}\right]
\nonu \\
&&
=
-\text{\tt Delta}[\text{\tt AA1},{\tt P}] \,
\text{\tt kappa}\left[{\tt \frac{3}{2},-\frac{3}{2},2}\right]\,
\text{\tt kappa}\left[{\tt 2,-\frac{3}{2},\frac{3}{2}}\right] \,
\text{\tt Phi}\left[\text{\tt h1}+\text{\tt h2}+\text{\tt h3},
{\tt -2}\right].
\label{Ceight}
\eea
The final relation of (\ref{Ceight})
is consistent with the corresponding one in Appendix A.

\subsection{The fourth order pole of the OPEs
of the gravitons of helicity $-2$ and the quadratic operators }

\bea
&& \text{\tt OPESimplify}
\left[\text{\tt OPEPole}[{\tt 4}][\text{\tt Phi}[\text{\tt h1},{\tt -2}],
\text{\tt comp}[[{\tt 1}]]],
\text{\tt Factor}\right]
\nonu \\
&& = \text{\tt kappa}[{\tt 2,-2,2}]^2 \, \text{\tt Phi}[\text{\tt h1}+
\text{\tt h2}+\text{\tt h3},{\tt -2}].
\label{Cnine}
\eea
Therefore, we can compare the results of
Appendix A or Appendix B with the results of this Appendix C,
(\ref{Cone})-(\ref{Cnine}).
In particular, the celestial operators and the couplings
can be compared to each other
\footnote{
\label{quadrupleopepole6}
For the quadruple-collinear case,
the sixth-order pole for the soft gravitons with $+2$ helicities,
provides the following result inside the package  
\bea
&& \text{\tt OPESimplify}[\text{\tt OPEPole}[{\tt 6}][\text{\tt Phi}
[\text{\tt h1},{\tt +2}],\text{\tt NO}[\text{\tt Phi}[\text{\tt h2},
{\tt +2}],\text{\tt NO}[\text{\tt Phi}[\text{\tt h3},{\tt +2}],
\text{\tt Phi}[\text{\tt h4},{\tt +2}]]]],\text{\tt Factor}]
\nonu \\
&& =\text{\tt kappa}[{\tt 2,2,-2}]^{\tt 3}
(\text{\tt h1}+{\tt 5} \text{\tt h2}+{\tt 6})
(\text{\tt h1}+\text{\tt h2}+{\tt 3} \text{\tt h3}+{\tt 4})
(\text{\tt h1}+\text{\tt h2}+\text{\tt h3}+\text{\tt h4}+{\tt 2})
\text{\tt Phi}[\text{\tt h1}+\text{\tt h2}+\text{\tt h3}+
\text{\tt h4},{\tt 2}].
\nonu 
\eea
}.

\subsection{The second-order pole of the OPEs
of the gravitons of helicity $+2$ and the quadratic operators }

We can compute the second-order poles of the following OPEs
inside the package for the first twenty-five OPEs
in Appendix A \footnote{This subsection corresponds to the first-order
poles (of the holomorphic coordinates) in Appendix A.}
\bea
&& \text{\tt OPESimplify}[\text{\tt OPEPole}[{\tt 2}][\text{\tt Phi}
[\text{\tt h1},{\tt +2}],
\text{\tt comp}[[{\tt 1}]]]
\nonu \\
&& -(\text{\tt kappa}[{\tt 2,2,-2}]
(\text{\tt h1}+\text{\tt h3}+{\tt 2})
\text{\tt NO}[\text{\tt Phi}[\text{\tt h2},{\tt 2}],
\text{\tt Phi}[\text{\tt h1}+\text{\tt h3},{\tt 2}]]
\nonu \\
&& +\text{\tt kappa}[{\tt 2,2,-2}]
(\text{\tt h1}+\text{\tt h2}+{\tt 2})
\text{\tt NO}[\text{\tt Phi}[\text{\tt h1}+\text{\tt h2},{\tt 2}],
\text{\tt Phi}[\text{\tt h3},{\tt 2}]]),\text{\tt Factor}] {\tt =}
{\tt 0},
\nonu \\
&& \text{\tt OPESimplify}[\text{\tt OPEPole}[{\tt 2}][\text{\tt Phi}
[\text{\tt h1},{\tt +2}],
\text{\tt comp}[[{\tt 2}]]]
\nonu \\
&&-\left({\tt \frac{1}{2}} \text{\tt kappa}
\left[{\tt 2,\frac{3}{2},-\frac{3}{2}}\right]
({\tt 2} \text{\tt h1}+ {\tt 2} \text{\tt h3}+{\tt 3})
\text{\tt NO}\left[\text{\tt Phi}[\text{\tt h2},{\tt 2}],
\text{\tt Phi}\left[\text{\tt h1}+\text{\tt h3},\{\text{\tt AA1}\},
{\tt \frac{3}{2}} \right]\right] \right.
\nonu \\
&& \left. +\text{\tt kappa}[{\tt 2,2,-2}]
(\text{\tt h1}+\text{\tt h2}+{\tt 2})
\text{\tt NO}\left[\text{\tt Phi}[\text{\tt h1}+\text{\tt h2},{\tt 2}],
\text{\tt Phi}\left[\text{\tt h3},\{\text{\tt AA1}\},{\tt \frac{3}{2}}
\right]\right]\right),\text{\tt Factor}] {\tt =} {\tt 0},
\nonu \\
&& \text{\tt OPESimplify}[\text{\tt OPEPole}[{\tt 2}][\text{\tt Phi}
[\text{\tt h1},{\tt +2}],
\text{\tt comp}[[{\tt 3}]]]
\nonu \\
&&
-(\text{\tt kappa}[{\tt 2,1,-1}]
(\text{\tt h1}+\text{\tt h3}+{\tt 1})
\text{\tt NO}[\text{\tt Phi}[\text{\tt h2},{\tt 2}],
\text{\tt Phi}[\text{\tt h1}+\text{\tt h3},\{\text{\tt AA1},
\text{\tt BB1}\},{\tt 1}]]
\nonu \\
&& +\text{\tt kappa}[{\tt 2,2,-2}]
(\text{\tt h1}+\text{\tt h2}+{\tt 2})
\text{\tt NO}[\text{\tt Phi}[\text{\tt h1}+\text{\tt h2},{\tt 2}],
\text{\tt Phi}[\text{\tt h3},\{\text{\tt AA1},\text{\tt BB1}\},{\tt 1}]]),
\text{\tt Factor}]{\tt =} {\tt 0},
\nonu \\
&& \text{\tt OPESimplify}[\text{\tt OPEPole}[{\tt 2}][\text{\tt Phi}
[\text{\tt h1},{\tt +2}],
\text{\tt comp}[[{\tt 4}]]]
-\left({\tt \frac{1}{2}} \text{\tt kappa}\left[{\tt 2,\frac{1}{2},
-\frac{1}{2}}\right] ({\tt 2} \text{\tt h1}+{\tt 2} \text{\tt h3}+{\tt 1})
\right. \nonu \\
&& \times \text{\tt NO}\left[\text{\tt Phi}[\text{\tt h2},{\tt 2}],
\text{\tt Phi}\left[\text{\tt h1}+\text{\tt h3},
\{\text{\tt AA1},\text{\tt BB1},\text{\tt CC1}\},
{ \tt \frac{1}{2}} \right]\right]  + {\tt kappa}[{ \tt 2,2,-2}]
(\text{\tt h1}+\text{\tt h2}+{\tt 2})
\nonu \\
&& \left. \left. \times
\text{\tt NO}\left[\text{\tt Phi}[\text{\tt h1}+\text{\tt h2},{\tt 2}],
\text{\tt Phi}\left[\text{\tt h3},\{\text{\tt AA1},\text{\tt BB1},
\text{\tt CC1}\},{\tt \frac{1}{2}}\right] \right]\right),
\text{\tt Factor} \right] {\tt =} {\tt 0},
\nonu \\
&& \text{\tt OPESimplify}[\text{\tt OPEPole}[{\tt 2}][\text{\tt Phi}
[\text{\tt h1},{\tt +2}],
\text{\tt comp}[[{\tt 5}]]]
\nonu \\
&&
-\left(\text{\tt kappa}[{\tt 2,0,0}] (\text{\tt h1}+\text{\tt h3})
\text{\tt NO}[\text{\tt Phi}[\text{\tt h2},{\tt 2}],
\text{\tt Phi}[\text{\tt h1}+\text{\tt h3},
\{\text{\tt AA1},\text{\tt BB1},\text{\tt CC1},\text{\tt DD1}\},
{\tt 0}]] \right. \nonu \\
&& \left.
+\text{\tt kappa}[{\tt 2,2,-2}]
(\text{\tt h1}+\text{\tt h2}+{\tt 2})
\text{\tt NO}[\text{\tt Phi}(\text{\tt h1}+\text{\tt h2},{\tt 2}],
\text{\tt Phi}[\text{\tt h3},\{\text{\tt AA1},\text{\tt BB1},
\text{\tt CC1},\text{\tt DD1}\},{\tt 0}]] \right),
\text{\tt Factor}] \nonu \\
&& {\tt =} {\tt 0},
\nonu \\
&& \text{\tt OPESimplify}[\text{\tt OPEPole}[{\tt 2}][\text{\tt Phi}
[\text{\tt h1},{\tt +2}],
\text{\tt comp}[[{\tt 6}]]]
-\left({\tt \frac{1}{2}} \text{\tt kappa}
\left[
{\tt   2,-\frac{1}{2},\frac{1}{2}}\right]
({\tt 2} \text{\tt h1}+ { \tt 2} \text{\tt h3}-{\tt 1})
\right. \nonu \\
&& \times
\text{\tt NO}
\left[\text{\tt Phi}[\text{\tt h2},{\tt 2}],
\text{\tt Phi}
\left[\text{\tt h1}+\text{\tt h3},\{\text{\tt AA1},
\text{\tt BB1},\text{\tt CC1}\},{\tt -\frac{1}{2}}
\right] \right]
+\text{\tt kappa}[{\tt 2,2,-2}] (\text{\tt h1}+\text{\tt h2}+{\tt 2})
\nonu \\
&& \left. \left. \times
\text{\tt NO}
\left[\text{\tt Phi}[\text{\tt h1}+\text{\tt h2},{\tt 2}],
\text{\tt Phi}
\left[\text{\tt h3},\{\text{\tt AA1},\text{\tt BB1},
\text{\tt CC1}\},{\tt -\frac{1}{2}}\right]\right] \right),
\text{\tt Factor}\right]  {\tt =} {\tt 0},
\nonu \\
&& \text{\tt OPESimplify}[\text{\tt OPEPole}[{\tt 2}][\text{\tt Phi}
[\text{\tt h1},{\tt +2}],
\text{\tt comp}[[{\tt 7}]]]
\nonu \\
&&
-(\text{\tt kappa}[{\tt 2,-1,1}]
(\text{\tt h1}+\text{\tt h3}-{\tt 1})
\text{\tt NO}[\text{\tt Phi}[\text{\tt h2},{\tt 2}],
\text{\tt Phi}(\text{\tt h1}+\text{\tt h3},
\{\text{\tt AA1},\text{\tt BB1}\},{\tt -1}]]
\nonu \\
&& +\text{\tt kappa}[{\tt 2,2,-2}]
(\text{\tt h1}+\text{\tt h2}+{\tt 2})
\text{\tt NO}[\text{\tt Phi}[\text{\tt h1}+\text{\tt h2},{\tt 2}],
\text{\tt Phi}[\text{\tt h3},\{\text{\tt AA1},\text{\tt BB1}\},{\tt -1}
]] ),\text{\tt Factor}] {\tt =} {\tt 0},
\nonu \\
&& \text{\tt OPESimplify}[\text{\tt OPEPole}[{\tt 2}][\text{\tt Phi}
[\text{\tt h1},{\tt +2}],
\text{\tt comp}[[{\tt 8}]]]
\nonu \\
&&
-\left({\tt \frac{1}{2}} \text{\tt kappa}
\left[{\tt 2,-\frac{3}{2},\frac{3}{2}}\right]
({\tt 2} \text{\tt h1}+{\tt 2} \text{\tt h3}-{\tt 3})
\text{\tt NO}\left[\text{\tt Phi}[\text{\tt h2},{\tt 2}],
\text{\tt Phi}\left[\text{\tt h1}+\text{\tt h3},
\{\text{\tt AA1}\},{\tt -\frac{3}{2}}\right]\right]
\right.  \nonu \\
&& \left. \left. +\text{\tt kappa}[{\tt 2,2,-2}]
(\text{\tt h1}+\text{\tt h2}+{\tt 2})
\text{\tt NO}
\left[\text{\tt Phi}[\text{\tt h1}+\text{\tt h2},{\tt 2}],
\text{\tt Phi}
\left[\text{\tt h3},\{\text{\tt AA1}\},
{\tt -\frac{3}{2}} \right]\right]
\right),\text{\tt Factor}\right] {\tt =} {\tt 0},
\nonu \\
&& \text{\tt OPESimplify}[\text{\tt OPEPole}[{\tt 2}][\text{\tt Phi}
[\text{\tt h1},{\tt +2}],
\text{\tt comp}[[{\tt 9}]]]
\nonu \\
&&
-(\text{\tt kappa}[{\tt 2,-2,2}]
(\text{\tt h1}+\text{\tt h3}-{\tt 2})
\text{\tt NO}[\text{\tt Phi}[\text{\tt h2},{\tt 2}],
\text{\tt Phi}[\text{\tt h1}+\text{\tt h3},{\tt -2}]]
\nonu \\
&&  +\text{\tt kappa}[{\tt 2,2,-2}]
[\text{\tt h1}+\text{\tt h2}+{\tt 2}]
\text{\tt NO}[\text{\tt Phi}[\text{\tt h1}+\text{\tt h2},{\tt 2}],
\text{\tt Phi}[\text{\tt h3},{\tt -2}]]),
\text{\tt Factor}] {\tt =} {\tt 0},
\nonu \\
&& \text{\tt OPESimplify}[\text{\tt OPEPole}[{\tt 2}][\text{\tt Phi}
[\text{\tt h1},{\tt +2}],
\text{\tt comp}[[{\tt 10}]]]
-\left({\tt \frac{1}{2}} \text{\tt kappa}\left[{\tt 2,\frac{3}{2},
-\frac{3}{2}}\right] ({\tt 2} \text{\tt h1}+{\tt 2} \text{\tt h3}+
{\tt 3}) \right.
\nonu \\
&& \times
\text{\tt NO}\left[\text{\tt Phi}\left[\text{\tt h2},
\{\text{\tt AA1}\},{\tt \frac{3}{2}}\right],
\text{\tt Phi}\left[\text{\tt h1}+\text{\tt h3},
\{\text{\tt BB1}\},{\tt \frac{3}{2}}\right]\right] 
\nonu \\
&&    +{\tt \frac{1}{2}} \text{\tt kappa}
\left[{\tt 2,\frac{3}{2},-\frac{3}{2}}\right]
({\tt 2} \text{\tt h1}+{\tt 2} \text{\tt h2}+{\tt 3})
\nonu \\
&& \left. \left. \times
\text{\tt NO}\left[\text{\tt Phi}\left[\text{\tt h1}+\text{\tt h2},
\{\text{\tt AA1}\},{\tt \frac{3}{2}}\right],
\text{\tt Phi}\left[\text{\tt h3},\{\text{\tt BB1}\},
{\tt \frac{3}{2}} \right]\right]\right),  \text{\tt Factor}\right]
{\tt = } { \tt 0},
\nonu \\
&& \text{\tt OPESimplify}[\text{\tt OPEPole}[{\tt 2}][\text{\tt Phi}
[\text{\tt h1},{\tt +2}],
\text{\tt comp}[[{\tt 11}]]]
\nonu    \\
&&
-\left(\text{\tt kappa}[{\tt 2,1,-1}]
(\text{\tt h1}+\text{\tt h3}+{\tt 1})
\text{\tt NO}\left[\text{\tt Phi}\left[\text{\tt h2},
\{\text{\tt AA1}\},{\tt \frac{3}{2}}\right],
\text{\tt Phi}[\text{\tt h1}+\text{\tt h3},\{\text{\tt BB1},
\text{\tt CC1}\},{\tt 1}]\right] \right.
\nonu \\
&&
+{\tt \frac{1}{2}} \text{\tt kappa}\left[{\tt 2,\frac{3}{2},
-\frac{3}{2}}\right]
({\tt 2} \text{\tt h1}+{\tt 2} \text{\tt h2}+{\tt 3})
\nonu \\
&&
\left. \left.
\times \text{\tt NO}\left[\text{\tt Phi}\left[\text{\tt h1}
+\text{\tt h2},\{\text{\tt AA1}\},{\tt \frac{3}{2}}\right],
\text{\tt Phi}[\text{\tt h3},\{\text{\tt BB1},\text{\tt CC1}\},
{\tt 1}]\right]\right),
\text{\tt Factor}\right] {\tt =} {\tt 0},
\nonu \\
&& \text{\tt OPESimplify}[\text{\tt OPEPole}[{\tt 2}][\text{\tt Phi}
[\text{\tt h1},{\tt +2}],
\text{\tt comp}[[{\tt 12}]]]
-\left({\tt \frac{1}{2}} \text{\tt kappa}
\left[{\tt 2,\frac{1}{2},-\frac{1}{2}}\right]
({\tt 2} \text{\tt h1}+{\tt 2} \text{\tt h3}+{\tt 1})
\right.
\nonu \\
&& \times \text{\tt NO}\left[\text{\tt Phi}
\left[\text{\tt h2},\{\text{\tt AA1}\},{\tt \frac{3}{2}}\right],
\text{\tt Phi}\left[\text{\tt h1}+\text{\tt h3},
\{\text{\tt BB1},\text{\tt CC1},\text{\tt DD1}\},{\tt \frac{1}{2}}
\right]\right] \nonu \\
&& 
+{\tt \frac{1}{2}} \text{\tt kappa}
\left[{\tt 2,\frac{3}{2},-\frac{3}{2}}\right]
({\tt 2} \text{\tt h1}+{\tt 2} \text{\tt h2}+{\tt 3})
\nonu \\
&& \left. \left. \times 
\text{\tt NO}\left[\text{\tt Phi}
\left[\text{\tt h1}+\text{\tt h2},\{\text{\tt AA1}\},
{\tt \frac{3}{2}} \right],
\text{\tt Phi}\left[\text{\tt h3},\{\text{\tt BB1},\text{\tt CC1},
\text{\tt DD1}\},{\tt \frac{1}{2}} \right]\right]\right),
\text{\tt Factor}\right] {\tt =} {\tt 0},
\nonu \\
&& \text{\tt OPESimplify}[\text{\tt OPEPole}[{\tt 2}][\text{\tt Phi}
[\text{\tt h1},{\tt +2}],
\text{\tt comp}[[{\tt 13}]]] \nonu \\
&&-\left(\text{\tt kappa}[{\tt 2,0,0}]
(\text{\tt h1}+\text{\tt h3})
\text{\tt NO}\left[\text{\tt Phi}
\left[\text{\tt h2},\{\text{\tt AA1}\},{\tt \frac{3}{2}}\right],
\text{\tt Phi}[\text{\tt h1}+\text{\tt h3},
\{\text{\tt BB1},\text{\tt CC1},\text{\tt DD1},\text{\tt EE1}\},{\tt 0}]
\right] \right. \nonu \\
&&  +{\tt \frac{1}{2}} \text{\tt kappa}\left[{\tt
2,\frac{3}{2},-\frac{3}{2}}\right]
({\tt 2} \text{\tt h1}+{\tt 2} \text{\tt h2}+{\tt 3})
\nonu \\
&& \left. \left.
\times \text{\tt NO}\left[\text{\tt Phi}\left[\text{\tt h1}+\text{\tt h2},
\{\text{\tt AA1}\},{\tt \frac{3}{2}}\right],
\text{\tt Phi}[\text{\tt h3},\{\text{\tt BB1},\text{\tt CC1},
\text{\tt DD1},\text{\tt EE1}\},{\tt 0}]\right]\right),
\text{\tt Factor}\right] {\tt =} {\tt 0},
\nonu \\
&& \text{\tt OPESimplify}[\text{\tt OPEPole}[{\tt 2}][\text{\tt Phi}
[\text{\tt h1},{\tt +2}],
\text{\tt comp}[[{\tt 14}]]]
-\left({\tt \frac{1}{2}} \text{\tt kappa}
\left[{\tt 2,-\frac{1}{2},\frac{1}{2}}\right]
({\tt 2} \text{\tt h1}+{\tt 2} \text{\tt h3}-{\tt 1}) \right.
\nonu \\ && \times
\text{\tt NO}\left[\text{\tt Phi}\left[\text{\tt h2},
\{\text{\tt AA1}\},{\tt \frac{3}{2}}\right],
\text{\tt Phi}\left[\text{\tt h1}+\text{\tt h3},
\{\text{\tt BB1},\text{\tt CC1},\text{\tt DD1}\},
{\tt -\frac{1}{2}}\right]\right] \nonu \\
&& +{\tt \frac{1}{2}}
\text{\tt kappa}\left[{\tt 2,\frac{3}{2},-\frac{3}{2}}\right]
({\tt 2} \text{\tt h1}+{\tt 2} \text{\tt h2}+{\tt 3})
\nonu \\
&&  \left. \left. \times
\text{\tt NO}\left[\text{\tt Phi}\left[\text{\tt h1}+\text{\tt h2},
\{\text{\tt AA1}\},{\tt \frac{3}{2}}\right],
\text{\tt Phi}\left[\text{\tt h3},\{\text{\tt BB1},
\text{\tt CC1},\text{\tt DD1}\},{\tt -\frac{1}{2}}\right]\right]\right),
\text{\tt Factor}\right] {\tt =} {\tt 0},
\nonu \\
&& \text{\tt OPESimplify}[\text{\tt OPEPole}[{\tt 2}][\text{\tt Phi}
[\text{\tt h1},{\tt +2}],
\text{\tt comp}[[{\tt 15}]]]
\nonu \\
&&-\left(\text{\tt kappa}[{\tt 2,-1,1}]
(\text{\tt h1}+\text{\tt h3}-{\tt 1})
\text{\tt NO}\left[\text{\tt Phi}\left[\text{\tt h2},
\{\text{\tt AA1}\},{\tt \frac{3}{2}}\right],
\text{\tt Phi}[\text{\tt h1}+\text{\tt h3},
\{\text{\tt BB1},\text{\tt CC1}\},{\tt -1}]\right] \right.\nonu \\
&&+{\tt \frac{1}{2}} \text{\tt kappa}
\left[{\tt 2,\frac{3}{2},-\frac{3}{2}}\right]
({\tt 2} \text{\tt h1}+{\tt 2} \text{\tt h2}+{\tt 3})
\nonu \\
&& \left. \left.
\times \text{\tt NO}\left[\text{\tt Phi}\left[\text{\tt h1}+\text{\tt h2},
\{\text{\tt AA1}\},{\tt \frac{3}{2}}\right],
\text{\tt Phi}[\text{\tt h3},\{\text{\tt BB1},\text{\tt CC1}\},{\tt -1}]
\right]\right),\text{\tt Factor}\right] {\tt =}{\tt 0 },
\nonu  \\
&& \text{\tt OPESimplify}[\text{\tt OPEPole}[{\tt 2}][\text{\tt Phi}
[\text{\tt h1},{\tt +2}],
\text{\tt comp}[[{\tt 16}]]]
\nonu \\
&&-\left({\tt \frac{1}{2}} \text{\tt kappa}
\left[{\tt 2,-\frac{3}{2},\frac{3}{2}}\right]
({\tt 2} \text{\tt h1}+{\tt 2} \text{\tt h3}-{\tt 3})
\text{\tt NO}\left[\text{\tt Phi}\left[\text{\tt h2},\{\text{\tt AA1}\},
{\tt \frac{3}{2}} \right],
 \text{\tt Phi}\left[\text{\tt h1}+\text{\tt h3},
\{\text{\tt BB1}\},{\tt -\frac{3}{2}}\right]\right] \right.\nonu \\
&& +{\tt \frac{1}{2}} \text{\tt kappa}
\left[{\tt 2,\frac{3}{2},-\frac{3}{2}}\right]
({\tt 2} \text{\tt h1}+{\tt 2} \text{\tt h2}+{\tt 3})
\nonu \\
&& \left. \left.  \times
\text{\tt NO}\left[\text{\tt Phi}
\left[\text{\tt h1}+\text{\tt h2},\{\text{\tt AA1}\},
{\tt \frac{3}{2}}\right],
\text{\tt Phi}\left[\text{\tt h3},\{\text{\tt BB1}\},
{\tt -\frac{3}{2}}\right]\right]\right),
\text{\tt Factor}\right] {\tt =}{\tt 0},
\nonu \\
&& \text{\tt OPESimplify}[\text{\tt OPEPole}[{\tt 2}][\text{\tt Phi}
[\text{\tt h1},{\tt +2}],
\text{\tt comp}[[{\tt 17}]]]
\nonu \\
&& -(\text{\tt kappa}[{\tt 2,1,-1}]
(\text{\tt h1}+\text{\tt h3}+{\tt 1})
\text{\tt NO}[\text{\tt Phi}[\text{\tt h2},
\{\text{\tt AA1},\text{\tt BB1}\},{\tt 1}],
\text{\tt Phi}[\text{\tt h1}+\text{\tt h3},
\{\text{\tt CC1},\text{\tt DD1}\},{\tt 1}]]
\nonu \\
&& +\text{\tt kappa}[{\tt 2,1,-1}]
(\text{\tt h1}+\text{\tt h2}+{\tt 1})
\nonu \\
&& \times
\text{\tt NO}[\text{\tt Phi}[\text{\tt h1}+\text{\tt h2},
\{\text{\tt AA1},\text{\tt BB1}\},{\tt 1}],
\text{\tt Phi}[\text{\tt h3},\{\text{\tt CC1},\text{\tt DD1}\},
{\tt 1}]]),\text{\tt Factor}] {\tt =}{\tt 0},
\nonu  \\
&& \text{\tt OPESimplify}[\text{\tt OPEPole}[{\tt 2}][\text{\tt Phi}
[\text{\tt h1},{\tt +2}],
\text{\tt comp}[[{\tt 18}]]]
-\left({\tt \frac{1}{2}} \text{\tt kappa}
\left[{\tt 2,\frac{1}{2},-\frac{1}{2}}\right]
({\tt 2} \text{\tt h1}+{\tt 2} \text{\tt h3}+{\tt 1})
\right.
\nonu \\
&& \times \text{\tt NO}\left[\text{\tt Phi}[\text{\tt h2},\{\text{\tt AA1},
\text{\tt BB1}\},{\tt 1}],
\text{\tt Phi}\left[\text{\tt h1}+\text{\tt h3},
\{\text{\tt CC1},\text{\tt DD1},\text{\tt EE1}\},
{\tt \frac{1}{2}} \right]\right]  \nonu \\
&&+\text{\tt kappa}[{\tt 2,1,-1}]
(\text{\tt h1}+\text{\tt h2}+{\tt 1})
\nonu \\
&& \left. \left. \times 
\text{\tt NO}\left[\text{\tt Phi}[\text{\tt h1}+\text{\tt h2},
\{\text{\tt AA1},\text{\tt BB1}\},{\tt 1}],
\text{\tt Phi}\left[\text{\tt h3},\{\text{\tt CC1},\text{\tt DD1},
 \text{\tt EE1}\},{\tt \frac{1}{2}} \right]\right]\right),
\text{\tt Factor}\right] {\tt =}{\tt 0},
\nonu \\
&& \text{\tt OPESimplify}[\text{\tt OPEPole}[{\tt 2}][\text{\tt Phi}
[\text{\tt h1},{\tt +2}],
\text{\tt comp}[[{\tt 19}]]]
\nonu \\
&&  -(\text{\tt kappa}[{\tt 2,0,0}]
(\text{\tt h1}+\text{\tt h3})
\text{\tt NO}[\text{\tt Phi}[\text{\tt h2},\{\text{\tt AA1},
\text{\tt BB1}\},{\tt 1}],
\text{\tt Phi}[\text{\tt h1}+\text{\tt h3},
\{\text{\tt CC1},\text{\tt DD1},\text{\tt EE1},\text{\tt FF1}\},{\tt 0}]]
\nonu \\
&& +\text{\tt kappa}[{\tt 2,1,-1}]
(\text{\tt h1}+\text{\tt h2}+{\tt 1})
\nonu \\
&&  \times
\text{\tt NO}[\text{\tt Phi}[\text{\tt h1}+\text{\tt h2},
\{\text{\tt AA1},\text{\tt BB1}\},{\tt 1}],
\text{\tt Phi}[\text{\tt h3},\{\text{\tt CC1},\text{\tt DD1},
\text{\tt EE1},\text{\tt FF1}\},{\tt 0}]]),
\text{\tt Factor}] {\tt =}{\tt 0},
\nonu \\
&& \text{\tt OPESimplify}[\text{\tt OPEPole}[{\tt 2}][\text{\tt Phi}
[\text{\tt h1},{\tt +2}],
\text{\tt comp}[[{\tt 20}]]]
\nonu \\
&&-\left({\tt \frac{1}{2}} \text{\tt kappa}
\left[{\tt 2,-\frac{1}{2},\frac{1}{2}}\right]
({\tt 2} \text{\tt h1}+{\tt 2} \text{\tt h3}-{\tt 1})
\right. \nonu \\
&& \times
\text{\tt NO}\left[\text{\tt Phi}[\text{\tt h2},\{\text{\tt AA1},
\text{\tt BB1}\},{\tt 1}],\text{\tt Phi}\left[\text{\tt h1}+
\text{\tt h3},\{\text{\tt CC1},\text{\tt DD1},\text{\tt EE1}\},
{\tt -\frac{1}{2}} \right]\right] \nonu \\
&& +\text{\tt kappa}[{\tt 2,1,-1}] (\text{\tt h1}+\text{\tt h2}+{\tt 1})
\nonu \\
&& \left. \left.
\times \text{\tt NO}\left[\text{\tt Phi}[\text{\tt h1}+\text{\tt h2},
\{\text{\tt AA1},\text{\tt BB1}\},{\tt 1}],
\text{\tt Phi}\left[\text{\tt h3},\{\text{\tt CC1},
\text{\tt DD1},\text{\tt EE1}\},{\tt -\frac{1}{2}}\right]\right]
\right),\text{\tt Factor}\right] {\tt = }{\tt 0},
\nonu \\
&& \text{\tt OPESimplify}[\text{\tt OPEPole}[{\tt 2}][\text{\tt Phi}
[\text{\tt h1},{\tt +2}],
\text{\tt comp}[[{\tt 21}]]]
\nonu \\
&& -(\text{\tt kappa}[{\tt 2,-1,1}]
(\text{\tt h1}+\text{\tt h3}-{\tt 1})
\text{\tt NO}[\text{\tt Phi}[\text{\tt h2},\{\text{\tt AA1},
\text{\tt BB1}\},{\tt 1}],\text{\tt Phi}[\text{\tt h1}+
\text{\tt h3},\{\text{\tt CC1},\text{\tt DD1}\},{\tt -1}]]
\nonu \\
&&+\text{\tt kappa}[{\tt 2,1,-1}]
(\text{\tt h1}+\text{\tt h2}+{\tt 1})
\nonu \\
&& \times \text{\tt NO}[\text{\tt Phi}[\text{\tt h1}+\text{\tt h2},
\{\text{\tt AA1},\text{\tt BB1}\},{\tt 1}],
\text{\tt Phi}[\text{\tt h3},\{\text{\tt CC1},\text{\tt DD1}\},
{\tt -1}]]),\text{\tt Factor}] {\tt =}{\tt 0},
\nonu \\
&& \text{\tt OPESimplify}[\text{\tt OPEPole}[{\tt 2}][\text{\tt Phi}
[\text{\tt h1},{\tt +2}],
\text{\tt comp}[[{\tt 22}]]]
-\left({\tt \frac{1}{2}} \text{\tt kappa}
\left[{\tt 2,\frac{1}{2},-\frac{1}{2}}\right]
({\tt 2} \text{\tt h1}+{\tt 2} \text{\tt h3}+{\tt 1})
\right.
\nonu \\
&& \times \text{\tt NO}\left[\text{\tt Phi}\left[\text{\tt h2},
\{\text{\tt AA1},\text{\tt BB1},\text{\tt CC1}\},
{\tt \frac{1}{2}}\right],
\text{\tt Phi}\left[\text{\tt h1}+\text{\tt h3},
\{\text{\tt DD1},\text{\tt EE1},\text{\tt FF1}\},
{\tt \frac{1}{2}} \right]\right]
\nonu \\
&& +{\tt \frac{1}{2}} \text{\tt kappa}\left[{\tt 2,\frac{1}{2},
-\frac{1}{2}}\right]
({\tt 2} \text{\tt h1}+{\tt 2} \text{\tt h2}+{\tt 1})
\nonu \\
&& \left. \left. \times \text{\tt NO}\left[\text{\tt Phi}\left[\text{\tt h1}
+\text{\tt h2},\{\text{\tt AA1},\text{\tt BB1},\text{\tt CC1}\},
{\tt \frac{1}{2}}\right],
\text{\tt Phi}\left[\text{\tt h3},\{\text{\tt DD1},\text{\tt EE1},
\text{\tt FF1}\},{\tt \frac{1}{2}}\right]\right]\right),
\text{\tt Factor}\right] {\tt =}{\tt 0},
\nonu   \\
&& \text{\tt OPESimplify}[\text{\tt OPEPole}[{\tt 2}][\text{\tt Phi}
[\text{\tt h1},{\tt +2}],
\text{\tt comp}[[{\tt 23}]]]
-\left(\text{\tt kappa}[{\tt 2,0,0}]
(\text{\tt h1}+\text{\tt h3}) \right.
\nonu \\
&& \times \text{\tt NO}\left[\text{\tt Phi}
\left[\text{\tt h2},\{\text{\tt AA1},\text{\tt BB1},\text{\tt CC1}\},
{\tt \frac{1}{2}} \right],
\text{\tt Phi}[\text{\tt h1}+\text{\tt h3},\{\text{\tt DD1},
\text{\tt EE1},\text{\tt FF1},\text{\tt GG1}\},{\tt 0}]\right]
\nonu \\
&& +{\tt \frac{1}{2}}
\text{\tt kappa}\left[{\tt 2,\frac{1}{2},-\frac{1}{2}}\right]
({\tt 2} \text{\tt h1}+{\tt 2} \text{\tt h2}+{\tt 1})
\nonu \\
&&\left. \left.
\times \text{\tt NO}\left[\text{\tt Phi}\left[\text{\tt h1}+\text{\tt h2},
\{\text{\tt AA1},\text{\tt BB1},\text{\tt CC1}\},
{\tt \frac{1}{2}} \right],
\text{\tt Phi}[\text{\tt h3},\{\text{\tt DD1},\text{\tt EE1},
\text{\tt FF1},\text{\tt GG1}\},{\tt 0}]\right]\right),
\text{\tt Factor}\right] {\tt =}{\tt 0},
\nonu  \\
&& \text{\tt OPESimplify}[\text{\tt OPEPole}[{\tt 2}][\text{\tt Phi}
[\text{\tt h1},{\tt +2}],
\text{\tt comp}[[{\tt 24}]]]
-\left(\frac{1}{2}
\text{\tt kappa}\left[{\tt 2,-\frac{1}{2},\frac{1}{2}}\right]
({\tt 2} \text{\tt h1}+{\tt 2} \text{\tt h3}-{\tt 1})
\right.
\nonu \\
&& \times \text{\tt NO}\left[\text{\tt Phi}\left[\text{\tt h2},
\{\text{\tt AA1},\text{\tt BB1},\text{\tt CC1}\},
{\tt \frac{1}{2}} \right],
\text{\tt Phi}\left[\text{\tt h1}+\text{\tt h3},
\{\text{\tt DD1},\text{\tt EE1},\text{\tt
FF1}\},{ \tt -\frac{1}{2}} \right]\right]
\nonu \\
&& +{\tt \frac{1}{2}} \text{\tt kappa}\left[{\tt 2,\frac{1}{2},
-\frac{1}{2}}\right]
({\tt 2} \text{\tt h1}+{\tt 2} \text{\tt h2}+{\tt 1})
\nonu \\
&& \left. \left. \times
\text{\tt NO}\left[\text{\tt Phi}\left[\text{\tt h1}+\text{\tt h2},
\{\text{\tt AA1},\text{\tt BB1},\text{\tt CC1}\},
{\tt \frac{1}{2}}\right],
\text{\tt Phi}\left[\text{\tt h3},\{\text{\tt DD1},
\text{\tt EE1},\text{\tt FF1}\},
{\tt -\frac{1}{2}} \right]\right]\right),
\text{\tt Factor}\right] {\tt =}{\tt 0},
\nonu \\
&& \text{\tt OPESimplify}[\text{\tt OPEPole}[{\tt 2}][\text{\tt Phi}
[\text{\tt h1},{\tt +2}],
\text{\tt comp}[[{\tt 25}]]]
-(\text{\tt kappa}[{\tt 2,0,0}]
(\text{\tt h1}+\text{\tt h3})
\nonu \\
&& \times \text{\tt NO}[\text{\tt Phi}[\text{\tt h2},
\{\text{\tt AA1},\text{\tt BB1},\text{\tt CC1},
\text{\tt DD1}\},{\tt 0}],
\text{\tt Phi}[\text{\tt h1}+\text{\tt h3},
\{\text{\tt EE1},\text{\tt FF1},\text{\tt GG1},\text{\tt HH1}\},
{\tt 0}]]\nonu \\
&& +\text{\tt kappa}[{\tt 2,0,0}]
(\text{\tt h1}+\text{\tt h2})
\label{Second} 
\\
&& \times \text{\tt NO}[\text{\tt Phi}[\text{\tt h1}+\text{\tt h2},
\{\text{\tt AA1},\text{\tt BB1},\text{\tt CC1},\text{\tt DD1}\},
{\tt 0}],
\text{\tt Phi}[\text{\tt h3},\{\text{\tt EE1},\text{\tt FF1},
\text{\tt GG1},\text{\tt HH1}\},{\tt 0}]]),\text{\tt Factor}]
{\tt =}{\tt 0}.
\nonu
\eea
In (\ref{Second}), we have calculated the subtractions
between the second-order poles and the correct
expressions appearing on the right-hand sides.
The reason for this is that
when we compute the second-order poles directly,
then there are derivative terms due to the
ordering of the two operators on the right-hand sides.
Inside the package, 
these derivative terms appear inevitably \footnote{
For example, in \cite{AK2509},
there are defining twenty-five OPEs. When
any normal-ordered product of two operators, which are reversed,
compared to the above defining OPEs, appears, then the package
performs to change this ordering automatically together with
the derivatives.}. 
Therefore, we can read off
all the second-order poles by looking at the left-hand sides
of (\ref{Second}) \footnote{As described before, because
the conditons $s_1+s_2 \geq 0$ and $s_1+s_3 \geq 0$ are satisfied,
there exist two kinds of quadratic terms on the right-hand sides of
(\ref{Second}). Due to the single contractions, compared to the
linear terms on the right-hand sides of the fourth-order poles,
we see only linear coupling dependence on the right-hand sides.
Moreover, the $h_i$ ($i=1,2,3$) dependence appear on the right-hand
sides according to the defining
relations on the OPEs of single-particle operators and themselves
in \cite{AK2509}. }.
The remaining second-order poles we do not write down
explicitly in this paper can be computed from the inside of
the package and we expect that they should be consistent with
the constructions given in Appendix A
\footnote{
The fourth-order pole for the OPE
corresponding to the footnote \ref{quadrupleopepole6} is 
\bea
&& \text{\tt OPESimplify}[\text{\tt OPEPole}[{\tt 4}]
[\text{\tt Phi}[\text{\tt h1},{\tt +2}],
\text{\tt NO}[\text{\tt Phi}[\text{\tt h2},{\tt +2}],
\text{\tt NO}[\text{\tt Phi}[\text{\tt h3},{\tt +2}],
\text{\tt Phi}[\text{\tt h4},{\tt +2}]]]],\text{\tt Factor}]
\nonu \\
&&=\text{\tt kappa}[{\tt 2,2,-2}]^{\tt 2}
(\text{\tt h1}+{\tt 3} \text{\tt h3}+{\tt 4})
(\text{\tt h1}+\text{\tt h3}+\text{\tt h4}+{\tt 2})
\text{\tt NO}[\text{\tt Phi}[\text{\tt h2},{\tt 2}],
\text{\tt Phi}[\text{\tt h1}+\text{\tt h3}+\text{\tt h4},{\tt 2}]]
\nonu \\
&& +\text{\tt kappa}[{\tt 2,2,-2}]^{\tt 2}
(\text{\tt h1}+{\tt 3} \text{\tt h2}+{\tt 4})
(\text{\tt h1}+\text{\tt h2}+\text{\tt h4}+{\tt 2})
\text{\tt NO}[\text{\tt Phi}[\text{\tt h3},{\tt 2}],
\text{\tt Phi}[\text{\tt h1}+\text{\tt h2}+\text{\tt h4},{\tt 2}]]
\nonu \\
&& +\text{\tt kappa}[{\tt 2,2,-2}]^{\tt 2}
(\text{\tt h1}+{\tt 3} \text{\tt h2}+{\tt 4})
(\text{\tt h1}+\text{\tt h2}+\text{\tt h3}+{\tt 2})
\text{\tt NO}[\text{\tt Phi}[\text{\tt h1}+\text{\tt h2}+
\text{\tt h3},{\tt 2}],\text{\tt Phi}[\text{\tt h4},{\tt 2}]]
\nonu
\eea
where the quadratic operators appear.
The cubic terms appear at the second-order pole as follows:
\bea
&&
\text{\tt OPESimplify}[\text{\tt OPEPole}[{\tt 2}]
[\text{\tt Phi}[\text{\tt h1},{\tt +2}],
\text{\tt NO}[\text{\tt Phi}[\text{\tt h2},{\tt +2}],
\text{\tt NO}[\text{\tt Phi}[\text{\tt h3},{\tt +2}],
\text{\tt Phi}[\text{\tt h4},{\tt +2}]]]],\text{\tt Factor}]
\nonu \\
&& =
\text{\tt kappa}[{\tt 2,2,-2}]
(\text{\tt h1}+\text{\tt h4}+{\tt 2})
\text{\tt NO}[\text{\tt Phi}[\text{\tt h2},{\tt 2}],
\text{\tt NO}[\text{\tt Phi}[\text{\tt h3},{\tt 2}],
\text{\tt Phi}[\text{\tt h1}+\text{\tt h4},{\tt 2}]]]
\nonu \\
&& +\text{\tt kappa}[{\tt 2,2,-2}]
(\text{\tt h1}
+\text{\tt h3}+{\tt 2})
\text{\tt NO}[\text{\tt Phi}[\text{\tt h2},{\tt 2}],
\text{\tt NO}[\text{\tt Phi}[\text{\tt h1}+\text{\tt h3},{\tt 2}],
\text{\tt Phi}[\text{\tt h4},{\tt 2}]]]
\nonu \\
&&
+\text{\tt kappa}[{\tt 2,2,-2}]
(\text{\tt h1}+\text{\tt h2}+{\tt 2})
\text{\tt NO}[\text{\tt Phi}[\text{\tt h1}+\text{\tt h2},{\tt 2}],
\text{\tt NO}[\text{\tt Phi}[\text{\tt h3},{\tt 2}],
\text{\tt Phi}[\text{\tt h4},{\tt 2}]]].
\nonu 
\eea}.

\section{The splitting functions for triple-collinear limits}

In this Appendix, we list the splitting functions in Tables~4--11.
By using the definitions for the coefficients in (\ref{threed}),
we can compute these splitting functions for each helicity.
In these Tables, the overall energy-dependent terms with common
factor are obtained.
As described before, they can be written as a single expression in~(\ref{SPLIT}).

\begin{table}[tbp]
\centering
\renewcommand{\arraystretch}{1.7}
\begin{tabular}{|c|c|c|c|c| }
\hline
$s_1$ & $s_2$ & $s_3$  & $
\Big(d_{1,2}\big|_{s_I=s_1+s_2-2}+
d_{2,3}\big|_{s_I=s_2+s_3-2} + d_{1,3}\big|_{s_I=s_1+s_3-2}\Big)
\Big|_{s_J=s_1+s_2+s_3-4}$  
\\
\hline
\hline
$+2$   & $+2$
& $+2$ 
& $\frac{1}{\ep^2}\,
\frac{(\omega_1+\omega_2+\omega_3)^2}{\omega_1 \, \omega_2 \,
\omega_3}\, \Big( \omega_1 \,
\frac{\bar{z}_{12} \, \bar{z}_{13}}{(1-\eta)} +
\omega_2 \,
\frac{\bar{z}_{12} \, \bar{z}_{23}}{\eta \, (1-\eta)}
+\omega_3 \,
\frac{\bar{z}_{13} \, \bar{z}_{23}}{\eta}\Big)$ 
\\
\hline
$+2$ & $+2$
& $+\frac{3}{2}$  
& $\frac{1}{\ep^2}\,
\frac{(\omega_1+\omega_2+\omega_3)^\frac{3}{2}}{\omega_1 \, \omega_2 \,
\omega_3^{\frac{1}{2}}}\, \Big( \omega_1 \,
\frac{\bar{z}_{12} \, \bar{z}_{13}}{(1-\eta)} +
\omega_2 \,
\frac{\bar{z}_{12} \, \bar{z}_{23}}{\eta \, (1-\eta)}
+\omega_3 \,
\frac{\bar{z}_{13} \, \bar{z}_{23}}{\eta}\Big)$ 
\\
\hline
$+2$ & $+2$
& $+1$  
& $\frac{1}{\ep^2}\,
\frac{(\omega_1+\omega_2+\omega_3)}{\omega_1 \, \omega_2 }\, \Big( \omega_1 \,
\frac{\bar{z}_{12} \, \bar{z}_{13}}{(1-\eta)} +
\omega_2 \,
\frac{\bar{z}_{12} \, \bar{z}_{23}}{\eta \, (1-\eta)}
+\omega_3 \,
\frac{\bar{z}_{13} \, \bar{z}_{23}}{\eta}\Big)$ 
\\
\hline
$+2$ & $+2$
& $+\frac{1}{2}$  
& $\frac{1}{\ep^2}\,
\frac{\omega_3^{\frac{1}{2}}\,
(\omega_1+\omega_2+\omega_3)^{\frac{1}{2}}}{\omega_1 \, \omega_2 }\, \Big( \omega_1 \,
\frac{\bar{z}_{12} \, \bar{z}_{13}}{(1-\eta)} +
\omega_2 \,
\frac{\bar{z}_{12} \, \bar{z}_{23}}{\eta \, (1-\eta)}
+\omega_3 \,
\frac{\bar{z}_{13} \, \bar{z}_{23}}{\eta}\Big)$ 
\\
\hline
$+2$ & $+2$
& $0$  
& $\frac{1}{\ep^2}\,
\frac{\omega_3 \,
}{\omega_1 \, \omega_2 }\, \Big( \omega_1 \,
\frac{\bar{z}_{12} \, \bar{z}_{13}}{(1-\eta)} +
\omega_2 \,
\frac{\bar{z}_{12} \, \bar{z}_{23}}{\eta \, (1-\eta)}
+\omega_3 \,
\frac{\bar{z}_{13} \, \bar{z}_{23}}{\eta}\Big)$ 
\\
\hline
$+2$ & $+2$
& $-\frac{1}{2}$  
& $\frac{1}{\ep^2}\,
\frac{\omega_3^{\frac{3}{2}} \,
}{\omega_1 \, \omega_2 \,
(\omega_1+\omega_2+\omega_3)^{\frac{1}{2}}}\, \Big( \omega_1 \,
\frac{\bar{z}_{12} \, \bar{z}_{13}}{(1-\eta)} +
\omega_2 \,
\frac{\bar{z}_{12} \, \bar{z}_{23}}{\eta \, (1-\eta)}
+\omega_3 \,
\frac{\bar{z}_{13} \, \bar{z}_{23}}{\eta}\Big)$ 
\\
\hline
$+2$ & $+2$
& $-1$  
& $\frac{1}{\ep^2}\,
\frac{\omega_3^{2} \,
}{\omega_1 \, \omega_2 \,
(\omega_1+\omega_2+\omega_3)}\, \Big( \omega_1 \,
\frac{\bar{z}_{12} \, \bar{z}_{13}}{(1-\eta)} +
\omega_2 \,
\frac{\bar{z}_{12} \, \bar{z}_{23}}{\eta \, (1-\eta)}
+\omega_3 \,
\frac{\bar{z}_{13} \, \bar{z}_{23}}{\eta}\Big)$ 
\\
\hline
$+2$ & $+2$
& $-\frac{3}{2}$  
& $\frac{1}{\ep^2}\,
\frac{\omega_3^{\frac{5}{2}} \,
}{\omega_1 \, \omega_2 \,
(\omega_1+\omega_2+\omega_3)^{\frac{3}{2}}}\, \Big( \omega_1 \,
\frac{\bar{z}_{12} \, \bar{z}_{13}}{(1-\eta)} +
\omega_2 \,
\frac{\bar{z}_{12} \, \bar{z}_{23}}{\eta \, (1-\eta)}
+\omega_3 \,
\frac{\bar{z}_{13} \, \bar{z}_{23}}{\eta}\Big)$ 
\\
\hline
$+2$ & $+2$
& $-2$  
& $\frac{1}{\ep^2}\,
\frac{\omega_3^{3} \,
}{\omega_1 \, \omega_2 \,
(\omega_1+\omega_2+\omega_3)^{2}}\, \Big( \omega_1 \,
\frac{\bar{z}_{12} \, \bar{z}_{13}}{(1-\eta)} +
\omega_2 \,
\frac{\bar{z}_{12} \, \bar{z}_{23}}{\eta \, (1-\eta)}
+\omega_3 \,
\frac{\bar{z}_{13} \, \bar{z}_{23}}{\eta}\Big)$ 
\\
\hline
$+2$ & $+\frac{3}{2}$
& $+\frac{3}{2}$  
& $\frac{1}{\ep^2}\,
\frac{(\omega_1+\omega_2+\omega_3) \,
}{\omega_1 \, \omega_2^{\frac{1}{2}} \, \omega_3^{\frac{1}{2}}
}\, \Big( \omega_1 \,
\frac{\bar{z}_{12} \, \bar{z}_{13}}{(1-\eta)} +
\omega_2 \,
\frac{\bar{z}_{12} \, \bar{z}_{23}}{\eta \, (1-\eta)}
+\omega_3 \,
\frac{\bar{z}_{13} \, \bar{z}_{23}}{\eta}\Big)$ 
\\
\hline
\end{tabular}
\caption{The first ten splitting functions of
twenty-five ones corresponding to (\ref{25comm}).
We are using the definitions in (\ref{threed}).
}
\end{table}
%

\begin{table}[tbp]
\centering
\renewcommand{\arraystretch}{1.7}
\begin{tabular}{|c|c|c|c|c| }
\hline
$s_1$ & $s_2$ & $s_3$  & $
\Big(d_{1,2}\big|_{s_I=s_1+s_2-2}+
d_{2,3}\big|_{s_I=s_2+s_3-2} + d_{1,3}\big|_{s_I=s_1+s_3-2}\Big)
\Big|_{s_J=s_1+s_2+s_3-4}$  
\\
\hline
\hline
$+2$ & $+\frac{3}{2}$
& $+1$  
& $\frac{1}{\ep^2}\,
\frac{(\omega_1+\omega_2+\omega_3)^{\frac{1}{2}} \,
}{\omega_1 \, \omega_2^{\frac{1}{2}} 
}\, \Big( \omega_1 \,
\frac{\bar{z}_{12} \, \bar{z}_{13}}{(1-\eta)} +
\omega_2 \,
\frac{\bar{z}_{12} \, \bar{z}_{23}}{\eta \, (1-\eta)}
+\omega_3 \,
\frac{\bar{z}_{13} \, \bar{z}_{23}}{\eta}\Big)$ 
\\
\hline
$+2$ & $+\frac{3}{2}$
& $+\frac{1}{2}$  
& $\frac{1}{\ep^2}\,
\frac{\omega_3^{\frac{1}{2}} \,
}{\omega_1 \, \omega_2^{\frac{1}{2}} 
}\, \Big( \omega_1 \,
\frac{\bar{z}_{12} \, \bar{z}_{13}}{(1-\eta)} +
\omega_2 \,
\frac{\bar{z}_{12} \, \bar{z}_{23}}{\eta \, (1-\eta)}
+\omega_3 \,
\frac{\bar{z}_{13} \, \bar{z}_{23}}{\eta}\Big)$ 
\\
\hline
$+2$ & $+\frac{3}{2}$
& $0$  
& $\frac{1}{\ep^2}\,
\frac{\omega_3 \,
}{\omega_1 \, \omega_2^{\frac{1}{2}} 
\, (\omega_1+\omega_2+\omega_3)^{\frac{1}{2}}}\, \Big( \omega_1 \,
\frac{\bar{z}_{12} \, \bar{z}_{13}}{(1-\eta)} +
\omega_2 \,
\frac{\bar{z}_{12} \, \bar{z}_{23}}{\eta \, (1-\eta)}
+\omega_3 \,
\frac{\bar{z}_{13} \, \bar{z}_{23}}{\eta}\Big)$ 
\\
\hline
$+2$ & $+\frac{3}{2}$
& $-\frac{1}{2}$  
& $\frac{1}{\ep^2}\,
\frac{\omega_3^{\frac{3}{2}} \,
}{\omega_1 \, \omega_2^{\frac{1}{2}} 
\, (\omega_1+\omega_2+\omega_3)}\, \Big( \omega_1 \,
\frac{\bar{z}_{12} \, \bar{z}_{13}}{(1-\eta)} +
\omega_2 \,
\frac{\bar{z}_{12} \, \bar{z}_{23}}{\eta \, (1-\eta)}
+\omega_3 \,
\frac{\bar{z}_{13} \, \bar{z}_{23}}{\eta}\Big)$ 
\\
\hline
$+2$ & $+\frac{3}{2}$
& $-1$  
& $\frac{1}{\ep^2}\,
\frac{\omega_3^{2} \,
}{\omega_1 \, \omega_2^{\frac{1}{2}} 
\, (\omega_1+\omega_2+\omega_3)^{\frac{3}{2}}}\, \Big( \omega_1 \,
\frac{\bar{z}_{12} \, \bar{z}_{13}}{(1-\eta)} +
\omega_2 \,
\frac{\bar{z}_{12} \, \bar{z}_{23}}{\eta \, (1-\eta)}
+\omega_3 \,
\frac{\bar{z}_{13} \, \bar{z}_{23}}{\eta}\Big)$ 
\\
\hline
$+2$   & $+\frac{3}{2}$
& $-\frac{3}{2}$ 
& $\frac{1}{\ep^2}\,
\frac{\omega_3^{\frac{5}{2}}}{\omega_1 \, \omega_2^{\frac{1}{2}} \,
(\omega_1+\omega_2+\omega_3)^2}\, \Big( \omega_1 \,
\frac{\bar{z}_{12} \, \bar{z}_{13}}{(1-\eta)} +
\omega_2 \,
\frac{\bar{z}_{12} \, \bar{z}_{23}}{\eta \, (1-\eta)}
+\omega_3 \,
\frac{\bar{z}_{13} \, \bar{z}_{23}}{\eta}\Big)$ 
\\
\hline
$+2$ & $+1$
& $+1$  
& $\frac{1}{\ep^2}\,
\frac{1}{\omega_1  }\, \Big( \omega_1 \,
\frac{\bar{z}_{12} \, \bar{z}_{13}}{(1-\eta)} +
\omega_2 \,
\frac{\bar{z}_{12} \, \bar{z}_{23}}{\eta \, (1-\eta)}
+\omega_3 \,
\frac{\bar{z}_{13} \, \bar{z}_{23}}{\eta}\Big)$ 
\\
\hline
$+2$ & $+1$
& $+\frac{1}{2}$  
& $\frac{1}{\ep^2}\,
\frac{\omega_3^{\frac{1}{2}}\,
}{\omega_1 \, (\omega_1+\omega_2+\omega_3)^{\frac{1}{2}} }\, \Big( \omega_1 \,
\frac{\bar{z}_{12} \, \bar{z}_{13}}{(1-\eta)} +
\omega_2 \,
\frac{\bar{z}_{12} \, \bar{z}_{23}}{\eta \, (1-\eta)}
+\omega_3 \,
\frac{\bar{z}_{13} \, \bar{z}_{23}}{\eta}\Big)$ 
\\
\hline
$+2$ & $+1$
& $0$  
& $\frac{1}{\ep^2}\,
\frac{\omega_3 \,
}{\omega_1 \, (\omega_1+\omega_2+\omega_3) }\, \Big( \omega_1 \,
\frac{\bar{z}_{12} \, \bar{z}_{13}}{(1-\eta)} +
\omega_2 \,
\frac{\bar{z}_{12} \, \bar{z}_{23}}{\eta \, (1-\eta)}
+\omega_3 \,
\frac{\bar{z}_{13} \, \bar{z}_{23}}{\eta}\Big)$ 
\\
\hline
$+2$ & $+1$
& $-\frac{1}{2}$  
& $\frac{1}{\ep^2}\,
\frac{\omega_3^{\frac{3}{2}} \,
}{\omega_1  \,
(\omega_1+\omega_2+\omega_3)^{\frac{3}{2}}}\, \Big( \omega_1 \,
\frac{\bar{z}_{12} \, \bar{z}_{13}}{(1-\eta)} +
\omega_2 \,
\frac{\bar{z}_{12} \, \bar{z}_{23}}{\eta \, (1-\eta)}
+\omega_3 \,
\frac{\bar{z}_{13} \, \bar{z}_{23}}{\eta}\Big)$ 
\\
\hline
$+2$ & $+1$
& $-1$  
& $\frac{1}{\ep^2}\,
\frac{\omega_3^{2} \,
}{\omega_1  \,
(\omega_1+\omega_2+\omega_3)^2}\, \Big( \omega_1 \,
\frac{\bar{z}_{12} \, \bar{z}_{13}}{(1-\eta)} +
\omega_2 \,
\frac{\bar{z}_{12} \, \bar{z}_{23}}{\eta \, (1-\eta)}
+\omega_3 \,
\frac{\bar{z}_{13} \, \bar{z}_{23}}{\eta}\Big)$ 
\\
\hline
$+2$ & $+\frac{1}{2}$
& $+\frac{1}{2}$  
& $\frac{1}{\ep^2}\,
\frac{\omega_2^{\frac{1}{2}}\, \omega_3^{\frac{1}{2}} \,
}{\omega_1  \,
(\omega_1+\omega_2+\omega_3)}\, \Big( \omega_1 \,
\frac{\bar{z}_{12} \, \bar{z}_{13}}{(1-\eta)} +
\omega_2 \,
\frac{\bar{z}_{12} \, \bar{z}_{23}}{\eta \, (1-\eta)}
+\omega_3 \,
\frac{\bar{z}_{13} \, \bar{z}_{23}}{\eta}\Big)$ 
\\
\hline
$+2$ & $+\frac{1}{2}$
& $0$  
& $\frac{1}{\ep^2}\,
\frac{\omega_2^{\frac{1}{2}} \, \omega_3
}{\omega_1  \,
(\omega_1+\omega_2+\omega_3)^{\frac{3}{2}}}\, \Big( \omega_1 \,
\frac{\bar{z}_{12} \, \bar{z}_{13}}{(1-\eta)} +
\omega_2 \,
\frac{\bar{z}_{12} \, \bar{z}_{23}}{\eta \, (1-\eta)}
+\omega_3 \,
\frac{\bar{z}_{13} \, \bar{z}_{23}}{\eta}\Big)$ 
\\
\hline
$+2$ & $+\frac{1}{2}$
& $-\frac{1}{2}$  
& $\frac{1}{\ep^2}\,
\frac{\omega_2^{\frac{1}{2}} \, \omega_3^{\frac{3}{2}} 
}{\omega_1 \, (\omega_1+\omega_2+\omega_3)^2
}\, \Big( \omega_1 \,
\frac{\bar{z}_{12} \, \bar{z}_{13}}{(1-\eta)} +
\omega_2 \,
\frac{\bar{z}_{12} \, \bar{z}_{23}}{\eta \, (1-\eta)}
+\omega_3 \,
\frac{\bar{z}_{13} \, \bar{z}_{23}}{\eta}\Big)$ 
\\
\hline
$+2$ & $0$
& $0$  
& $\frac{1}{\ep^2}\,
\frac{ \omega_2 \,
\omega_3}{\omega_1 \,(\omega_1+\omega_2+\omega_3)^{2}}\, \Big( \omega_1 \,
\frac{\bar{z}_{12} \, \bar{z}_{13}}{(1-\eta)} +
\omega_2 \,
\frac{\bar{z}_{12} \, \bar{z}_{23}}{\eta \, (1-\eta)}
+\omega_3 \,
\frac{\bar{z}_{13} \, \bar{z}_{23}}{\eta}\Big)$ 
\\
\hline
\end{tabular}
\caption{
The remaining fifteen splitting functions of twenty-five ones
corresponding to (\ref{25comm}).
Again, the relation (\ref{threed}) is used.}
\end{table}

\begin{table}[tbp]
\centering
\renewcommand{\arraystretch}{1.7}
\begin{tabular}{|c|c|c|c|c| }
\hline
$s_1$ & $s_2$ & $s_3$  & $
\Big(d_{1,2}\big|_{s_I=s_1+s_2-2}+
d_{2,3}\big|_{s_I=s_2+s_3-2} + d_{1,3}\big|_{s_I=s_1+s_3-2}\Big)
\Big|_{s_J=s_1+s_2+s_3-4}$  
\\
\hline
\hline
$+\frac{3}{2}$ & $+2$
& $+2$  
& $\frac{1}{\ep^2}\,
\frac{(\omega_1+\omega_2+\omega_3)^{\frac{3}{2}} \,
}{\omega_1^{\frac{1}{2}} \, \omega_2 \, \omega_3 
}\, \Big( \omega_1 \,
\frac{\bar{z}_{12} \, \bar{z}_{13}}{(1-\eta)} +
\omega_2 \,
\frac{\bar{z}_{12} \, \bar{z}_{23}}{\eta \, (1-\eta)}
+\omega_3 \,
\frac{\bar{z}_{13} \, \bar{z}_{23}}{\eta}\Big)$ 
\\
\hline
$+\frac{3}{2}$ & $+2$
& $+\frac{3}{2}$  
& $\frac{1}{\ep^2}\,
\frac{ (\omega_1+\omega_2+\omega_3)
}{\omega_1^{\frac{1}{2}} \,\omega_2\,  \omega_3^{\frac{1}{2}} 
}\, \Big( \omega_1 \,
\frac{\bar{z}_{12} \, \bar{z}_{13}}{(1-\eta)} +
\omega_2 \,
\frac{\bar{z}_{12} \, \bar{z}_{23}}{\eta \, (1-\eta)}
+\omega_3 \,
\frac{\bar{z}_{13} \, \bar{z}_{23}}{\eta}\Big)$ 
\\
\hline
$+\frac{3}{2}$ & $+2$
& $+1$  
& $\frac{1}{\ep^2}\,
\frac{ (\omega_1+\omega_2+\omega_3)^{\frac{1}{2}}
}{\omega_1^{\frac{1}{2}} 
\,\omega_2}\, \Big( \omega_1 \,
\frac{\bar{z}_{12} \, \bar{z}_{13}}{(1-\eta)} +
\omega_2 \,
\frac{\bar{z}_{12} \, \bar{z}_{23}}{\eta \, (1-\eta)}
+\omega_3 \,
\frac{\bar{z}_{13} \, \bar{z}_{23}}{\eta}\Big)$ 
\\
\hline
$+\frac{3}{2}$ & $+2$
& $+\frac{1}{2}$  
& $\frac{1}{\ep^2}\,
\frac{\omega_3^{\frac{1}{2}} \,
}{\omega_1^{\frac{1}{2}} 
\, \omega_2}\, \Big( \omega_1 \,
\frac{\bar{z}_{12} \, \bar{z}_{13}}{(1-\eta)} +
\omega_2 \,
\frac{\bar{z}_{12} \, \bar{z}_{23}}{\eta \, (1-\eta)}
+\omega_3 \,
\frac{\bar{z}_{13} \, \bar{z}_{23}}{\eta}\Big)$ 
\\
\hline
$+\frac{3}{2}$ & $+2$
& $0$  
& $\frac{1}{\ep^2}\,
\frac{\omega_3 \,
}{\omega_1^{\frac{1}{2}} 
\, \omega_2 \, (\omega_1+\omega_2+\omega_3)^{\frac{1}{2}}}\, \Big( \omega_1 \,
\frac{\bar{z}_{12} \, \bar{z}_{13}}{(1-\eta)} +
\omega_2 \,
\frac{\bar{z}_{12} \, \bar{z}_{23}}{\eta \, (1-\eta)}
+\omega_3 \,
\frac{\bar{z}_{13} \, \bar{z}_{23}}{\eta}\Big)$ 
\\
\hline
$+\frac{3}{2}$   & $+2$
& $-\frac{1}{2}$ 
& $\frac{1}{\ep^2}\,
\frac{\omega_3^{\frac{3}{2}}}{\omega_1^{\frac{1}{2}} \,
\omega_2\, (\omega_1+\omega_2+\omega_3)}\, \Big( \omega_1 \,
\frac{\bar{z}_{12} \, \bar{z}_{13}}{(1-\eta)} +
\omega_2 \,
\frac{\bar{z}_{12} \, \bar{z}_{23}}{\eta \, (1-\eta)}
+\omega_3 \,
\frac{\bar{z}_{13} \, \bar{z}_{23}}{\eta}\Big)$ 
\\
\hline
$+\frac{3}{2}$ & $+2$
& $-1$  
& $\frac{1}{\ep^2}\,
\frac{\omega_3^2}{\omega_1^{\frac{1}{2}}\, \omega_2\,
(\omega_1+\omega_2+\omega_3)^{\frac{3}{2}}}\, \Big( \omega_1 \,
\frac{\bar{z}_{12} \, \bar{z}_{13}}{(1-\eta)} +
\omega_2 \,
\frac{\bar{z}_{12} \, \bar{z}_{23}}{\eta \, (1-\eta)}
+\omega_3 \,
\frac{\bar{z}_{13} \, \bar{z}_{23}}{\eta}\Big)$ 
\\
\hline
$+\frac{3}{2}$ & $+2$
& $-\frac{3}{2}$  
& $\frac{1}{\ep^2}\,
\frac{\omega_3^{\frac{5}{2}}\,
}{\omega_1^{\frac{1}{2}}\, \omega_2 \,
(\omega_1+\omega_2+\omega_3)^{2} }\, \Big( \omega_1 \,
\frac{\bar{z}_{12} \, \bar{z}_{13}}{(1-\eta)} +
\omega_2 \,
\frac{\bar{z}_{12} \, \bar{z}_{23}}{\eta \, (1-\eta)}
+\omega_3 \,
\frac{\bar{z}_{13} \, \bar{z}_{23}}{\eta}\Big)$ 
\\
\hline
$+\frac{3}{2}$ & $+\frac{3}{2}$
& $+\frac{3}{2}$  
& $\frac{1}{\ep^2}\,
\frac{(\omega_1+\omega_2+\omega_3)^{\frac{1}{2}}
}{\omega_1^{\frac{1}{2}} \, \omega_2^{\frac{1}{2}} \, \omega_3^{\frac{1}{2}} }\, \Big( \omega_1 \,
\frac{\bar{z}_{12} \, \bar{z}_{13}}{(1-\eta)} +
\omega_2 \,
\frac{\bar{z}_{12} \, \bar{z}_{23}}{\eta \, (1-\eta)}
+\omega_3 \,
\frac{\bar{z}_{13} \, \bar{z}_{23}}{\eta}\Big)$ 
\\
\hline
$+\frac{3}{2}$ & $+\frac{3}{2}$
& $+1$  
& $\frac{1}{\ep^2}\,
\frac{1
}{\omega_1^{\frac{1}{2}}  \,\omega_2^{\frac{1}{2}} 
}\, \Big( \omega_1 \,
\frac{\bar{z}_{12} \, \bar{z}_{13}}{(1-\eta)} +
\omega_2 \,
\frac{\bar{z}_{12} \, \bar{z}_{23}}{\eta \, (1-\eta)}
+\omega_3 \,
\frac{\bar{z}_{13} \, \bar{z}_{23}}{\eta}\Big)$ 
\\
\hline
$+\frac{3}{2}$ & $+\frac{3}{2}$
& $+\frac{1}{2}$  
& $\frac{1}{\ep^2}\,
\frac{\omega_3^{\frac{1}{2}} \,
}{\omega_1^{\frac{1}{2}}  \,\omega_2^{\frac{1}{2}}\,
(\omega_1+\omega_2+\omega_3)^{\frac{1}{2}}}\, \Big( \omega_1 \,
\frac{\bar{z}_{12} \, \bar{z}_{13}}{(1-\eta)} +
\omega_2 \,
\frac{\bar{z}_{12} \, \bar{z}_{23}}{\eta \, (1-\eta)}
+\omega_3 \,
\frac{\bar{z}_{13} \, \bar{z}_{23}}{\eta}\Big)$ 
\\
\hline
$+\frac{3}{2}$ & $+\frac{3}{2}$
& $0$  
& $\frac{1}{\ep^2}\,
\frac{\omega_3
}{\omega_1^{\frac{1}{2}}  \,\omega_2^{\frac{1}{2}}  \,
(\omega_1+\omega_2+\omega_3)}\, \Big( \omega_1 \,
\frac{\bar{z}_{12} \, \bar{z}_{13}}{(1-\eta)} +
\omega_2 \,
\frac{\bar{z}_{12} \, \bar{z}_{23}}{\eta \, (1-\eta)}
+\omega_3 \,
\frac{\bar{z}_{13} \, \bar{z}_{23}}{\eta}\Big)$ 
\\
\hline
$+\frac{3}{2}$ & $+\frac{3}{2}$
& $-\frac{1}{2}$  
& $\frac{1}{\ep^2}\,
\frac{\omega_3^{\frac{3}{2}} 
}{\omega_1^{\frac{1}{2}}  \,\omega_2^{\frac{1}{2}}  \,
(\omega_1+\omega_2+\omega_3)^{\frac{3}{2}}}\, \Big( \omega_1 \,
\frac{\bar{z}_{12} \, \bar{z}_{13}}{(1-\eta)} +
\omega_2 \,
\frac{\bar{z}_{12} \, \bar{z}_{23}}{\eta \, (1-\eta)}
+\omega_3 \,
\frac{\bar{z}_{13} \, \bar{z}_{23}}{\eta}\Big)$ 
\\
\hline
$+\frac{3}{2}$ & $+\frac{3}{2}$
& $-1$  
& $\frac{1}{\ep^2}\,
\frac{ \omega_3^{2} 
}{\omega_1^{\frac{1}{2}} \,
\omega_2^{\frac{1}{2}} \,
(\omega_1+\omega_2+\omega_3)^2
}\, \Big( \omega_1 \,
\frac{\bar{z}_{12} \, \bar{z}_{13}}{(1-\eta)} +
\omega_2 \,
\frac{\bar{z}_{12} \, \bar{z}_{23}}{\eta \, (1-\eta)}
+\omega_3 \,
\frac{\bar{z}_{13} \, \bar{z}_{23}}{\eta}\Big)$ 
\\
\hline
$+\frac{3}{2}$ & $+1$
& $+1$  
& $\frac{1}{\ep^2}\,
\frac{1}{\omega_1^{\frac{1}{2}} \,
(\omega_1+\omega_2+\omega_3)^{\frac{1}{2}}}\, \Big( \omega_1 \,
\frac{\bar{z}_{12} \, \bar{z}_{13}}{(1-\eta)} +
\omega_2 \,
\frac{\bar{z}_{12} \, \bar{z}_{23}}{\eta \, (1-\eta)}
+\omega_3 \,
\frac{\bar{z}_{13} \, \bar{z}_{23}}{\eta}\Big)$ 
\\
\hline
\end{tabular}
\caption{
The first fifteen splitting functions for (\ref{20case}). }
\end{table}

\begin{table}[tbp]
\centering
\renewcommand{\arraystretch}{1.7}
\begin{tabular}{|c|c|c|c|c| }
\hline
$s_1$ & $s_2$ & $s_3$  & $
\Big(d_{1,2}\big|_{s_I=s_1+s_2-2}+
d_{2,3}\big|_{s_I=s_2+s_3-2} + d_{1,3}\big|_{s_I=s_1+s_3-2}\Big)
\Big|_{s_J=s_1+s_2+s_3-4}$  
\\
\hline
\hline
$+\frac{3}{2}$ & $+1$
& $+\frac{1}{2}$  
& $\frac{1}{\ep^2}\,
\frac{\omega_3^{\frac{1}{2}} \,
}{\omega_1^{\frac{1}{2}}  \,
(\omega_1+\omega_2+\omega_3)}\, \Big( \omega_1 \,
\frac{\bar{z}_{12} \, \bar{z}_{13}}{(1-\eta)} +
\omega_2 \,
\frac{\bar{z}_{12} \, \bar{z}_{23}}{\eta \, (1-\eta)}
+\omega_3 \,
\frac{\bar{z}_{13} \, \bar{z}_{23}}{\eta}\Big)$ 
\\
\hline
$+\frac{3}{2}$ & $+1$
& $0$  
& $\frac{1}{\ep^2}\,
\frac{\omega_3
}{\omega_1^{\frac{1}{2}}  \,
(\omega_1+\omega_2+\omega_3)^{\frac{3}{2}}}\, \Big( \omega_1 \,
\frac{\bar{z}_{12} \, \bar{z}_{13}}{(1-\eta)} +
\omega_2 \,
\frac{\bar{z}_{12} \, \bar{z}_{23}}{\eta \, (1-\eta)}
+\omega_3 \,
\frac{\bar{z}_{13} \, \bar{z}_{23}}{\eta}\Big)$ 
\\
\hline
$+\frac{3}{2}$ & $+1$
& $-\frac{1}{2}$  
& $\frac{1}{\ep^2}\,
\frac{\omega_3^{\frac{3}{2}} 
}{\omega_1^{\frac{1}{2}}  \,
(\omega_1+\omega_2+\omega_3)^{2}}\, \Big( \omega_1 \,
\frac{\bar{z}_{12} \, \bar{z}_{13}}{(1-\eta)} +
\omega_2 \,
\frac{\bar{z}_{12} \, \bar{z}_{23}}{\eta \, (1-\eta)}
+\omega_3 \,
\frac{\bar{z}_{13} \, \bar{z}_{23}}{\eta}\Big)$ 
\\
\hline
$+\frac{3}{2}$ & $+\frac{1}{2}$
& $+\frac{1}{2}$  
& $\frac{1}{\ep^2}\,
\frac{\omega_2^{\frac{1}{2}} \, \omega_3^{\frac{1}{2}} 
}{\omega_1^{\frac{1}{2}} \, (\omega_1+\omega_2+\omega_3)^{\frac{3}{2}}
}\, \Big( \omega_1 \,
\frac{\bar{z}_{12} \, \bar{z}_{13}}{(1-\eta)} +
\omega_2 \,
\frac{\bar{z}_{12} \, \bar{z}_{23}}{\eta \, (1-\eta)}
+\omega_3 \,
\frac{\bar{z}_{13} \, \bar{z}_{23}}{\eta}\Big)$ 
\\
\hline
$+\frac{3}{2}$ & $+\frac{1}{2}$
& $0$  
& $\frac{1}{\ep^2}\,
\frac{ \omega_2^{\frac{1}{2}} \,
\omega_3}{\omega_1^{\frac{1}{2}}
\,(\omega_1+\omega_2+\omega_3)^{2}}\, \Big( \omega_1 \,
\frac{\bar{z}_{12} \, \bar{z}_{13}}{(1-\eta)} +
\omega_2 \,
\frac{\bar{z}_{12} \, \bar{z}_{23}}{\eta \, (1-\eta)}
+\omega_3 \,
\frac{\bar{z}_{13} \, \bar{z}_{23}}{\eta}\Big)$ 
\\
\hline
\end{tabular}
\caption{
The remaining five splitting functions for (\ref{20case}). }
\end{table}

\begin{table}[tbp]
\centering
\renewcommand{\arraystretch}{1.7}
\begin{tabular}{|c|c|c|c|c| }
\hline
$s_1$ & $s_2$ & $s_3$  & $
\Big(d_{1,2}\big|_{s_I=s_1+s_2-2}+
d_{2,3}\big|_{s_I=s_2+s_3-2} + d_{1,3}\big|_{s_I=s_1+s_3-2}\Big)
\Big|_{s_J=s_1+s_2+s_3-4}$  
\\
\hline
\hline
$+1$ & $+2$
& $+2$  
& $\frac{1}{\ep^2}\,
\frac{(\omega_1+\omega_2+\omega_3)
}{\omega_2 \, \omega_3 
}\, \Big( \omega_1 \,
\frac{\bar{z}_{12} \, \bar{z}_{13}}{(1-\eta)} +
\omega_2 \,
\frac{\bar{z}_{12} \, \bar{z}_{23}}{\eta \, (1-\eta)}
+\omega_3 \,
\frac{\bar{z}_{13} \, \bar{z}_{23}}{\eta}\Big)$ 
\\
\hline
$+1$ & $+2$
& $+\frac{3}{2}$  
& $\frac{1}{\ep^2}\,
\frac{ (\omega_1+\omega_2+\omega_3)^{\frac{1}{2}}
}{\omega_2\,  \omega_3^{\frac{1}{2}} 
}\, \Big( \omega_1 \,
\frac{\bar{z}_{12} \, \bar{z}_{13}}{(1-\eta)} +
\omega_2 \,
\frac{\bar{z}_{12} \, \bar{z}_{23}}{\eta \, (1-\eta)}
+\omega_3 \,
\frac{\bar{z}_{13} \, \bar{z}_{23}}{\eta}\Big)$ 
\\
\hline
$+1$ & $+2$
& $+1$  
& $\frac{1}{\ep^2}\,
\frac{1
}{\omega_2}\, \Big( \omega_1 \,
\frac{\bar{z}_{12} \, \bar{z}_{13}}{(1-\eta)} +
\omega_2 \,
\frac{\bar{z}_{12} \, \bar{z}_{23}}{\eta \, (1-\eta)}
+\omega_3 \,
\frac{\bar{z}_{13} \, \bar{z}_{23}}{\eta}\Big)$ 
\\
\hline
$+1$ & $+2$
& $+\frac{1}{2}$  
& $\frac{1}{\ep^2}\,
\frac{\omega_3^{\frac{1}{2}} \,
}{
\omega_2 \, (\omega_1+\omega_2+\omega_3)^{\frac{1}{2}}}\, \Big( \omega_1 \,
\frac{\bar{z}_{12} \, \bar{z}_{13}}{(1-\eta)} +
\omega_2 \,
\frac{\bar{z}_{12} \, \bar{z}_{23}}{\eta \, (1-\eta)}
+\omega_3 \,
\frac{\bar{z}_{13} \, \bar{z}_{23}}{\eta}\Big)$ 
\\
\hline
$+1$ & $+2$
& $0$  
& $\frac{1}{\ep^2}\,
\frac{\omega_3 \,
}{\omega_2 \, (\omega_1+\omega_2+\omega_3)}\, \Big( \omega_1 \,
\frac{\bar{z}_{12} \, \bar{z}_{13}}{(1-\eta)} +
\omega_2 \,
\frac{\bar{z}_{12} \, \bar{z}_{23}}{\eta \, (1-\eta)}
+\omega_3 \,
\frac{\bar{z}_{13} \, \bar{z}_{23}}{\eta}\Big)$ 
\\
\hline
$+1$   & $+2$
& $-\frac{1}{2}$ 
& $\frac{1}{\ep^2}\,
\frac{\omega_3^{\frac{3}{2}}}{
\omega_2\, (\omega_1+\omega_2+\omega_3)^{\frac{3}{2}}}\, \Big( \omega_1 \,
\frac{\bar{z}_{12} \, \bar{z}_{13}}{(1-\eta)} +
\omega_2 \,
\frac{\bar{z}_{12} \, \bar{z}_{23}}{\eta \, (1-\eta)}
+\omega_3 \,
\frac{\bar{z}_{13} \, \bar{z}_{23}}{\eta}\Big)$ 
\\
\hline
$+1$ & $+2$
& $-1$  
& $\frac{1}{\ep^2}\,
\frac{\omega_3^2}{\omega_2\,
(\omega_1+\omega_2+\omega_3)^{2}}\, \Big( \omega_1 \,
\frac{\bar{z}_{12} \, \bar{z}_{13}}{(1-\eta)} +
\omega_2 \,
\frac{\bar{z}_{12} \, \bar{z}_{23}}{\eta \, (1-\eta)}
+\omega_3 \,
\frac{\bar{z}_{13} \, \bar{z}_{23}}{\eta}\Big)$ 
\\
\hline
$+1$ & $+\frac{3}{2}$
& $+\frac{3}{2}$  
& $\frac{1}{\ep^2}\,
\frac{1
}{\omega_2^{\frac{1}{2}}\, \omega_3^{\frac{1}{2}} 
 }\, \Big( \omega_1 \,
\frac{\bar{z}_{12} \, \bar{z}_{13}}{(1-\eta)} +
\omega_2 \,
\frac{\bar{z}_{12} \, \bar{z}_{23}}{\eta \, (1-\eta)}
+\omega_3 \,
\frac{\bar{z}_{13} \, \bar{z}_{23}}{\eta}\Big)$ 
\\
\hline
$+1$ & $+\frac{3}{2}$
& $+1$  
& $\frac{1}{\ep^2}\,
\frac{1
}{\omega_2^{\frac{1}{2}} \,
(\omega_1+\omega_2+\omega_3)^{\frac{1}{2}}
 }\, \Big( \omega_1 \,
\frac{\bar{z}_{12} \, \bar{z}_{13}}{(1-\eta)} +
\omega_2 \,
\frac{\bar{z}_{12} \, \bar{z}_{23}}{\eta \, (1-\eta)}
+\omega_3 \,
\frac{\bar{z}_{13} \, \bar{z}_{23}}{\eta}\Big)$ 
\\
\hline
$+1$ & $+\frac{3}{2}$
& $+\frac{1}{2}$  
& $\frac{1}{\ep^2}\,
\frac{\omega_3^{\frac{1}{2}} 
}{\omega_2^{\frac{1}{2}} \, (\omega_1+\omega_2+\omega_3) 
}\, \Big( \omega_1 \,
\frac{\bar{z}_{12} \, \bar{z}_{13}}{(1-\eta)} +
\omega_2 \,
\frac{\bar{z}_{12} \, \bar{z}_{23}}{\eta \, (1-\eta)}
+\omega_3 \,
\frac{\bar{z}_{13} \, \bar{z}_{23}}{\eta}\Big)$ 
\\
\hline
$+1$ & $+\frac{3}{2}$
& $0$  
& $\frac{1}{\ep^2}\,
\frac{\omega_3
}{\omega_2^{\frac{1}{2}}  \,
(\omega_1+\omega_2+\omega_3)^{\frac{3}{2}}}\, \Big( \omega_1 \,
\frac{\bar{z}_{12} \, \bar{z}_{13}}{(1-\eta)} +
\omega_2 \,
\frac{\bar{z}_{12} \, \bar{z}_{23}}{\eta \, (1-\eta)}
+\omega_3 \,
\frac{\bar{z}_{13} \, \bar{z}_{23}}{\eta}\Big)$ 
\\
\hline
$+1$ & $+\frac{3}{2}$
& $-\frac{1}{2}$  
& $\frac{1}{\ep^2}\,
\frac{\omega_3^{\frac{3}{2}} 
}{\omega_2^{\frac{1}{2}}  \,
(\omega_1+\omega_2+\omega_3)^{2}}\, \Big( \omega_1 \,
\frac{\bar{z}_{12} \, \bar{z}_{13}}{(1-\eta)} +
\omega_2 \,
\frac{\bar{z}_{12} \, \bar{z}_{23}}{\eta \, (1-\eta)}
+\omega_3 \,
\frac{\bar{z}_{13} \, \bar{z}_{23}}{\eta}\Big)$ 
\\
\hline
$+1$ & $+1$
& $+1$  
& $\frac{1}{\ep^2}\,
\frac{1}{
(\omega_1+\omega_2+\omega_3)}\, \Big( \omega_1 \,
\frac{\bar{z}_{12} \, \bar{z}_{13}}{(1-\eta)} +
\omega_2 \,
\frac{\bar{z}_{12} \, \bar{z}_{23}}{\eta \, (1-\eta)}
+\omega_3 \,
\frac{\bar{z}_{13} \, \bar{z}_{23}}{\eta}\Big)$ 
\\
\hline
$+1$ & $+1$
& $+\frac{1}{2}$  
& $\frac{1}{\ep^2}\,
\frac{\omega_3^{\frac{1}{2}} }{
(\omega_1+\omega_2+\omega_3)^{\frac{3}{2}}}\, \Big( \omega_1 \,
\frac{\bar{z}_{12} \, \bar{z}_{13}}{(1-\eta)} +
\omega_2 \,
\frac{\bar{z}_{12} \, \bar{z}_{23}}{\eta \, (1-\eta)}
+\omega_3 \,
\frac{\bar{z}_{13} \, \bar{z}_{23}}{\eta}\Big)$ 
\\
\hline
$+1$ & $+1$
& $0$  
& $\frac{1}{\ep^2}\,
\frac{\omega_3}{
(\omega_1+\omega_2+\omega_3)^{2}}\, \Big( \omega_1 \,
\frac{\bar{z}_{12} \, \bar{z}_{13}}{(1-\eta)} +
\omega_2 \,
\frac{\bar{z}_{12} \, \bar{z}_{23}}{\eta \, (1-\eta)}
+\omega_3 \,
\frac{\bar{z}_{13} \, \bar{z}_{23}}{\eta}\Big)$ 
\\
\hline
$+1$ & $+\frac{1}{2}$
& $+\frac{1}{2}$  
& $\frac{1}{\ep^2}\,
\frac{\omega_2^{\frac{1}{2}} \, \omega_3^{\frac{1}{2}}}{
(\omega_1+\omega_2+\omega_3)^{2}}\, \Big( \omega_1 \,
\frac{\bar{z}_{12} \, \bar{z}_{13}}{(1-\eta)} +
\omega_2 \,
\frac{\bar{z}_{12} \, \bar{z}_{23}}{\eta \, (1-\eta)}
+\omega_3 \,
\frac{\bar{z}_{13} \, \bar{z}_{23}}{\eta}\Big)$ 
\\
\hline
\end{tabular}
\caption{
The sixteen splitting functions for (\ref{16case}). }
\end{table}

\begin{table}[tbp]
\centering
\renewcommand{\arraystretch}{1.7}
\begin{tabular}{|c|c|c|c|c| }
\hline
$s_1$ & $s_2$ & $s_3$  & $
\Big(d_{1,2}\big|_{s_I=s_1+s_2-2}+
d_{2,3}\big|_{s_I=s_2+s_3-2} + d_{1,3}\big|_{s_I=s_1+s_3-2}\Big)
\Big|_{s_J=s_1+s_2+s_3-4}$  
\\
\hline
\hline
$+\frac{1}{2}$ & $+2$
& $+2$  
& $\frac{1}{\ep^2}\,
\frac{\omega_1\,(\omega_1+\omega_2+\omega_3)^{\frac{1}{2}}
}{\omega_2 \, \omega_3 
}\, \Big( \omega_1 \,
\frac{\bar{z}_{12} \, \bar{z}_{13}}{(1-\eta)} +
\omega_2 \,
\frac{\bar{z}_{12} \, \bar{z}_{23}}{\eta \, (1-\eta)}
+\omega_3 \,
\frac{\bar{z}_{13} \, \bar{z}_{23}}{\eta}\Big)$ 
\\
\hline
$+\frac{1}{2}$ & $+2$
& $+\frac{3}{2}$  
& $\frac{1}{\ep^2}\,
\frac{ \omega_1^{\frac{1}{2}}
}{\omega_2\,  \omega_3^{\frac{1}{2}} 
}\, \Big( \omega_1 \,
\frac{\bar{z}_{12} \, \bar{z}_{13}}{(1-\eta)} +
\omega_2 \,
\frac{\bar{z}_{12} \, \bar{z}_{23}}{\eta \, (1-\eta)}
+\omega_3 \,
\frac{\bar{z}_{13} \, \bar{z}_{23}}{\eta}\Big)$ 
\\
\hline
$+\frac{1}{2}$ & $+2$
& $+1$  
& $\frac{1}{\ep^2}\,
\frac{\omega_1^{\frac{1}{2}}
}{\omega_2 \, (\omega_1+\omega_2+\omega_3)^{\frac{1}{2}}}\, \Big( \omega_1 \,
\frac{\bar{z}_{12} \, \bar{z}_{13}}{(1-\eta)} +
\omega_2 \,
\frac{\bar{z}_{12} \, \bar{z}_{23}}{\eta \, (1-\eta)}
+\omega_3 \,
\frac{\bar{z}_{13} \, \bar{z}_{23}}{\eta}\Big)$ 
\\
\hline
$+\frac{1}{2}$ & $+2$
& $+\frac{1}{2}$  
& $\frac{1}{\ep^2}\,
\frac{\omega_1^{\frac{1}{2}}\, \omega_3^{\frac{1}{2}} \,
}{
\omega_2 \, (\omega_1+\omega_2+\omega_3)}\, \Big( \omega_1 \,
\frac{\bar{z}_{12} \, \bar{z}_{13}}{(1-\eta)} +
\omega_2 \,
\frac{\bar{z}_{12} \, \bar{z}_{23}}{\eta \, (1-\eta)}
+\omega_3 \,
\frac{\bar{z}_{13} \, \bar{z}_{23}}{\eta}\Big)$ 
\\
\hline
$+\frac{1}{2}$ & $+2$
& $0$  
& $\frac{1}{\ep^2}\,
\frac{\omega_1^{\frac{1}{2}}\, \omega_3 \,
}{\omega_2 \, (\omega_1+\omega_2+\omega_3)^{\frac{3}{2}}}\, \Big( \omega_1 \,
\frac{\bar{z}_{12} \, \bar{z}_{13}}{(1-\eta)} +
\omega_2 \,
\frac{\bar{z}_{12} \, \bar{z}_{23}}{\eta \, (1-\eta)}
+\omega_3 \,
\frac{\bar{z}_{13} \, \bar{z}_{23}}{\eta}\Big)$ 
\\
\hline
$+\frac{1}{2}$   & $+2$
& $-\frac{1}{2}$ 
& $\frac{1}{\ep^2}\,
\frac{\omega_1^{\frac{1}{2}}\,\omega_3^{\frac{3}{2}}}{
\omega_2\, (\omega_1+\omega_2+\omega_3)^{2}}\, \Big( \omega_1 \,
\frac{\bar{z}_{12} \, \bar{z}_{13}}{(1-\eta)} +
\omega_2 \,
\frac{\bar{z}_{12} \, \bar{z}_{23}}{\eta \, (1-\eta)}
+\omega_3 \,
\frac{\bar{z}_{13} \, \bar{z}_{23}}{\eta}\Big)$ 
\\
\hline
$+\frac{1}{2}$ & $+\frac{3}{2}$
& $+\frac{3}{2}$  
& $\frac{1}{\ep^2}\,
\frac{\omega_1^{\frac{1}{2}}
}{\omega_2^{\frac{1}{2}} \, \omega_3^{\frac{1}{2}}
(\omega_1+\omega_2+\omega_3)^{\frac{1}{2}}  
 }\, \Big( \omega_1 \,
\frac{\bar{z}_{12} \, \bar{z}_{13}}{(1-\eta)} +
\omega_2 \,
\frac{\bar{z}_{12} \, \bar{z}_{23}}{\eta \, (1-\eta)}
+\omega_3 \,
\frac{\bar{z}_{13} \, \bar{z}_{23}}{\eta}\Big)$ 
\\
\hline
$+\frac{1}{2}$ & $+\frac{3}{2}$
& $+1$  
& $\frac{1}{\ep^2}\,
\frac{\omega_1^{\frac{1}{2}}
}{\omega_2^{\frac{1}{2}} \,
(\omega_1+\omega_2+\omega_3)
}\, \Big( \omega_1 \,
\frac{\bar{z}_{12} \, \bar{z}_{13}}{(1-\eta)} +
\omega_2 \,
\frac{\bar{z}_{12} \, \bar{z}_{23}}{\eta \, (1-\eta)}
+\omega_3 \,
\frac{\bar{z}_{13} \, \bar{z}_{23}}{\eta}\Big)$ 
\\
\hline
$+\frac{1}{2}$ & $+\frac{3}{2}$
& $+\frac{1}{2}$  
& $\frac{1}{\ep^2}\,
\frac{ \omega_1^{\frac{1}{2}} \, \omega_3^{\frac{1}{2}} 
}{\omega_2^{\frac{1}{2}} \, (\omega_1+\omega_2+\omega_3)^{\frac{3}{2}} 
}\, \Big( \omega_1 \,
\frac{\bar{z}_{12} \, \bar{z}_{13}}{(1-\eta)} +
\omega_2 \,
\frac{\bar{z}_{12} \, \bar{z}_{23}}{\eta \, (1-\eta)}
+\omega_3 \,
\frac{\bar{z}_{13} \, \bar{z}_{23}}{\eta}\Big)$ 
\\
\hline
$+\frac{1}{2}$ & $+\frac{3}{2}$
& $0$  
& $\frac{1}{\ep^2}\,
\frac{\omega_1^{\frac{1}{2}}\,\omega_3
}{\omega_2^{\frac{1}{2}}  \,
(\omega_1+\omega_2+\omega_3)^{2}}\, \Big( \omega_1 \,
\frac{\bar{z}_{12} \, \bar{z}_{13}}{(1-\eta)} +
\omega_2 \,
\frac{\bar{z}_{12} \, \bar{z}_{23}}{\eta \, (1-\eta)}
+\omega_3 \,
\frac{\bar{z}_{13} \, \bar{z}_{23}}{\eta}\Big)$ 
\\
\hline
$+\frac{1}{2}$ & $+1$
& $+1$  
& $\frac{1}{\ep^2}\,
\frac{\omega_1^{\frac{1}{2}}
}{
(\omega_1+\omega_2+\omega_3)^{\frac{3}{2}}}\, \Big( \omega_1 \,
\frac{\bar{z}_{12} \, \bar{z}_{13}}{(1-\eta)} +
\omega_2 \,
\frac{\bar{z}_{12} \, \bar{z}_{23}}{\eta \, (1-\eta)}
+\omega_3 \,
\frac{\bar{z}_{13} \, \bar{z}_{23}}{\eta}\Big)$ 
\\
\hline
$+\frac{1}{2}$ & $+1$
& $+\frac{1}{2}$  
& $\frac{1}{\ep^2}\,
\frac{\omega_1^{\frac{1}{2}}\,\omega_3^{\frac{1}{2}}
}{
(\omega_1+\omega_2+\omega_3)^{2}}\, \Big( \omega_1 \,
\frac{\bar{z}_{12} \, \bar{z}_{13}}{(1-\eta)} +
\omega_2 \,
\frac{\bar{z}_{12} \, \bar{z}_{23}}{\eta \, (1-\eta)}
+\omega_3 \,
\frac{\bar{z}_{13} \, \bar{z}_{23}}{\eta}\Big)$ 
\\
\hline
\end{tabular}
\caption{
The twelve splitting functions for (\ref{12case}). }
\end{table}

\begin{table}[tbp]
\centering
\renewcommand{\arraystretch}{1.7}
\begin{tabular}{|c|c|c|c|c| }
\hline
$s_1$ & $s_2$ & $s_3$  & $
\Big(d_{1,2}\big|_{s_I=s_1+s_2-2}+
d_{2,3}\big|_{s_I=s_2+s_3-2} + d_{1,3}\big|_{s_I=s_1+s_3-2}\Big)
\Big|_{s_J=s_1+s_2+s_3-4}$  
\\
\hline
\hline
$0$ & $+2$
& $+2$  
& $\frac{1}{\ep^2}\,
\frac{\omega_1
}{\omega_2 \, \omega_3 
}\, \Big( \omega_1 \,
\frac{\bar{z}_{12} \, \bar{z}_{13}}{(1-\eta)} +
\omega_2 \,
\frac{\bar{z}_{12} \, \bar{z}_{23}}{\eta \, (1-\eta)}
+\omega_3 \,
\frac{\bar{z}_{13} \, \bar{z}_{23}}{\eta}\Big)$ 
\\
\hline
$0$ & $+2$
& $+\frac{3}{2}$  
& $\frac{1}{\ep^2}\,
\frac{\omega_1 
}{\omega_2\,  \omega_3^{\frac{1}{2}} \,
(\omega_1+\omega_2+\omega_3)^{\frac{1}{2}}
}\, \Big( \omega_1 \,
\frac{\bar{z}_{12} \, \bar{z}_{13}}{(1-\eta)} +
\omega_2 \,
\frac{\bar{z}_{12} \, \bar{z}_{23}}{\eta \, (1-\eta)}
+\omega_3 \,
\frac{\bar{z}_{13} \, \bar{z}_{23}}{\eta}\Big)$ 
\\
\hline
$0$ & $+2$
& $+1$  
& $\frac{1}{\ep^2}\,
\frac{\omega_1
}{\omega_2 \, (\omega_1+\omega_2+\omega_3)}\, \Big( \omega_1 \,
\frac{\bar{z}_{12} \, \bar{z}_{13}}{(1-\eta)} +
\omega_2 \,
\frac{\bar{z}_{12} \, \bar{z}_{23}}{\eta \, (1-\eta)}
+\omega_3 \,
\frac{\bar{z}_{13} \, \bar{z}_{23}}{\eta}\Big)$ 
\\
\hline
$0$ & $+2$
& $+\frac{1}{2}$  
& $\frac{1}{\ep^2}\,
\frac{\omega_1\, \omega_3^{\frac{1}{2}} \,
}{
\omega_2 \, (\omega_1+\omega_2+\omega_3)^{\frac{3}{2}}}\, \Big( \omega_1 \,
\frac{\bar{z}_{12} \, \bar{z}_{13}}{(1-\eta)} +
\omega_2 \,
\frac{\bar{z}_{12} \, \bar{z}_{23}}{\eta \, (1-\eta)}
+\omega_3 \,
\frac{\bar{z}_{13} \, \bar{z}_{23}}{\eta}\Big)$ 
\\
\hline
$0$ & $+2$
& $0$  
& $\frac{1}{\ep^2}\,
\frac{\omega_1\, \omega_3 
}{\omega_2 \, (\omega_1+\omega_2+\omega_3)^2}\, \Big( \omega_1 \,
\frac{\bar{z}_{12} \, \bar{z}_{13}}{(1-\eta)} +
\omega_2 \,
\frac{\bar{z}_{12} \, \bar{z}_{23}}{\eta \, (1-\eta)}
+\omega_3 \,
\frac{\bar{z}_{13} \, \bar{z}_{23}}{\eta}\Big)$ 
\\
\hline
$0$   & $+\frac{3}{2}$
& $+\frac{3}{2}$ 
& $\frac{1}{\ep^2}\,
\frac{\omega_1}{
\omega_2^{\frac{1}{2}}\,
\omega_3^{\frac{1}{2}} \, (\omega_1+\omega_2+\omega_3)}\, \Big( \omega_1 \,
\frac{\bar{z}_{12} \, \bar{z}_{13}}{(1-\eta)} +
\omega_2 \,
\frac{\bar{z}_{12} \, \bar{z}_{23}}{\eta \, (1-\eta)}
+\omega_3 \,
\frac{\bar{z}_{13} \, \bar{z}_{23}}{\eta}\Big)$ 
\\
\hline
$0$ & $+\frac{3}{2}$
& $+1$  
& $\frac{1}{\ep^2}\,
\frac{\omega_1}{\omega_2^{\frac{1}{2}}\,
(\omega_1+\omega_2+\omega_3)^{\frac{3}{2}}}\, \Big( \omega_1 \,
\frac{\bar{z}_{12} \, \bar{z}_{13}}{(1-\eta)} +
\omega_2 \,
\frac{\bar{z}_{12} \, \bar{z}_{23}}{\eta \, (1-\eta)}
+\omega_3 \,
\frac{\bar{z}_{13} \, \bar{z}_{23}}{\eta}\Big)$ 
\\
\hline
$0$ & $+\frac{3}{2}$
& $+\frac{1}{2}$  
& $\frac{1}{\ep^2}\,
\frac{\omega_1 \, \omega_3^{\frac{1}{2}}
}{\omega_2^{\frac{1}{2}}\, (\omega_1+\omega_2+\omega_3)^2 
 }\, \Big( \omega_1 \,
\frac{\bar{z}_{12} \, \bar{z}_{13}}{(1-\eta)} +
\omega_2 \,
\frac{\bar{z}_{12} \, \bar{z}_{23}}{\eta \, (1-\eta)}
+\omega_3 \,
\frac{\bar{z}_{13} \, \bar{z}_{23}}{\eta}\Big)$ 
\\
\hline
$0$ & $+1$
& $+1$  
& $\frac{1}{\ep^2}\,
\frac{\omega_1 
}{
(\omega_1+\omega_2+\omega_3)^{2}
 }\, \Big( \omega_1 \,
\frac{\bar{z}_{12} \, \bar{z}_{13}}{(1-\eta)} +
\omega_2 \,
\frac{\bar{z}_{12} \, \bar{z}_{23}}{\eta \, (1-\eta)}
+\omega_3 \,
\frac{\bar{z}_{13} \, \bar{z}_{23}}{\eta}\Big)$ 
\\
\hline
\hline
$-\frac{1}{2}$ & $+2$
& $+2$  
& $\frac{1}{\ep^2}\,
\frac{\omega_1^{\frac{3}{2}} 
}{\omega_2 \, \omega_3\, (\omega_1+\omega_2+\omega_3)^{\frac{1}{2}} 
}\, \Big( \omega_1 \,
\frac{\bar{z}_{12} \, \bar{z}_{13}}{(1-\eta)} +
\omega_2 \,
\frac{\bar{z}_{12} \, \bar{z}_{23}}{\eta \, (1-\eta)}
+\omega_3 \,
\frac{\bar{z}_{13} \, \bar{z}_{23}}{\eta}\Big)$ 
\\
\hline
$-\frac{1}{2}$ & $+2$
& $+\frac{3}{2}$  
& $\frac{1}{\ep^2}\,
\frac{\omega_1^{\frac{3}{2}}
}{\omega_2\, \omega_3^{\frac{1}{2}}  \,
(\omega_1+\omega_2+\omega_3)}\, \Big( \omega_1 \,
\frac{\bar{z}_{12} \, \bar{z}_{13}}{(1-\eta)} +
\omega_2 \,
\frac{\bar{z}_{12} \, \bar{z}_{23}}{\eta \, (1-\eta)}
+\omega_3 \,
\frac{\bar{z}_{13} \, \bar{z}_{23}}{\eta}\Big)$ 
\\
\hline
$-\frac{1}{2}$ & $+2$
& $+1$  
& $\frac{1}{\ep^2}\,
\frac{\omega_1^{\frac{3}{2}} 
}{\omega_2  \,
(\omega_1+\omega_2+\omega_3)^{\frac{3}{2}}}\, \Big( \omega_1 \,
\frac{\bar{z}_{12} \, \bar{z}_{13}}{(1-\eta)} +
\omega_2 \,
\frac{\bar{z}_{12} \, \bar{z}_{23}}{\eta \, (1-\eta)}
+\omega_3 \,
\frac{\bar{z}_{13} \, \bar{z}_{23}}{\eta}\Big)$ 
\\
\hline
$-\frac{1}{2}$ & $+2$
& $+\frac{1}{2}$  
& $\frac{1}{\ep^2}\,
\frac{\omega_1^{\frac{3}{2}}\, \omega_3^{\frac{1}{2}}}{\omega_2\,
(\omega_1+\omega_2+\omega_3)^2}\, \Big( \omega_1 \,
\frac{\bar{z}_{12} \, \bar{z}_{13}}{(1-\eta)} +
\omega_2 \,
\frac{\bar{z}_{12} \, \bar{z}_{23}}{\eta \, (1-\eta)}
+\omega_3 \,
\frac{\bar{z}_{13} \, \bar{z}_{23}}{\eta}\Big)$ 
\\
\hline
$-\frac{1}{2}$ & $+\frac{3}{2}$
& $+\frac{3}{2}$  
& $\frac{1}{\ep^2}\,
\frac{\omega_1^{\frac{3}{2}} }{\omega_2^{\frac{1}{2}} \,
\omega_3^{\frac{1}{2}} \,
(\omega_1+\omega_2+\omega_3)^{\frac{3}{2}}}\, \Big( \omega_1 \,
\frac{\bar{z}_{12} \, \bar{z}_{13}}{(1-\eta)} +
\omega_2 \,
\frac{\bar{z}_{12} \, \bar{z}_{23}}{\eta \, (1-\eta)}
+\omega_3 \,
\frac{\bar{z}_{13} \, \bar{z}_{23}}{\eta}\Big)$ 
\\
\hline
$-\frac{1}{2}$ & $+\frac{3}{2}$
& $+1$  
& $\frac{1}{\ep^2}\,
\frac{\omega_1^{\frac{3}{2}}}{
\omega_2^{\frac{1}{2}}\,
(\omega_1+\omega_2+\omega_3)^{2}}\, \Big( \omega_1 \,
\frac{\bar{z}_{12} \, \bar{z}_{13}}{(1-\eta)} +
\omega_2 \,
\frac{\bar{z}_{12} \, \bar{z}_{23}}{\eta \, (1-\eta)}
+\omega_3 \,
\frac{\bar{z}_{13} \, \bar{z}_{23}}{\eta}\Big)$ 
\\
\hline
\end{tabular}
\caption{
The nine  splitting functions for (\ref{9case})
and the six ones for (\ref{6case}). }
\end{table}

\begin{table}[tbp]
\centering
\renewcommand{\arraystretch}{1.7}
\begin{tabular}{|c|c|c|c|c| }
\hline
$s_1$ & $s_2$ & $s_3$  & $
\Big(d_{1,2}\big|_{s_I=s_1+s_2-2}+
d_{2,3}\big|_{s_I=s_2+s_3-2} + d_{1,3}\big|_{s_I=s_1+s_3-2}\Big)
\Big|_{s_J=s_1+s_2+s_3-4}$  
\\
\hline
\hline
$-1$ & $+2$
& $+2$  
& $\frac{1}{\ep^2}\,
\frac{\omega_1^2
}{\omega_2 \, \omega_3 
\, (\omega_1+\omega_2+\omega_3)}\, \Big( \omega_1 \,
\frac{\bar{z}_{12} \, \bar{z}_{13}}{(1-\eta)} +
\omega_2 \,
\frac{\bar{z}_{12} \, \bar{z}_{23}}{\eta \, (1-\eta)}
+\omega_3 \,
\frac{\bar{z}_{13} \, \bar{z}_{23}}{\eta}\Big)$ 
\\
\hline
$-1$ & $+2$
& $+\frac{3}{2}$  
& $\frac{1}{\ep^2}\,
\frac{\omega_1^2 
}{\omega_2\,  \omega_3^{\frac{1}{2}} \,
(\omega_1+\omega_2+\omega_3)^{\frac{3}{2}}
}\, \Big( \omega_1 \,
\frac{\bar{z}_{12} \, \bar{z}_{13}}{(1-\eta)} +
\omega_2 \,
\frac{\bar{z}_{12} \, \bar{z}_{23}}{\eta \, (1-\eta)}
+\omega_3 \,
\frac{\bar{z}_{13} \, \bar{z}_{23}}{\eta}\Big)$ 
\\
\hline
$-1$ & $+2$
& $+1$  
& $\frac{1}{\ep^2}\,
\frac{\omega_1^2
}{\omega_2 \, (\omega_1+\omega_2+\omega_3)^2}\, \Big( \omega_1 \,
\frac{\bar{z}_{12} \, \bar{z}_{13}}{(1-\eta)} +
\omega_2 \,
\frac{\bar{z}_{12} \, \bar{z}_{23}}{\eta \, (1-\eta)}
+\omega_3 \,
\frac{\bar{z}_{13} \, \bar{z}_{23}}{\eta}\Big)$ 
\\
\hline
$-1$ & $+\frac{3}{2}$
& $+\frac{3}{2}$  
& $\frac{1}{\ep^2}\,
\frac{\omega_1^2
}{
\omega_2^{\frac{1}{2}} \, \omega_3^{\frac{1}{2}}\,
(\omega_1+\omega_2+\omega_3)^{2}}\, \Big( \omega_1 \,
\frac{\bar{z}_{12} \, \bar{z}_{13}}{(1-\eta)} +
\omega_2 \,
\frac{\bar{z}_{12} \, \bar{z}_{23}}{\eta \, (1-\eta)}
+\omega_3 \,
\frac{\bar{z}_{13} \, \bar{z}_{23}}{\eta}\Big)$ 
\\
\hline
\hline
$-\frac{3}{2}$ & $+2$
& $+2$  
& $\frac{1}{\ep^2}\,
\frac{\omega_1^{\frac{5}{2}} 
}{\omega_2 \, \omega_3\,
(\omega_1+\omega_2+\omega_3)^{\frac{3}{2}}}\, \Big( \omega_1 \,
\frac{\bar{z}_{12} \, \bar{z}_{13}}{(1-\eta)} +
\omega_2 \,
\frac{\bar{z}_{12} \, \bar{z}_{23}}{\eta \, (1-\eta)}
+\omega_3 \,
\frac{\bar{z}_{13} \, \bar{z}_{23}}{\eta}\Big)$ 
\\
\hline
$-\frac{3}{2}$   & $+2$
& $+\frac{3}{2}$ 
& $\frac{1}{\ep^2}\,
\frac{\omega_1^{\frac{5}{2}}}{
\omega_2\,
\omega_3^{\frac{1}{2}} \, (\omega_1+\omega_2+\omega_3)^2}\, \Big( \omega_1 \,
\frac{\bar{z}_{12} \, \bar{z}_{13}}{(1-\eta)} +
\omega_2 \,
\frac{\bar{z}_{12} \, \bar{z}_{23}}{\eta \, (1-\eta)}
+\omega_3 \,
\frac{\bar{z}_{13} \, \bar{z}_{23}}{\eta}\Big)$ 
\\
\hline
\hline
$-2$ & $+2$
& $+2$  
& $\frac{1}{\ep^2}\,
\frac{\omega_1^3}{\omega_2\, \omega_3\,
(\omega_1+\omega_2+\omega_3)^{2}}\, \Big( \omega_1 \,
\frac{\bar{z}_{12} \, \bar{z}_{13}}{(1-\eta)} +
\omega_2 \,
\frac{\bar{z}_{12} \, \bar{z}_{23}}{\eta \, (1-\eta)}
+\omega_3 \,
\frac{\bar{z}_{13} \, \bar{z}_{23}}{\eta}\Big)$ 
\\
\hline
\end{tabular}
\caption{
The four  splitting functions for (\ref{4case}),
the two ones for (\ref{2case})
and the one for (\ref{1case}). }
\end{table}

\section{The redundant twenty (anti)commutators}

For convenience, we present the redundant twenty (anti)commutators,
which were not present in \cite{AK2509}, as follows:
\bea
\comm{(\Phi^{(h_1),A}_{+\frac{3}{2}})_r}{(\Phi^{(h_2)}_{+2})_m}&=&
\kappa_{+\frac{3}{2},+2,-\frac{3}{2}}\,\Big((h_2+1)r-(h_1+\tfrac{1}{2})m\Big)\,(\Phi^{(h_1+h_2),A}_{+\frac{3}{2}})_{r+m}\,\,\, :\text{eq.2},
\nonu \\
\comm{(\Phi^{(h_1),AB}_{+1})_m}{(\Phi^{(h_2)}_{+2})_n}&=&
\kappa_{+1,+2,-1}\,\Big((h_2+1)m-h_1\,n\Big)\,(\Phi^{(h_1+h_2),AB}_{+1})_{m+n}
\,\,\, :\text{eq.3},
\nonu \\
\comm{(\Phi^{(h_1),ABC}_{+\frac{1}{2}})_r}{(\Phi^{(h_2)}_{+2})_m}&=&
\kappa_{+\frac{1}{2},+2,-\frac{1}{2}}\,\Big((h_2+1)r-(h_1-\tfrac{1}{2})m\Big)\,(\Phi^{(h_1+h_2),ABC}_{+\frac{1}{2}})_{r+m}
\nonu \\
&: & \text{eq.4},
\nonu \\
\comm{(\Phi^{(h_1),ABCD}_{0})_m}{(\Phi^{(h_2)}_{+2})_n}&=&
\kappa_{0,+2,0}\,\Big((h_2+1)m-(h_1-1)n\Big)\,(\Phi^{(h_1+h_2),ABCD}_{0})_{m+n}
\,\,\, :\text{eq.5},
\nonu \\
\comm{(\Phi^{(h_1)}_{ABC,-\frac{1}{2}})_r}{(\Phi^{(h_2)}_{+2})_m}&=&
\kappa_{-\frac{1}{2},+2,+\frac{1}{2}}\,\Big((h_2+1)r-(h_1-\tfrac{3}{2})m\Big)\,(\Phi^{(h_1+h_2)}_{ABC,-\frac{1}{2}})_{r+m}\,\,\, :\text{eq.6},
\nonu \\
\comm{(\Phi^{(h_1)}_{AB,-1})_m}{(\Phi^{(h_2)}_{+2})_n}&=&
\kappa_{-1,+2,+1}\,\Big((h_2+1)m-(h_1-2)n\Big)\,(\Phi^{(h_1+h_2)}_{AB,-1})_{m+n}
\,\,\, :\text{eq.7},
\nonu \\
\comm{(\Phi^{(h_1)}_{A,-\frac{3}{2}})_r}{(\Phi^{(h_2)}_{+2})_m}&=&
\kappa_{-\frac{3}{2},+2,+\frac{3}{2}}\,\Big((h_2+1)r-(h_1-\tfrac{5}{2})m\Big)\,(\Phi^{(h_1+h_2)}_{A,-\frac{3}{2}})_{r+m}
\,\,\, :\text{eq.8},
\nonu \\
\comm{(\Phi^{(h_1)}_{-2})_m}{(\Phi^{(h_2)}_{+2})_n}&=&
\kappa_{-2,+2,+2}\,\Big((h_2+1)m-(h_1-3)n\Big)\,(\Phi^{(h_1+h_2)}_{-2})_{m+n}
\,\,\, :\text{eq.9},
\nonu \\
\comm{(\Phi^{(h_1),AB}_{+1})_m}{(\Phi^{(h_2),C}_{+\frac{3}{2}})_r}
&=& \kappa_{+1,+\frac{3}{2},-\frac{1}{2}}\,\Big((h_2+\tfrac{1}{2})m-h_1\,r\Big)\,(\Phi^{(h_1+h_2),ABC}_{+\frac{1}{2}})_{r+m}
\,\,\, :\text{eq.11},
\nonu \\
\acomm{(\Phi^{(h_1),ABC}_{+\frac{1}{2}})_r}{(\Phi^{(h_2),D}_{+\frac{3}{2}})_s}
&=& \kappa_{+\frac{1}{2},+\frac{3}{2},0}\,\Big((h_2+\tfrac{1}{2})r-(h_1-\tfrac{1}{2})s\Big) \nonu \\
& \times & (\Phi^{(h_1+h_2),ABCD}_0)_{r+s}
\,\,\, :\text{eq.12},
\nonu \\
\comm{(\Phi^{(h_1),ABCD}_0)_m}{(\Phi^{(h_2),E}_{+\frac{3}{2}})_r}&=&
\kappa_{0,+\frac{3}{2},+\frac{1}{2}}\,\Big((h_2+\tfrac{1}{2})m-(h_1-1)r\Big)\,\frac{1}{3!}\epsilon^{ABCDEFGH}
\nonu \\
& \times & (\Phi^{(h_1+h_2)}_{FGH,-\frac{1}{2}})_{r+m}
\,\,\, : 
\text{eq.13},
\nonu \\
\acomm{(\Phi^{(h_1)}_{ABC,-\frac{1}{2}})_r}{(\Phi^{(h_2),D}_{+\frac{3}{2}})_s}
&
=& \kappa_{-\frac{1}{2},+\frac{3}{2},+1}\,
\Big((h_2+\tfrac{1}{2})r-(h_1-\tfrac{3}{2})s\Big)\,3\,(\Phi^{(h_1+h_2)}_{[AB,-1})_{r+s}\delta^{\,\,\,D}_{C]}
\nonu \\
&: & \text{eq.14},
\nonu \\
\comm{(\Phi^{(h_1)}_{AB,-1})_m}{(\Phi^{(h_2),C}_{+\frac{3}{2}})_r}
&=&
\kappa_{-1,+\frac{3}{2},+\frac{3}{2}}\,\Big((h_2+\tfrac{1}{2})m-(h_1-2)r\Big)\,2\,\delta^{\,\,\,C}_{[A}
  (\Phi^{(h_1+h_2)}_{B],-\frac{3}{2}})_{r+m}
\nonu \\
&: & \text{eq.15},
\nonu \\
\acomm{(\Phi^{(h_1)}_{A,-\frac{3}{2}})_r}{(\Phi^{(h_2),B}_{+\frac{3}{2}})_s}
&=&
\kappa_{-\frac{3}{2},+\frac{3}{2},+2}\,\Big((h_2+\tfrac{1}{2})r-(h_1-\tfrac{5}{2})s\Big)\,\delta^{\,\,\,B}_{A}\,(\Phi^{(h_1+h_2)}_{-2})_{r+s}
\nonu \\
&: & \text{eq.16},
\nonu \\
\comm{(\Phi^{(h_1),ABC}_{+\frac{1}{2}})_r}{(\Phi^{(h_2),DE}_{+1})_m}&=&
\kappa_{+\frac{1}{2},+1,+\frac{1}{2}}\,\Big(h_2\,r-(h_1-\tfrac{1}{2})m\Big)\frac{1}{3!}\,\varepsilon^{ABCDEFGH}(\Phi^{(h_1+h_2)}_{FGH,-\frac{1}{2}})_{m+r}
\nonu \\
&: & \text{eq.18},
\nonu \\
\comm{(\Phi^{(h_1),ABCD}_0)_m}{(\Phi^{(h_2),EF}_{+1})_n}&=&
\kappa_{0,+1,+1}\,\Big(h_2\,m-(h_1-1)n\Big)\,\epsilon^{ABCDEFGH}\,\frac{1}{2!}\,(\Phi^{(h_1+h_2)}_{GH,-1})_{m+n}
\nonu \\
&:& \text{eq.19},
\nonu \\
\comm{(\Phi^{(h_1)}_{ABC,-\frac{1}{2}})_r}{(\Phi^{(h_2),DE}_{+1})_m}&=&
\kappa_{-\frac{1}{2},+1,+\frac{3}{2}}\,\Big(h_2\,r-(h_1-\tfrac{3}{2})m\Big)\,3!\,\delta^{D}_{\,\,\,[A}(\Phi^{(h_1+h_2)}_{B,-\frac{3}{2}})_{m+r}\delta_{C]}^{\,\,\,E}
\nonu \\
&: & \text{eq.20},
\nonu \\
\comm{(\Phi^{(h_1)}_{AB,-1})_m}{(\Phi^{(h_2),CD}_{+1})_n}&=&
\kappa_{-1,+1,+2}\,\Big(h_2\,m-(h_1-2)n\Big)\,\delta^{CD}_{AB}
\,(\Phi^{(h_1+h_2)}_{-2})_{m+n}
\,\,\, :\text{eq.21},
\nonu \\
\comm{(\Phi^{(h_1),ABCD}_0)_m}{(\Phi^{(h_2),EFG}_{+\frac{1}{2}})_r}&=&
\kappa_{0,+\frac{1}{2},+\frac{3}{2}}\,\Big((h_2-\tfrac{1}{2})m-(h_1-1)r\Big)\,\epsilon^{ABCDEFGH}\,\nonu \\
& \times & (\Phi^{(h_1+h_2)}_{H,-\frac{3}{2}})_{r+m}
\,\,\, :  \text{eq.23},
\nonu \\
\acomm{(\Phi^{(h_1)}_{ABC,-\frac{1}{2}})_r}{(\Phi^{(h_2),DEF}_{+\frac{1}{2}})_s}
&=& \kappa_{-\frac{1}{2},+\frac{1}{2},+2}\,\Big((h_2-\tfrac{1}{2})r-(h_1-\tfrac{3}{2})s\Big)\,
\delta^{DEF}_{ABC}\,(\Phi^{(h_1+h_2)}_{-2})_{r+s}
\nonu \\
&: & \text{eq.24}.
\label{20expression}
\eea
We can easily check the relations between the couplings explicitly
\footnote{
That is,
there are
\bea 
\kappa_{+\frac{3}{2},-\frac{1}{2},+1}=-\kappa_{-\frac{1}{2},+\frac{3}{2},+1},
\qquad
\kappa_{+\frac{3}{2},-\frac{3}{2},+2}=-\kappa_{-\frac{3}{2},+\frac{3}{2},+2}, \qquad
\kappa_{+\frac{1}{2},-\frac{1}{2},+2}=-\kappa_{-\frac{1}{2},+\frac{1}{2},+2}
\label{ferfer}
\eea
and the remaining seventeen couplings satisfy
$\kappa_{s_1,s_2,-s_3}=\kappa_{s_2,s_1,-s_3}$.
In particular, we have
$\kappa_{+\frac{1}{2},+\frac{3}{2},0}=\kappa_{+\frac{3}{2},+\frac{1}{2},0}$
corresponding to eq. $12$.
If we define this anticommutator having the $SU(8)$ indices
$DABC$ on the right-hand side, then there is an extra minus sign
for the couplings like as (\ref{ferfer}).
Note that in \cite{AK2509},
the commutator corresponding to eq. $15$ of (\ref{20expression})
was defined as
$\comm{(\Phi^{(h_1)}_{AB,-1})_m}{(\Phi^{(h_2),C}_{+\frac{3}{2}})_r}
=
\kappa_{-1,+\frac{3}{2},+\frac{3}{2}}\,\Big((h_2+\tfrac{1}{2})m-(h_1-2)r\Big)\,2\,
(\Phi^{(h_1+h_2)}_{[A,-\frac{3}{2}})_{r+m}\,\delta^{\,\,\,C}_{B]}$
with the condition
$\kappa_{+\frac{3}{2},-1,+\frac{3}{2}}=-\kappa_{-1,+\frac{3}{2},+\frac{3}{2}}$.
If one uses the right-hand side of the commutator corresponding to
eq. $15$ of (\ref{20expression}) (the tensorial structure
is different from the above), then there is no sign difference between
the couplings:$\kappa_{+\frac{3}{2},-1,+\frac{3}{2}}=
\kappa_{-1,+\frac{3}{2},+\frac{3}{2}}$.}

%

\end{document}